\documentclass[letter,11pt]{article}

\usepackage{jheppub} 

\usepackage[T1]{fontenc} 

\usepackage{mciteplus}

\usepackage{amsmath,amssymb,bm,mathrsfs}
\usepackage{times}
\usepackage{srcltx}
\usepackage{epsfig}

\usepackage{subfigure}
\usepackage{graphicx}
\usepackage{multirow}

\usepackage{tikz}
\usetikzlibrary{positioning,decorations.pathreplacing,decorations.markings,shapes}
\usetikzlibrary{calc}
\usetikzlibrary{arrows,shapes,backgrounds}
\usetikzlibrary{decorations.pathmorphing}
\usetikzlibrary{decorations.markings}

\usepackage{braket}

\usepackage{ulem} 
\usepackage{slashed}
\usepackage{cancel}

\usepackage{feynmp}
\usepackage{mathtools}

\makeatletter
\newsavebox\myboxA
\newsavebox\myboxB
\newlength\mylenA
\newcommand*\xoverline[2][0.75]{%
    \sbox{\myboxA}{$\m@th#2$}%
    \setbox\myboxB\null
    \ht\myboxB=\ht\myboxA%
    \dp\myboxB=\dp\myboxA%
    \wd\myboxB=#1\wd\myboxA
    \sbox\myboxB{$\m@th\overline{\copy\myboxB}$}
    \setlength\mylenA{\the\wd\myboxA}
    \addtolength\mylenA{-\the\wd\myboxB}%
    \ifdim\wd\myboxB<\wd\myboxA%
       \rlap{\hskip 0.5\mylenA\usebox\myboxB}{\usebox\myboxA}%
    \else
        \hskip -0.5\mylenA\rlap{\usebox\myboxA}{\hskip 0.5\mylenA\usebox\myboxB}%
    \fi}
\makeatother

\newcommand{\drawsquare}[2]{\hbox{%
\rule{#2pt}{#1pt}\hskip-#2pt
\rule{#1pt}{#2pt}\hskip-#1pt
\rule[#1pt]{#1pt}{#2pt}}\rule[#1pt]{#2pt}{#2pt}\hskip-#2pt
\rule{#2pt}{#1pt}}

\newcommand{\Yfund}{\raisebox{-.5pt}{\drawsquare{6.5}{0.4}}}

\def\pd{\partial}
\def\avg#1{\langle #1 \rangle}
\def\abs#1{\left| #1 \right|}

\def\a{\alpha}
\def\b{\beta}
\def\d{\delta}
\def\D{\Delta}
\def\g{\gamma}
\def\G{\Gamma}
\def\e{\epsilon}
\def\et{\eta}
\def\k{\kappa}
\def\l{\lambda}
\def\L{\Lambda}
\def\m{\mu}
\def\n{\nu}
\def\s{\sigma}
\def\r{\rho}
\def\ps{\psi}
\def\ph{\phi}
\def\Ph{\Phi}

\def\t{\tau}

\def\x{\xi}

\def\scL{\mathcal{L}}

\def\scO{\mathcal{O}}

\def\b4{\xoverline{4}}

\def\b{\beta}

\def\Tr{\text{Tr}}
\def\tr{\text{tr}}

\def\Seff{S_{\text{eff}}}
\def\Leff{\scL_{\text{eff}}}

\usepackage{accents}

\DeclareMathSymbol{\widehatsym}{\mathord}{largesymbols}{"62}
\newcommand\lowerwidehatsym{%
  \text{\footnotesize{\smash{\raisebox{-1.3ex}{
    $\widehatsym$}}}}}
\newcommand\lowerwidehatsymscript{%
  \text{\scriptsize{\smash{\raisebox{-1.3ex}{
    $\widehatsym$}}}}}

\newcommand\fixwidehat[1]{%
  \mathchoice
    {\accentset{\displaystyle\lowerwidehatsym}{#1}}
    {\accentset{\textstyle\lowerwidehatsym}{#1}}
    {\accentset{\scriptstyle\lowerwidehatsymscript}{#1}}
    {\accentset{\scriptscriptstyle\lowerwidehatsymscript}{#1}}
}

\def\wh#1{\fixwidehat{#1}}

\def\hq{\wh{q}}
\def\hx{\wh{x}}

\def\hxo{\wh{x}^{}_1}
\def\hxt{\wh{x}^{}_2}
\def\hqo{\wh{q}^{}_1}
\def\hqt{\wh{q}^{}_2}

\DeclareMathSymbol{\widetildesym}{\mathord}{largesymbols}{"65}
\newcommand\lowerwidetildesym{%
  \text{\smash{\raisebox{-1.3ex}{%
    $\widetildesym$}}}}
\newcommand\fixwidetilde[1]{%
  \mathchoice
    {\accentset{\displaystyle\lowerwidetildesym}{#1}}
    {\accentset{\textstyle\lowerwidetildesym}{#1}}
    {\accentset{\scriptstyle\lowerwidetildesym}{#1}}
    {\accentset{\scriptscriptstyle\lowerwidetildesym}{#1}}
}

\def\wt#1{\fixwidetilde{#1}}

\def\tG{\wt{G}}
\def\tU{\wt{U}}

\def\tH{\wt{H}}
\def\tk{\wt{k}}

\def\uxo{1_{x^{}_1}}

\def\uqt{1_{q^{}_2}}

\def\bps{\xoverline{\psi}}

\def\sP{\slashed{P}}

\def\scI{\mathcal{I}}

\newcommand{\Dfb}{\mathord{\buildrel{\lower3pt\hbox{$\scriptscriptstyle{\leftrightarrow \tiny{ \ \ \ } }$}}\over {D^{\mu}}}}

\newcommand{\Dfbd}{\mathord{\buildrel{\lower3pt\hbox{$\scriptscriptstyle\leftrightarrow$}}\over {D}_{\mu}}}

\def\mt{m_{\tilde{t}}}

\begin{document}
\unitlength = 1mm

\title{How to use the Standard Model effective field theory}

\author[a,b]{Brian Henning,}
\author[a,b]{Xiaochuan Lu,}
\author[a,b,c]{and Hitoshi Murayama}


\affiliation[a]{Department of Physics, University of California,
  Berkeley,\\ Berkeley, California 94720, USA}
\affiliation[b]{Theoretical Physics Group, Lawrence Berkeley National
  Laboratory,\\ Berkeley, California 94720, USA}
\affiliation[c]{Kavli Institute for the Physics and Mathematics of the
  Universe (WPI), Todai Institutes for Advanced Study, University of Tokyo,\\
  Kashiwa 277-8583, Japan}

\emailAdd{bhenning@berkeley.edu}
\emailAdd{luxiaochuan123456@berkeley.edu}
\emailAdd{hitoshi@berkeley.edu, hitoshi.murayama@ipmu.jp}

%
%
%
\abstract{
We present a practical three-step procedure of using the Standard Model effective field theory (SM EFT) to connect ultraviolet (UV) models of new physics with weak scale precision observables. With this procedure, one can interpret precision measurements as constraints on a given UV model. We give a detailed explanation for calculating the effective action up to one-loop order in a manifestly gauge covariant fashion. This covariant derivative expansion method dramatically simplifies the process of matching a UV model with the SM EFT, and also makes available a universal formalism that is easy to use for a variety of UV models. A few general aspects of RG running effects and choosing operator bases are discussed. Finally, we provide mapping results between the bosonic sector of the SM EFT and a complete set of precision electroweak and Higgs observables to which present and near future experiments are sensitive. Many results and tools which should prove useful to those wishing to use the SM EFT are detailed in several appendices.
}
\preprint{UCB-PTH 14/40, IPMU14-0353}
\maketitle
\flushbottom

\newpage

\section{Introduction} \label{sec:intro}

The discovery of a Standard Model (SM)-like Higgs boson~\cite{Aad:2012tfa,Chatrchyan:2012ufa} is a milestone in particle physics. Direct study of this boson {\it will} shed light on the mysteries surrounding the origin of the Higgs boson and the electroweak (EW) scale. Additionally, it will potentially provide insight into some of the many long standing experimental observations that remain unexplained (see, {\it e.g.},~\cite{Murayama:2014ita}) by the SM. In attempting to answer questions raised by the EW sector and these presently unexplained observations, a variety of new physics models have been proposed, with little clue which---if any---Nature actually picks.

It is exciting that ongoing and possible near future experiments can achieve an estimated per mille sensitivity on precision Higgs and EW observables~\cite{Behnke:2013xla,*Baer:2013cma,*Adolphsen:2013jya,*Adolphsen:2013kya,*Behnke:2013lya,Gomez-Ceballos:2013zzn,Dawson:2013bba,Baak:2012kk,Baak:2013fwa,Fan:2014vta}. This level of precision provides a window to indirectly explore the theory space of BSM physics and place constraints on specific UV models. For this purpose, an efficient procedure of connecting new physics models with precision Higgs and EW observables is clearly desirable.

In this paper, we make use of the Standard Model effective field theory (SM EFT) as a bridge to connect models of new physics with experimental observables. The SM EFT consists of the renormalizable SM Lagrangian supplemented with higher-dimension interactions:
\begin{equation}
{\cal L}_\text{eff} = {\cal L}_\text{SM} + \sum_i \frac{1}{\Lambda^{d_i-4}} c_i {\cal O}_i . \label{eqn:LeffSM}
\end{equation}
In the above, $\Lambda$ is the cutoff scale of the EFT, ${\cal O}_i$ are a set of dimension $d_i$ operators that respect the $SU(3)_c \times SU(2)_L \times U(1)_Y$ gauge invariance of ${\cal L}_\text{SM}$, and $c_i$ are their Wilson coefficients that run as functions $c_i(\mu)$ of the renormalization group (RG) scale $\mu$. The estimated per-mille sensitivity of future precision Higgs measurements justifies truncating the above expansion at dimension-six operators.

It is worth noting that the SM EFT parameterized by the \(c_i\) of Eq.~\eqref{eqn:LeffSM} is totally different from the widely used seven-$\kappa$ parametrization ({\it e.g.},~\cite{LHCHiggsCrossSectionWorkingGroup:2012nn}), which captures only a change in size of each of the SM-type Higgs couplings. In fact, the seven $\kappa$'s parameterize models that do not respect the electroweak gauge symmetry, and hence, violate unitarity. As a result, future precision programs can show spuriously high sensitivity to the \(\k\). The SM EFT of Eq.~\eqref{eqn:LeffSM}, on the other hand, parameterizes new physics in directions that respect the SM gauge invariance and are therefore free from unitarity violations.\footnote{Equation~\eqref{eqn:LeffSM} is a linear-realization of EW gauge symmetry. An EFT constructed as a non-linear realization of EW gauge symmetry is, of course, perfectly acceptable.}

\begin{figure}[t]
 \centering
 \includegraphics[height=8cm]{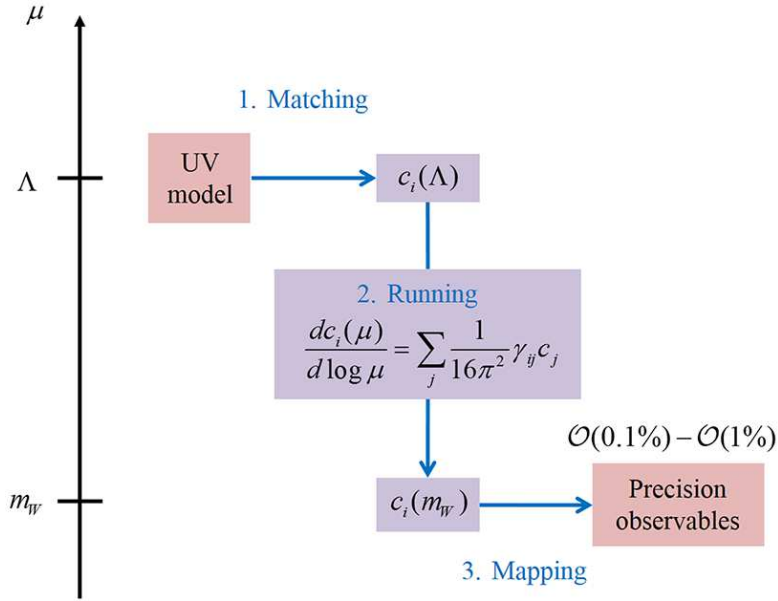}
 \caption{SM EFT as a bridge to connect UV models and weak scale precision observables.} \label{fig:connection}
\end{figure}

In an EFT framework, the connection of UV models\footnote{In this work we take ``UV model'' to generically mean the SM supplemented with new states that couple to the SM. In particular, the UV model does not need to be UV complete; it may itself be an effective theory of some other, unknown description.} with low-energy observables is accomplished through a three-step procedure schematically described in Fig.~\ref{fig:connection}.\footnote{For an introduction to the basic techniques of effective field theories see, for example,~\cite{Manohar:1996cq,*Burgess:2007pt,*Georgi:1985kw}.} First, the UV model is matched onto the SM EFT at a high-energy scale $\Lambda$. This matching is performed order-by-order in a loop expansion. At each loop order, $c_i(\Lambda)$ is determined
such that the \(S\)-matrix elements in the EFT and the UV model are the same at the RG scale \(\m = \Lambda\). Next, the \(c_i(\L)\) are run down to the weak scale \(c_i(m_W)\) according to the RG equations of the SM EFT. The leading order solution to these RG equations is determined by the anomalous dimension matrix \(\g_{ij}\). Finally, we use the effective Lagrangian at \(\m = m_W\) to compute weak scale observables in terms of the \(c_i(m_W)\) and SM parameters of \(\scL_{\text{SM}}\). We refer to this third step as mapping the Wilson coefficients onto observables.

In the rest of this paper we consider each of these three steps---matching, running, and mapping---in detail for the SM EFT. In the SM EFT, the main challenge presented at each step is complexity: truncating the expansion in~\eqref{eqn:LeffSM} at dimension-six operators leaves us with \(\scO(10^2)\) independent deformations of the Standard Model.\footnote{This counting excludes flavor. With flavor, this number jumps to \(\scO(10^3)\).} This large number of degrees of freedom can obscure the incredible simplicity and utility that the SM EFT has to offer. One of the main purposes of the present work is to provide tools and results to help a user employ the SM EFT and take advantage of the many benefits it can offer.

A typical scenario that we imagine is one where a person has some UV model containing massive BSM states and she wishes to understand how these states affect Higgs and EW observables. With a UV model in hand she can, of course, compute these effects using the UV model itself. This option sounds more direct and can, in principle, be more accurate since it does not require an expansion in powers of \(\L^{-1}\). However, performing a full computation with the UV model is typically quite involved, especially at loop-order and beyond, and needs to be done on a case-by-case basis for each UV model. Among the great advantages of using an EFT is that the computations related to running and mapping, being intrinsic to the EFT, only need to be done once; in other words, once the RG evolution and physical effects of the \(\scO_i\) are known (to a given order), the results can be tabulated for general use.

Moreover, for many practical purposes, a full computation in the UV model does not offer considerable improvement in accuracy over the EFT approach when one considers future experimental resolution. The difference between an observable computed using the UV theory versus the (truncated) EFT will scale in powers of \(E_{\text{obs}}/\L\), typically beginning at \((E_{\text{obs}}/\L)^2\), where \(E_{\text{obs}} \sim m_W\) is the energy scale at which the observable is measured. The present lack of evidence for BSM physics coupled to the SM requires in many cases \(\L\) to be at least a factor of a few above the weak scale. With an estimated per mille precision of future Higgs and EW observables, this means that the leading order calculation in the EFT will rapidly converge with the calculation from the UV model, providing essentially the same result for \(\L \gtrsim (\text{several}\times E_{\text{obs}})\).\footnote{For example, in considering the impact of scalar tops on the associated \(Zh\) production cross-section at an \(e^+e^-\) collider, Craig {\it et. al.} recently compared~\cite{Craig:2014una} the result of a full NLO calculation versus the SM EFT calculation. They found that the results were virtually indistinguishable for stop masses above 500 GeV. In their calculation, they used Wilson coefficients previously obtained by us in~\cite{Henning:2014gca}; in section~\ref{sec:CDE} of this paper we explain the details of how these Wilson coefficients are easily computed using the covariant derivative expansion.} For the purpose of determining the physics reach of future experiments on specific UV models---{\it i.e.} estimating the largest values of \(\L\) in a given model that experiments can probe---the EFT calculation is sufficiently accurate in almost all cases.

As mentioned above, the steps of RG running the \(\scO_i\) and mapping these operators to observables are done within the EFT; once these results are known they can be applied to any set of \(\{c_i(\L)\}\) obtained from matching a given UV model onto the SM EFT. Therefore, an individual wishing to study the impact of some UV model on weak scale observables ``only'' needs to obtain the \(c_i(\L)\) at the matching scale \(\L\). We put ``only'' in quotes because this step, while straightforward, can also be computationally complex owing to the large number of operators in the SM EFT.

A large amount of literature pertaining to the SM EFT already exists, some of which dates back a few decades, and is rapidly growing and evolving. Owing to the complexity of the SM EFT, many results are scattered throughout the literature at varying levels of completeness. This body of research can be difficult to wade through for a newcomer (or expert) wishing to use the SM EFT to study the impact of BSM physics on Higgs and EW observables. We believe an explication from a UV perspective, oriented to consider how one uses the SM EFT as a bridge to connect UV models with weak-scale precision observables, is warranted. We have strived to give such a perspective by providing new results and tools with the full picture of matching, running, and mapping in mind. Moreover, our results are aimed to be complete and systematic---especially in regards to the mapping onto observables---as well as usable and self-contained. These goals have obviously contributed to the considerable length of this paper. In the rest of this introduction, we summarize more explicitly our results in order to provide an overview for what is contained where in this paper.

In section~\ref{sec:CDE}, we present a method to considerably ease the matching of a UV model onto the SM EFT. The SM EFT is obtained by taking a given UV model and integrating out the massive BSM states. The resultant effective action is given by~\eqref{eqn:LeffSM}, where the higher dimension operators are suppressed by powers of \(\L = m\), the mass of the heavy BSM states. Although every \(\scO_i\) respects SM gauge invariance, traditional methods of evaluating the effective action, such as Feynman diagrams, require working with gauge non-invariant pieces at intermediate steps, so that the process of arranging an answer back into the gauge invariant \(\scO_i\) can be quite tedious. Utilizing techniques introduced in~\cite{Gaillard:1985,Cheyette:1987} and termed the covariant derivative expansion (CDE), we present a method of computing the effective action through one-loop order in a manifestly gauge-invariant manner. By working solely with gauge-covariant quantities, an expansion of the effective action is obtained that immediately produces the gauge-invariant operators \(\scO_i\) of the EFT and their associated Wilson coefficients.

At one-loop order, the effective action that results when integrating out a heavy field \(\Ph\) of mass \(m\) is generally of the form
\begin{equation}
\D S_{\text{eff,1-loop}} \propto i \Tr \log \Big[ D^2 + m^2 + U(x)\Big],
\label{eqn:intro_TrLog}
\end{equation}
where \(D^2 = D_{\m}D^{\m}\) with \(D_{\m}\) a gauge covariant derivative and \(U(x)\) depends on the light, SM fields. The typical method for evaluating the functional trace relies on splitting the covariant derivative into its component parts, \(D_{\m} = \pd_{\m} - i A_{\m}\) with \(A_{\m}\) a gauge field, and performing a derivative expansion in \(\pd^2 - m^2\). This splitting clearly causes intermediate steps of the calculation to be gauge non-covariant. Many years ago, Gaillard found a transformation~\cite{Gaillard:1985} that allows the functional trace to be evaluated while keeping gauge covariance manifest at every step of the calculation, which we derive and explain in detail in section~\ref{sec:CDE}. In essence, the argument of the logarithm in Eq.~\eqref{eqn:intro_TrLog} is transformed such that the covariant derivative only appears in a series of commutators with itself and \(U(x)\). The effective action is then evaluated in a series of ``free propagators'' of the form \((q^2-m^2)^{-1}\) with \(q_{\m}\) a momentum parameter that is integrated over. The coefficients of this expansion are the commutators of \(D_{\m}\) with itself and \(U(x)\) and correspond to the \(\scO_i\) of the EFT. Thus, one immediately obtains the gauge-invariant \(\scO_i\) of the effective action.

In our discussion, we clarify and streamline certain aspects of the derivation and use of the covariant derivative expansion of~\cite{Gaillard:1985,Cheyette:1987}. Moreover, we generalize the results of~\cite{Gaillard:1985,Cheyette:1987} and provide explicit formulas for scalars, fermions, and massless as well as massive vector bosons. As a sidenote, for massive gauge bosons it is known that the magnetic dipole coefficient is universal~\cite{Weinberg:1970,Ferrara:1992yc}; in appendix~\ref{sec:app_gkpiece} we present a new, completely algebraic proof of this fact. In addition to addressing the one-loop effective action, we present a method for obtaining the tree-level effective action using a covariant derivative expansion. While this tree-level evaluation is very straightforward, to the best of our knowledge, it has not appeared elsewhere in the literature. 

We believe the CDE to be quite useful in general, but especially so when used to match a UV model onto the SM EFT. It is perhaps not widely appreciated that an inverse mass expansion of the one-loop effective action is essentially universal; one of the benefits of the CDE is that this fact is transparent at all stages of the computation. Therefore, the results of the inverse mass expansion, Eq.~\eqref{eqn:CDE_universal_lag}, can be applied to a large number of UV models, allowing one to calculate one-loop matched Wilson coefficients with ease. To demonstrate this, we compute the Wilson coefficients of a handful of non-trivial examples that could be relevant for Higgs physics, including an electroweak triplet scalar, an electroweak scalar doublet (the two Higgs doublet model), additional massive gauge bosons, and several others.

In section~\ref{sec:operatorbasis} we consider the step of running Wilson coefficients from the matching scale \(\L\) to the electroweak scale \(m_W\) where measurements are made. Over the past few years, the RG evolution of the SM EFT has been investigated quite intensively~\cite{Grojean:2013kd,Elias-Miro:2013gya,Jenkins:2013zja,Jenkins:2013wua,Alonso:2013hga,Alonso:2014zka,Elias-Miro:2013mua,Elias-Miro:2013eta,Jenkins:2013fya,Manohar:2013rga,Jenkins:2013sda}. It is a great accomplishment that the entire one-loop anomalous dimension matrix within a complete operator basis has been obtained~\cite{Jenkins:2013zja,Jenkins:2013wua,Alonso:2013hga,Alonso:2014zka},\footnote{Not only is the computation of \(\g_{ij}\) practically useful, its structure may be hinting at something deep in regards to renormalization and effective actions~\cite{Alonso:2014rga}.} as well as components of \(\g_{ij}\) in other operator bases~\cite{Elias-Miro:2013mua,Elias-Miro:2013eta}. As the literature has been quite thorough on the subject, we have little to contribute in terms of new calculations; instead, our discussion on RG running primarily concerns determining when this step is important to use and how to use it. Since future precision observables have a sensitivity of \(\scO(0.1\%)\)-\(\scO(1\%)\), they will generically be able to probe new physics at one-loop order. RG evolution introduces a loop factor; therefore, as a rule of thumb, RG running of the \(c_i(\L)\) to \(c_i(m_W)\) is usually only important if the \(c_i(\L)\) are tree-level generated. RG evolution includes a logarithm which may serve to counter its loop suppression; however, from \(v^2/\L^2 \sim 0.1\%\), we see that \(\L\) can be probed at most to a few TeV, so that the logarithm is not large, \(\log (\L/m_W) \sim 3\). We note that this estimate also means that in a perturbative expansion a truncation by loop-order counting is reasonable.

A common theme in the literature on the SM EFT is the choice of an operator basis. We will discuss this in detail in section~\ref{sec:operatorbasis}, but we would like to comment here on relevance of choosing an operator basis to the steps of matching and running. One does not need to choose an operator basis at the stage of matching a UV model onto the effective theory. The effective action obtained by integrating out some massive modes will simply produce a set of higher-dimension operators. One can then decide to continue to work with this UV generated operator set as it is, or to switch to a different set due to some other considerations. An operator basis needs to be picked once one RG evolves the Wilson coefficients using the anomalous dimension matrix $\gamma_{ij}$, as the anomalous dimension matrix is obviously basis dependent. When RG running is relevant, it is crucial that the operator basis be complete or overcomplete~\cite{Jenkins:2013zja}.

In section~\ref{sec:mapping} we consider the mapping step, {\it i.e.} obtaining Higgs and EW precision observables as functions of the Wilson coefficients at the weak scale, \(c_i(m_W)\). While there have been a variety of studies concerning the mapping of operators onto weak-scale observables in the literature~\cite{Han:2004az,Cacciapaglia:2006pk,Bonnet:2011yx,*Bonnet:2012nm,delAguila:2011zs,*deBlas:2013qqa,*Blas:2013ana,Dumont:2013wma,Elias-Miro:2013mua,Elias-Miro:2013eta,Pomarol:2013zra,Alonso:2013hga,Chen:2013kfa,Willenbrock:2014bja,Gupta:2014rxa,*Gupta:2014toa,Masso:2014xra,Englert:2014cva,Ciuchini:2014dea,Ellis:2014jta,Falkowski:2014tna,Craig:2014una}, to the best of our knowledge, a complete and systematic list does not exist yet. In this paper, we study a complete set of the Higgs and EW precision observables that present and possible near future experiments can have a decent $\big($1\%$\text{ or better}\big)$ sensitivity on. These include the seven Electroweak precision observables (EWPO) $S, T, U, W, Y, X, V$ up to $p^4$ order in the vacuum polarization functions, the three independent triple gauge couplings (TGC), the deviation in Higgs decay widths $\{\Gamma_{h\to f {\bar f}}, \Gamma_{h\to gg}, \Gamma_{h\to\gamma\gamma}, \Gamma_{h\to\gamma Z}, \Gamma_{h\to WW^*}, \Gamma_{h\to ZZ^*}\}$, and the deviation in Higgs production cross sections at both lepton and hadron colliders $\{\sigma_{ggF}^{},\sigma_{WWh}^{}, \sigma_{Wh}^{}, \sigma_{Zh}^{}\}$. We write these precision observables up to linear power and tree-level order in the Wilson coefficients $c_i(m_W)$ of a complete set of dimension-six CP-conserving bosonic operators\footnote{In this paper, we use the term ``bosonic operators'' to refer to the operators that contain only bosonic fields, {\it i.e.} Higgs and gauge bosons. Other operators will be referred to as ``fermionic operators''.} shown in Table~\ref{tbl:operators}. Quite a bit calculation steps are also listed in Appendix~\ref{sec:app_mapping}. These include a list of two-point and three-point Feynman rules (appendix~\ref{subsec:FeynmanRules}) from operators in Table~\ref{tbl:operators}, interference corrections to Higgs decay widths (appendix~\ref{subsec:GammaIdetails}) and production cross sections (appendix~\ref{subsec:SigmaIdetails}), and general analysis on residue modifications (appendix~\ref{subsec:RMdetails}) and Lagrangian parameter modifications (appendix~\ref{subsec:LPMdetails}). With a primary interest in new physics that only couples with bosons in the SM, we have taken the Wilson coefficients of all the fermionic operators to be zero while calculating the mapping results. However, the general analysis we present for calculating the Higgs decay widths and production cross sections completely applies to fermionic operators.


\section{Covariant derivative expansion} \label{sec:CDE}

In this paper, we advocate the use of the Standard Model EFT from a UV perspective. Let's recapitulate this program. First, match a given UV theory onto the EFT: integrate out heavy physics from the UV model to obtain the Wilson coefficients of the higher dimension operators in the EFT. Second, run the Wilson coefficients down to weak scale using their RG equations. Third, use the EFT at the weak scale to calculate the contribution of new physics, in the form of non-zero Wilson coefficients, to physical observables. In this section, we present tools that considerably ease the step of matching the UV model onto the EFT. We take up the task of running and mapping in later sections.

The process of matching the UV theory onto the EFT is done order-by-order in perturbation theory. As present and future tests of the Standard Model Higgs and gauge sector are typically only sensitive to one-loop order effects, for most purposes it is sufficient to do this matching only up to one-loop order. In this case, the contribution of the UV physics to the low-energy effective action consists of a tree level piece and a one-loop piece.

The point of this section is to present a method for computing the one-loop effective action that leaves gauge invariance manifest {\it at every step of the calculation}. By this we mean that one only works with gauge covariant quantities, such as the covariant derivative. We find it somewhat surprising that this method---developed in the 80s by Gaillard~\cite{Gaillard:1985} (see also her summer school lectures~\cite{Gaillard:1986} and the work by Cheyette~\cite{Cheyette:1987})---is not widely known considering the incredible simplifications it provides. Therefore, in order to spread the good word so to speak, we will explain the method of the covariant derivative expansion (CDE) as developed in~\cite{Gaillard:1985,Cheyette:1987}. Along the way, we will make more rigorous and clear a few steps in the derivation, present a more transparent expansion method to evaluate the CDE, and provide generalized results for scalars, fermions, and massless as well as massive gauge bosons. We also show how to evaluate the tree-level effective action in manifestly gauge-covariant manner. In order to explicitly demonstrate the utility of the CDE, we take up a handful of non-trivial examples and compute their Wilson coefficients in the SM EFT.

Besides providing an easier computational framework, the CDE illuminates a certain universality in computing Wilson coefficients from different UV theories. This occurs because individual terms in the expansion split into a trace over internal indices (gauge, flavor, {\it etc.}) involving covariant derivatives times low energy fields---these are the operators in the EFT---times a simple momentum integral whose value corresponds to the Wilson coefficient of the operator. The UV physics is contained in the specific form of the covariant derivatives and low energy fields, but the momentum integral is independent of these details and therefore can be considered universal.

So far our discussion has been centered around the idea of integrating out some heavy mode to get an effective action, to which we claim the CDE is a useful tool. More precisely, the CDE is a technique for evaluating functional determinants of a generalized Laplacian operator, \(\det [D^2 + U(x)]\), where \(D\) is some covariant derivative. Therefore the technique is not limited to gauge theories; in fact, the CDE was originally introduced in~\cite{Gaillard:1985} primarily as a means for computing the one-loop effective action of non-linear sigma models. In these applications, the use of the CDE keeps the geometric structure of the target manifold and its invariance to field redefinitions manifest~\cite{Gaillard:1985}. Moreover, functional determinants are prolific in the computation of the (1PI or Wilsonian) effective action to one-loop order. Therefore, the use of the CDE extends far beyond integrating out some heavy field and can be used as a tool to, for example, renormalize a (effective) field theory or compute thermal effects.

The 1980s saw considerable effort in developing methods to compute the effective action with arbitrary background fields. While we cannot expect to do justice to this literature, let us provide a brief outline of some relevant works. The CDE developed in~\cite{Gaillard:1985,Cheyette:1987} built upon the derivative expansion technique of~\cite{Chan:1985,Cheyette:1985}. A few techniques for covariant calculation of the one-loop effective action were developed somewhat earlier in~\cite{NSVZ:1983}. While these techniques do afford considerable simplification over traditional methods, they are less systematic and more cumbersome than the CDE presented here~\cite{Gaillard:1985}. In using a heat kernel to evaluate the effective action, a covariant derivative expansion has also been developed, see, {\it e.g.},~\cite{Ball:1988}. This method utilizes a position space representation and is significantly more involved than the approach presented here, where we work in Fourier space.

An outline for this section is as follows. In Sec.~\ref{sec:CDE_summary} we consider the tree and one-loop contributions to the effective action in turn and show how to evaluate each using a covariant derivative expansion. The tree-level result is very simple, as well as useful, and, to the best of our knowledge, has not been appeared in the literature before.   In Sec.~\ref{sec:CDE_gen_consider} we examine evaluation of the functional trace at the more abstract matrix level, thereby clarifying a few steps in the derivation of the CDE. These results are somewhat tangential towards our main focus and can be safely omitted in a first reading. The explicit extension to fermions and gauge bosons is provided in Sec.~\ref{sec:CDE_fermions} together with summary formulas of the CDE for different spin particles. In Sec.~\ref{sec:CDE_universal} we demonstrate how to explicitly evaluate terms in the CDE. Following this, universal formulas for terms in the expansion are presented. As a first example using these results, we derive the \(\b\) function for non-abelian gauge theory and present the Wilson coefficients for the purely gluonic dimension six operators for massive spin 0, 1/2, and 1 particles transforming under some representation of the gauge group. The universal formulas can also immediately be used to obtain the one-loop effective action for a wide variety of theories, as we show in Sec.~\ref{sec:CDE_examples} with a variety of explicit examples. The examples considered are non-trivial demonstrations of the power of the CDE; moreover, they are models that may be relevant to Higgs and other BSM physics: they are related to supersymmetry, extended Higgs sectors, Higgs portal operators, little Higgs theories, extra-dimensional theories, and kinetic mixing of gauge bosons.

We have strived to make accessible the results of this section to a wide audience, primarily because we believe the CDE and its results to be so useful for practical and presently relevant computations. In doing so, however, this section is quite long and it may be helpful to provide a readers guide of sorts in addition to the above outline. Readers mainly interested in the basic idea of the CDE can consider reading the first section, Sec.~\ref{sec:CDE_summary}, then looking over the universal results in Sec.~\ref{sec:CDE_universal} (and equation~\eqref{eqn:CDE_universal_lag} in particular), and skimming a few of the examples in Sec.~\ref{sec:CDE_examples}.

\subsection{Covariant evaluation of the tree-level and one-loop effective action}\label{sec:CDE_summary}
\subsubsection*{Setting up the problem}
Consider \(\Ph\) to be a heavy, real scalar field of mass \(m\) that we wish to integrate out. Let \(S[\ph,\Ph]\) denote the piece of the action in the full theory consisting of \(\Ph\) and its interactions with Standard Model fields \(\ph\). The effective action resultant from integrating out \(\Ph\) is given by
\begin{equation}
e^{i\Seff[\ph](\m)} = \int \mathcal{D} \Ph \, e^{i S[\ph,\Ph](\m)}.
\label{eqn:int_out_phi}
\end{equation}
The above defines the effective action at the scale \(\m \sim m\), where we have matched the UV theory onto the effective theory. In the following we do not write the explicit \(\m\) dependence and it is to be implicitly understood that the effective action is being computed at \(\m \sim m\).

Following standard techniques, \(\Seff\) can be computed to one-loop order by a saddle point approximation to the above integral. To do this, expand \(\Ph\) around its minimum value, \(\Ph = \Ph_c + \et\), where \(\Ph_c\) is determined by
\begin{equation}
\frac{\d S[\ph,\Ph]}{\d \Ph} = 0 \Rightarrow \Ph_c[\ph] .
\label{eqn:tree_solve}
\end{equation}
Expanding the action around this minimum,
\[
S[\ph,\Ph_c + \et] = S[\Ph_c] + \frac{1}{2} \left. \frac{\d^2 S}{\d\Ph^2}\right|_{\Ph_c} \et^2 + \scO(\et^3),
\]
the integral is computed as\footnote{The minus sign inside the logarithm comes from Wick rotating to Euclidean space, computing the path integral using the method of steepest descent, and then Wick rotating back to Minkowski space.}
\begin{align*}
e^{iS_{\text{eff}}[\ph]} &= \int \mathcal{D} \et \, e^{iS[\ph, \Ph_c + \et]} \\
&\approx e^{i S[\Ph_c]}\left[\det\left(-\left. \frac{\d^2 S}{\d\Ph^2}\right|_{\Ph_c} \right)\right]^{-1/2} ,
\end{align*}
so that the effective action is given by
\begin{equation}
S_{\text{eff}} \approx S[\Ph_c] + \frac{i}{2} \Tr\log\left(-\left. \frac{\d^2 S}{\d\Ph^2}\right|_{\Ph_c} \right) .
\label{eqn:Seff_realscalar}
\end{equation}
The first term in the above is the tree-level piece when integrating out a field, {\it i.e.} solving for a field's equation of motion and plugging it back into the action, while the second term is the one-loop piece.

As is clear in the defining equation of the effective action, Eq.~\eqref{eqn:int_out_phi}, the light fields \(\ph\) are held fixed while the path integral over \(\Ph\) is computed. The \(\ph(x)\) fields are therefore referred to as background fields. The fact that the background fields are held fixed while only \(\Ph\) varies in Eq.~\eqref{eqn:int_out_phi} leads to an obvious diagrammatic interpretation of the effective action: the effective action is the set of all Feynman diagrams with \(\ph\) as external legs and only \(\Ph\) fields as internal lines. The number of loops in these diagrams correspond to a loop expansion of the effective action.

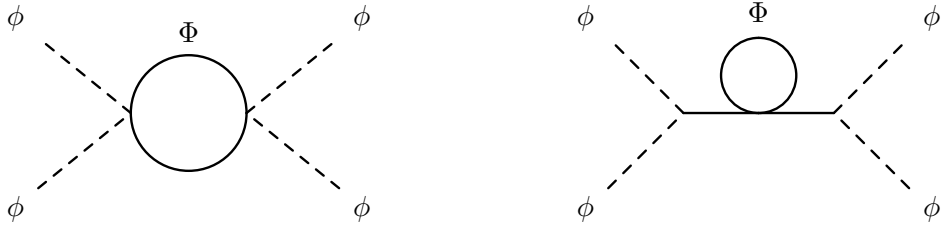
\begin{figure}[t]
\centering
 \subfigure{
 \centering
\begin{fmffile}{CDE1}
\begin{fmfgraph*}(50,20)
\fmfleft{i1,i2}
\fmfright{o1,o2}
\fmftop{t}
\fmfbottom{b}
\fmflabel{$\phi$}{i2}
\fmflabel{$\phi$}{i1}
\fmflabel{$\phi$}{o2}
\fmflabel{$\phi$}{o1}
\fmflabel{$\Phi$}{p}
\fmf{phantom,tension=1}{t,p}
\fmf{phantom,tension=0.15}{p,b}
\fmf{dashes}{i1,i,i2}
\fmf{dashes}{o1,o,o2}
\fmf{plain,left,tension=0.8}{i,o,i}
\end{fmfgraph*}
\end{fmffile}
 }\hspace{2cm}
 \subfigure{
 \centering
\begin{fmffile}{CDE2}
\begin{fmfgraph*}(50,20)
\fmfleft{i1,i2}
\fmfright{o1,o2}
\fmftop{t}
\fmfbottom{b}
\fmflabel{$\phi$}{i2}
\fmflabel{$\phi$}{i1}
\fmflabel{$\phi$}{o2}
\fmflabel{$\phi$}{o1}
\fmflabel{$\Phi$}{t}
\fmf{dashes}{i1,i,i2}
\fmf{dashes}{o1,o,o2}
\fmf{phantom,tension=1.2}{p,b}
\fmf{plain,tension=2}{i,p,o}
\fmf{plain,left,tension=0.6}{p,t,p}
\end{fmfgraph*}
\end{fmffile}
 }
 \caption{Example diagrams that arise in the one-loop effective action.} \label{fig:CDEdiagrams}
\end{figure}

The diagrams with external \(\ph\) and internal \(\Ph\) are sometimes referred to as one-light-particle irreducible (1LPI) in the sense that no lines of the light particle \(\ph\) can be cut to obtain disjoint diagrams. Note, however, that some the diagrams may not be 1PI in the traditional sense. Figure~\ref{fig:CDEdiagrams} shows two example diagrams that could arise in the evaluation of the one-loop effective action; the diagram on the left is 1PI in the traditional sense, while the one on the right is not. The origin of non-1PI diagrams is \(\Ph_c[\ph] \ne 0\). Moreover, these non-1PI diagrams are related to renormalization of the UV Lagrangian parameters, as is clear in the second diagram of Fig.~\ref{fig:CDEdiagrams}. One can find more details on this in the explicit examples considered in Sec.~\ref{sec:CDE_examples}.

\subsubsection{Covariant evaluation of the tree-level effective action}\label{sec:CDE_tree}

First, we show how to evaluate the tree-level piece to the effective action in a covariant fashion. The most na\"ive guess of how to do this turns out to be correct: in the exact same way one would do a derivative expansion, one can do a covariant derivative expansion.

To have a tree-level contribution to the effective action there needs to be a term in the UV Lagrangian that is linear in the heavy field \(\Ph\). We take a Lagrangian,
\begin{equation}
\scL[\Ph,\ph] \supset \big(\Ph^{\dag}B(x) + \text{h.c.}\big) + \Ph^{\dag}\big(-D^2 - m^2 - U(x) \big) \Ph + \scO(\Ph^3),
\label{eqn:tree_UV_Lag}
\end{equation}
where \(B(x)\) and \(U(x)\) are generically functions of the light fields \(\ph(x)\) and we have not specified the interaction terms that are cubic or higher in \(\Ph\). To get the tree-level effective action, one simply solves the equation of motion for \(\Ph\), and plugs it back into the action. The equation of motion for \(\Ph\) is
\begin{equation*}
\big(P^2 - m^2 - U(x)\big) \Ph = -B(x) + \scO(\Ph^2) ,
\end{equation*}
where \(P_{\m} \equiv i D_{\m} = i\pd_{\m} + A_{\m}(x)\) is the covariant derivative\footnote{\(A_{\m} = A_{\m}^aT^a\) with \(T^a\) in the representation of \(\Ph\). We do not specify the coupling constant in the covariant derivative. Of course, the coupling constant can be absorbed into the gauge field; however, unless otherwise stated, for calculations in this paper we implicitly assume the coupling constant to be in the covariant derivative. The primary reason we have not explicitly written the coupling constant is because \(\Ph\) may carry multiple gauge quantum numbers. For example, if \(\Ph\) is charged under \(SU(2)_L\times U(1)_Y\) then we will take \(D_{\m} = \pd_{\m} - ig W_{\m} -i g' Y B_{\m}\).} that acts on \(\Ph\). The solution of this gives \(\Ph_c[\ph]\) denoted in Eq.~\eqref{eqn:tree_solve}. To leading approximation, we can linearize the above equation to solve for \(\Ph_c\),
\begin{equation}
\Ph_c = -\frac{1}{P^2 - m^2 - U(x)}B(x).
\label{eqn:phic_lin}
\end{equation}
If the covariant derivative were replaced with the partial derivative, \(P^2 = -\pd^2\), one would evaluate the above in an inverse-mass expansion producing a series in \(\pd^2/m^2\). The exact same inverse-mass expansion can be used with the covariant derivative as well to obtain\footnote{This is trivially true. In the case of a partial derivative, \(-\pd^2 - m^2 - U(x)\), the validity of the expansion relies not only on \(\pd^2/m^2 \ll 1\) but also on \(U(x)/m^2 \ll 1\), {\it i.e.} momenta in the EFT need to be less than \(m\) which also means the fields in the EFT need to be slowly varying on distance scales of order \(m^{-1}\). Obviously, the same conditions can be imposed on the covariant derivative as a whole.}
\begin{align}
\Ph_c &= \left[1 - \frac{1}{m^2}\big(P^2 - U\big)\right]^{-1}\frac{1}{m^2}B \nonumber \\
&= \frac{1}{m^2} B + \frac{1}{m^2}\big(P^2 - U\big)\frac{1}{m^2}B + \frac{1}{m^2}\big(P^2 - U\big)\frac{1}{m^2}\big(P^2 - U\big)\frac{1}{m^2}B + \dots \hspace{2mm}.
\label{eqn:phic_expand}
\end{align}
In general, the mass-squared matrix need not be proportional to the identity, so that \(1/m^2\) should be understood as the inverse of the matrix \(m^2\). In this case, \(1/m^2\) would not necessarily commute with \(U\) and hence we used the matrix expansion from Eq.~\eqref{eqn:sum_matrix_exp} in the above equation.

Plugging \(\Ph_c\) back into the Lagrangian gives the tree-level effective action. Using the linearized solution to the equation of motion, Eq.~\eqref{eqn:phic_lin}, we have
\begin{equation}
\scL_{\text{eff,tree}} = -B^{\dag} \frac{1}{P^2 - m^2 - U(x)} B + \scO(\Ph_c^3).
\end{equation}
Although we have not specified the interactions in Eq.~\eqref{eqn:tree_UV_Lag} that are cubic or higher in \(\Ph\), one needs to also substitute \(\Ph_c\) for these pieces as well, as indicated in the above equation. The first few terms in the inverse mass expansion are
\begin{equation}
\scL_{\text{eff,tree}} = B^{\dag}\frac{1}{m^2}B +B^{\dag}\frac{1}{m^2}\big(P^2 - U\big)\frac{1}{m^2}B + \dots + \scO(\Ph_c^3) .
\label{eqn:tree_cde_terms}
\end{equation}

\subsubsection{CDE of the one-loop effective action}
\label{sec:CDE_one-loop_deriv}

Now let us discuss the one-loop piece of the effective action. Let \(\Ph\) be field of mass \(m\) that we wish to integrate out to obtain a low-energy effective action in terms of light fields. Assume that \(\Ph\) has quantum numbers under the low-energy gauge groups. The one-loop contribution to the effective action that results from integrating out \(\Ph\) is
\begin{equation}
\D\Seff = i c_s \Tr \log\Big( -P^2 + m^2 + U(x) \Big) ,
\label{eqn:sum_seff}
\end{equation}
where \(c_s = +1/2, +1,\text{ or } -1/2\) for \(\Ph\) a real scalar, complex scalar, or fermion, respectively.\footnote{The reason fermions have \(c_s=-1/2\) instead of the usual \(-1\) is because we have squared the usual argument of the logarithm, \(\D\Seff = -\frac{i}{2} \Tr \log (i \slashed{D} + \dots)^2\), to bring it to the form in Eq.~\eqref{eqn:sum_seff}. See Appendix~\ref{sec:app_cde_ferm} for details.}

We evaluate the trace in the usual fashion by inserting a complete set of momentum and spatial states to arrive at
\begin{equation}
\D\Seff = i c_s \int d^4x \int \frac{d^4q}{(2\pi)^4} \tr \, e^{iq\cdot x} \log\Big(-P^2 + m^2+ U(x)\Big) e^{-iq\cdot x} ,
\end{equation}
where the lower case ``tr'' denotes a trace on internal indices, {\it e.g.} gauge, spin, flavor, {\it etc.} For future shorthand we define \(dx \equiv d^4x\) and \(dq \equiv d^4q/(2\pi)^4\). Using the Baker-Campbell-Hausdorff (BCH) formula,
\begin{equation}
e^BAe^{-B} = \sum_{n=0}^{\infty} \frac{1}{n!}L_B^nA  , \quad L_BA = [B,A],
\label{eqn:sum_BCH}
\end{equation}
together with the fact that we can bring the \(e^{\pm iq\cdot x}\) into the logarithm, we see that the \(P_{\m} \to P_{\m} + q_{\m}\). Then, after changing variables \(q \to -q\), the one-loop effective action is given by
\begin{equation}
\D\Seff = i c_s \int dx\, dq \, \tr \, \log\Big[ -\big(P_{\m} - q_{\m}\big)^2 +m^2 + U(x) \Big].
\label{eqn:sum_shiftq}
\end{equation}
Following~\cite{Gaillard:1985,Cheyette:1987}, we sandwich the above by \(e^{\pm P^{\m}\pd/\pd q^{\m}}\)
\begin{equation}
\D\Seff = i c_s \int dx \, dq \, \tr \, e^{P \cdot \frac{\pd}{\pd q}} \log\Big[ -\big(P_{\m} - q_{\m}\big)^2 + m^2 + U(x) \Big]e^{-P \cdot \frac{\pd}{\pd q}}.
\label{eqn:sum_trans}
\end{equation}
In the above it is to be understood that the derivatives \(\pd/\pd q\) and \(\pd/\pd x \subset P\) act on unity to the right (for \(e^{-P\cdot \pd/\pd q}\)) and, by integration by parts, can be made to act on unity to the left (for \(e^{P\cdot \pd/\pd q}\)). Since the derivative of one is zero, the above insertion is allowed. We emphasize that the ability to insert \(e^{\pm P \cdot \pd/\pd q}\) in Eq.~\eqref{eqn:sum_trans} {\it does not} rely on cyclic property of the trace: the ``tr'' trace in Eq.~\eqref{eqn:sum_trans} is over internal indices only and we therefore cannot cyclically permute the infinite dimensional matrices in Eq.~\eqref{eqn:sum_trans}.\footnote{
While the above arguments leading to Eq.~\eqref{eqn:sum_trans} are correct, they may seem slightly unclear because we have, in fact, brushed over some subtle steps: Why could we use the BCH formula in Eq.~\eqref{eqn:sum_shiftq}? Where does this magical unity on the right and left come from? In section~\ref{sec:CDE_gen_consider} we provide a more abstract and general treatment that answers these questions and makes clear what transformations in general we can make on the argument of the trace.}

One advantage of this choice of insertion is that it makes the linear term in $P_\mu$ vanish when transforming the combination $(P_\mu-q_\mu)$, and so the expansion starts from a commutator $\left[P_\mu,P_\nu\right]$, which is the field strength. Indeed, by making use of the BCH formula and the fact $\left(L_{P\cdot\partial/\partial q}\right)q_\mu=[P\cdot\partial/\partial q,q_\mu]=P_\mu$, we get
\begin{eqnarray}
{e^{P \cdot \frac{\partial}{\partial q}}}({P_\mu } - {q_\mu }){e^{ - P \cdot \frac{\partial}{\partial q}}} &=& \sum\limits_{n = 0}^\infty  {\frac{1}{{n!}}{{\left( {{L_{P \cdot \partial/\partial q}}} \right)}^n}{P_\mu }}  - \sum\limits_{n = 0}^\infty  {\frac{1}{{n!}}{{\left( {{L_{P \cdot \partial/\partial q}}} \right)}^n}{q_\mu }} \nonumber \\
 &=&  - {q_\mu } + \sum\limits_{n = 1}^\infty  {\frac{n}{{(n + 1)!}}{{\left( {{L_{P \cdot \partial/\partial q}}} \right)}^n}{P_\mu }} \nonumber \\
 &=&  - {q_\mu } - \sum\limits_{n = 0}^\infty  {\frac{{n + 1}}{{(n + 2)!}}\bigg[ {{P_{{\alpha _1}}},\Big[ { \ldots \big[ {{P_{{\alpha _n}}},\left[ {{D_\nu },{D_\mu }} \right]} \big]} \Big]} \bigg]\frac{{{\partial ^n}}}{{\partial {q_{{\alpha _1}}} \ldots \partial {q_{{\alpha _n}}}}}\frac{\partial }{{\partial {q_\nu }}}} \nonumber \\
 &\equiv&  - \left( {{q_\mu } + {{\tG}_{\nu \mu }}\frac{\partial }{{\partial {q_\nu }}}} \right) , \label{eqn:asdf}
\end{eqnarray}
and similarly,
\begin{equation}
{e^{P \cdot {\frac{\partial}{\partial q}}}}U{e^{ - P \cdot \frac{\partial}{\partial q}}} = \sum\limits_{n = 0}^\infty  {\frac{1}{{n!}}\bigg[ {{P_{{\alpha _1}}},\Big[ {{P_{{\alpha _2}}},\big[ { \ldots \left[ {{P_{{\alpha _n}}},U} \right]} \big]} \Big]} \bigg]\frac{{{\partial ^n}}}{{\partial {q_{{\alpha _1}}} \ldots \partial {q_{{\alpha _n}}}}}} \equiv \tU .
\end{equation}
Bringing the \(e^{\pm P \cdot \pd/\pd q}\) into the logarithm to compute the transformation of the integrand in Eq.~\eqref{eqn:sum_trans}, one gets the results obtained in~\cite{Gaillard:1985,Cheyette:1987}
\begin{equation}
\D\Seff = \int dx \, \D\Leff = i c_s \int dx \int dq \, \tr \, \log\bigg[ -\Big(q_{\m} + \tG_{\n\m}\frac{\pd}{\pd q_{\n}}\Big)^2 + m^2 + \tU \bigg]
, \label{eqn:sum_CDE_lag}
\end{equation}
where we have defined
\begin{subequations}
\label{eqn:sum_tilde_defs}
\begin{align}
\tG_{\n\m} &= \sum_{n=0}^{\infty} \frac{n+1}{(n+2)!} \Bigg[P_{\a_1},\bigg[P_{\a_2},\Big[ \dots \big[P_{\a_n},[D_{\n},D_{\m}]\big]\Big]\bigg]\Bigg] \frac{\pd^n}{\pd q_{\a_1} \pd q_{\a_2} \dots \pd q_{\a_n}} , \label{eqn:sum_tG_def} \\
\tU &= \sum_{n=0}^{\infty} \frac{1}{n!} \bigg[P_{\a_1},\Big[P_{\a_2},\big[\dots [P_{\a_n},U]\big]\Big]\bigg]\frac{\pd^n}{\pd q_{\a_1} \pd q_{\a_2} \dots \pd q_{\a_n}} . \label{eqn:sum_tU_def}
\end{align}
\end{subequations}
The commutators in the above correspond to manifestly gauge invariant higher dimension operators: In Eq.~\eqref{eqn:sum_tG_def} the commutators of \(P\)'s with \([D_{\n},D_{\m}] = -i G_{\n\m}\), where \(G_{\n\m}\) is the gauge field strength, correspond to higher dimension operators of the field strength and its derivatives. In Eq.~\eqref{eqn:sum_tU_def}, the commutators will generate higher dimension derivative operators on the fields inside \(U(x)\).

While it should be clear, it is worth emphasizing that \(x\) and \(\pd/\pd x\) commute with \(q\) and \(\pd/\pd q\), {\it i.e.} \(P = i\pd/\pd x + A(x)\) and \(U(x)\) commute with \(q\) and \(\pd/\pd q\). This, together with the fact that the commutators in Eq.~\eqref{eqn:sum_tilde_defs} correspond to higher dimension operators, allows us to develop a simple expansion of Eq.~\eqref{eqn:sum_CDE_lag} in terms of higher dimension operators whose coefficients are determined from easy to compute momentum integrals, which we now describe.

Instead of working with the logarithm, we work with its derivative with respect to \(m^2\). Using \(\pd_{\m}\) to denote the derivative {\it with respect to q}, \(\pd_{\m} \equiv \pd/\pd q^{\m}\), and defining \(\D \equiv (q^2 -m^2)^{-1}\), the effective Lagrangian is
\begin{equation}
\D \Leff = -i c_s \int dq \int dm^2 \, \tr \, \frac{1}{\D^{-1}\Big[1 + \D\Big(\big\{q_{\m},\tG_{\n\m}\pd^{\n}\big\} + \tG_{\s\m}\tG^\s_\n\pd^{\m}\pd^{\n} - \tU\Big) \Big]}.
\label{eqn:sum_Leff_begin}
\end{equation}
In the above, \(\D\) is a free propagator for a massive particle; we can develop an expansion of powers of \(\D\) and its derivatives (from the \(q\) derivatives inside \(\tG\) and \(\tU\)) where the coefficients are the higher dimension operators. The derivatives and integrals in \(q\) are then simple, albeit tedious, to compute and correspond to the Wilson coefficient of the higher dimension operator. Explicitly, using
\begin{equation}
[A^{-1}(1+AB)]^{-1} = A - A B A + ABABA - \dots,
\label{eqn:sum_matrix_exp}
\end{equation}
we have (using obvious shorthand notation)
\begin{align}
\D \Leff = -ic_s \int dq \, dm^2 \, &\tr \bigg[ \D - \D\Big(\{q,\tG\} + \tG^2 - \tU\Big)\D \nonumber \\
&+ \D\Big(\{q,\tG\} + \tG^2 - \tU\Big)\D\Big(\{q,\tG\} + \tG^2 - \tU\Big)\D +\dots \bigg] . \label{eqn:sum_Leff_exp}
\end{align}
There are two points that we would like to draw attention to:
\begin{description}
\item[Power counting] Power counting is very transparent in the expansion in Eq.~\eqref{eqn:sum_Leff_exp}. This makes it simple to identify the dimension of the operators in the resultant EFT and to truncate the expansion at the desired order. For example, the lowest dimension operator in \(\tG_{\m\n}\) is the field strength \([D_{\m},D_{\n}] = -i G_{\m\n}\); each successive term in \(\tG\) increases the EFT operator dimension by one through an additional \(P_{\a}\). The dimension increase from additional \(P\)'s is compensated by additional \(q\) derivatives which, by acting on \(\D\), increase the numbers of propagators.
\item[Universality] When the mass squared matrix \(m^2\) is proportional to the identity then \(\D\) commutes with the matrices in \(\tG\) and \(\tU\). In this case, for any given term in the expansion in Eq.~\eqref{eqn:sum_Leff_exp}, the \(q\) integral trivially factorizes out of the trace and can be calculated separately. Because of this, there is a certain universality of the expansion in Eq.~\eqref{eqn:sum_Leff_exp}: specifics of a given UV theory are contained in \(P_{\m}\) and \(U(x)\), but the coefficients of EFT operators are determined by the \(q\) integrals and can be calculated without any reference to the UV model.
\end{description}

Before we end this section, let us introduce a more tractable notation that we use in later calculations and results. We provide the notation here for the reader who wishes to skim ahead to results. As we already have used, \(\pd_{\m} \equiv \pd/\pd q^{\m}\). The action of the covariant derivative on matrix is defined as a commutator and we use as shorthand \(P_{\m}A \equiv [P_{\m},A]\). We also define \(G_{\m\n}' \equiv [D_{\m},D_{\n}]\).\footnote{If \(D_{\m}= \pd/\pd x^{\m} - i g A_{\m}\), then \(G'_{\m\n}\) is related to the usual field strength as \(G'_{\m\n}  = [D_{\m},D_{\n}] = -i g G_{\m\n}\). In the case where we have integrated out multiple fields with possibly multiple and different gauge numbers, it is easier to just work with \(D_{\m}\), hence the definition of \(G'_{\m\n}\).} To summarize and repeat ourselves:
\begin{equation}
\pd_{\m} \equiv \frac{\pd}{\pd q^{\m}}, \quad P_{\m}A \equiv [P_{\m},A], \quad G_{\m\n}' \equiv [D_{\m},D_{\n}] .
\label{eqn:notation_short}
\end{equation}
Finally, as everything is explicitly Lorentz invariant, we will typically not bother with raised and lowered indices. With this notation, \(\tG\) and \(\tU\) as defined in Eq.~\eqref{eqn:sum_tilde_defs} are given by
\begin{subequations}
\label{eqn:sum_tilde_short}
\begin{align}
\tG_{\n\m} &= \sum_{n=0}^{\infty} \frac{n+1}{(n+2)!} \big(P_{\a_1}\dots P_{\a_n}G_{\n\m}' \big) \pd^n_{\a_1\dots \a_n} , \label{eqn:sum_tG} \\
\tU &= \sum_{n=0}^{\infty} \frac{1}{n!}\big(P_{\a_1}\dots P_{\a_n}U\big) \pd^n_{\a_1\dots \a_n} .   \label{eqn:sum_tU}
\end{align}
\end{subequations}

\subsection{General considerations}\label{sec:CDE_gen_consider}
Here we look at the covariant evaluation of the one-loop effective action at the operator level, to clarify a few steps presented in the derivation of the previous section. These results are not essential to the rest of this paper and can be omitted in a first reading.

For the one-loop effective action, we are interested in evaluating the functional trace,
\[
\Tr \log\left(-\left.\frac{\d^2S[\ph,\Ph]}{\d\Ph^2}\right|_{\Ph = \Ph_c}\right) \equiv \Tr \, f.
\]
\(\d^2 S/\d \Ph^2\), and hence \(f\), is Hermitian.\footnote{The usual care should be taken when defining the functional determinant: we go to Euclidean space and take \(K\equiv\d^2S_E/\d\Ph^2\) to be Hermitian, positive definite. For general background fields the matrix is non-singular, although specific field configurations may make \(K\) singular, in which case the zero eigenvalues have to be handled with care. These properties allow us to define the functional determinant, \(\det K\), as well as the functional trace \(\Tr \log K\) where the Hermiticity of \(\log K\) follows from that of \(K\). We assume there is no issue with Wick rotation and work in Minkowski space.} Since \(f\) is Hermitian, its eigenvectors lie in a Hilbert space. Since we are working in a Hilbert space, we will use notation familiar from quantum mechanics. Unfortunately, we cannot diagonalize \(f\) and compute its spectrum in general because \(f\) depends on arbitrary functions (the background fields). However, we can still develop a perturbative approximation of the trace.

For our purposes, \(f\) derives from a Lagrangian and is therefore a function of the position and momentum operators, \(f(\wh{x}_{\m},\wh{q}_{\m})\). For example, a particular form of \(f\) of interest to us in this work is
\begin{equation}
f = \log \Big[ -\big(\wh{q}_{\m} + A_{\m}(\wh{x}) \big)^2 + U(\wh{x})\Big].
\label{eqn:example_f}
\end{equation}
For notational simplicity, in the following we will typically not write the Lorentz indices explicitly.

To explicitly evaluate \(\Tr\, f\), we will need to give a representation to \(f\). Recall that operators take on a given representation when acting on some basis vector, {\it e.g.}\footnote{With a metric \(g_{\m\n} = \text{diag}(+,-,-,-)\) the position representation of \(\wh{q}\) is \(\wh{q}_{\m} = i\pd/\pd x^{\m}\). In this convention, the commutation relation is \([\wh{x},\wh{q}] = -i\) and a plane wave is given by \(\braket{x|q} = e^{-iq\cdot x}\).}
\begin{equation}
\bra{x} f(\wh{x},\wh{q}) = f(x,i\pd_x)\bra{x} \quad \text{or} \quad \bra{q} f(\wh{x},\wh{q}) = f(-i\pd_q,q)\bra{q},
\label{eqn:QMrep}
\end{equation}
Note that the derivative acts on the eigenvalue of the basis vector which gave the operator that particular representation.

To evaluate \(\Tr f\) we begin by inserting the identity and resolving the identity in momentum space,
\begin{equation}
\Tr \, f(\wh{x},\wh{q}) = \int dq \, \tr \bra{q} f(\wh{x},\wh{q}) \ket{q}.
\label{eqn:Trf_start}
\end{equation}
As before, \(dq \equiv d^4q/(2\pi)^4\), \(dx \equiv d^4x\), and the lower case ``tr'' denotes a trace over internal indices only. For the rest of this subsection we will leave the trace on internal indices implicit and drop the ``tr'' in expressions.

The momentum states \(\ket{q}\) can be written in a particularly useful way. Define the unit function in \(x\)-space as
\begin{equation}
\ket{1_x} \equiv \int dy \ket{y}.
\label{eqn:unit_def}
\end{equation}
Since a constant function has zero momentum, obviously the unit function in \(x\)-space is equivalent to the zero momentum state:
\begin{equation*}
\ket{1_x} = \int dp \ket{p}\braket{p|1_x} = \int dp \, dy \ket{p}e^{ip\cdot y} = \int dp \ket{p}\d(p) = \ket{0_q}.
\end{equation*}
While we could just work with the zero momentum state \(\ket{0_q}\), when explicitly evaluating the functional determinant it will be conceptually more convenient to think of it as the unit function \(\ket{1_x}\). This state possesses the following properties which are easily checked
\begin{equation}
\braket{x|1_x} = 1, \quad \wh{q}\ket{1_x} = 0, \quad \braket{1_x|1_x} = \int dx \, .
\end{equation}

With the use of the unit function, the plane wave \(\ket{q}\) can be written as
\begin{equation}
\ket{q} = e^{-iq\cdot \wh{x}} \ket{1_x} .
\label{eqn:q_as_unit_fxn}
\end{equation}
This is easily seen by using the eigen-decomposition \(e^{-iq\cdot \wh{x}} = \int dy \, e^{-iq\cdot y}\ket{y}\bra{y}\), or even more simply by noting that Eq.~\eqref{eqn:q_as_unit_fxn} is obviously consistent with \(\braket{x|q} = e^{-iq\cdot x}\).

Using the decomposition for the momentum states in Eq.~\eqref{eqn:q_as_unit_fxn}, the trace in Eq.~\eqref{eqn:Trf_start} is
\begin{equation}
\Tr\, f = \int dq \bra{1_x} e^{iq\cdot\wh{x}} f(\wh{x},\wh{q}) e^{-iq\cdot \wh{x}} \ket{1_x}.
\label{eqn:Trf_qshift}
\end{equation}
By making use of the Baker-Cambell-Hausdorff (BCH) formula,
\begin{equation}
e^BAe^{-B} = \sum_{n=0}^{\infty} \frac{1}{n!}L_B^nA, \quad L_BA \equiv [B,A],
\label{eqn:BCH}
\end{equation}
we see that \(\wh{q} \to \wh{q} + q\) in Eq.~\eqref{eqn:Trf_qshift}
 \begin{equation}
\Tr\, f = \int dq \bra{1_x}f(\wh{x},\wh{q}+q) \ket{1_x}.
\label{eqn:Trf_qshift2}
\end{equation}
Inserting a complete set of position states,
\begin{eqnarray}
\Tr\, f &=& \int dx\, dq \braket{1_x|x}\bra{x}f(\wh{x},\wh{q}+q) \ket{1_x} \nonumber \\
&=& \int dx\, dq\, \braket{1_x|x} f(x,i\pd_x +q) \braket{x|1_x} = \int dx\, dq\, f(x,i\pd_x +q), \nonumber
\end{eqnarray}
Taking \(f\) as in Eq.~\eqref{eqn:example_f}, we see that we recover~\eqref{eqn:sum_shiftq} where now it is clear why we could use the BCH formula to get~\eqref{eqn:sum_shiftq}. Moreover, it is explicitly clear what it means for the derivative \(i \pd_x\) to be acting on unity to the right; in the above \(\wh{q}\) takes a representation from \(\bra{x}\), \(\bra{x}\wh{q} = i\pd_x\bra{x}\), and acts upon the eigenvalue of \(\bra{x}\). When \(\bra{x}\) hits \(\ket{1_x}\), \(\braket{x|1_x} = 1\), it is to be understood that the derivative \(i\pd_x\) then acts on unity when it gets all the way to the right.

Let us consider more general transformations that can be made within the inner product of Eq.~\eqref{eqn:Trf_qshift}. Note that since \(q\) is simply a parameter, it commutes with everything. Let us promote this parameter to a second momentum operator, \(q \to \hqt\), that acts on a second position-momentum space. Denoting the original \(\hx\) and \(\hq\) as \(\hxo\) and \(\hqo\), the commutation relations are
\begin{equation}
[\hx^{}_i,\hq^{}_j] = -i \d_{ij}, \quad [\hx^{}_i,\hx^{}_j] = [\hq^{}_i,\hq^{}_j] = 0, \quad i,j=1,2 \, .
\label{eqn:comm_rel}
\end{equation}
\(\hxt\) and \(\hqt\) are operators on a second Hilbert space; the entire Hilbert space is the direct product \(\mathcal{H} = \mathcal{H}_1 \otimes \mathcal{H}_2\). We denote states in \(\mathcal{H}_1 \otimes \mathcal{H}_2\) with a single bra or ket with a semi-colon separating labels between the \(\mathcal{H}_i\) and the state in \(\mathcal{H}_1\) always to the left of the semi-colon. For example,
\begin{equation}
\ket{x^{}_1;q^{}_2} = \ket{x^{}_1} \otimes \ket{q^{}_2} \quad , \quad \bra{x^{}_1; q^{}_2} = \bra{x^{}_1} \otimes \bra{q^{}_2}
\label{eqn:ex_kets}
\end{equation}

Making use of the property
\begin{equation}
\braket{\uqt|g(\hqt)|\uqt} = \int dq \, g(q),
\end{equation}
where \(\ket{1_{q^{}_i}} = \int dp_i \ket{p_i}\) is the unit function in \(q_i\)-space, we see that that we can rewrite the trace in Eq.~\eqref{eqn:Trf_qshift} as
\begin{eqnarray}
\Tr\, f(\hxo ,\hqo) &=& \int dq \bra{\uxo} e^{i\hxo \cdot q} f(\hxo,\hqo) e^{-i \hxo \cdot q} \ket{\uxo} \nonumber \\
&=& \bra{\uxo; \uqt} e^{i\hxo \cdot \hqt} f(\hxo,\hqo) e^{-i \hxo \cdot \hqt} \ket{\uxo;\uqt} \nonumber \\
&=& \bra{\uxo; \uqt} f(\hxo,\hqo+\hqt) \ket{\uxo;\uqt} \label{eqn:q_promote}
\end{eqnarray}
where in the last line we used BCH to shift \(\hqo\to \hqo + \hqt\).

What have we gained by going through this more abstract way of writing the trace? The point is that Eq.~\eqref{eqn:q_promote} makes it clear that we can make many transformations on \(f(\hxo,\hqo+\hqt)\) that leave the trace invariant: A large number of operators leave the unit function \(\ket{\uxo;\uqt}\) invariant; by inserting these into the inner product in Eq.~\eqref{eqn:q_promote} we can then regard them as transformations on \(f(\hxo,\hqo+\hqt)\). Moreover, by promoting the parameter \(q\) to be operator valued, \(q \to \hqt\), it is clear that we can consider transformations on \(\hqt\) as well. The idea, of course, is that some of these transformations may bring \(f\) to a particularly convenient form.

Let us consider the operators which leave the state \(\ket{\uxo;\uqt}\) invariant. Let \(h(\hx^{}_i,\hq^{}_i)\) be an analytic function of the position and momentum operators and we ask
\begin{equation}
e^{i h(\hx^{}_i,\hq^{}_i)} \ket{\uxo;\uqt} \overset{\text{\normalsize{?}}}{=} \ket{\uxo;\uqt}.
\label{eqn:leave_unity_inv}
\end{equation}
We have put \(h\) in the exponential for convenience, from which clearly the above condition is satisfied when \(h\) annihilates \(\ket{\uxo;\uqt}\). We are not particularly interested in general considerations on the form of \(h\), but rather concern ourselves with pointing out some classes of \(h\) that satisfy Eq.~\eqref{eqn:leave_unity_inv} which will prove useful in explicit calculations. Recalling that \(\hq \ket{1_x} = \hx \ket{1_q} = 0\), we see that if \(h\) only depends on \(\hqo\) and \(\hxt\) then any function \(h(\hqo,\hxt)\) such that \(h(\hqo,0) = 0\) {\it or} \(h(0,\hxt)=0\) will annihilate \(\ket{\uxo;\uqt}\). If we consider \(h\) to depend on \(\hxo\) as well, then any function \(h(\hxo,\hqo,\hxt)\) such that \(h(\hxo,\hqo,0) = 0\) will annihilate \(\ket{\uxo;\uqt}\). This follows from that fact that since \(\hxt\) commutes with \(\hxo\) and \(\hqo\), we can always bring it to the right where it will annihilate \(\ket{\uqt}\).

Let \(h(\hx^{}_i,\hq^{}_i)\) and \(h'(\hx^{}_i,\hq^{}_i)\) be two Hermitian operators satisfying Eq.~\eqref{eqn:leave_unity_inv}. We can therefore insert these into the inner product in Eq.~\eqref{eqn:q_promote} and consider the properties of the transformed operator
\begin{equation}
e^{ih(\hx^{}_i,\hq^{}_i)} f(\hxo,\hqo + \hqt) e^{-ih'(\hx^{}_i,\hq^{}_i)}.
\end{equation}
When \(h' = h\), this amounts to a unitary transformation on \(f\). In this case, assuming \(f\) has a well-defined Taylor expansion, we have
\begin{equation}
e^{ih}f(\hxo,\hqo + \hqt)e^{-ih} = f\Big(e^{ih}\hxo e^{-ih}, e^{ih}(\hqo + \hqt)e^{-ih}\Big)
\end{equation}
and the transformations can be evaluated using the BCH formula Eq.~\eqref{eqn:BCH}. When \(h\) is not very complicated, these are not hard to compute. As an example, consider the case \(h = -\hqo \cdot \hxt\):
\begin{equation}
f(\hxo,\hqo+\hqt) \to e^{-i\hqo \cdot \hxt} f(\hxo,\hqo+\hqt) e^{i\hqo \cdot \hxt} = f(\hxo + \hxt, \hqt).
\label{eqn:trans_chan}
\end{equation}
This transformation takes us from the starting point of a derivative expansion of Eq.~\eqref{eqn:Trf_qshift2} to the form used in~\cite{Chan:1985,Cheyette:1985}.

Finally, let us consider the case where \(f\) contains the covariant derivative:
\[
f = \log \Big[ -\big(\hq + A(\hx)\big)^2 + U(\hx) \Big] = \log \Big[ -\wh{P}^2 + U(\hx) \Big].
\]
From the above discussion we have,
\begin{equation}
\Tr f = \bra{\uxo; \uqt} \log \Big[ - \big( \wh{P}_1 + \hqt)^2 + U(\hxo) \Big]\ket{\uxo;\uqt}.
\label{eqn:cde_f_trans}
\end{equation}
We consider the unitary transformation \(e^{ih}\) with \(h = -\wh{P}_1\cdot \hxt\), which is the operator statement of the transformation introduced by~\cite{Gaillard:1985} and used in the previous subsection in deriving the CDE. As per our discussion on the allowed forms of \(h\), while \(\wh{P}_1\) does not annihilate \(\ket{\uxo}\), \(\hxt\) does annihilate \(\ket{\uqt}\) and therefore \(h \ket{\uxo;\uqt} = 0\). The nice property of this \(h\) is that it shifts \(\hqt\) by the covariant derivative: \(\hqt \to \hqt - \wh{P}_1 + \dots\) where the higher order terms are commutators of the covariant derivative with itself times powers of \(\hxt\), {\it i.e.},
\[
e^{-i\wh{P}_1\cdot \wh{x}_2}\big(\wh{P}_1 + \hqt\big)e^{i\wh{P}_1\cdot \wh{x}_2} = \hqt + \sum_{n=0}\frac{n+1}{(n+2)!} \big(\wh{P}_1^n [\wh{P}_1,\wh{P}_1]\big) (-i\wh{x}_2)^{n+1},
\]
just as in Eq.~\eqref{eqn:asdf}. Upon using this shift and inserting the complete set of states,
\[
\int dx \, dq \, \ket{x;q}\bra{x;q},
\]
into Eq. \eqref{eqn:cde_f_trans}, it is straightforward to see that we recover the covariant derivative expansion in formula~\eqref{eqn:sum_CDE_lag}.

\subsection{CDE for fermions, gauge bosons, and summary formulas}\label{sec:CDE_fermions}

The CDE as presented in section~\ref{sec:CDE_one-loop_deriv} is for evaluating functional determinants of the form
\begin{equation*}
\log \det \big(-P^2 + W(x) \big) = \Tr \log \big(-P^2 + W(x) \big) ,
\end{equation*}
where \(P_{\m} = iD_{\m}\) is a covariant derivative. As such, the results of section~\ref{sec:CDE_one-loop_deriv} apply for any generalized Laplacian operator of the form \(-P^2 + W(x)\).\footnote{This is loosely speaking, but applies to many of the cases physicists encounter. More correctly, the functional determinant should exist and so we actually work in Euclidean space and consider elliptic operators of the form \(+P^2 + W(x)\) with \(W\) hermitian, positive-definite. The transformations leading to the CDE in section~\ref{sec:CDE_one-loop_deriv} then apply to these elliptic operators as well. In the cases we commonly encounter in physics, these properties are satisfied by the fact that operator is the second variation of the Euclidean action which is typically taken to be Hermitian and positive-definite.} The lightning summary is
\begin{align}
 \Tr \log \big(-P^2 + W \big) &= \int dx\, dq \, \tr \, e^{P \cdot \pd_q}e^{iq\cdot x} \log \big( -P^2 + W \big) e^{-iq\cdot x} e^{-P\cdot \pd_q} \nonumber \\
&= \int dx \, dq \, \tr\, \log \Big[ - \Big(q_{\m} +\tG_{\n\m}\pd_{\n}\Big)^2 + \wt{W} \Big] ,
\end{align}
where we \(\tG\) and \(\wt{W}\) are given in Eq.~\eqref{eqn:sum_tilde_short} with \(U\) replaced by \(W\) and we are using the notation defined in Eq.~\eqref{eqn:notation_short}. In section~\ref{sec:CDE_one-loop_deriv} we took \(W(x) = m^2 + U(x)\) for its obvious connection to massive scalar fields.

When we integrate out fermions and gauge bosons, at one-loop they also give functional determinants of generalized Laplacian operators of the form \(-P^2 + W(x)\). It is straightforward to apply the steps of section~\ref{sec:CDE_one-loop_deriv} to these cases. Nevertheless, it is useful to tabulate these results for easy reference. Therefore, in this subsection we summarize the results for integrating out massive scalars, fermions, and gauge bosons. We also include the result of integrating out the high energy modes of a massless gauge field. We relegate detailed derivations of the fermion and gauge boson results to appendix~\ref{sec:app_cde_ferm}. The results for fermions were first obtained in~\cite{Gaillard:1985}\footnote{We note that there is an error in the results for fermions in~\cite{Gaillard:1985} (see appendix~\ref{sec:app_cde_ferm}).} and for gauge bosons in~\cite{Cheyette:1987}.

Let us state the general result and then specify how it specializes to the various cases under consideration. The one-loop effective action is given by
\begin{equation}
\D S_{\text{eff,1-loop}} = i c_s \Tr \log \big(-P^2 + m^2 + U(x)\big),
\label{eqn:gen_Seff_Tr}
\end{equation}
where the constant \(c_s\) and the form of \(U\) depend on the species we integrate out, as we explain below. After evaluating the trace and using the transformations introduced in~\cite{Gaillard:1985} and explained in section~\ref{sec:CDE_one-loop_deriv}, the one-loop effective Lagrangian is given by
\begin{equation}
\D \scL_{\text{eff,1-loop}} = i c_s \int dq \, \tr\, \log \Big[ - \Big(q_{\m} +\tG_{\n\m}\pd_{\n}\Big)^2 + m^2 + \tU \Big],
\label{eqn:gen_Leff}
\end{equation}
where the lower case trace, ``tr'', is over internal indices and
\begin{subequations}
\begin{align}
\tG_{\n\m} &= \sum_{n=0}^{\infty}\frac{n+1}{(n+2)!} \big(P_{\a_1}\dots P_{\a_n}G'_{\n\m}\big) \pd^n_{\a_1 \dots \a_n}, \\
\tU &= \sum_{n=0}^{\infty}\frac{1}{n!} \big(P_{\a_1}\dots P_{\a_n}U\big) \pd^n_{\a_1 \dots \a_n}, \\
P_{\m} &= i D_{\m}, \ \pd_{\m} \equiv \frac{\pd}{\pd q^{\m}}, \ G'_{\n\m} \equiv [D_{\n},D_{\m}].
\end{align}
\end{subequations}

\begin{description}
\item[Real scalars] The effective action originates from the Gaussian integral
\[
\exp\big(i \D S_{\text{eff,1-loop}}\big) = \int \mathcal{D} \Ph \exp \bigg[ i \int dx \, \frac{1}{2} \Ph^T \big(P^2 - m^2 - M^2(x) \big)\Ph \bigg].
\]
For this case, in Eqs.~\eqref{eqn:gen_Seff_Tr} and~\eqref{eqn:gen_Leff} we have
\begin{equation}
c_s =1/2, \quad U(x) = M^2(x) .
\label{eqn:realscal_cs_U}
\end{equation}

\item[Complex scalars] The effective action originates from the Gaussian integral
\[
\exp\big(i \D S_{\text{eff,1-loop}}\big) = \int \mathcal{D} \Ph\mathcal{D}\Ph^* \exp \bigg[ i \int dx \, \Ph^{\dag} \big(P^2 - m^2 - M^2(x) \big)\Ph \bigg].
\]
For this case, in Eqs.~\eqref{eqn:gen_Seff_Tr} and~\eqref{eqn:gen_Leff} we have
\begin{equation}
c_s =1, \quad U(x) = M^2(x)
\label{eqn:comscal_cs_U}
\end{equation}

\item[Massive fermions] We work with Dirac fermions. The effective action originates from the Gaussian integral
\[
\exp\big(i \D S_{\text{eff,1-loop}}\big) = \int \mathcal{D} \ps\mathcal{D}\bps \exp \bigg[ i \int dx \, \bps\big(\slashed{P} - m - M(x) \big) \ps\bigg],
\]
where \(\slashed{P} = \g^{\m}P_{\m}\) with \(\g^{\m}\) the usual gamma matrices. As shown in appendix~\ref{sec:app_cde_ferm}, in Eqs.~\eqref{eqn:gen_Seff_Tr} and~\eqref{eqn:gen_Leff} we have
\begin{equation}
c_s =-1/2, \quad U = U_{\text{ferm}} \equiv -\frac{i}{2}\s^{\m\n}G'_{\m\n} + 2mM + M^2 + \sP M,
\label{eqn:ferm_cs_U}
\end{equation}
where \(\s^{\m\n} = i [\g^{\m},\g^{\n}]/2\) and, by definition, \(\slashed{P}M = [\slashed{P},M]\). Note that the trace in~\eqref{eqn:gen_Leff} includes tracing over the spinor indices. The $2mM$ and $M^2$ terms in \(U_{\text{ferm}}\) and the \(-P^2\) term are proportional to the identity matrix in the spinor indices which, since we use the \(4\times 4\) gamma matrices, is the \(4\times 4\) identity matrix \(\mathbf{1}_4\).
\item[Massless gauge fields] We take pure Yang-Mills theory for non-abelian gauge group \(G\),
\[
\scL_{YM} = -\frac{1}{2g^2\m(G)} \tr\, F_{\m\n} F^{\m\n}, \quad F_{\m\n} = F_{\m\n}^a t^a_G ,
\]
where \(t^a_G\) are generators in the adjoint representation and \(\m(G)\) is the Dynkin index for the adjoint representation.\footnote{For representation \(R\), the Dynkin index is given by \(\tr \, T^a_RT^b_R = \m(R)\d^{ab}\). For \(SU(N)\), \(\m(G) = N\) while the fundamental representation has \(\m(\Yfund) = 1/2\). In the adjoint representation \((t^b_G)_{ac} = i f^{abc}\) where \(f^{abc}\) are the structure constants, \([T^a,T^b] = if^{abc}T^c\).} We are considering the 1PI effective action, \(\G[A]\), of the gauge field \(A_{\m}\).

We explain the essential details here and explicate them in full in appendix~\ref{sec:app_cde_ferm}. The 1PI effective action is evaluated using the background field method: the gauge field is expanded around a background piece and a fluctuating piece, \(A_{\m}(x) = A_{B,\m}(x) + Q_{\m}\), and we integrate out \(Q_{\m}\). The field \(Q_{\m}\) is gauge-fixed in such a way as to preserve the background field gauge invariance. The gauge-fixed functional integral we evaluate is,
\begin{align}
\exp\big(i \G_{\text{1-loop}}[A_B]\big) = &\int \mathcal{D} Q_{\m}^a \mathcal{D}c^a\mathcal{D}\xoverline{c}^a\nonumber \\
&\times \exp\bigg[ i \int dx\, -\frac{1}{2g^2} Q_{\r}^a \big(P^2 + i \mathcal{J}^{\m\n}G'_{\m\n}\big)^{\r,ab}_{\s} Q^{\s,b} + \xoverline{c}^a\big(P^2)^{ab} c^b \bigg], \nonumber
\end{align}
where \(c^a\) are Fadeev-Popov ghosts. In the above, \(G'_{\m\n} = [D_{\m},D_{\n}]\) where \(D_{\m} = \pd_{\m} - i A_{B,\m}\) is the covariant derivative with respect to the background field, \(\mathcal{J}^{\m\n}\) is the generator of Lorentz transformations on four-vectors,\footnote{Note the similarity with the fermion case, where \(\s^{\m\n}/2\) is the generator of Lorentz transformations on spinors. Explicitly, the components of \(\mathcal{J}^{\m\n}\) are given by \((\mathcal{J}^{\m\n})_{\r\s} = i (\d^{\m}_{\r}\d^{\n}_{\s} - \d^{\m}_{\s}\d^{\n}_{\r})\).} and we have taken Feynman gauge (\(\x = 1\)).

The effective Lagrangian is composed of two-pieces of the form in Eqs.~\eqref{eqn:gen_Seff_Tr} and~\eqref{eqn:gen_Leff} with \(m^2 =0\). The first is the ghost piece, for which \(c_s = -1\) since the ghost fields are anti-commuting and \(m^2 = U = 0\):
\begin{equation}
\text{Ghost piece: } \quad c_s = -1, \quad m^2 = U = 0.
\label{eqn:gh_cs_U}
\end{equation}
The second piece is from the gauge field \(Q_{\m}^a\) which gives Eqs.~\eqref{eqn:gen_Seff_Tr} and~\eqref{eqn:gen_Leff} with \(m^2=0\), \(c_s = 1/2\) since each component of \(Q_{\m}^a\) is a real boson, and \(U = -i \mathcal{J}\cdot G'\)
\begin{equation}
\text{Gauge piece: } \quad c_s = 1/2, \quad U = U_{\text{gauge}} \equiv - i \mathcal{J}^{\m\n}G'_{\m\n}, \quad m^2 =0.
\label{eqn:gauge_cs_U}
\end{equation}

With \(m^2 =0\), Eqs.~\eqref{eqn:gen_Seff_Tr} and~\eqref{eqn:gen_Leff} contain IR divergences. These IR divergences can be regulated by adding a mass term for \(Q_{\m}^a\) and \(c^a\) (essentially keeping \(m^2\) in Eqs.~\eqref{eqn:gen_Seff_Tr} and~\eqref{eqn:gen_Leff}).

\item[Massive vector bosons] We consider a UV model with gauge group $G$ that is spontaneously broken into $H$. A set of gauge bosons $Q_\mu^i$, $i=1,2,...,\dim(G)-\dim(H)$ that correspond to the broken generators obtain mass $m_Q$ by ``eating'' the Nambu-Goldstone bosons $\chi^i$. Here, we restrict ourselves to the degenerate mass spectrum of all $Q_\mu^i$ for simplicity. These heavy gauge bosons form a representation of the unbroken gauge group. As we show in appendix~\ref{sec:app_gkpiece}, the general gauge-kinetic piece of the Lagrangian up to quadratic term in $Q_\mu^i$ is
    \begin{eqnarray}
    {\cal L}_\text{g.k.} \supset \frac{1}{2} Q_\mu^i \left\{ - P^2 g^{\mu\nu} + P^\nu P^\mu - \left[ P^\mu, P^\nu \right] \right\}^{ij} Q_\nu^j , \label{eqn:Lgk}
    \end{eqnarray}
    where $P_\mu=iD_\mu$, with $D_\mu$ denotes the covariant derivative that contains only the unbroken gauge fields. One remarkable feature of this general gauge-kinetic term is that the coefficient of the ``magnetic dipole term'' $\frac{1}{2} Q_\mu^i \left\{-\left[P^\mu, P^\nu\right] \right\}^{ij} Q_\nu^j$ is universal, namely that its coefficient is fixed to $1$ relative to the ``curl'' terms $\frac{1}{2} Q_\mu^i \left\{ -P^2 g^{\mu\nu} + P^\nu P^\mu\right\}^{ij} Q_\nu^j$, regardless of the details of the symmetry breaking. In appendix~\ref{sec:app_gkpiece}, we will give both an algebraic derivation and a physical argument to prove Eq.~\eqref{eqn:Lgk}.

    The piece shown in Eq.~\eqref{eqn:Lgk} is to be combined with a gauge boson mass term due to the symmetry breaking, a generalized $R_\xi$ gauge fixing term which preserves the unbroken gauge symmetry, an appropriate ghost term, and a possible generic interaction term. More details about all these terms are in appendix~\ref{sec:app_cde_ferm}. The resultant one-loop effective action is given by computing
    \begin{eqnarray}
    \exp \left( i\Delta S_\text{eff,1-loop} \right) &=& \int {{\cal D} Q_\mu^i {\cal D} \chi^i {\cal D} c^i {\cal D}{\bar c}^i} \nonumber \\
    && \times \exp \Bigg\{ i\int dx \Bigg[ \frac{1}{2} Q_\mu^i \left( -P^2 g^{\mu\nu} + m_Q^2{g^{\mu \nu }} - 2[P^\mu, P^\nu] + M^{\mu\nu} \right)^{ij} Q_\nu^j \nonumber \\
    && + \frac{1}{2} \chi^i (P^2-m_Q^2)^{ij} \chi^j + {\bar c}^i (P^2-m_Q^2)^{ij} c^j \Bigg] \Bigg\} , \label{eqn:Sgauge}
    \end{eqnarray}
    where $c^i$, ${\bar c}^i$ denote the ghosts, $M^{\mu\nu}$ parameterizes the possible generic interaction term, and we have taken Feynman gauge $\xi=1$. Clearly, the effective Lagrangian is composed of three-pieces of the form in Eqs.~\eqref{eqn:gen_Seff_Tr} and~\eqref{eqn:gen_Leff}
    \begin{subequations}
    \begin{eqnarray}
    \text{Gauge piece: } \quad c_s &=& 1/2, \quad U = - i \mathcal{J}^{\m\n}\left(G'_{\m\n}+\frac{1}{2} M_{\mu\nu}\right), \quad m^2 = m_Q^2 . \label{eqn:U_massive_gauge}\\
    \text{Goldstone piece: } \quad c_s &=& 1/2, \quad U=0, \quad m^2 = m_Q^2 . \\
    \text{Ghost piece: } \quad c_s &=& -1, \quad U=0, \quad m^2 = m_Q^2 .
    \end{eqnarray}
    \end{subequations}
\end{description}

\subsection{Evaluating the CDE and universal results}\label{sec:CDE_universal}
In the present subsection we explicitly show how to evaluate terms in covariant derivative expansion of the one-loop effective action in Eqs.~\eqref{eqn:sum_Leff_begin} and~\eqref{eqn:sum_Leff_exp}. Following this, we provide the results of the expansion through a given order in covariant derivatives. Specifically, for an effective action of the form \(\Seff \propto \Tr \log (-P^2 +m^2 + U)\), we provide the results of the CDE through dimension-six operators assuming \(U\) is at least linear in background fields. These results make no explicit reference to a specific UV model and therefore they are, in a sense, universal. This universal result is tabulated in Eq.~\eqref{eqn:CDE_universal_lag} and can be immediately used to compute the effective action of a given UV model.

\subsubsection{Evaluating terms in CDE}\label{sec:CDE_evaluate}
Let us consider how to evaluate expansion terms from the effective Lagrangian of Eq.~\eqref{eqn:sum_Leff_begin}, which we reproduce here for convenience
\begin{equation*}
\D \scL_{\text{eff,1-loop}} = -i c_s \int dq \int dm^2 \, \tr \, \frac{1}{\D^{-1}\Big[1 - \D\Big(-\big\{q_{\m},\tG_{\n\m}\big\}\pd_{\n} - \tG_{\m\s}\tG_{\n\s}\pd_{\m}\pd_{\n} + \tU\Big) \Big]}.
\end{equation*}
In the above, \(\tG\) and \(\tU\) are as defined in Eq.~\eqref{eqn:sum_tilde_short}, \(dq \equiv d^4q/(2\pi)^4\), \(\D \equiv 1/(q^2 -m^2)\), and we employ the shorthand notation defined in~\eqref{eqn:notation_short}. We also used the fact that \(\{q_{\m},\tG_{\n\m}\pd_{\n}\} = \{q_{\m},\tG_{\n\m}\}\pd_{\n}\) which follows from \(\{A,BC\} = \{A,B\}C + B[C,A]\) and the antisymmetry of \(\tG_{\n\m}\), \(\tG_{\n\m} = - \tG_{\m\n}\). Using the matrix expansion
\begin{equation*}
\frac{1}{A^{-1}(1-AB)} = \sum_{n=0}^{\infty}(AB)^nA ,
\end{equation*}
we define the integrals
\begin{equation*}
\scI_n \equiv \tr\, \int dq \, dm^2\,\Big[ \D\Big( -\{q,\tG\}\pd -\tG^2\pd^2 + \tU \Big)\Big]^n\D.
\label{eqn:scI_n_def}
\end{equation*}
The effective action from a given \(\scI_n\) integral is given by \(\D \scL_{\scI_n} = -i c_s \scI_n\).

\(\tG_{\n\m}\) and \(\tU\) are infinite expansions in covariant derivatives of \(G_{\n\m}'\) and \(U\), and thus contain higher-dimension operators (HDOs). Therefore, each \(\scI_n\) is an infinite expansion containing these HDOs. For this work, motivated by present and future precision measurements, we are interested in corrections up to dimension-six operators. This dictates how many \(\scI_n\) we have to calculate as well as what order in \(\tG_{\n\m}\) and \(\tU\) we need to expand within a given \(\scI_n\).

As a typical example to demonstrate how to evaluate the \(\scI_n\), we consider \(\scI_1\),
\begin{equation}
\scI_1 = \tr\, \int dq \, dm^2\,\D\Big( -\{q,\tG\}\pd -\tG^2\pd^2 + \tU \Big)\D.
\label{eqn:scI_1}
\end{equation}
This term is fairly easy to compute and captures the basic steps to evaluate any of the \(\scI_n\) while also highlighting a few features that are unique to low order terms in the expansion. We remind the reader that \(q_{\m}\) and \(\pd_{\m}\) commute with \(P_{\m}\) and \(U\), which is what makes the \(\scI_n\) very simple to compute. We also assume that the mass-squared matrix \(m^2\) commutes with \(G_{\m\n}'\) and \(\tU\).\footnote{
This is always the case if \(m^2\) is proportional to identity, {\it i.e.} if every particle integrated out has the same mass. If we integrate out multiple particles with different masses, typically \(m^2\) commutes with \(G'_{\m\n}\) but, in general, will not commute with \(U\). For \(m^2\) to commute with \(G'_{\m\n}\), in the operator \(P^2 - m^2 - U(x)\), it amounts to assuming \(P_{\m}\) and \(m^2\) are block diagonal of the form \(P_{\m} = \text{diag}(P_{\m}^{(1)}, \dots, P_{\m}^{(n)})\) and \(m^2 = \text{diag}(m_1^2,\dots,m_n^2)\). Physically, this means we are integrating out \(n\) particles, where the \(i\)th particle has mass-squared \(m_i^2\) and a covariant derivative \(P_{\m}^{(i)}\) associated to its gauge interactions. The block-diagonal mass matrix means we diagonalized the mass matrix before integrating out the particles. If \(U\) happens to have the same block-diagonal structure, then of course \(m^2\) commutes with \(U\) as well.
}
In this case, \( \D \) commutes with the HDOs in \(\tG\) and \(\tU\), {\it i.e.} \([\D,P_{\a_1} \dots P_{\a_n}G'_{\m\n}] =0\) and similarly for the HDOs in \(\tU\). This allows us to separate the \(q\)-integral from the trace over the HDOs.

Let us now evaluate \(\scI_1\) in~\eqref{eqn:scI_1}. We consider the \(\tU\) term first,
\[
\scI_1 \supset \tr \, \int dq \, dm^2 \, \D \, \tU \, \D = \sum_{n=0}^{\infty}\frac{1}{n!}\tr \big(P_{\a_1} \dots P_{\a_n} U\big) \times \int dq \, \D \, \pd^n_{\a_1 \dots \a_n} \, \D.
\]
Recall that the covariant derivative action on a matrix is defined as the commutator, {\it e.g.} \(P_{\a}U = [P_{\a},U]\). Since the trace of a commutator vanishes, all the \(n \ge 1\) terms become total derivatives after the evaluation of the trace, and therefore do not contribute to the effective action. Thus,
\[
\tr \, \int dq \, dm^2 \, \D \, \tU \, \D  = \tr \, U \times \int dq \, dm^2 \, \D^2.
\]
The above term is divergent. It may be the case---as in the above integral---that the order of integration does not commute and changes the divergent structure of the integral. In these cases, to properly capture the divergent structure (and therefore define counter-terms) the integral on \(m^2\) should be performed first since we are truly evaluating \(\int dq \int dm^2 \frac{\pd}{\pd m^2} \tr \, \log(\dots)\).\footnote{Simple power counting easily shows that divergences in \(\scI_n\) can only occur for \(n= 0,1,\) and \(2\). In the expansions of \(\tG\) and \(\tU\) within \(\scI_{0,1,2}\), it is not difficult to see that there are only four non-vanishing divergent terms: \(\scI_0\), in \(\scI_1\) they are the \(\tr \, U\) and \(\tr \, G'_{\m\n}G'_{\r\s}\) terms, and in \(\scI_2\) it is the \(\tr\, U^2\) term.} In this paper, we use dimensional regularization with \(\xoverline{\text{MS}}\) for our renormalization scheme, in which case
\[
\tr\, U \int dq \, dm^2 \, \D^2 = \tr \, U \int dq \, \D = -\frac{i}{(4\pi)^2} m^2 \Big(\log \frac{m^2}{\m^2} - 1\Big) \tr \, U,
\]
where \(\m\) is the renormalization scale.

We now turn our attention to the pieces in \(\scI_1\) involving \(\tG_{\m\n}\). The term linear in \(\tG\) in \(\scI_1\) vanishes since it is the trace of a commutator, as was the case for the higher derivative terms in \(\tU\) discussed above. Thus, only the \(\tG^2\) term in non-zero and we seek to evaluate
\[
\scI_1 \supset - \tr \, \int dq \, dm^2 \, \D \, \tG_{\m\s} \tG_{\n\s}\pd^2_{\m\n} \, \D .
\]
We evaluate the above up to dimension-six operators. Since $G_{\mu\nu}'=-\left[P_\mu,P_\nu\right]$ is ${\cal O}(P^2)$, we need the expansion of \(\tG\tG\) to \(\scO(P^6)\):
\begin{align*}
\tG_{\m\s}\tG_{\n\s} \pd^2_{\m\n} &= \frac{1}{4} G'_{\m\s}G'_{\n\s}\pd^2_{\m\n} + \frac{1}{9}(P_{\a}G'_{\m\s})(P_{\b}G'_{\n\s})\pd^4_{\a\b\m\n} \\
& \quad + \frac{1}{16}\Big[ G'_{\m\s}(P_{\b_1}P_{\b_2}G'_{\n\s})\pd^4_{\b_1\b_2\m\n} + (P_{\a_1}P_{\a_2}G'_{\m\s})G'_{\n\s}\pd^4_{\a_1\a_2\m\n} \Big],
\end{align*}
where we dropped the \(\scO(P^5)\) terms since they vanish as required by Lorentz invariance. It is straightforward to plug the above back into \(\scI_1\) and compute the \(q\)-derivatives and integrals. For example, the \(G'^2\pd^2\) requires computing
\begin{align*}
\int dq\, dm^2 \, \D \, \pd^2_{\m\n} \, \D &= \int dq \, dm^2\, \D \big(-2g_{\m\n}\D^2 + 8 q_{\m}q_{\n}\D^3\big) \\
&=2g_{\m\n}\int dq\, dm^2 \, \big(-\D^3 + q^2\D^4\big)  \\
&=2g_{\m\n} \int dq \, \Big(-\frac{1}{2}\D^2 + \frac{1}{3}q^2\D^4\Big) \\
&= 2g_{\m\n} \cdot \frac{i}{(4\pi)^2}\cdot \frac{1}{6}\cdot \Big(\log \frac{m^2}{\m^2} - 1\Big),
\end{align*}
where we computed the \(m^2\) integral first and used dimensional regularization with \(\xoverline{\text{MS}}\). Thus, we see that
\[
\scI_1 \supset -\frac{1}{4} \tr \big(G'_{\m\s}G'_{\n\s}\big) \int dq\, dm^2 \, \D \, \pd^2_{\m\n} \, \D  =  -\frac{i}{(4\pi)^2}\cdot \Big(\log \frac{m^2}{\m^2} - 1\Big)\cdot \frac{1}{12} \cdot \tr \big(G'_{\m\n}G'_{\m\n}\big),
\]
which we clearly recognize as a contribution to the \(\b\) function of the gauge coupling constant.

The other \(\scO(P^6)\) terms in the expansion of \(\tG^2\) are computed similarly. In appendix~\ref{sec:app_cde_identity} we tabulate several useful identities that frequently occur, such as \(\pd^n_{\a_1\dots \a_n} \D\) and what this becomes under the $q$-integral. For example, in the above computation we used
\[
\pd^2_{\m\n} \D = -2g_{\m\n}\D^2 + 8 q_{\m}q_{\n}\D^3 \Rightarrow \text{under }q\text{-integral: } \pd^2_{\m\n} \D= 2g_{\m\n}\big(-\D^2 +q^2\D^3\big).
\]
The end result of computing the \(q\)-integrals for the \(\scO(P^6)\) terms in \(\scI_1\) gives
\begin{align*}
-\tr \int dq\, dm^2 \, &\D \, \tG_{\m\s}\tG_{\n\s}\pd^2_{\m\n} \, \D \supset -\frac{i}{(4\pi)^2} \frac{1}{30} \, \frac{1}{m^2} \, \tr \, \bigg\{\\
& \frac{4}{9}\Big[ \big(P_{\m}G_{\m\n}'\big)^2 + \big(P_{\m}G_{\n\s}'\big) \big(P_{\m}G'_{\n\s}\big) + \big(P_{\m}G_{\n\s}'\big) \big(P_{\n}G'_{\m\s}\big)\Big] \\
& +\frac{1}{2} \Big[G_{\m\n}'\big(P^2 G_{\m\n} + P_{\m}P_{\s}G_{\s\n}' + P_{\s}P_{\m}G'_{\s\n}\big) \Big] \bigg\}.
\end{align*}
There are only two possible dimension-six operators involving just \(P_{\m}\) and \(G_{\m\n}'\), namely \(\tr \, (P_{\m}G'_{\m\n})^2\) and \(\tr \, (G_{\m\n}'G_{\n\s}'G_{\s\m}')\). Using the Bianchi identity and integration by parts, \(\tr\, [A (P_{\m}B)] = -\tr \, [(P_{\m}A)B] + \text{total deriv.}\), the above can be arranged into just these two dimension-six operators:
\begin{equation*}
-\frac{i}{(4\pi)^2} \frac{1}{m^2} \bigg[ \frac{1}{135} \tr \, \big(P_{\m}G'_{\m\n}\big)^2 + \frac{1}{90} \tr \, \big(G_{\m\n}'G_{\n\s}'G_{\s\m}'\big) \bigg].
\end{equation*}
Combining all these terms together, we find the contribution to the effective Lagrangian from \(\scI_1\) is
\begin{align*}
\D \scL_{\scI_1} = -i c_s \scI_1 = -\frac{c_s}{(4\pi)^2} \Bigg[& \Big(\log \frac{m^2}{\m^2} - 1\Big)\, \frac{1}{12} \, \tr\, \big(G'_{\m\n}G'_{\m\n}\big) + \frac{1}{m^2}\frac{1}{135} \tr \, \big(P_{\m}G'_{\m\n}\big)^2 \\
&+ \frac{1}{m^2} \, \frac{1}{90} \tr \, \big(G_{\m\n}'G_{\n\s}'G_{\s\m}'\big) \Bigg] + \text{dim-8 operators}.
\end{align*}
For the reader following closely, we note that the only contribution to \(\tr \, (G_{\m\n}')^2\) is the above term from \(\scI_1\), while \(\tr\, (P_{\m}G_{\m\n}')^2\) and \(\tr \, G'^3\) also receive contributions from \(\scI_2\).

In a similar fashion, one can compute the other \(\scI_n\). In the next subsection we tabulate the result of all possible contributions to dimension-six operators from the \(\scI_n\); in appendix~\ref{sec:app_CDE_universal} the results for each individual \(\scI_n\) are listed.
\vspace{4.5cm}

\subsubsection{Universal results}\label{sec:CDE_universal_results}

We just showed how to evaluate terms in the CDE to a given order. Here we tabulate the results that allow one to compute the one-loop effective action through dimension-six operators. In the next subsection we use these results to obtain the dimension-six Wilson coefficients of the SM EFT for several non-trivial BSM models.

The one-loop effective action is given by
\[
\D S_{\text{eff,1-loop}} = i c_s \Tr \log\Big(-P^2 + m^2 + U(x)\Big) ,
\]
where, as discussed these in section~\ref{sec:CDE_fermions}, \(c_s\) and \(U(x)\) depend on the species we integrate out. We assume that the mass-squared matrix \(m^2\) commutes with \(U\) and \(G'_{\m\n}\). Under this assumption, we tabulate results of the CDE through dimension-six operators. In general, \(U\) may have terms which are linear in the background fields.\footnote{For example, a Yukawa interaction \(y \ph \bps \ps\) for massive fermions leads to a term linear in the light field \(\ph\): from Eq.~\eqref{eqn:ferm_cs_U}, \(U_{\text{ferm}}\supset 2m  M(x) = y m \ph\).} In this case, although the scaling dimension of \(U\) is two, its operator dimension may be one. Simple power counting tells us that we will have to evaluate terms in the \(\scI_n\) integrals of Eq.~\eqref{eqn:scI_n_def} through \(\scI_6\).\footnote{While this is tedious, it isn't too hard. Moreover, there are many terms within each \(\scI_n\) that we don't need to compute since they lead to too large of an operator dimension. For example, the only term in \(\scI_6\) that we need to compute is
\[
\scI_6 = \tr\, \int dq \, dm^2 \, \Big[\D \big(-\{q,\tG\}\pd - \tG^2\pd^2 + \tU\big)\Big]^6 \D \supset \tr \, U^6\,  \int dq \, dm^2 \D^7 = \tr \, U^6 \cdot \frac{i}{(4\pi)^2}\cdot \frac{1}{120} \cdot \frac{1}{m^8}.
\]
All other terms in \(\scI_6\) have too large of operator dimension and can be dropped.} In appendix~\ref{sec:app_CDE_universal}, we give the result of this calculation for each of the relevant terms in \(\scI_1\)-\(\scI_6\). Gathering all of the terms together, the one-loop effective action is:

\begin{align}
\D &\scL_{\text{eff,1-loop}} =\frac{c_s}{(4\pi)^2} \, \tr \, \Bigg\{ \nonumber \\
&\qquad +m^4\bigg[-\frac{1}{2} \Big(\log \frac{m^2}{\m^2} -\frac{3}{2}\Big) \bigg]  \nonumber\\
&\qquad +m^2\bigg[-\Big(\log \frac{m^2}{\m^2} - 1\Big) \, U\bigg]  \nonumber\\
&\qquad +m^0\Bigg[-\frac{1}{12}\Big(\log \frac{m^2}{\m^2} - 1\Big) \, G_{\m\n}'^2 - \frac{1}{2} \log \frac{m^2}{\m^2} \, U^2 \Bigg]\nonumber \\
&\qquad + \frac{1}{m^2}\Bigg[ -\frac{1}{60} \, \big(P_{\m}G_{\m\n}'\big)^2 - \frac{1}{90} \, G_{\m\n}'G_{\n\s}'G_{\s\m}' -\frac{1}{12} \, (P_{\m}U)^2 - \frac{1}{6}\, U^3 - \frac{1}{12}\, U G_{\m\n}'G_{\m\n}' \Bigg] \nonumber \\
&\qquad +\frac{1}{m^4} \Bigg[\frac{1}{24} \, U^4 + \frac{1}{12}\, U \big(P_{\m}U\big)^2 + \frac{1}{120}\, \big(P^2U\big)^2 +\frac{1}{24} \, \Big( U^2 G'_{\m\n}G'_{\m\n} \Big) \nonumber \\
&\qquad \qquad \qquad - \frac{1}{120} \, \big[(P_{\m}U),(P_{\n}U)\big] G'_{\m\n} - \frac{1}{120}\, \big[U[U,G'_{\m\n}]\big] G'_{\m\n} \Bigg] \nonumber \\
&\qquad +\frac{1}{m^6} \Bigg[-\frac{1}{60} \, U^5 - \frac{1}{20} \, U^2\big(P_{\m}U\big)^2 - \frac{1}{30} \, \big(UP_{\m}U\big)^2 \Bigg] \nonumber \\
&\qquad + \frac{1}{m^8} \bigg[ \frac{1}{120} \, U^6 \bigg]
 \Bigg\} .
\label{eqn:CDE_universal_lag}
\end{align}

Equation~\eqref{eqn:CDE_universal_lag} is one of the central results that we present, so let us make a few comments about it:
\begin{itemize}
\item This formula is the expansion of a functional trace of the form \(i c_s \Tr \log \big[-P^2 + m^2 + U(x)\big]\) where \(P_{\m} = i D_{\m}\) is a covariant derivative and \(U(x)\) is an arbitrary function of spacetime. We have worked in Minkowski space and defined the one-loop action and Lagrangian from \(i c_s \Tr \log\big[-P^2 + m^2 + U\big] = \D S_{\text{eff,1-loop}} = \int d^4x \D \scL_{\text{eff,1-loop}}\).
\item The results of Eq.~\eqref{eqn:CDE_universal_lag} are valid when the mass-squared matrix \(m^2\) commutes with \(U(x)\) and \(G'_{\m\n} = [D_{\m},D_{\n}]\).
\item The lower case ``tr'' in~\eqref{eqn:CDE_universal_lag} is over internal indices. These indices may include gauge indices, Lorentz indices (spinor, vector, {\it etc.}), flavor indices, {\it etc.}.
\item \(c_s\) is a constant which relates the functional trace to the effective action, \'a la the first bullet point above. For example, for real scalars, complex scalars, Dirac fermions, gauge bosons, and Fadeev-Popov ghosts \(c_s = 1/2, 1, -1/2, 1/2,\) and \(-1\), respectively. \(U(x)\) is a function of the background fields. In section~\ref{sec:CDE_fermions} we discussed the form of \(U(x)\) for various particle species, namely scalars, fermions, and gauge bosons.
\item Given the above statements, it is clear that~\eqref{eqn:CDE_universal_lag} is universal in the sense that it applies to any effective action of the form \(\Tr \log \big(-P^2 + m^2 +U\big)\).\footnote{Under the assumption \(m^2\) commutes with \(U\) and \(G'_{\m\n}\); see the second bullet point.} For any specific theory, one only needs to determine the form of the covariant derivative \(P_{\m}\) and the matrix \(U(x)\) and then~\eqref{eqn:CDE_universal_lag} may be used. We provide several examples in the next subsection.
\item Equation~\eqref{eqn:CDE_universal_lag} is an expansion of the effective Lagrangian through dimension-six operators. \(U\) has scaling dimension two, but its operator dimension may be one or greater. In the case \(U\) contains a term with unit operator dimension, one needs all the terms in~\eqref{eqn:CDE_universal_lag} to capture all dimension-six operators.
\item The lines proportional to \(m^4\), \(m^2\), and \(m^0\) in~\eqref{eqn:CDE_universal_lag} come from UV divergences in the evaluation of the trace;  \(\m\) is a renormalization scale and we used dimensional regularization and \(\xoverline{\text{MS}}\) scheme.
\item The lines proportional to \(m^2\) and \(m^0\) can always be absorbed by renormalization. They can also be used to find the contribution of the particles we integrate out to the \(\b\)-functions of operators.
\end{itemize}

\begin{center} \textbf{Evaluation of the pure glue pieces} \end{center}

The operators involving only gauge bosons, \(G'^2\) at dimension four and \((PG')^2\) and \(G'^3\) at dimension six, are determined solely by stating the field content and their representations under the gauge groups. As such, we can evaluate these terms more generally. For the dimension four term \(G'^2\) we will immediately produce the \(\b\) function of Yang-Mills coupling constant.

We take a simple gauge group and evaluate the contribution of different particle species to these pure glue operators. For a semi-simple group, the following results apply to each individual gauge group. The covariant derivative is given by \(D_{\m} = \pd_{\m} - i g A_{\m}\) so that \(G'_{\m\n} = [D_{\m},D_{\n}] = -igG_{\m\n}\) where \(G_{\m\n}\) is the Yang-Mills field strength.

All particle species contribute to renormalization of the Yang-Mills kinetic term, \(-(G_{\m\n}^a)^2/4\), through the \(\tr \, G_{\m\n}'^2\) term in~\eqref{eqn:CDE_universal_lag}. In addition, the magnetic moment coupling for fermions and gauge bosons is contained within \(U\), \(U\supset -i S^{\m\n}G'_{\m\n}\) where \(S^{\m\n}\) is the Lorentz generator in a given representation---see Eqs.~\eqref{eqn:ferm_cs_U} and~\eqref{eqn:gauge_cs_U}. This term then contributes to the Yang-Mills kinetic term through \(\tr \, U^2\). Evaluating these terms for a particle with spin \(j\) particle and representation \(R\) under the gauge group we have
\[
-c_s\frac{1}{12}\tr \, G_{\m\n}'^2 = \frac{g^2}{3}\cdot c_s \cdot d(j) \cdot \m(R) \times \Big(\frac{1}{4} G_{\m\n}^aG^{a\, \m\n}\Big),
\]
where \(d(j)\) is the number of components of the spin \(j\) particle\footnote{\(d = 1,\ 4, \text{ and }4\) for scalars, Dirac fermions, and vectors, respectively.} and \(\m(R)\) is the Dynkin index of the \(R^{\text{th}}\) representation, \(\tr \, T_R^aT_R^b = \m(R) \d^{ab}\). For the \(\tr \, U^2\) term we have
\[
-c_s\frac{1}{2} \tr \, U^2 \supset -c_s\frac{g^2}{2}\tr\, \Big( S^{\m\n}G_{\m\n}S^{\r\s}G_{\r\s}\Big) = -4g^2\cdot c_s \cdot k(j) \cdot \m(R) \times \Big(\frac{1}{4} G_{\m\n}^aG^{a\, \m\n}\Big),
\]
where \(k = 1\) (\(k=2\)) for Dirac spinors (vectors).\footnote{In the spinor representation and vector representations \(S^{\m\n} = \s^{\m\n}/2\) and \(S^{\m\n} = \mathcal{J}^{\m\n}\), respectively. With this, \(\tr \, S^{\m\n}S^{\r\s} = k(j)(g^{\m\r}g^{\n\s} - g^{\m\s}g^{\n\r})\) with \(k(j =1/2) =1\) for spinors, \(k(j=1) = 2\) for vectors, and, obviously, \(k(j=0) = 0\) for scalars.} Combining these terms together, we see that a given species that we integrate out produces
\begin{equation}
\D S_{\text{eff,1-loop}} \supset \frac{g^2}{(4\pi)^2}\Big[c_s\m(R)\Big(\frac{1}{3}d(j) - 4 k(j)\Big)\Big] \log \frac{\m^2}{m^2}\times \Big(-\frac{1}{4} G_{\m\n}^aG^{a\, \m\n}\Big).
\end{equation}
We recognize the term in square brackets as the contribution to the one-loop \(\b\) function coefficient.\footnote{For massless particles, the \(m^2\) inside the logarithm should be interpreted as an IR regulator. Note that interpreting this result as the contribution to the running of the coupling constant means we are regarding this as the 1PI effective action or an EFT where the particle of mass \(m\) remains in the spectrum, its mass small compared to the cutoff of the EFT. In the case where we are integrating out a heavy particle of mass \(m\), as is well known, we are still picking up the massive particle's contribution to the \(\b\) function since dimensional regularization is a mass-independent renormalization scheme. Of course, since we have integrated out the massive species we should not include its contribution to the running of the coupling constant in the low-energy EFT.} In particular, for scalars, fermions, and vector bosons (including the ghost contribution, Eq.~\eqref{eqn:gh_cs_U}), we have
\begin{equation*}
c_s \m(R) \Big(\frac{1}{3}d(j) - 4 k(j)\Big) = \m(R)\left\{ \begin{array}{cr} \frac{1}{3} & \text{complex scalars} \\ -\frac{2}{3} + 2 =\frac{4}{3} & \text{Dirac fermions} \\ \frac{1}{2}\big(\frac{4}{3} - 8\big) - \frac{1}{3} = -\frac{11}{3} & \text{vector bosons} \end{array} \right. .
\end{equation*}

In a similar fashion, we can compute the dimension-six pure glue operators. In Eq.~\eqref{eqn:CDE_universal_lag}, these come from \(\tr (P_{\m}G_{\m\n}')^2\) and \(\tr \, (G_{\m\n}'G_{\n\s}'G_{\s\m}')\) as well as \(\tr \, U^3\) and \(\tr (P_{\m}U)^2\) when \(U\) contains the magnetic moment coupling. These traces are straightforward to compute. Defining the dimension-six operators
\begin{equation}
\scO_{2G} \equiv -\frac{1}{2} \big(D_{\m}G_{\m\n}^a\big)^2 \quad , \quad \scO_{3G} \equiv \frac{g}{3!}f^{abc}G_{\m\n}^aG_{\n\s}^bG_{\s\m}^c,
\label{eqn:dim6_pureglueOps}
\end{equation}
we find
\begin{eqnarray}
-\frac{c_s}{60}\tr \, (P_{\m}G_{\m\n}')^2 & = & \frac{g^2}{30} \cdot c_s \cdot d(j) \cdot \m(R) \times \scO_{2G}, \nonumber\\
-\frac{c_s}{90} \tr\, (G_{\m\n}'G_{\n\s}'G_{\s\m}') & = & \frac{g^2}{30} \cdot c_s \cdot d(j) \cdot \m(R) \times \scO_{3G},  \nonumber\\
-\frac{c_s}{6} \tr \, U^3 & = & 2 g^2 \cdot c_s \cdot k(j) \cdot \m(R) \times \scO_{3G}, \nonumber \\
-\frac{c_s}{12} \tr \, \big(P_{\m}U\big)^2 & = & 2 g^2 \cdot c_s \cdot k(j) \cdot \m(R) \times \big( -\scO_{3G} - \frac{1}{3} \scO_{2G} \big). \nonumber
\end{eqnarray}
Adding these terms up we have
\begin{equation}
\Delta\scL_{\text{eff,1-loop}} \supset \frac{1}{(4\pi)^2} \, \frac{1}{m^2} \, \frac{g^2}{30} \, c_s \, \m(R) \bigg[d(j) \times \scO_{3G} + \Big(d(j) - 20 k(j)\Big) \times \scO_{2G} \bigg].
\label{eqn:dim6_pureglue}
\end{equation}
In Table~\ref{tbl:dim6_pureglue} we tabulate these coefficients for different species, where in the massive gauge boson case, proper contributions from Goldstone and ghosts are already included.

\begin{table}
\centering
\[
\scL_{\text{eff,1-loop}} \supset \frac{1}{(4\pi)^2}\, \frac{1}{m^2}\, \frac{g^2}{60} \, \m(R) \Big( a_{2\, s} \scO_{2G} + a_{3\, s} \scO_{3G} \Big) \quad \renewcommand\arraystretch{1.}
\begin{array}{crrrrr}
& a_{2\, s}&  & a_{3\, s}&  \\
\hline
 & 2 & &2 & &\text{complex scalar}\\
 &  16 & & -4 & &\text{Dirac fermion}\\
 & -37 & & 3 & &\text{massive vector}\\
\end{array}
\]
\vspace{-15pt}
\caption{\label{tbl:dim6_pureglue}  Contribution of different massive species to the purely gluonic dimension-six operators, computed from~\eqref{eqn:dim6_pureglue}. The operators \(\scO_{2G}\) and \(\scO_{3G}\) are defined in Eq.~\eqref{eqn:dim6_pureglueOps}. The particle has mass \(m\) and transforms in the \(R^{\text{th}}\) representation of the group, with \(\m(R)\) its index. Real scalars are half the value of complex scalars. For \(U(1)\) gauge groups, \(\m(R)\) is replaced by \(Q^2\) and \(a_{2s}\) by the number of degrees of freedom transforming under the \(U(1)\), where \(Q\) the charge of the massive particle under the \(U(1)\). Note that, by anti-symmetry of the Lorentz indices, \(\scO_{3G}\) vanishes for abelian groups.}
\vspace{-5pt}
\end{table}

\subsection{Example calculations}\label{sec:CDE_examples}

In this subsection, we give several example models where we calculate the effective action using the covariant derivative expansion. As we will explicitly see, computing the Wilson coefficients for a given model proceeds in an essentially algorithmic fashion. If there is a tree-level contribution to the effective action, we use Eq.~\eqref{eqn:tree_cde_terms}. For the one-loop contribution, we use Eq.~\eqref{eqn:CDE_universal_lag}. Given a model, the brunt of the work is to identify the appropriate \(U\) to plug into Eqs.~\eqref{eqn:tree_cde_terms} and~\eqref{eqn:CDE_universal_lag} and then to evaluate the traces in these equations. In the following matching calculations, it should be understood that all the Wilson coefficients obtained are at the matching scale $\Lambda$, namely that all our results are actually about $c_i(\Lambda)$. That said, throughout this subsection we drop the specification of RG scale.

A note on terminology. We frequently, and somewhat inappropriately, refer to the use of Eqs.~\eqref{eqn:tree_cde_terms} and~\eqref{eqn:CDE_universal_lag} as ``using the CDE''. If we are just using the results in these equations, then such a statement is technically incorrect. The expansion of the effective action in these two equations can be obtained from \textit{any} consistent method to compute the effective action. The CDE is a \textit{particular} method which considerably eases obtaining these results, but, nevertheless, is still just a means to the end. With this clarification, we hope the reader can forgive our sloppy language in this section.

In demonstrating how to use the CDE to compute the effective action, we would also like to pick models that are of phenomenological interest. As such, we focus on models that couple to the bosonic sector of the SM, with particular attention towards those models which generate tree-level Wilson coefficients. UV models that generate tree-level Wilson coefficients of the bosonic operators in Table~\ref{tbl:operators} may substantially contribute to precision observables. As a result, these models are typically either already tightly constrained or will be probed in future. Note that RG running may be of practical relevance when the Wilson coefficient is generated at tree-level (see the discussion in section~\ref{sec:operatorbasis}). 

With the above motivations, we would like to make a list of possible UV models that have tree-level contributions to the effective action. Let us limit this list to heavy scalars which can couple at tree-level to the Higgs sector via renormalizable interactions. There are only four such theories:

\begin{enumerate}
  \item A real singlet scalar $\Phi$
  \begin{equation}
  \Delta \mathcal{L} \supset \Phi \left|H\right|^2 .
  \end{equation}

  \item A real (complex) $SU(2)_L$ triplet scalar $\Phi_0=\Phi_0^a\tau^a$ ($\Phi_1=\Phi_1^a\tau^a$) with hypercharge $Y_\Phi=0$ ($Y_\Phi=1$)
  \begin{eqnarray}
  \Delta \mathcal{L} &\supset& H^\dagger \Phi_0 H , \\
  \Delta \mathcal{L} &\supset& H^\dagger \Phi_1 {\tilde H} +c.c. ,
  \end{eqnarray}
  where ${\tilde H}=i\sigma^2 H^*$.

  \item A complex $SU(2)_L$ doublet scalar $\Phi$ with $U(1)_Y$ hypercharge $Y_\Phi=\frac{1}{2}$
  \begin{equation}
  \Delta \mathcal{L} \supset \left|H\right|^2 (\Phi^\dagger H + c.c.) .
  \end{equation}

  \item A complex $SU(2)_L$ quartet scalar $\Phi_{3/2}$ ($\Phi_{1/2}$) with hypercharge $Y_\Phi=\frac{3}{2}$ ($Y_\Phi=\frac{1}{2}$)
  \begin{equation}
  \Delta \mathcal{L} \supset \Phi^\dag H^3 + c.c. ,
  \end{equation}

\end{enumerate}

We now show that the above list exhausts the possibilities of heavy scalars that couple via renormalizable interactions to the Higgs and produce tree-level Wilson coefficients. In order to have tree-level generated Wilson coefficients, the UV Lagrangian must contain a term that is linear in the heavy field. Therefore, we need to count all possible Lagrangian terms formed by $\Phi$ and $H$ that are linear in $\Phi$. After appropriate diagonalization of $\Phi$ and $H$, we do not need to consider the quadratic terms. Then there are only two types of renormalizable interactions $H^aH^b\Phi^{ab}$ and $H^aH^bH^c\Phi^{abc}$, where we have written the SM Higgs field $H$ in terms of its four real components $H^a$ with $a=1,2,3,4$. Because only symmetric combinations are non-vanishing, it is clear that there are in total $10$ real components $\Phi^{ab}$ that are enumerated by No.1 and No.2 in the above list, and $20$ real components $\Phi^{abc}$ that are enumerated by No.3 and No.4.

In the rest of this subsection, we will discuss in detail the examples above and compute their effective actions through one-loop order. Additionally, we will compute the one-loop effective action of three other examples: (1) degenerate scalar tops in the MSSM, (2) a heavy \(U(1)\) gauge boson that kinetically mixes with hypercharge, and (3) massive vector bosons that transform in the triplet of (unbroken) \(SU(2)_L\) and couple universally to fermions. The latter model can arise in extra-dimension and little Higgs theories.

When there is a non-zero tree-level contribution, \(\Ph_c \ne 0\), the dependence of the one-loop functional determinant on the classical configuration can introduce divergences into the Wilson coefficients of operators with dimension greater than four. These terms generically are associated with renormalization of parameters in the UV Lagrangian (see the discussion at the beginning of Sec.~\ref{sec:CDE_summary}, around Fig.~\ref{fig:CDEdiagrams}). Therefore, the effects of the contributions can be absorbed into a redefinition (renormalization scheme dependence) of the UV Lagrangian parameters, and hence dropped from the matching analysis. Another natural scheme choice is to use \(\xoverline{\text{MS}}\). In \(\xoverline{\text{MS}}\) scheme, from Eq.~\eqref{eqn:CDE_universal_lag}, there is a finite contribution to higher dimension operators from the \(\tr\, U\) piece. To show where this difference arises in doing calculations, in our examples of the triplet scalar and doublet scalar we will use the \(\xoverline{\text{MS}}\) renormalization scheme, while for all the other examples we will absorb the divergences of HDOs into the UV Lagrangian parameters. For the latter case, this essentially amounts to dropping \(\Ph_c\) from the one-loop calculation. 

\renewcommand\arraystretch{1.4}
\begin{table}[tb]
\centering
\begin{tabular}{|rcl|rcl|}\hline
 \(\scO_{GG}\) &\(=\)& \(g_s^2 \abs{H}^2G_{\mu \nu }^aG^{a,\mu \nu }\) & \(\scO_H\)   &\(=\)& \(\frac{1}{2}\big(\pd_{\mu} \abs{H}^2\big)^2\)\\
 \(\scO_{WW}\) &\(=\)& \(g^2  \abs{H}^2 W_{\mu \nu }^aW^{a,\mu \nu } \) &  \(\scO_T\)   &\(=\)& \(\frac{1}{2}\big( H^{\dag} \Dfbd H\big)^2\) \\
 \(\scO_{BB}\) &\(=\)& \(g'^2 \abs{H}^2 B_{\mu \nu }B^{\mu \nu }\) & \(\scO_R\)   &\(=\)& \(\abs{H}^2\abs{D_{\m}H}^2\) \\
 \(\scO_{WB}\) &\(=\)& \(2gg'H^\dag {\tau^a}H W_{\mu \nu }^a B^{\mu \nu }\) &  \(\scO_D\)   &\(=\)& \(\abs{D^2H}^2\) \\
 \(\scO_W\)   &\(=\)& \(ig\big(H^\dag \tau^a \Dfb H\big)D^\nu W_{\mu \nu }^a\) &  \(\scO_6\)   &\(=\)& \(\abs{H}^6\) \\
 \(\scO_B\)   &\(=\)& \(ig'Y_H\big(H^\dag \Dfb H\big)\pd^\nu B_{\mu \nu }\)  &  \(\scO_{2G}\) &\(=\)& \(-\frac{1}{2} \big(D^\mu G_{\mu \nu }^a\big)^2\) \\
 \(\scO_{3G}\) &\(=\)& \(\frac{1}{3!}g_sf^{abc}G_\rho ^{a\mu }G_\mu ^{b\nu }G_\nu ^{c\rho }\) &  \(\scO_{2W}\) &\(=\)& \(-\frac{1}{2} \big(D^\mu W_{\mu \nu }^a\big)^2\) \\
 \(\scO_{3W}\) &\(=\)& \(\frac{1}{3!}g \e^{abc}W_\rho ^{a\mu }W_\mu ^{b\nu }W_\nu ^{c\rho }\) & \(\scO_{2B}\) &\(=\)& \(-\frac{1}{2} \big(\pd^{\mu} B_{\mu \nu }\big)^2\) \\
  \hline
\end{tabular}
\caption{CP conserving dimension-six bosonic operators.} \label{tbl:operators}
\vspace{-10pt}
\end{table}
\renewcommand\arraystretch{0}

\subsubsection{Electroweak triplet scalar}

Let us consider an electroweak triplet scalar \(\Ph\) with neutral hypercharge. The Lagrangian contains the trilinear interaction \(H^{\dag}\Ph H\), where \(H\) is the electroweak Higgs doublet. This interaction, being linear in \(\Ph\), leads to a tree-level contribution to the effective action when we integrate out \(\Ph\).

While our main purpose here is to demonstrate how to use the CDE, we note that EW triplet scalars are phenomenologically interesting~\cite{Georgi:1985nv,*Chanowitz:1985ug} and well studied (for a recent study of triplet collider phenomenology and constraints see, \textit{e.g.},~\cite{Englert:2013wga}). As shown below, the electroweak \(T\) parameter is generated at tree-level due to the custodial violating interaction \(H^{\dag}\Ph H\). The strong constraints on the \(T\) parameter require the triplet scalar to have a large mass, \(m \gg v\). In this regime, the leading terms of the EFT are quite accurate.

For readers interested in comparing the CDE with traditional Feynman diagram techniques, we note that triplet scalars were studied within the EFT framework in~\cite{Khandker:2012,*Skiba:2010} where the Wilson coefficients were calculated using Feynman diagrams (see the appendices of~\cite{Khandker:2012,*Skiba:2010}). Tree-level Feynman diagrams involving scalar propagators are straightforward to deal with; yet, we believe that even in this simple case the CDE offers a significantly easier method of calculation. In particular, at no point do we (1) have to break the Lagrangian into gauge non-covariant pieces to obtain Feynman rules, (2) look up a table of higher dimension operators to know how to rearrange the answer back into a gauge-invariant form, or (3) consider various momenta configurations of external particles in order to extract which particular higher dimension operator is generated.

\begin{center} \small{\textbf{Tree-level matching}} \end{center}

Let \(\Ph = \Ph^aT^a\) be an electroweak, real scalar triplet with hypercharge \(Y_{\Ph}=0\).\footnote{For \(Y_{\Ph} \ne 0\), \(\Ph^a\) must be complex. Only for \(Y_{\Ph} = 0 \text{ or }1\) can \(\Ph\) have a trilinear interaction with \(H\).} We take the \(SU(2)_L\) generators in the fundamental representation, \(T^a = \t^a = \s^a/2\) with \(\s^a\) the Pauli matrices. The Lagrangian involving \(\Ph\) and its interactions with the Standard Model Higgs doublet is given by\footnote{The coupling names and normalization are chosen to coincide with those in~\cite{Khandker:2012,*Skiba:2010}.}
\begin{equation}
\scL[\Ph,H] = \frac{1}{2} \big( D_{\m}\Ph^a\big)^2 - \frac{1}{2} m^2\Ph^a\Ph^a +2\k H^{\dag}\t^a H \Ph^a - \et \abs{H}^2\Ph^a\Ph^a -\frac{1}{4}\l_{\Ph}(\Ph^a\Ph^a)^2,
\label{eqn:lag_triplet}
\end{equation}
where \(D_{\m}\Ph = [D_{\m},\Ph] = (\pd_{\m}\Ph^a + g\e^{abc}W_{\m}^b\Ph^c)T^a = (D_{\m}\Ph^a)T^a\). The interaction \(H^{\dag}\Ph H\), being linear in \(\Ph\), leads to a tree-level contribution to the effective action. To calculate this contribution, we follow the steps outlined in section~\ref{sec:CDE_tree}. Introducing an obvious vector notation and writing the Lagrangian as in Eq.~\eqref{eqn:tree_UV_Lag},
\begin{equation}
\scL = \frac{1}{2} \vec{\Ph}^T \big(P^2 - m^2 - U\big) \vec{\Ph} + \vec{\Ph}\cdot\vec{B} + \scO(\Ph^3), \quad U = 2 \et \abs{H}^2 \text{ and } \vec{B} = 2\k H^{\dag}\vec{\t} H,
\label{eqn:lag_triplet2}
\end{equation}
we solve the equation of motion for \(\Ph\) and plug it back into the action. Linearizing the equation of motion, we have
\begin{equation}
\vec{\Ph}_c = -\frac{1}{P^2 - m^2 - U} \vec{B}.
\label{eqn:trip_Ph_c}
\end{equation}
The tree-level effective action is given by \(\scL_{\text{eff,tree}}[H] = \scL[\Ph_c,H]\). Performing an inverse mass expansion on \(\Ph_c\), the effective action through dimension-six operators is, 
\[
\scL_{\text{eff,tree}} = \frac{1}{2m^2}\vec{B}\cdot\vec{B} +\frac{1}{2m^4}\vec{B}^T\big(P^2 - U\big)\vec{B} + \text{ dim 8 operators},
\]
where the factor of two difference from Eq.~\eqref{eqn:tree_cde_terms} occurs because \(\vec{\Ph}\) is real.

Now we need to evaluate the terms in the above. For the \(\vec{B}\cdot \vec{B}\) term we have\footnote{Here and below we use the following relation for generators \(T^a\) in the fundamental representation of \(SU(N)\): \((T^a)_{ij}(T^a)_{kl} = \frac{1}{2}(\d_{il}\d_{jk} - \frac{1}{N}\d_{ij}\d_{kl})\).}
\begin{equation*}
B^aB^a = 4\k^2(H^{\dag}\t^aH)(H^{\dag}\t^aH) = \k^2 \abs{H}^4,
\end{equation*}
from which it follows
\begin{equation*}
B^aUB^a = 2\et \k^2 \abs{H}^6.
\end{equation*}
Integrating by parts, the term in involving the covariant derivative is \(\vec{B}^{T} (-D^2) \vec{B} = (D_{\m}\vec{B})^2\) where
\begin{equation*}
D_{\m}B^a \propto D_{\m}(H^{\dag}\t^aH) = (D_{\m}H)^{\dag}\t^aH+ H^{\dag}\t^a(D_{\m}H) .
\end{equation*}
Squaring this, using the identity in the previous footnote and the one in Eq.~\eqref{eqn:cdeA_opsq} we have
\begin{align*}
(D_{\m}B^a)^2 &= \k^2 \big(H^{\dag}\Dfb H \big)^2 + 4\k^2 \abs{H}^2\abs{D_{\m}H}^2 \\
&= 2\k^2 \big(\scO_T + 2\scO_R \big) ,
\end{align*}
where \(H^{\dag}\Dfb H = H^{\dag}(D_{\m}H) - (D_{\m}H)^{\dag}H\) and the operators \(\scO_{T,R}\) are as defined in Table~\ref{tbl:operators}.

Putting it all together, we find
\begin{equation}
\scL_{\text{eff,tree}} = \frac{\k^2}{2m^2} \abs{H}^4 + \frac{\k^2}{m^4} \big(\scO_T + 2\scO_R \big) - \frac{\et \k^2}{m^4} \scO_6 ,
\label{eqn:triplet_tree}
\end{equation}
where \(\scO_6 = \abs{H}^6\). As mentioned previously, these results were also obtained in~\cite{Khandker:2012,*Skiba:2010} using Feynman diagrams.\footnote{The notation in the first reference of~\cite{Khandker:2012,*Skiba:2010} uses the three operators \(\scO_1\), \(\scO_2\), and \(\scO_T'\) where we added the prime since it is not the same as our \(\scO_T\). What they call \(\scO_T'\) is now more commonly called \(\scO_{HD}\). In our notation, \(\scO_T' = \abs{H^{\dag}D_{\m}H}^2 \equiv \scO_{HD} = (\scO_H - \scO_T)/2\), \(\scO_1 = -(\scO_R + \scO_H)\), and \(\scO_2 = \scO_R\).} The first term in the above can be absorbed into the renormalization of the Higgs quartic coupling. As we will discuss in section~\ref{sec:mapping}, \(\scO_T\) contributes to the electroweak \(T\) parameter. Thus, we see in the effective theory that the \(T\) parameter is generated at tree-level.

\begin{center} \small{\textbf{One-loop level matching}} \end{center}

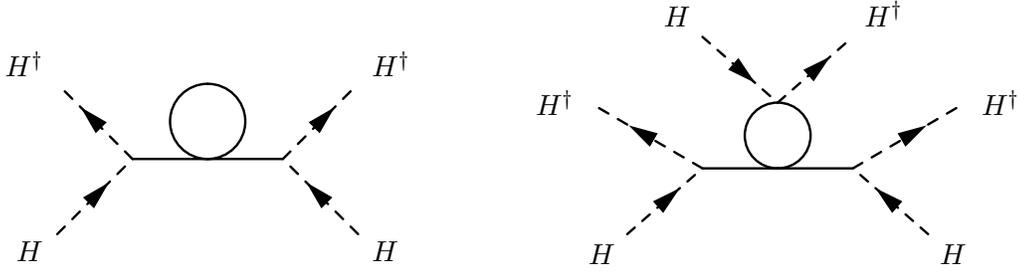
\begin{figure}[t]
\centering
 \subfigure{
 \centering
\begin{fmffile}{H4}
\begin{fmfgraph*}(50,20)
\fmfleft{i1,i2}
\fmfright{o1,o2}
\fmftop{t}
\fmfbottom{b}
\fmflabel{$H^\dagger$}{i2}
\fmflabel{$H$}{i1}
\fmflabel{$H^\dagger$}{o2}
\fmflabel{$H$}{o1}
\fmf{scalar}{i1,i,i2}
\fmf{scalar}{o1,o,o2}
\fmf{phantom,tension=1.2}{p,b}
\fmf{plain,tension=2}{i,p,o}
\fmf{plain,left,tension=0.6}{p,t,p}
\end{fmfgraph*}
\end{fmffile}
 }\hspace{2cm}
 \subfigure{
 \centering
\begin{fmffile}{H6}
\begin{fmfgraph*}(50,35)
\fmfleft{i1,i2,i3}
\fmfright{o1,o2,o3}
\fmfbottom{b}
\fmflabel{$H^\dagger$}{i2}
\fmflabel{$H$}{i1}
\fmflabel{$H^\dagger$}{o2}
\fmflabel{$H$}{o1}
\fmflabel{$H^\dagger$}{o4}
\fmflabel{$H$}{i4}
\fmf{phantom,tension=1.0}{i3,i4}
\fmf{phantom,tension=1.0}{o3,o4}
\fmf{phantom,tension=2}{p,b}
\fmf{scalar}{i1,i,i2}
\fmf{scalar}{o1,o,o2}
\fmf{scalar}{i4,v,o4}
\fmf{plain,tension=2.5}{i,p,o}
\fmf{plain,left,tension=1.0}{p,v,p}
\end{fmfgraph*}
\end{fmffile}
 }
 \caption{Feynman diagrams for $\vec{\Ph}_c \ne 0$ effects at one-loop.
} \label{fig:RGtermdiagrams}
\end{figure}

Let us also calculate the one-loop effective action from integrating out the scalar triplet. It is given by
\begin{equation*}
\D S_{\text{eff,1-loop}} = \frac{i}{2} \Tr \log \left[ - \left. \frac{\d^2S}{\d \Ph^2} \right|_{\Ph = \Ph_c} \right] = \frac{i}{2} \Tr \log \big[ -P^2 + m^2 + U' \big],
\end{equation*}
with
\begin{equation*}
U' = 2\et \abs{H}^2 \cdot \mathbf{1}_3+ \l_{\Ph} \Big[\big(\vec{\Ph}^T_c\cdot \vec{\Ph}_c\big)  \cdot \mathbf{1}_3 + 2 \vec{\Ph}_c \vec{\Ph}_c^T\Big],
\end{equation*}
where \(\mathbf{1}_3\) is the \(3\times 3\) identity matrix and we explicitly wrote it above to remind the reader that each piece in \(U'\) is a matrix. The term in square brackets above is due to the fact that there is a non-zero tree-level piece, \textit{i.e.} that \(\vec{\Ph}_c \ne 0\). Diagrammatically, this term leads to connected, but not 1PI, diagrams of the sort shown in Fig.~\ref{fig:RGtermdiagrams}. Such diagrams are clearly associated with renormalization of parameters in the UV Lagrangian, \textit{e.g.} \(\Ph\)'s mass \(m\) or the cross-quartic coupling \(\et\) in the left and right panels of Fig.~\ref{fig:RGtermdiagrams}, respectively. We recall that \(\Ph_c\) is given by Eq.~\eqref{eqn:trip_Ph_c}, 
\[
\vec{\Ph}_c = \frac{1}{m^2}\vec{B} + \frac{1}{m^4}(P^2 - U)\vec{B} + \dots.
\]

To evaluate the one-loop effective action, we take the universal results from Eq.~\eqref{eqn:CDE_universal_lag} with \(c_s = 1/2\) since \(\Ph^a\) is a real scalar. As \(U'\) contains no term that is linear in fields, for dimension-six and less operators we take the \(m^2\), \(m^0\), and \(m^{-2}\) terms from Eq.~\eqref{eqn:CDE_universal_lag}
\begin{align}
32\pi^2&\D \scL_{\text{eff,1-loop}} = -m^2\Big(\log \frac{m^2}{\m^2} - 1\Big) \tr \, U' -\frac{1}{12}\Big(\log \frac{m^2}{\m^2} - 1\Big) \tr \, G_{\m\n}'^2 - \frac{1}{2} \log \frac{m^2}{\m^2} \tr \, U'^2 \nonumber \\
& +\frac{1}{m^2} \bigg[- \frac{1}{60} \tr \big(P_{\m}G_{\m\n}'\big)^2 - \frac{1}{90} \tr \, G_{\m\n}'G_{\n\s}'G_{\s\m}' -\frac{1}{12}\tr \, (P_{\m}U')^2 - \frac{1}{6}\tr \, U'^3 - \frac{1}{12} \tr \, U' G_{\m\n}'G_{\m\n}' \bigg].
\label{eqn:trip_oneloop_gen}
\end{align}
We are interested in the dimension-six operators generated by integrating out \(\Ph\); since the \(\scO(\Phi_c^2)\) term in \(U'\) is minimally quartic in SM fields, \(\scO(\Ph_c^2) \sim \scO(H^4)+\dots\), we can set \(U' \approx U = 2\et\abs{H}^2\) in the second line of the above equation. In the first line of~\eqref{eqn:trip_oneloop_gen}, higher dimension operators arise because \(\Ph_c \ne 0\); by simple power counting, to capture the dim-6 operators we need to take \(\Ph_c \approx \vec{B}/m^2 + (P^2 - U) \vec{B}/m^4\) in the \(\tr \, U'\) term and \(\Ph_c \approx \vec{B}/m^2\) in the \(\tr \, U'^2\) term.
\footnote{As a side comment, we note that the terms in the first line of Eq.~\eqref{eqn:trip_oneloop_gen} can be used to find the contribution of \(\Ph\) to the beta functions of SM couplings. In particular, the triplet contributes to the running of the Higgs' mass and quartic coupling and also to the \(SU(2)_L\) gauge coupling \(g\). This is easy to see since
\begin{align*}
\tr \, U' &= 3 U + 5 \frac{\l_{\Ph}}{m^4} \vec{B}^T\vec{B} + \text{ dim-six ops } = 6\et \abs{H}^2 + 5\frac{\l_{\Ph}\k^2}{m^4}\abs{H}^4 \\
\tr \, U'^2 &= 3 U^2 + \text{ dim-six ops } = 12 \et^2 \abs{H}^4 + \dots \\
\tr G_{\m\n}'^2 &= -2 g^2 (W_{\m\n}^a)^2.
\end{align*}
} 

To evaluate the traces in~\eqref{eqn:trip_oneloop_gen}, recall that \(G_{\m\n}' = [D_{\m},D_{\n}]\). Since \(\Ph\) is in the adjoint of \(SU(2)_L\), \(G_{\m\n}' = [D_{\m},D_{\n}] = -ig W_{\m\n}^at^a_G\) where the generators \(t^a_G\) are in the adjoint representation, so \(\tr(t^a_Gt^b_G) = 2 \d^{ab}\). 
Keeping only up to dimension-six operators and using the operator definitions given in table~\ref{tbl:operators}, the traces evaluate to\footnote{For example,
\begin{align*}
\tr \, U'^2 &= \tr \, \Big[U \cdot \mathbf{1}_3 + \frac{\l_{\Ph}}{m^4}\Big(\vec{B}^T\vec{B} \cdot 1_{3 \times 3} + 2 \vec{B}\vec{B}^T\Big) \Big]^2 \\
&\supset 2 \frac{\l_{\Ph}}{m^4}\, U\cdot  \tr \, \Big(\vec{B}^T\vec{B} \cdot 1_{3 \times 3} + 2 \vec{B}\vec{B}^T\Big) \\
&= 2 \frac{\l_{\Ph}}{m^4}\, U \cdot \big(5 \vec{B}^T\vec{B} \big) \Rightarrow U = 2\et \abs{H}^2, \ \vec{B}^T\vec{B} = \k^2 \abs{H}^4 \Rightarrow\\
&= 20 \, \frac{\k^2 \et \l_{\Ph} }{m^4} \, \abs{H}^6
\end{align*}
}
\[
\def\arraystretch{1.4}
\begin{array}{rclll}
\tr \, U' & \supset & 5 \l_{\Ph}\vec{\Ph}_c^2 & \supset & 20 \dfrac{\l_{\Ph} \k^2}{m^6}\big(-\et \scO_6 + \scO_T + 2\scO_R\big) \\
\tr \, U'^2 & \supset & 20 \, \dfrac{\l_{\Ph} \k^2 \et }{m^4} \, \abs{H}^6 & = & 20 \, \dfrac{\k^2 \et \l_{\Ph} }{m^4} \scO_6\\
\tr \, U^3 &=& 3 \big(2\et \abs{H})^3 & = & +24\et^3 \scO_6 \\
\tr \big(P_{\m}U\big)^2 & = & -3 \big(2\et \pd_{\m} \abs{H}^2\big)^2 &=& -24 \et^2 \scO_H \\
\tr UG_{\m\n}'G_{\m\n}' & = & -4\et g^2 \abs{H}^2 \big(W_{\m\n}^a\big)^2 &=& -4\et \scO_{WW}\\
 \tr G'^3 & = & -g^3 \e_{abc}W_{\m\n}^aW_{\n\s}^bW_{\s\m}^c& = & -6g^2 \scO_{3W} \\
\tr \big(P_{\m}G_{\m\n}'\big)^2 & = & 2g^2 \big(D_{\m}W_{\m\n}^a)^2 &=& -4 g^2 \scO_{2W}
\end{array}
\]
Plugging these back into~\eqref{eqn:trip_oneloop_gen}, the dimension-six operators in the one-loop effective action are
\begin{align}
\D \scL_{\text{eff,1-loop,dim 6}} = \frac{1}{32\pi^2} \frac{1}{m^2} \bigg[ \frac{g^2}{15}\Big(\scO_{2W} + \scO_{3W}\Big) &+ 2 \et^2 \,\scO_H + \frac{\et}{3} \scO_{WW} - 4\et^3 \scO_6 \nonumber \\
&+ 20 \frac{\l_{\Ph}\k^2}{m^2} \Big(-\et \scO_6 + \scO_T + 2\scO_R\Big) \bigg].
\end{align}
Note that for the present example we use \(\xoverline{\text{MS}}\) renormalization scheme, whose scheme-dependent finite pieces manifest as the terms proportional to \(\l_{\Ph}\) in the above. These terms are associated to the renormalization of the \(\Ph\) mass and the cross-quartic coupling \(\et\), see Fig.~\ref{fig:RGtermdiagrams}; one can in principle choose a different scheme so that these contributions vanish. Finally, we reiterate that the above effective Lagrangian is at the matching scale \(\m = m\), hence why the logarithm pieces from Eq.~\eqref{eqn:trip_oneloop_gen} vanish (this is scheme-independent).

\subsubsection{Extra EW scalar doublet}
Here we integrate out an additional electroweak scalar doublet \(\Ph\) with hypercharge \(Y_\Ph = -1/2\) and mass \(m^2 \gg v^2\). This is essentially the two Higgs doublet model (2HDM) where the mass term for the extra scalar is taken large compared to the EW symmetry breaking scale.

The general Lagrangian for a 2HDM model can be rather complex; often, if the UV model doesn't already impose some restriction on the 2HDM model (as it does in, \textit{e.g.}, supersymmetry), then some other simplifying approximation is made to make more tractable the study of the second doublet. Below, we will consider the most general scalar sector for the second EW doublet; this is rather easy to handle within our EFT framework and requires little additional effort.\footnote{Of course, a large reason why this is much easier in the EFT framework is because we have made the simplifying assumption that the second doublet is heavy.}

The most general Lagrangian consisting of an extra EW scalar doublet \(\Ph\) with \(Y_{\Ph} = -1/2\) interacting with the Higgs sector is given by
\begin{align}
\scL \supset &\abs{D_{\m}\Ph}^2 - m^2 \abs{\Ph}^2 - \frac{\l_{\Ph}}{4}\abs{\Ph}^4 \nonumber\\
&+ \big(\et_H \abs{\tH}^2+\et_{\Ph}\abs{\Ph}^2\big)\big(\tH^{\dag}\Ph + \Ph^{\dag}\tH\big) \nonumber\\
&-\l_1\abs{\tH}^2\abs{\Ph}^2 - \l_2 \abs{\tH^{\dag}\Ph}^2 -\l_3 \big[\big(\tH^{\dag}\Ph\big)^2 + \big(\Ph^{\dag}\tH\big)^2 \big].
\label{eqn:EW_doub_lag}
\end{align}
where \(D_{\m}\Ph = (\pd_{\m} - i g W_{\m}^a \t^a - ig' Y_{\Ph} B_{\m})\Ph\), \(\t^a = \s^a/2\) are the \(SU(2)_L\) generators in the fundamental representation, and \(\tH \equiv i \s_2 H^*\) so that \(\e^{\a\b}\Ph_{\a}H_{\b} = \tH^{\dag}\Ph\). The first line of the above is the potential of \(\Ph\) alone, the second line contains a linear term in \(\Ph\) which leads to a tree-level contribution to the effective action, while the last line contains interactions with the Higgs doublet \(H\) that appear in the effective action at one-loop order.

The main purpose of this section is to show how to use the covariant derivative expansion; in this regard, we remain agnostic to restrictions specific 2HDM models might impose on the Lagrangian in~\eqref{eqn:EW_doub_lag}. However, let us make a few, brief comments. Here we focus on the Higgs sector and have not included a Yukawa sector with couplings to \(\Ph\); these would lead to tree-level generated dimension-six operators involving only fermions. If a parity \(\Ph \to -\Ph, H \to H\) is imposed, then the terms in the second line of~\eqref{eqn:EW_doub_lag} and extra Yukawa terms are forbidden. This parity prevents \(\Ph\) from developing a vacuum expectation value\footnote{Since we assume \(m^2 >0\), \(\Ph\) can only get a vacuum expectation value via the term linear in \(\Ph\) in~\eqref{eqn:EW_doub_lag}, \textit{i.e.} the \(\et_{\Ph}\abs{H}^2\big(\tH^{\dag} \Ph + \text{h.c.}\big)\) term.} and \(\Ph\) in this case is sometimes known as an ``inert Higgs''~\cite{Barbieri:2006}. Finally, imposing an exact or approximate global \(U(1)\) on \(\Ph\) eliminates the second line in~\eqref{eqn:EW_doub_lag}, the term proportional to \(\l_3\) in~\eqref{eqn:EW_doub_lag}, and any potential Yukawa terms involving \(\Ph\).

\begin{center} \small{\textbf{Tree-level matching}} \end{center}

When we integrate out the massive doublet the term linear in \(\Ph\) in~\eqref{eqn:EW_doub_lag}, \(\et_{H}\abs{H}^2\big(\tH^{\dag} \Ph + \text{h.c.}\big)\), leads to a tree-level contribution to the effective action. As this interaction is cubic in the Higgs field, it is simple to see that the only dimension-six operator will be \(\scO_6 = \abs{H}^6\). Concretely, \(B\) from the general tree-level formula Eq.~\eqref{eqn:tree_cde_terms} is given by \(B = \et_{H} \abs{H}^2 \tH\). The solution to the linearized equation of motion is
\begin{equation}
\Ph_c = -\frac{1}{P^2 - m^2 - \l_1 \abs{H}^2 - \l_2 \tH\tH^{\dag}}B \approx \frac{1}{m^2}B = \frac{\et_H}{m^2}\abs{H}^2 \tH ,
\label{eqn:doub_Ph_c}
\end{equation}
and the tree-level effective action through dimension-six operators is
\begin{equation}
\D \scL_{\text{eff,tree,dim-6}} = \frac{1}{m^2} B^{\dag}B = \frac{\et_{H}^2}{m^2} \abs{H}^6 =  \frac{\et_{H}^2}{m^2} \scO_6 .
\label{eqn:EW_doub_tree}
\end{equation}

\begin{center} \small{\textbf{One-loop-level matching}} \end{center}

Let us now find the one-loop effective action from integrating out the massive scalar doublet \(\Ph\) in Eq.~\eqref{eqn:EW_doub_lag}. One of the main reasons we provide these examples is to show how to use the covariant derivative expansion. All the couplings in Eq.~\eqref{eqn:EW_doub_lag} make the effective action calculation complicated, but not very difficult. For the moment, however, let us make several simplifying assumptions on the couplings simply so that the basic setup and use of the CDE is not obscured. After we show the CDE for the simpler Lagrangian, we will return to the full Lagrangian in Eq.~\eqref{eqn:EW_doub_lag} and use the CDE to compute the one-loop effective action.\\

\noindent \textit{Simplifying case}\\

For the simplifying assumptions, let us impose a global \(U(1)\) on \(\Ph\) so that \(\et_H = \et_{\Ph} = \l_3 = 0\) in the Lagrangian. Again, we will come back and let these terms be non-zero shorty. In this case, there is no tree-level effective action. We integrate \(\Ph\) out of the Lagrangian
\[
\scL \supset \Ph^{\dag}\big( -D^2 -m^2 - \l_1 \abs{H}^2 - \l_2 \tH \tH^{\dag} \big) \Ph.
\]
After performing the gaussian integral we are left with the effective action
\[
\D S_{\text{eff,1-loop}} = i\text{Tr}\log\big[ -P^2 + m^2 + A \big],
\]
where we defined
\begin{equation}
A \equiv \l_1 \abs{H}^2 + \l_2 \tH \tH^{\dag}.
\label{eqn:doub_A_def}
\end{equation}
From here, we can use the univeral formula in Eq.~\eqref{eqn:CDE_universal_lag} with \(c_s = 1\) since \(\Ph\) is a complex boson and \(A\) substituted for \(U\) in~\eqref{eqn:CDE_universal_lag}. At this point, we are essentially done; all that is left is to compute the traces.

Let us give a few examples of trace computations by considering \(\tr \big( G_{\m\n}' G_{\n\s}'G_{\s\m}'\big)\) and \(\tr \big( A G_{\m\n}'G_{\m\n}'\big)\). The covariant derivative acting on \(\Ph\) is \(D_{\m} = \pd_{\m} - i g W_{\m} - i g' Y_{\Ph} B_{\m}\cdot \mathbf{1}_2\) where we have explicitly denoted the \(2\times 2\) identity matrix by \(\mathbf{1}_2\). Therefore,
\[
G_{\m\n}' = [D_{\m},D_{\n}] = - ig W_{\m\n}^a \t^a - ig' Y_{\Ph}B_{\m\n} \cdot \mathbf{1}_2.
\]
In \(\tr G'^3\) the anti-symmetry on the Lorentz indices only leaves \(\tr W^3\) non-vanishing. Thus,\footnote{We used
\[
\tr \big(T^a T^b T^c\big) = \frac{1}{2} \tr \Big( [T^a,T^b]T^c + \underbrace{\{T^a,T^b\}T^c}_{\text{vanishes by }W^aW^bW^c\text{ anti-symm}} \Big) = \frac{i}{2}f^{abd}\tr T^dT^c = \frac{i}{2} \m(R) f^{abc}
\]
}
\[
\tr \, G_{\m\n}' G_{\n\s}'G_{\s\m}' = i g^3 \tr \, W_{\m\n}W_{\n\s}W_{\s\m} = -\frac{g^3}{2} \m(R) \e_{abc} W_{\m\n}^aW_{\n\s}^bW_{\s\m}^c = -3g^2 \m(R) \scO_{3W} ,
\]
where \(\m(R)\) is the Dynkin index for representation \(R\) and is equal to \(1/2\) for the fundamental representation and \(\scO_{3W}\) is as defined in table~\ref{tbl:operators}.

For \(\tr \big( A G_{\m\n}'G_{\m\n}'\big)\) we have
\begin{align*}
\tr \big( A G_{\m\n}'G_{\m\n}'\big) &= - \tr\Big[ A \times \big(g W_{\m\n}^a\t^a +g'Y_{\Ph}B_{\m\n} \cdot \mathbf{1}_2)^2 \Big]\\
&=  -g^2 \tr \big(A W_{\m\n}W_{\m\n} \big) - g'^2Y_{\Ph}^2B_{\m\n}B_{\m\n}\tr\, A - 2gg'Y_{\Ph}B_{\m\n}\tr\big(A W_{\m\n}\big) ,
\end{align*}
using \(\tr A \t^a = \l_2 \tH^{\dag}\t^a \tH = -\l_2 H^{\dag}\t^a H\) and a few other manipulations, it is straightforward to see that
\begin{align*}
\tr \big( A G_{\m\n}'G_{\m\n}'\big) &= - (2\l_1 + \l_2)\Big(\frac{g^2}{4} \abs{H}^2 W_{\m\n}^aW_{\m\n}^a   +g'^2 Y_{\Ph}^2\abs{H}^2 B_{\m\n}B_{\m\n}\Big) +2 g g' \l_2 Y_{\Ph} \big(H^{\dag}\t^a H\big) W_{\m\n}^aB_{\m\n} \\
&= -(2\l_1 + \l_2)\Big(\frac{1}{4} \scO_{WW} + Y_{\Ph}^2 \scO_{BB}\Big) + \l_2Y_{\Ph} \scO_{WB} .
\end{align*}

\noindent \textit{Returning to the full Lagrangian}\\

Now we return to the full Lagrangian in~\eqref{eqn:EW_doub_lag} and leave all couplings non-zero. This makes the calculation more complicated; however, it will not be too difficult---we will simply need to evaluate some traces which, while tedious, is very straightforward. In many regards, most of the work goes into setting up the matrix that we are tracing over.

To evaluate the one-loop effective action, we expand the action around the solution to the equation of motion, \(\Ph = \Ph_c + \s\). Because the interaction \((\tH^{\dag}\Ph)^2\) is holomorphic in \(\Ph\), it is easiest to treat \(\Ph\) and \(\Ph^*\) as separate variables. This is equivalent to splitting \(\Ph\) into its real and imaginary pieces, although more convenient to work with. Then, upon expanding \(\Ph = \Ph_c + \s\) and doing a little algebra, the terms quadratic in \(\s\) are
\begin{equation}
\scL[\Ph_c + \s] \supset \frac{1}{2}\begin{matrix} \begin{pmatrix} \s^{\dag} & \s^T \end{pmatrix} \\ \mbox{} \end{matrix} \begin{pmatrix} P^2-m^2 - A' & -2V \\ -2V^{\dag} & \big(P^T\big)^2-m^2 - A'^T \end{pmatrix} \left(\def\arraystretch{0.8}\begin{array}{c} \s \\ \s^* \end{array}\right) , 
\label{eqn:doub_matrix}
\end{equation}
where
\begin{align}
A' &= A - \et_{\Ph}\big(\tH^{\dag}\Ph_c + \Ph_c\tH^{\dag} + \text{h.c.}\big) + \frac{\l_{\Ph}}{2}\big(\abs{\Ph_c}^2 + \Ph_c\Ph_c^{\dag}\big) , \nonumber\\
V &= \l_3 \tH\tH^T - \et_{\Ph}\Ph_c\tH^T + \frac{\l_{\Ph}}{4} \Ph_c\Ph_c^T .
\label{eqn:doub_matrix_defs}
\end{align}
A few comments:
\begin{itemize}
\item We are treating \(\s\) and \(\s^*\) as separate variables, which is the same procedure as working with the real and imaginary parts of \(\s\).
\item The one-loop effective action is given by
\[
\D S_{\text{eff,1-loop}} = \frac{i}{2} \Tr \log \big( \dots \big) ,
\]
with the matrix in~\eqref{eqn:doub_matrix} inserted into the trace. Note the factor of \(1/2\); we take \(c_s = 1/2\) since we are treating \(\s\) and \(\s^*\) as separate, real variables.
\item The classical configuration is given by
\[
\Ph_c = \left[\frac{1}{m^2} + \frac{1}{m^4}\big(P^2 - A\big) + \dots\right] B.
\]
Recall that \(B \sim \scO(H^3)\) and \(A \sim \scO(H^2)\). Keeping up to dimension-six operators, for the traces below we need to keep the above two terms in \(\Ph_c\) for \(\tr\, U\), only the leading term for \(\tr \, U^2\), and we can drop \(\Ph_c\) from the other traces.

\end{itemize}

We now use the CDE to compute \(\D S_{\text{eff,1-loop}}\). In the CDE formulas, we take
\begin{equation}
P_{\m} = \begin{pmatrix} P_{\m} & 0 \\ 0 & P_{\m}^T \end{pmatrix}, \quad m^2 = \begin{pmatrix} m^2_{\textcolor{white}{\m}}\mathbf{1}_2 & 0 \\ 0 & m^2_{\textcolor{white}{\m}}\mathbf{1}_2 \end{pmatrix}, \quad U = \begin{pmatrix}  A'_{\textcolor{white}{\m}} & 2V \\ 2V^{\dag} & A'^T_{\textcolor{white}{\m}} \end{pmatrix} ,
\label{eqn:doub_CDE_defs}
\end{equation}
where \(\mathbf{1}_2\) is the \(2\times2\) identity matrix. The effective action is of the form \(\Tr \log \big(-P^2 + m^2 + U\big)\), so that the transformation \(e^{\pm P^{\m}\pd/\pd q^{\m}}\) in Eq.~\eqref{eqn:sum_trans} is still allowed and the CDE proceeds as discussed.

Thus, we can immediately use the universal results in Eq.~\eqref{eqn:CDE_universal_lag} with matrices \(P_{\m}\) and \(U\) defined as above in~\eqref{eqn:doub_CDE_defs}, and all that is left to do is evaluate some traces. Tabulating only dim-6 operators, using the operator definitions in Table~\ref{tbl:operators}, and including a factor of \(1/2\) for convenience, we find
\begin{equation}
\def\arraystretch{1.4}
\begin{array}{rclll}
\frac{1}{2}\tr \, U^{\ } & = & \tr \, A' & \supset & \big[\frac{3}{2} \l_{\Ph}\et_H^2 + 6\et_{\Ph}(\l_1+\l_2)\big] \, \scO_6/m^4 \\
& & & & - 6\et_{\Ph}\et_{H}\big(\scO_R + \scO_H\big)/m^4\\
\frac{1}{2} \tr \, U^2 & = & \tr \, A'^2 + 4 \tr \, VV^{\dag} & \supset & -4(3\l_1 + 3\l_2 + \l_3)\et_{H}\et_{\Ph} \, \scO_6/m^2\\
\frac{1}{2}\tr \, U^3 & = & \tr \, A^3 + 6 \tr \, \big(A V V^{\dag} + A^T V^{\dag}V \big) &= & \big(2\l_1^3 + 3\l_1^2\l_2 + 3\l_1\l_2^2+\l_2^3 +12(\l_1+\l_2)\l_3^2\big)\scO_6\\
\frac{1}{2}\tr \big(P_{\m}U\big)^2 & = & \tr \big(P_{\m}A\big)^2 + 4\tr \big(P_{\m}V P_{\m}^TV^{\dag}\big)&=& -\big(4\l_1^2 + 4 \l_1\l_2 + \l_2^2 + 4\l_3^2\big)\scO_H - 2\big(\l_2^2 + 4\l_3^2\big)\scO_R \\
 & & & & \quad - \big(\l_2^2 - 4\l_3^2\big)\scO_T  \\
\frac{1}{2}\tr \, UG_{\m\n}'G_{\m\n}' & = & \tr \, A G_{\m\n}'G_{\m\n}'  &=&  -(2\l_1 + \l_2)\Big(\frac{1}{4} \scO_{WW} + Y_{\Ph}^2 \scO_{BB}\Big) + \l_2 Y_{\Ph} \scO_{WB} \\
\frac{1}{2}\tr \, G'^3 & = & ig^3\tr \, W_{\m\n}W_{\n\s}W_{\s\m}& = & -\frac{3}{2}g^2 \scO_{3W} \\
\frac{1}{2} \tr \big(P_{\m}G_{\m\n}'\big)^2 & = & \frac{g^2}{2} \big(D_{\m}W_{\m\n}^a)^2 +2 g'^2 Y_{\Ph}^2 \big(\pd_{\m}B_{\m\n}\big)^2&=& -g^2 \scO_{2W} -4 g'^2Y_{\Ph}^2 \scO_{2B}\\
\end{array}
\label{eqn:doub_loop_traces}
\end{equation}
Plugging these traces into Eq.~\eqref{eqn:CDE_universal_lag} we obtain the one-loop effective Lagrangian. We summarize the results below.

\begin{center} \small{\textbf{Electroweak scalar doublet summary}} \end{center}

We took an electroweak scalar doublet \(\Ph\) with hypercharge \(Y_\Ph = -1/2\) and Lagrangian
\begin{align}
\scL \supset &\abs{D_{\m}\Ph}^2 - m^2 \abs{\Ph}^2 - \frac{\l_{\Ph}}{4}\abs{\Ph}^4 +\big(\et_H \abs{H}^2+\et_{\Ph}\abs{\Ph}^2\big)\big(\Ph\cdot H + \text{h.c}\big) \nonumber\\
&-\l_1\abs{H}^2\abs{\Ph}^2 - \l_2 \abs{\Ph\cdot H}^2 -\l_3 \big[\big(\Ph\cdot H\big)^2 + \text{h.c.} \big],
\label{eqn:EW_doub_lag_sum}
\end{align}
and integrated out \(\Ph\) to find the dimension-six operators of the effective action matched at one-loop order.

The tree-level effective action, given in Eq.~\eqref{eqn:EW_doub_tree}, only contains \(\scO_6 = \abs{H}^6\). The one-loop effective action is obtained from plugging the traces in Eq.~\eqref{eqn:doub_loop_traces} into~\eqref{eqn:CDE_universal_lag}. This piece contains a host of dimension-six operators that affect electroweak and Higgs physics. In summary, the effective Lagrangian at the matching scale is given by
\begin{align}
\scL_{\text{eff}} = \scL_{\text{SM}} &+ \frac{1}{m^2}\bigg(c_6 \scO_6 + c_H \scO_H + c_T \scO_T + c_R \scO_R + c_{BB}\scO_{BB} + c_{WW}\scO_{WW} \nonumber\\ &+ c_{WB}\scO_{WB} + c_{3W}\scO_{3W} +c_{2W}\scO_{2W} + c_{2B}\scO_{2B}\bigg) , 
\end{align}
where the Wilson coefficients are given in Table~\ref{tbl:EW_doublet}.  As in the previous example with the triplet scalar, we have used \(\xoverline{\text{MS}}\) renormalization scheme. In this scheme, the non-zero finite pieces at the matching scale \(\m = m\) are given by the terms in Table~\ref{tbl:EW_doublet} involving the parameters \(\et_{\Ph}\), \(\et_H\), and \(\l_{\Ph}\).  

\begin{table}
\centering
\[\renewcommand\arraystretch{1.7}
\begin{array}{|rcl|rcl|rcl|}
\hline
c_H &=& \frac{1}{(4\pi)^2} \big[6\et_{\Ph}\et_{H} + \frac{1}{12} \big( 4\l_1^2 + 4 \l_1\l_2 + \l_2^2 + 4\l_3^2\big)\big]
& c_{BB} &=& \frac{1}{(4\pi)^2}  \frac{1}{12} Y_\Ph^2(2\l_1+\l_2)
& c_{3W} &=& \frac{1}{(4\pi)^2}  \frac{1}{60} g^2
\\
c_T &=& \frac{1}{(4\pi)^2}\frac{1}{12}  \big(\l_2^2-4\l_3^2\big)
& c_{WW} &=& \frac{1}{(4\pi)^2}  \frac{1}{48}  (2\l_1+\l_2)
& c_{2W} &=& \frac{1}{(4\pi)^2}  \frac{1}{60} g^2
\\
c_R &=& \frac{1}{(4\pi)^2}\big[ 6\et_{\Ph}\et_{H} + \frac{1}{6} \big(\l_2^2+4\l_3^2\big)\big]
& c_{WB} &=& - \frac{1}{(4\pi)^2}  \frac{1}{12} \l_2 Y_\Ph
& c_{2B} &=& \frac{1}{(4\pi)^2}  \frac{1}{60} 4g'^2Y_\Ph^2
\\
\hline
\end{array}
\]
\[
\begin{array}{|l|}
\hline
c_6 = \et_{H}^2 + \frac{1}{(4\pi)^2} \bigg[\frac{3}{2} \l_{\Ph}\et_H^2 + 6\et_{\Ph}(\l_1+\l_2) -\frac{1}{6}\big( 2\l_1^3 + 3\l_1^2\l_2 + 3\l_1\l_2^2+\l_2^3\big) -2\big(\l_1+\l_2\big)\l_3^2\bigg]\\
\hline
\end{array}
\]
\vspace{-15pt}
\caption{\label{tbl:EW_doublet}  Wilson coefficients \(c_i\) for the operators \(\scO_i\) in Table~\ref{tbl:operators} generated from integrating out a massive electroweak scalar doublet \(\Ph\) with hypercharge \(Y_{\Ph} = -1/2\). \(g\) and \(g'\) denote the gauge couplings of \(SU(2)_L\) and \(U(1)_Y\), respectively. The couplings \(\l_{1,2,3}\) and \(\et_{\Ph,H}\) are defined by the Lagrangian in Eq.~\eqref{eqn:EW_doub_lag_sum}; they are associated with various interactions between \(\Ph\) and the SM Higgs doublet \(H\).}
\vspace{-5pt}
\end{table}
\renewcommand\arraystretch{1.0}

\subsubsection{A $SU(2)_L$ quartet scalar}

In this example, we consider a heavy complex $SU(2)_L$ quartet scalar $\Phi$ with mass $m$ and SM hypercharge $Y_\Phi=\frac{3}{2}$. An allowed \(\Phi H^3\) coupling to the Higgs leads to tree-level contributions in the effective action. For brevity, we will ignore other interaction terms with the Higgs, \textit{e.g.} \(\abs{\Ph}^2\abs{H}^2\), as well as the quartet's self-couplings---they can be easily included as in previous examples. This amounts to taking \(U =0\) in Eq.~\ref{eqn:CDE_universal_lag}. Thus, we consider the following Lagrangian
\begin{equation}
\Delta \mathcal{L} = {\Phi ^\dag }\left( - {D^2} - {m^2}\right)\Phi  - \left(\Phi^\dag B + c.c.\right) , \label{eqn:Lquartet}
\end{equation}
where $\Phi=\left(\Phi_1,\Phi_2,\Phi_3,\Phi_4\right)^T$, with each component being eigenstate of the third $SU(2)_L$ generator $t_\Phi^3=\text{diag}\left(\frac{3}{2}, \frac{1}{2},-\frac{1}{2},-\frac{3}{2}\right)$, and \(B\sim H^3\). Specifically,
\begin{equation}
B = \left( {\begin{array}{*{20}{c}}
{H_1^3}\\
{\sqrt 3 H_1^2{H_2}}\\
{\sqrt 3 {H_1}H_2^2}\\
{H_2^3}
\end{array}} \right) ,
\end{equation}
where $H_1$ and $H_2$ are components of the SM Higgs field $H=(H_1,H_2)^T$.\footnote{For quartet scalar $\Phi$ with $Y_\Phi=1/2$, $B$ would be given by (\(\tH \equiv i\s_2 H\))
\begin{equation}
B_{Y=1/2} = \left( {\begin{array}{*{20}{c}}
{H_1^2 {\tilde H}_1}\\
{\frac{1}{\sqrt 3} H_1^2{\tilde H}_2 + \frac{2}{\sqrt 3} H_1H_2{\tilde H}_1}\\
{\frac{1}{\sqrt 3} H_2^2{\tilde H}_1 + \frac{2}{\sqrt 3} H_1H_2{\tilde H}_2}\\
{H_2^2 {\tilde H}_2}
\end{array}} \right) .
\end{equation}
}

Again, we follow the procedure described in Section~\ref{sec:CDE_tree} to compute the tree-level effective Lagrangian. We first get the equation of motion
\begin{equation}
\left(-D^2-m^2\right)\Phi_c = B , \nonumber
\end{equation}
which gives the solution
\begin{equation}
\Phi_c = -\frac{1}{D^2+m^2} B \approx -\frac{1}{m^2} B . \nonumber
\end{equation}
Plugging this solution back to Eq.~\eqref{eqn:Lquartet}, we get
\begin{equation}
\Delta \mathcal{L}_\text{eff,tree} = -B^\dag \Phi_c \approx \frac{1}{m^2} B^\dag B = \frac{1}{m^2}\left|H\right|^6 = \frac{1}{m^2}\mathcal{O}_6 .
\end{equation}

Because we are ignoring other interactions that \(\Ph\) may have, at one-loop we only get dimension-six operators solely involving gauge fields. The general contribution of particles to the pure glue Wilson coefficients was given in Table~\ref{tbl:dim6_pureglue}. The quartet is the spin \(3/2\) representation of \(SU(2)\) and has Dynkin index \(\m(R) = 5\). Therefore, for \(\scO_{2W}\) and \(\scO_{3W}\), we find
\begin{equation}
\D \scL_{\text{eff,1-loop}} \supset \frac{1}{(4\pi)^2} \frac{1}{m^2}\, \frac{g^2}{6} \big( \scO_{2W} + \scO_{3W} \big).
\end{equation}
For \(U(1)\) gauge groups we can also use the results of Table~\ref{tbl:dim6_pureglue}: replace \(a_{2s}\m(R)\) with \(n_{\Ph}Q^2\), where \(Q\) is the charge of \(\Ph\) under the \(U(1)\) and \(n_{\Ph}\) is the number of real-degrees of freedom in \(\Ph\). (Note that, by anti-symmetry on the Lorentz indices, \(\scO_{3G}\) vanishes if the group is abelian.)  For the case at hand, the quartet has hypercharge \(3/2\) and four complex (eight real) degrees of freedom. Therefore,
\begin{equation}
\D \scL_{\text{eff,1-loop}} \supset \frac{1}{(4\pi)^2} \frac{1}{m^2}\, \frac{3}{10}\, g'^2\scO_{2B}.
\end{equation}

\subsubsection{A real singlet scalar}

In this example, we consider a heavy real singlet scalar field $\Phi$ with mass $m$ that couples to the SM through the following Lagrangian
\begin{equation}
\Delta \mathcal{L} = \frac{1}{2} \left(\partial_\mu \Phi\right)^2 - \frac{1}{2} m^2 \Phi^2 -A \left| H \right|^2 \Phi - \frac{1}{2} k\left| H \right|^2 \Phi^2 - \frac{1}{3!}\mu \Phi^3 -\frac{1}{4!}\lambda_\Phi \Phi^4 . \label{eqn:LSinglet}
\end{equation}
We previously computed the tree-level Wilson coefficients in~\cite{Henning:2014gca}; here we demonstrate how to perform this calculation using the CDE as well as provide the one-loop values of the Wilson coefficients. We note that this real scalar can have interesting phenomenological consequences, such as generating a first order EW phase transition~\cite{Grojean:2004xa}; see the discussion and references in~\cite{Henning:2014gca}.

To compute the tree-level effective Lagrangian we follow the procedure described in Section~\ref{sec:CDE_tree}, taking \(P_{\m} = i \pd_{\m}\). The solution to the linearized equation of motion is,
\begin{equation}
\Phi_c \approx - \frac{1}{\partial^2 + m^2 + k\left| H \right|^2} A\left| H \right|^2 \approx -\frac{1}{m^2} A\left| H \right|^2 + \frac{1}{m^4}\left(\partial^2 + k\left| H \right|^2\right) A\left| H \right|^2 . \nonumber
\end{equation}
Plugging this solution back to Eq.~\eqref{eqn:LSinglet}, we get the tree-level effective Lagrangian
\begin{eqnarray}
\Delta \mathcal{L}_\text{eff,tree} &=&  -A\left| H \right|^2 \Phi_c + \frac{1}{2}\Phi_c\left( -\partial^2 - m^2 - k\left| H \right|^2\right) \Phi_c - \frac{1}{3!}\mu \Phi_c^3 - \frac{1}{4!}\lambda_\Phi \Phi_c^4 \nonumber \\
 &\approx& \frac{1}{2m^2} A^2\left| H \right|^4 + \frac{A^2}{m^4}\mathcal{O}_H + \left( -\frac{kA^2}{2m^4} + \frac{1}{3!}\frac{\mu A^3}{m^6}\right)\mathcal{O}_6 .
\end{eqnarray}

Next let us compute the 1-loop piece of the effective Lagrangian, which according to Eq.~\eqref{eqn:Seff_realscalar}, is
\begin{eqnarray}
\Delta S_\text{eff,1-loop} &=& \frac{i}{2}\text{Tr}\log\left( -{\left. {\frac{{{\delta ^2}S}}{{\delta {\Phi ^2}}}} \right|_{{\Phi _c}}}\right) = \frac{i}{2}\text{Tr}\log\left({\partial ^2} + {m^2} + k{\left| H \right|^2} + \mu {\Phi _c} + \frac{1}{2}{\lambda _\Phi }\Phi _c^2\right) \nonumber \\
 &=& \frac{i}{2}\text{Tr}\log\left({\partial ^2} + {m^2} + k{\left| H \right|^2}\right). 
\end{eqnarray}
Recall that for \(\Ph_c \ne 0\), terms in the functional trace involving \(\Ph_c\) are related to renormalization of parameters in the UV Lagrangian. At the matching scale, they can only lead to scheme-dependent finite terms. In going to the second line, we have picked a renormalization scheme where these effects are absorbed, and hence \(\Ph_c\) is dropped from the analysis. The above is clearly in a form of Eq.~\eqref{eqn:sum_seff}, with \(P_{\m} = i \pd_{\m}\), $U=k\left|H\right|^2$, and $G'_{\mu\nu}=[D_\mu, D_\nu]=[\partial_\mu, \partial_\nu]=0$. Plugging these specific values of $U$ and $G'_{\mu\nu}$ into Eq.~\eqref{eqn:CDE_universal_lag}, we obtain
\begin{eqnarray}
\Delta {\mathcal{L}_\text{eff,1-loop}} &=& \frac{1}{{2{{(4\pi )}^2}}}\frac{1}{{m^2}}\left[ { - \frac{1}{{12}}{{\left({P_\mu }U\right)}^2} - \frac{1}{6}U^3} \right] \nonumber \\
 &=& \frac{1}{{{{(4\pi )}^2}}}\frac{1}{{m^2}}\left(\frac{{{k^2}}}{{12}}{\mathcal{O}_H} - \frac{{{k^3}}}{{12}}{\mathcal{O}_6}\right) .
\end{eqnarray}

\subsubsection{Supersymmetry and light scalar tops}
Supersymmetric states at or near the electroweak scale could explain the origin of this scale and its radiative stability. Scalar tops (stops) hold a privileged position in providing a natural explanation to origin of the EW scale. This motivated us in a previous work~\cite{Henning:2014gca} to study the low-energy EFT that results when stops are integrated out. In that work, we considered a supersymmetric spectrum with light stops and other superpartners decoupled and computed the Wilson coefficients of the one-loop effective action. Here we provide details of how to obtain the Wilson coefficients using the covariant derivative expansion.

As stops carry all SM gauge quantum numbers, every operator in Table~\ref{tbl:operators} is generated. In~\cite{Henning:2014gca}, we computed the Wilson coefficients separately using the CDE and traditional Feynman diagram techniques. The results agreed, providing a good consistency check of the calculation.\footnote{A recent paper by Craig \textit{et al.} \cite{Craig:2014una} computed the correction from scalar tops to the \(Zh\) associated production cross section $\sigma_{Zh}^{}$. They compared the result of the full NLO calculation versus the Wilson coefficients from the SM EFT. Excellent agreement was found, which also serves as a good consistency check.} More importantly, however, the two methods highlighted just how much effort the CDE saves over traditional techniques. No doubt the CDE computation is still complicated, as we will see below, but that is because stops have a large number of various interactions with the SM Higgs and gauge bosons. Nevertheless, it is extremely systematic.

We integrate out the multiplet $\Phi=(\tilde{Q}_3,{\tilde t}_R)^T$, the Lagrangian of which up to quadratic order is given by
\begin{equation}
\mathcal{L} = \Phi^\dagger \left( -D^2-m^2-U \right) \Phi ,
\end{equation}
where
\begin{equation}
{m^2} = \left( {\begin{array}{*{20}{c}}
{m_{{{\tilde Q}_3}}^2}&0\\
0&{m_{{{\tilde t}_R}}^2}
\end{array}} \right) ,
\end{equation}
and the matrix $U$ is
\begin{align*}
U &= \begin{pmatrix} \big(y_t^2s_{\b}^2 + \frac{1}{2}g^2c_{\b}^2\big)\tH\tH^{\dag} + \frac{1}{2} g^2 s_{\b}^2HH^{\dag} - \frac{1}{2}\big(g'^2 Y_Q c_{2\b} + \frac{1}{2}g^2\big) \abs{H}^2 & y_t s_{\b} X_t \tH \\
y_t s_{\b} X_t \tH^{\dag} & \big( y_t^2s_{\b}^2 - \frac{1}{2} g'^2 Y_{t_r} c_{2\b}\big) \abs{H}^2 \end{pmatrix} \\
&\equiv \begin{pmatrix} \tk \tH\tH^{\dag} + kH H^{\dag} + \l_L \abs{H}^2 & X_t \tH \\ X_t \tH^{\dag} & \l_R \abs{H}^2 \end{pmatrix} \\
&\equiv \begin{pmatrix} A_L & X_t \tH \\ X_t\tH^{\dag} & A_R \end{pmatrix}
\end{align*}
where we have defined
\[
\def\arraystretch{1.5}
\begin{array}{ccc}
A_L \equiv \tk \tH\tH^{\dag} + kH H^{\dag} + \l_L \abs{H}^2 \hspace{3mm} & A_R \equiv \l_R \abs{H}^2 \hspace{3mm} & y_ts_{\b} X_t \to X_t  \\
\tk \equiv y_t^2s_{\b}^2 + \frac{1}{2}g^2c_{\b}^2 & k \equiv \frac{1}{2} g^2 s_{\b}^2 & \l_L \equiv - \frac{1}{2}\big(g'^2 Y_Q c_{2\b} + \frac{1}{2}g^2\big) \\
\l_R \equiv y_t^2s_{\b}^2 - \frac{1}{2} g'^2 Y_{t_r} c_{2\b} & & \\
\end{array}
\]
Now with both the representation and the interaction matrix $U$ at hand, we are ready to make use of Eq.~\eqref{eqn:CDE_universal_lag} to compute the Wilson coefficients. However, in order for Eq.~\eqref{eqn:CDE_universal_lag} to be valid, we need $U$ to commute with the mass square matrix $m^2$, which limits us to the degenerate mass scenario $m_{{\tilde Q}_3}^2=m_{{\tilde t}_R}^2\equiv m_{\tilde t}^2$. It is also worth noting that due to the appearance of $X_t$, $U$ is no long quadratic in $H$, but also contains a linear term in $H$. This means that one has to keep all of the trace terms in Eq.~\eqref{eqn:CDE_universal_lag} in computing the Wilson coefficients of dimension-six operators. Another thing to keep in mind while evaluating the terms in Eq.~\eqref{eqn:CDE_universal_lag} is that ${\tilde Q}_3$ and ${\tilde t}_R$ have different charges under the SM gauge group, and the covariant derivative $D_\mu$ or $P_\mu=iD_\mu$ should take on the appropriate form for each,
\[
P_{\m} = \begin{pmatrix} P_{L\m} & 0 \\ 0 & P_{R\m} \end{pmatrix}.
\]
For example, the commutator $[P_\mu,U]$ is,
\[
[P_{\m},U] = \begin{pmatrix} [P_{L\m},A_L] & X_t (P_{\m}\tH) \\ X_t (P_{\m}\tH)^{\dag} & [P_{R\m},A_R] \end{pmatrix}.
\]
Through a straightforward, albeit tedious, use of Eq.~\eqref{eqn:CDE_universal_lag}, we obtain the final result of Wilson coefficients listed in Table~\ref{tbl:MSSM}.

\begin{table}\renewcommand\arraystretch{1.5}
\centering
{\footnotesize
\begin{tikzpicture}[>=latex]
\begin{scope}[xshift=.7cm, yshift = -3.4cm]
\node[anchor=east] at (-1.05,-.4) {
\scalebox{1.1}{\def\arraystretch{1.8}
\begin{tabular}{|rl|}
  \hline
  \(c_{3G}^{}=\)& \hspace{-3mm}\(\frac{g_s^2}{(4\pi )^2}\frac{1}{20}\) \\
  \(c_{3W}^{}=\)& \hspace{-3mm}\(\frac{g^2}{(4\pi)^2}\frac{1}{20}\)  \\
  \(c_{2G}^{}=\)& \hspace{-3mm}\(\frac{g_s^2}{(4\pi )^2}\frac{1}{20}\) \\
  \(c_{2W}^{}=\)& \hspace{-3mm}\(\frac{g^2}{(4\pi)^2}\frac{1}{20}\)  \\
  \(c_{2B}^{}=\)& \hspace{-3mm}\(\frac{g'^2}{(4\pi)^2}\frac{1}{20}\)  \\
  \hline
\end{tabular}
}
};
\end{scope}
\begin{scope}[xshift=-3.5cm]
\node[anchor=west] at (0,0) {
\scalebox{1.1}{\def\arraystretch{1.8}
\begin{tabular}{|rl|rl|}
  \hline
  \(c_{GG}^{}=\) & \hspace{-3mm}\(\frac{h_t^2}{(4\pi)^2} \frac{1}{12} \left[ \left( 1 + \frac{1}{12}\frac{g'^2c_{2\b} }{h_t^2} \right) - \frac{1}{2}\frac{X_t^2}{\mt^2} \right]\) &
  \(c_{WB}^{}=\) & \hspace{-3mm}\(-\frac{h_t^2}{(4\pi )^2}\frac{1}{24}\left[ \left( 1 + \frac{1}{2}\frac{g^2c_{2\b} }{h_t^2} \right) - \frac{4}{5}\frac{X_t^2}{\mt^2} \right]\)  \\
  \(c_{WW}^{}=\) & \hspace{-3mm}\(\frac{h_t^2}{(4\pi)^2} \frac{1}{16} \left[ \left( 1 - \frac{1}{6}\frac{g'^2c_{2\b} }{h_t^2} \right) - \frac{2}{5}\frac{X_t^2}{\mt^2} \right]\) &
  \(c_W^{}=\) & \hspace{-3mm}\(\frac{h_t^2}{(4\pi)^2}\frac{1}{40} \frac{X_t^2}{\mt^2}\) \\
  \(c_{BB}^{}=\) & \hspace{-3mm}\(\frac{h_t^2}{(4\pi)^2} \frac{17}{144} \left[ \left( 1 + \frac{31}{102}\frac{g'^2c_{2\b} }{h_t^2} \right) - \frac{38}{85}\frac{X_t^2}{\mt^2} \right]\) &
  \(c_B^{}=\) & \hspace{-3mm}\(\frac{h_t^2}{(4\pi)^2}\frac{1}{40} \frac{X_t^2}{\mt^2}\) \\[.2cm]
  \hline
\end{tabular}
}
};
\end{scope}

\begin{scope}[xshift=.9cm,yshift = -3.4cm]
\node[anchor=west] at (-1.3,-.4) {
\scalebox{1.1}{\def\arraystretch{1.8}
\begin{tabular}{|rl|}
  \hline
  \(c_H^{}=\) & \hspace{-3mm}\(\frac{h_t^4}{(4\pi)^2}\frac{3}{4}\left[ \left( 1 + \frac{1}{3}\frac{g'^2c_{2\b} }{h_t^2} + \frac{1}{12} \frac{g'^4c_{2\b}^2}{h_t^4} \right) - \frac{7}{6}\frac{X_t^2}{\mt^2} \left( 1 + \frac{1}{14}\frac{(g^2 + 2g'^2)c_{2\b} }{h_t^2} \right) + \frac{7}{30}\frac{X_t^4}{\mt^2} \right]\) \\
  \(c_T^{}=\) & \hspace{-3mm}\(\frac{h_t^4}{(4\pi )^2}\frac{1}{4}\left[ \left( 1 + \frac{1}{2}\frac{g^2c_{2\b} }{h_t^2} \right)^2 - \frac{1}{2}\frac{X_t^2}{\mt^2}\left( 1 + \frac{1}{2}\frac{g^2c_{2\b} }{h_t^2} \right) + \frac{1}{10}\frac{X_t^4}{\mt^4} \right]\) \\
  \(c_R^{}=\) & \hspace{-3mm}\(\frac{h_t^4}{(4\pi)^2}\frac{1}{2}\left[ \left( 1 + \frac{1}{2}\frac{g^2c_{2\b}}{h_t^2} \right)^2 - \frac{3}{2}\frac{X_t^2}{\mt^2}\left( 1 + \frac{1}{12}\frac{(3g^2 + g'^2)c_{2\b}}{h_t^2} \right) + \frac{3}{10}\frac{X_t^4}{\mt^4} \right]\) \\
  \(c_D^{}=\) & \hspace{-3mm}\(\frac{h_t^2}{(4\pi)^2}\frac{1}{20} \frac{X_t^2}{\mt^2}\) \\[0.2cm]
  \hline
\end{tabular}
}
};
\end{scope}

\begin{scope}[xshift=-3.5cm,yshift = -7.cm]
\node[anchor=west] at (0,-.6) {
\scalebox{1.1}{
\begin{tabular}{|rl|}
  \hline
  \(c_6^{}=\) & \hspace{-3mm}\(-\frac{h_t^6}{(4\pi )^2}\frac{1}{2}\left\{ \def\arraystretch{2.1} \begin{array}{l} \left[ 1 + \frac{1}{12}\frac{(3g^2 - g'^2)c_{2\beta }}{h_t^2} \right]^3 + \left[ - \frac{1}{12}\frac{(3g^2 + g'^2)c_{2\beta }}{h_t^2} \right]^3 + \left( 1 + \frac{1}{3}\frac{g'^2c_{2\beta }}{h_t^2} \right)^3 \\  - \frac{X_t^2}{\mt^2}\left[ 2\left( 1 + \frac{1}{12}\frac{(3g^2 - g'^2)c_{2\beta }}{h_t^2} \right) \left( 1 + \frac{1}{8}\frac{(g^2 + g'^2)c_{2\beta }}{h_t^2} \right) + \left(1 + \frac{1}{3}\frac{g'^2c_{2\beta }}{h_t^2} \right)^2 \right] \\+ \frac{X_t^4}{\mt^4}\left[1 + \frac{1}{8}\frac{(g^2 + g'^2)c_{2\beta }}{h_t^2} \right] - \frac{X_t^6}{\mt^6}\frac{1}{10} \\ \end{array} \right\}\) \\
  \hline
\end{tabular}
}
};
\end{scope}

\end{tikzpicture}
}
\vspace{-15pt}
\caption{\label{tbl:MSSM}  Wilson coefficients \(c_i\) for
  the operators \(\scO_i\) in Table~\ref{tbl:operators} generated
  from integrating out MSSM stops with degenerate soft mass
  \(\mt\). \(g_s,g,\) and \(g'\) denote the gauge couplings of
  \(SU(3),SU(2)_L,\) and \(U(1)_Y\), respectively, \(h_t = m_t/v\) with $v=174GeV$,
  and \(\tan \b = \langle
  H_u\rangle / \langle H_d \rangle\) in the MSSM.}
\vspace{-5pt}
\end{table}
\renewcommand\arraystretch{1.0}

\subsubsection{Kinetic mixing of gauge bosons}

In this example, we consider a heavy $U(1)$ gauge boson $K_\mu$ with mass $m_K$ that has a kinetic mixing with the SM $U(1)_Y$ gauge boson $B_\mu$,
\begin{equation}
\Delta\mathcal{L} = - \frac{1}{4}{K_{\mu \nu }}{K^{\mu \nu }} + \frac{1}{2}m_K^2{K_\mu }{K^\mu } - \frac{k}{2}{B^{\mu \nu }}{{K}_{\mu \nu }} , \label{eqn:Lmixing}
\end{equation}
where $K_{\mu\nu}$ denotes the field strength $K_{\mu\nu}=\partial_\mu K_\nu -\partial_\nu K_\mu$. Again, the tree-level effective Lagrangian can be obtained by following the procedure described in Section~\ref{sec:CDE_tree}. We first find the equation of motion of this heavy gauge boson $K_\mu$,
\begin{equation}
{\partial _\nu }{K^{\mu \nu }} + k({\partial _\nu }{B^{\mu \nu }}) = m_K^2{K^\mu } , \nonumber
\end{equation}
which, as usual for vector bosons, can be decomposed into two equations,
\begin{eqnarray}
{\partial _\mu }{K^\mu } &=& 0 , \nonumber \\
\left( -\partial^2 - m_K^2\right){K^\mu } &=&  - k({\partial _\nu }{B^{\mu \nu }}) . \nonumber
\end{eqnarray}
Solving these, we get the classical solution
\begin{equation}
{K_{c\mu} } = \frac{k}{{{\partial ^2} + m_K^2}}({\partial ^\nu }{B_{\mu \nu }}) \approx \frac{k}{{m_K^2}}({\partial ^\nu }{B_{\mu \nu }}). \label{eqn:mixingsolution}
\end{equation}
Next we plug this solution back into the UV model Lagrangian (Eq.~\eqref{eqn:Lmixing}) to get the tree-level effective Lagrangian. With ${B^{\mu \nu }}{K_{\mu \nu }} = 2({\partial _\nu }{B^{\mu \nu }}){K_\mu }$, we obtain
\begin{eqnarray}
\Delta\mathcal{L}_\text{eff,tree} &=& - \frac{1}{2}{K_{c\mu} }\left[ {\left( - {\partial ^2} - m_K^2\right){g^{\mu \nu }} + {\partial ^\mu }{\partial ^\nu }} \right]{K_{c\nu} } - \frac{k}{2}{B^{\mu \nu }}{K_{c\mu \nu }} \nonumber \\
 &=&  \frac{k}{2}({\partial _\nu }{B^{\mu \nu }}){K_{c\mu} } - k({\partial _\nu }{B^{\mu \nu }}){K_{c\mu} } \nonumber \\
 &=& -\frac{k}{2}({\partial _\nu }{B^{\mu \nu }}){K_{c\mu} } \nonumber \\
 &=&  \frac{{{k^2}}}{{m_K^2}}{\mathcal{O}_{2B}}. \label{eqn:mixingresult}
\end{eqnarray}
Note that this example has a trivial one-loop contribution to the effective action.

\subsubsection{Heavy vector bosons in the triplet representation of $SU(2)_L$}

Here we consider an example involving heavy vector bosons transforming under a low-energy (unbroken) non-abelian gauge symmetry. Massive vector bosons near the electroweak scale generically arise in, for example, extra-dimensional compactifications~\cite{Manton:1979kb} and little Higgs theories~\cite{ArkaniHamed:2002qx,*ArkaniHamed:2002qy}. We wish to draw attention to the comparative simplicity with the present covariant method versus traditional loop methods involving massive vector bosons. For example, this method could be readily employed to study massive vector bosons whose tree-level contributions are absent due to, \textit{e.g.}, KK-parity~\cite{Cheng:2002iz} in extra-dimensional models or T-parity~\cite{Cheng:2003ju,*Cheng:2004yc} in little Higgs models.

We consider an $SU(2)_1\times SU(2)_2$ gauge symmetry with a scalar \(\Ph\) transforming as a bifundamental. We take the Standard Model fermions and Higgs field to be localized to the \(SU(2)_1\) gauge group. (We suppress color and hypercharge; the full gauge symmetry is \(SU(2)_1\times SU(2)_2 \times U(1)_Y \times SU(3)_c\).) The scalar \(\Ph\) takes a vev, breaking the \(SU(2)\) groups down to their diagonal subgroup, which we identify with the weak interactions of the SM, \(SU(2)_1 \times SU(2)_2 \to SU(2)_L\). This is simply a deconstructed~\cite{ArkaniHamed:2001nc} version of an extra-dimensional model (\textit{e.g.} \cite{Randall:1999ee}), where the weak gauge bosons, being a diagonal combination of the \(SU(2)_1\times SU(2)_2\) gauge bosons, ``propagate in the bulk'', while the SM fermions and Higgs only transform under one gauge group and are therefore ``localized''.

The relevant kinetic terms of the Lagrangian are
\begin{equation}
\Delta \mathcal{L}_K = - \frac{1}{2}\tr{\big(F_1^{\mu\nu}\big)^2} - \frac{1}{2}\tr{\big(F_2^{\mu\nu}\big)^2} + \frac{1}{2}\tr{({\mathcal{D}_\mu }\Ph )^\dag }({\mathcal{D}^\mu }\Ph ) ,
\end{equation}
where the scalar $\Ph$ transforms as a bifundamental, $\Ph \to U_1 \Ph U_2^\dag$. The covariant derivative of the UV theory is given by\footnote{Note that the action of \(\mathcal{D}_{\m}\) on \(\Ph\) is \(\mathcal{D}_{\m}\Ph = \pd_{\m}\Ph - i g_1 A_{1\, \m}\Ph + i g_2 \Ph A_{2\, \m}\)}
\[
\mathcal{D}_{\m} = \pd_{\m} - i g_1 A_{1\, \m} - i g_2A_{2\, \m},
\]
where \(g_i\) and \(A_{i\, \m} = A_{i\, \m}^a\t_i^a\) are the gauge coupling and gauge bosons of the \(SU(2)_i\) with the generators \(\t_i^a\) taken in the fundamental representation. A vacuum expectation value for \(\Ph\),
\[
\avg{\Ph} = \frac{1}{\sqrt{2}}\begin{pmatrix} v& 0\\0 & v \end{pmatrix},
\]
breaks $SU{(2)_1} \times SU{(2)_2} \to SU{(2)_L}$. 
The mass eigenstates are
\begin{subequations}
\begin{eqnarray}
{Q^a} &\equiv& \frac{1}{{\sqrt {g_1^2 + g_2^2} }}({g_1}A_1^a - {g_2}A_2^a) , \\
{W^a} &\equiv& \frac{1}{{\sqrt {g_1^2 + g_2^2} }}({g_2}A_1^a + {g_1}A_2^a) ,
\end{eqnarray}
\end{subequations}
where $W^a$ are the SM gauge bosons corresponding to the unbroken symmetry $SU(2)_L$, and $Q^a$ obtain a mass \(m_Q^2 = (g_1^2+g_2^2)v^2/4\) from the Higgs mechanism. \(Q_{\m}\) transforms in the adjoint (triplet) representation of the unbroken \(SU(2)_L\).

In terms of the mass eigenstates, the covariant derivative becomes
\begin{equation}
{\mathcal{D}_\mu } = {\partial _\mu } - igW_\mu ^a\t_L^a - iQ_\mu ^a\bigg(\frac{{g_1^2}}{{\sqrt {g_1^2 + g_2^2} }}\t_1^a - \frac{{g_2^2}}{{\sqrt {g_1^2 + g_2^2} }}\t_2^a\bigg) ,
\end{equation}
where \(\t_L^a = \t_1^a + \t_2^a\) are the unbroken generators and we identify $g \equiv g_1g_2/\sqrt {g_1^2 + g_2^2}$ as the weak coupling constant of the SM. We expand \(\Ph\) around \(\avg{\Ph}\),
\begin{equation}
\Ph  = \frac{1}{{\sqrt 2 }}(v + h) \begin{pmatrix} 1& 0\\0 & 1 \end{pmatrix} + i\sqrt 2 {\chi ^a}{\t^a} .
\end{equation}
where the \(\chi^a\) are the Nambu-Goldstone bosons transforming in the adjoint of the unbroken \(SU(2)_L\) and \(h\) is the massive Higgs field.

Now we integrate out the massive \(Q_{\m}\). At tree-level, \(Q_{\m}\) couples to the \(SU(2)_L\) source current. At loop-level, we need to gauge fix---as summarized in section~\ref{sec:CDE_fermions}, we take a generalized \(R_{\x}\) gauge which preserves the unbroken \(SU(2)_L\) gauge symmetry (simply promote \(\pd_{\m}\) in the usual \(R_{\x}\) gauge to \(D_{\m}\)). Expanding out $\mathcal{L}_K$ in terms of $W_\mu$, $Q_\mu$, $\chi$, and adding the gauge fixing piece $\mathcal{L}_\text{g.f.}$, the ghost term $\mathcal{L}_\text{ghost}$, and the interaction $\mathcal{L}_\text{I}$ between $Q_\mu$ and the SM fields,
\begin{subequations}
\begin{eqnarray}
\Delta \mathcal{L}_K &\supset& - \frac{1}{2}\tr{({D^\mu }{Q^\nu } - {D^\nu }{Q^\mu })^2} + ig\tr([{Q^\mu },{Q^\nu }]{W_{\mu \nu }}) + \tr{({D_\mu }\chi - {m_Q}{Q_\mu })^2} , \\
{\mathcal{L}_\text{g.f.}} &=&  - \frac{1}{\xi }\tr{(\xi {m_Q}\chi  + {D^\mu }{Q_\mu })^2} , \\
{\mathcal{L}_\text{ghost}} &\supset& {{\bar c}^a}{( - {D^2} - \xi m_Q^2)^{ab}}{c^b} , \\
{\mathcal{L}_\text{I}} &=& \frac{{g_1^4}}{{4(g_1^2 + g_2^2)}}{\left| H \right|^2}Q_\mu ^a{Q^{a\mu }} + \frac{{g_1^2}}{\sqrt{g_1^2 + g_2^2}}Q_\mu ^aJ_W^{a\mu} ,
\end{eqnarray}
\end{subequations}
we find the Lagrangian up to quadratic terms in $Q_\mu^a$ to be,
\begin{eqnarray}
\Delta \mathcal{L} &=& \frac{1}{2}Q_\mu ^a{\left\{ {{D^2}{g^{\mu \nu }} - {D^\nu }{D^\mu } + m_Q^2{g^{\mu \nu }} + [{D^\mu },{D^\nu }] + \frac{1}{\xi }{D^\mu }{D^\nu } + \frac{{g_1^4}}{{2(g_1^2 + g_2^2)}}{{\left| H \right|}^2}} g^{\mu\nu} \right\}^{ab}}Q_\nu ^b \nonumber \\
 && + \frac{{g_1^2}}{{\sqrt {g_1^2 + g_2^2} }}Q_\mu ^aJ_W^{a\mu } + \frac{1}{2}{\chi ^a}{( - {D^2} - \xi m_Q^2)^{ab}}{\chi ^b} + {{\bar c}^a}{( - {D^2} - \xi m_Q^2)^{ab}}{c^b} ,
\end{eqnarray}
where $H$ denotes the SM Higgs field and $J_W^{a\mu}$ is the source current of the SM $W_\mu^a$. Working with Feynman gauge $\xi=1$, we get
\begin{eqnarray}
\Delta \mathcal{L} &=& \frac{1}{2}Q_\mu ^a{\left\{ {{D^2}{g^{\mu \nu }} + m_Q^2{g^{\mu \nu }} + 2[{D^\mu },{D^\nu }] + \frac{{g_1^4}}{{2(g_1^2 + g_2^2)}}{{\left| H \right|}^2}} g^{\mu\nu} \right\}^{ab}}Q_\nu ^b \nonumber \\
 && + \frac{{g_1^2}}{{\sqrt {g_1^2 + g_2^2} }}Q_\mu ^aJ_W^{a\mu } + \frac{1}{2}{\chi ^a}{( - {D^2} - m_Q^2)^{ab}}{\chi ^b} + {{\bar c}^a}{( - {D^2} - m_Q^2)^{ab}}{c^b} . \label{eqn:LSU2}
\end{eqnarray}
This Lagrangian is clearly in the form of Eq.~\eqref{eqn:Sgauge}, supplemented by a linear interaction term.

Although the heavy fields $Q_\mu^a$ couple directly to the fermions in SM, upon using the equation of motion \(D_{\m}W^{a\m\n} = J_W^{a\n}\), the tree-level effective Lagrangian can be written in a way such that it only contains bosonic operators. To see this, we first solve the equation of motion for \(Q_{\m}\) at leading order,
\begin{equation}
Q_c^{a\mu} =  - \frac{{g_1^2}}{{\sqrt {g_1^2 + g_2^2} }}\frac{1}{{m_Q^2}}J_W^{a\mu } .
\end{equation}
Then we plug this back into Eq.~\eqref{eqn:LSU2} and obtain the tree-level effective Lagrangian:
\begin{equation}
\Delta {\mathcal{L}_{{\text{eff,tree}}}} = \frac{1}{2}\frac{{g_1^2}}{{\sqrt {g_1^2 + g_2^2} }}Q_{c\mu} ^aJ_W^{a\mu } =  - \frac{1}{{2m_Q^2}}\frac{{g_1^4}}{{g_1^2 + g_2^2}}J_{W\mu }^aJ_W^{a\mu } = \frac{{g_1^4}}{{g_1^2 + g_2^2}}\frac{1}{{m_Q^2}}{\mathcal{O}_{2W}} .
\end{equation}

The one-loop effective Lagrangian can be read off from Table~\ref{tbl:dim6_pureglue} and Eq.~\eqref{eqn:CDE_universal_lag} using \(U\) as in Eq.~\eqref{eqn:U_massive_gauge} with $M^{\mu\nu} = \frac{g_1^4}{2(g_1^2+g_2^2)}{\left| H \right|^2}g^{\mu\nu}$:
\begin{eqnarray}
\Delta {\mathcal{L}_{{\rm{eff,1 - loop}}}} &=& \frac{1}{{{{(4\pi )}^2}}}\frac{1}{{m_Q^2}}\Bigg[ \frac{{{g^2}}}{{20}}\left( {3{\mathcal{O}_{3W}} - 37{\mathcal{O}_{2W}}} \right) + \frac{1}{4}{{\left( {\frac{{g_1^4}}{{g_1^2 + g_2^2}}} \right)}^2}{\mathcal{O}_H} \nonumber \\
 && - \frac{1}{{24}}{{\left( {\frac{{g_1^4}}{{g_1^2 + g_2^2}}} \right)}^3}{\mathcal{O}_6} \Bigg] .
\end{eqnarray}








\section{Running of Wilson coefficients and choosing an operator set} \label{sec:operatorbasis}

To connect with measurements, the Wilson coefficients \(c_i(\L)\) determined at the matching scale \(\L\) need to be evolved down to the weak scale \(m_W\) according to their renormalization group (RG) equations. From the perspective of {\it using} the SM EFT, the most important question surrounding RG running is whether or not it is relevant. In other words, when is it sufficient to simply take the zeroth order solution \(c_i(m_W) = c_i(\Lambda)\) versus higher order corrections? This, of course, depends on the sensitivity of present and future precision measurements. We discuss details below, but a short rule of thumb is that RG running is relevant only if \(c_i(\L)\) is generated at tree-level.

If one needs to include RG running, it follows from the above rule of thumb that it is sufficient to take just the leading order correction. At leading order, the RG equations are governed by the anomalous dimension matrix $\gamma_{ij}$,
\begin{equation}
\frac{d c_i(\mu)}{d \log \mu} = \sum_j \frac{1}{16\pi^2} \gamma_{ij} c_j, \label{eqn:RGE}
\end{equation}
whose leading order solution is
\begin{equation}
c_i(m_W) = c_i(\Lambda) - \sum_j \frac{1}{16\pi^2} \gamma_{ij} c_j(\Lambda) \log \frac{\Lambda}{m_W} \, . \label{eqn:RGrelation}
\end{equation}
Computing \(\g_{ij}\) in the SM EFT is no small endeavor; fortunately, results for the one-loop anomalous dimension matrix are known~\cite{Jenkins:2013zja,Jenkins:2013wua,Alonso:2013hga,Alonso:2014zka,Elias-Miro:2013mua,Elias-Miro:2013eta}. To consistently make use of these results, the main issue concerns operator bases---as with any matrix, the components \(\g_{ij}\) depend on the basis in which the matrix is expressed! We will discuss how the choice of operator sets affects the expression and use of \(\g_{ij}\). Following this, we will give a short summary of common basis choices in the literature and how to go between them.

\subsection{When is RG running important?}

Although the running of Wilson coefficients is a conceptually important step, there turn out to be strong requirements on the class of UV models for it to be of practical relevance. Near future measurements have an estimated sensitivity at the per mille level: from $v^2/\Lambda^2 \sim 0.1\%$, we see that $\Lambda$ can be probed at most up to a few TeV. So the logarithm is not large, $\log (\Lambda/m_W) \sim 3$, and therefore loop order counting in perturbative expansions is reasonable.

Counting by loop order, per mille level precision means that we can truncate perturbative calculations at one-loop. Since RG evolution contributes a loop factor, the running of \(c_j(\L)\) into \(c_i(m_W)\), \(i\ne j\), will be of practical relevance if \(c_j(\L)\) is of tree-level size. In particular, if \(c_j(\L)\) is generated at one-loop level, then its contribution to \(c_i(m_W)\) from RG running is of two loop size and hence negligible. Additionally, even in the case that $c_j(\Lambda)$ is generated at tree level, its contribution to $c_i(m_W)$ is subdominant if $c_i(\Lambda)$ is also generated at tree level. Therefore, as a rule of thumb, one needs to take account for RG evolution of \(c_j\) into \(c_i\) only when both of the following conditions are satisfied:
\begin{enumerate}
  \item $c_j(\Lambda)$ is generated at tree level from the UV model.
  \item $c_i(\Lambda)$ is not generated at tree level from the UV model.
\end{enumerate}

The fact that \(c_j(\L)\) need to be generated at tree-level for RG running to be important is a strong requirement---many motivated models of new physics only generate Wilson coefficients at one-loop level. Familiar examples of such cases are SUSY with R-parity, extra dimensions with KK parity, and little Higgs models with T parity. The parity in all these examples is a discrete symmetry which forces the new particles to always come in pairs, hence leading only to loop-level contributions of Wilson coefficients.

Let us discuss this rule of thumb in the context of the examples in Sec.~\ref{sec:CDE_examples} where a heavy scalar couples at tree-level to the Higgs sector. There are only four such models! Among these, the \(SU(2)\) scalar doublet and quartet only generate \(\scO_6 = \abs{H}^6\) at tree-level. Since \(\scO_6\) does not run into other dimension-six operators, the RG running is trivial. Therefore, RG analysis is only relevant for the two other examples in the list. An explicit example of this RG analysis can be found in~\cite{Henning:2014gca}, where we found the RG-induced constraints on the singlet example of Sec.~\ref{sec:CDE_examples} to be quite constraining. 


\subsection{Choosing an operator set in light of RG running analysis}

As mentioned before, the anomalous dimension matrix $\gamma_{ij}$ has been computed in the literature~\cite{Jenkins:2013zja,Jenkins:2013wua,Alonso:2013hga,Alonso:2014zka,Elias-Miro:2013mua,Elias-Miro:2013eta}. When RG running analysis is relevant, one just needs to make use of the known $\gamma_{ij}$ appropriately.

There are many dimension-six operators that respect the SM gauge invariance. However, some of these operators are redundant in the sense that they lead to the same physical effects. The relations among these operators stem from group identities, integration by parts, and use of the equations of motion; the first two of these are obvious, the latter is a result of the fact that physical quantities are on-shell, and therefore respect the equations of motion. An operator set is said to be \textit{complete} if it can capture all possible physical effects stemming from the higher dimension operators. A complete operator set with a minimal number of operators is called an ``operator basis''. We will discuss specific operator basis for the SM EFT in the next subsection.

Note that when performing calculations (matching, RG running, {\it etc.}), the theory does not select for a particular operator set or basis---choosing an operator set is something imposed by hand. {\it A priori}, there is no clear criteria to tell which operator set is ``best'', or if using a non-redundant versus redundant set of operators is ``better''. In general, there are three types of operator sets: (1) an operator basis, (2) an overcomplete set that has some redundant operators, and (3) an incomplete set that lacks of some components compared to a complete operator basis. For a consistent RG analysis, one generically should choose a complete operator set such that the RG running (Eq.~\eqref{eqn:RGE}) is closed~\cite{Jenkins:2013zja}.

Before discussing the above three choices of operator sets, we would like to include a relevant technical remark that regard how the anomalous dimension matrix is computed. One first chooses an operator set and then computes the anomalous dimension matrix for this operator set. For a chosen operator set, there are generically two types of contributions to $\gamma_{ij}$: the \textit{direct contribution} where ${\cal O}_j$ generates ${\cal O}_i$ directly through a loop Feynman diagram, and the \textit{indirect contribution} where ${\cal O}_j$ generates some ${\cal O}_k$ outside the operator set chosen, whose elimination (through the equations of motion or an operator identity) in turn gives ${\cal O}_i$.

\begin{itemize}
  \item Working with a Complete Operator Basis

    This case is fairly straightforward. The full anomalous dimension matrix $\gamma_{ij}$ in the ``standard basis'' (see the next subsection for definition) has been computed~\cite{Jenkins:2013zja,Jenkins:2013wua,Alonso:2013hga}. One can simply carry out a basis transformation to obtain the $\gamma_{ij}$ in the new basis.

  \item Working with an Overcomplete Operator Set

  Sometimes it is helpful to use a redundant operator set because it can make the physics more transparent. For example, the matching from a UV model may generate an overcomplete set of effective operators. An obvious drawback of working with an overcomplete set of operators is that the size of $\gamma_{ij}$ would be larger than necessary, and that the value of $\gamma_{ij}$ would not be unique~\cite{Jenkins:2013zja}. However, this does not necessarily mean that $\gamma_{ij}$ is harder to calculate. For example, consider the extreme case of using all the dim-6 operators, before using equations of motion to remove any redundant combination. This is a super overcomplete set, and as a result the size of $\gamma_{ij}$ would be way larger than that in the standard basis. But with this choice of operator set, all the contributions to $\gamma_{ij}$ are direct contributions by definition. Some of these direct contributions would become indirect in a smaller operator set, and one has to accommodate them by using equations of motion or operator identities, which is a further step of calculation. Therefore, in some cases, it is the reduction from an over complete set to an exact complete set that requires more work. Note that the ambiguity in the explicit form of $\gamma_{ij}$ from using an over-complete basis does not cause any problem when computing physical effects.

  \item Working with an Incomplete Operator Set

  An operator basis contains 59 operators, which has $76$ ($2499$) real valued Wilson coefficients for the number of generation being one (three) \cite{Alonso:2013hga}. Practically, that is a very large basis to work with. In some cases, only a small number of operators are relevant to the physics considered and it is tempting to just focus on this small, incomplete set for the purpose of simplification. However, while a complete or overcomplete operator basis is obviously guaranteed to be RG closed, an incomplete operator set is typically not. When the incomplete operator set is not RG closed, Eq.~\eqref{eqn:RGE} no longer holds. To fix this problem, one can view the incomplete operator set $\{{\cal O}_i\}$ as a subset of a certain complete operator basis $\{{\cal O}_i,{\cal O}_a\}$. Once this full operator basis is specified, one has a clear definition of the sub matrix $\gamma_{ij}$ to compute the RG induced effects. Obviously, the off-diagonal block $\gamma_{ai}$ is generically nonzero, which means some operator ${\cal O}_a$ outside the chosen operator set $\{{\cal O}_i\}$ can also be RG induced. In this case, the generation of \(\scO_a\) could bring additional constraints on the UV model under consideration. Ignoring these effects makes the constraints over conservative (see also the discussion in section 2 of~\cite{Elias-Miro:2013eta}).
\end{itemize}

\subsection{Popular operator bases in the literature}

Here we summarize a few popular choices of dimension-six operator bases that are commonly used in the literature (see~\cite{Willenbrock:2014bja} for a recent review). These sets have been developed with two different types of motivations: (1) completeness, and (2) phenomenological relevance. In spite of that, however, they are actually not very different from each other. In this subsection, we will briefly describe each basis and then discuss the relation among them.

With a motivation of completeness, one starts with enumerating all the possible dim-6 operators that respect the Standard Model gauge symmetry. Some combinations of these operators are zero due to simple operator identities.\footnote{For example, \(0= 2 \abs{H^{\dag} D_{\m} H}^2 - \frac{1}{2} \big(\pd_{\m} \abs{H}^2\big)^2 + \frac{1}{2} \big( H^{\dag} \Dfb H \big)^2\) is an operator identity that makes use of integration by parts.} One can use these redundances to remove operators and shrink the operator set. In addition, many other combinations are zero upon using equation of motions, and hence would not contribute to physical observables which are on-shell quantities. These combinations can also be removed because they are redundant in respect of describing physics.\footnote{An example identity which makes use of the equations of motion is \(0= (\pd_{\m}B^{\m\n})^2 - j_{\m,Y}^2\), where \(B^{\m\n}\) is the hypercharge field strength and \(j_{\m,Y}\) is its associated current.} After all of these reductions, one arrives at an operator set that is non-redundant but still complete, in a sense that it has the full capability of describing the physical effects of any dim-6 operators. Clearly, the non-redundant, complete set of operators forms an ``operator basis''. There are, of course, multiple choices of operator bases, all related by usual basis transformations.

The first attempt of this completeness motivated construction dates back to~\cite{Buchmuller:1985jz}, where 80 dim-6 operators were claimed to be independent. However, it was later discovered that there were still some redundant combinations within the set of 80. The non-redundant basis was eventually found to contain only 59 dim-6 operators~\cite{Grzadkowski:2010es}. (There are also 5 baryon violating operators, bringing the total to 64, which are typically dropped from the analysis). To respect this first success, we will call the 59 dim-6 operators listed in~\cite{Grzadkowski:2010es} the ``standard basis''. During the past year, the full anomalous dimension matrix $\gamma_{ij}$ has been calculated in the standard basis~\cite{Jenkins:2013zja,Jenkins:2013wua,Alonso:2013hga,Alonso:2014zka}.

The second type of motivation in choosing an operator set is the relevance to phenomenology. With this kind of motivation, one usually starts with a quite small set of operators that are immediately relevant to the physics concerned. However, if RG running effects are important, a complete operator set is required for the analysis. As discussed in the previous subsection, one can then extend the initial operator set into a complete operator basis by adding enough non-redundant operators to it. Popular operator bases constructed along this line include the ``EGGM basis''~\cite{Elias-Miro:2013eta}, the ``HISZ basis''~\cite{Hagiwara:1993ck}, and the ``SILH basis''~\cite{Giudice:2007fh,Elias-Miro:2013mua,Pomarol:2013zra}. These three bases are all motivated by studying physics relevant to the Higgs boson and the electroweak bosons. As a result, they all maximize the use of bosonic operators. In fact, these bases are very closely related to each other. Consider the following seven operators $\{ {\cal O}_W, {\cal O}_B, {\cal O}_{WW}, {\cal O}_{WB}, {\cal O}_{BB}, {\cal O}_{HW}, {\cal O}_{HB} \}$, where ${\cal O}_{HW}$ and ${\cal O}_{HB}$ are defined as
\begin{eqnarray}
O_{HW} &\equiv& 2ig(D^\mu H)^\dag \tau^a (D^\nu H) W_{\mu\nu}^a , \\
O_{HB} &\equiv& ig'(D^\mu H)^\dag (D^\nu H) B_{\mu\nu} ,
\end{eqnarray}
and the other five are defined in Table~\ref{tbl:operators}. There are two identities among them as following
\begin{eqnarray}
O_W &=& O_{HW} + \frac{1}{4}(O_{WW} + O_{WB}) , \label{eqn:OW} \\
O_B &=& O_{HB} + \frac{1}{4}(O_{BB} + O_{WB}) . \label{eqn:OB}
\end{eqnarray}
So only five out of the seven are non-redundant. The difference among ``EGGM basis'', ``HISZ basis'', and ``SILH basis'' just lies in different ways of choosing five operators out of these seven: ``EGGM basis'' drops $\{ {\cal O}_{HW}, {\cal O}_{HB} \}$, ``HISZ basis'' drops $\{ {\cal O}_W, {\cal O}_B \}$, and ``SILH basis'' drops $\{ {\cal O}_{WW}, {\cal O}_{WB} \}$.

The three phenomenologically motivated bases are not that different from the standard basis either. As mentioned before, due to motivation difference, the second type maximizes the use of bosonic operators. It turns out that to obtain the ``EGGM basis'' from the standard basis, one only needs to do the following basis transformation (trading five fermionic operators into five bosonic operators using equation of motion):
\begin{eqnarray}
({H^\dag }{\tau^a} \Dfb H)({{\bar L}_1}{\gamma _\mu }{\tau^a}{L_1}) &\to& {O_W} = ig({H^\dag }{\tau^a}\Dfb H)({D^\nu }W_{\mu \nu }^a) , \\
({H^\dag }\Dfb H)(\bar e{\gamma _\mu }e) &\to& {O_B} = ig'{Y_H}({H^\dag }\Dfb H)({\partial ^\nu }{B_{\mu \nu }}) , \\
(\bar u{\gamma ^\mu }t_s^Au)(\bar d{\gamma _\mu }t_s^Ad) &\to& {O_{2G}} =  - \frac{1}{2}{({D^\mu }G_{\mu \nu }^a)^2} , \\
({{\bar L}_1}{\gamma ^\mu }{\tau^a}{L_1})({{\bar L}_1}{\gamma _\mu }{\tau^a}{L_1}) &\to& {O_{2W}} =  - \frac{1}{2}{({D^\mu }W_{\mu \nu }^a)^2} , \\
(\bar e{\gamma ^\mu }e)(\bar e{\gamma _\mu }e) &\to& {O_{2B}} =  - \frac{1}{2}{({\partial ^\mu }{B_{\mu \nu }})^2} .
\end{eqnarray}


\section{Mapping Wilson coefficients onto observables} \label{sec:mapping}
So far we have described how to compute the Wilson coefficients $c_i(\Lambda)$ from a given UV model and how to run them down to the weak scale $c_i(m_W)$ with the appropriate anomalous dimension matrix $\gamma_{ij}$. This section then is devoted to the last step in Fig.~\ref{fig:connection} --- mapping $c_i$ \footnote{Throughout this section, all the Wilson coefficients mentioned will be at the weak scale $\mu=m_W$. In order to reduce the clutter, we hence suppress this specification of the RG scale and use $c_i$ as a shorthand for $c_i(m_W)$.} onto the weak scale precision observables. The Wilson coefficients $c_i$ will bring various corrections to the precision observables at the weak scale. The goal of this section is to study the deviation of each weak scale precision observable as a function of $c_i$.

It is worth noting that our SM EFT parameterized by Eq.~\eqref{eqn:LeffSM} and $c_i$ is totally different from the widely used seven-$\kappa$ parametrization (for example see~\cite{LHCHiggsCrossSectionWorkingGroup:2012nn}), which parameterizes only a size change in each of the SM type Higgs couplings. The seven-$\kappa$ actually parameterize models that do not respect the electroweak gauge symmetry and hence violates unitarity. As a result, future precision programs show spuriously high sensitivity on them. Our SM EFT on the other hand, parameterize new physics in the direction that respects the SM gauge invariance and is therefore free from unitarity violations.

In order to provide a concrete mapping result, we need to specify a set of operators to work with. Keeping in mind a special interest in UV models in which new physics is CP preserving and couples with the SM only through the Higgs and gauge bosons, we choose the set of dim-6 operators that are purely bosonic and CP conserving. All the dim-6 operators satisfying these conditions are listed in Table~\ref{tbl:operators2}. This set of effective operators coincides with the set chosen in \cite{Elias-Miro:2013eta}, supplemented by the operators ${\cal O}_D$ and ${\cal O}_R$. Wilson coefficients of all the fermionic operators are assumed to be zero.

\renewcommand\arraystretch{1.4}
\begin{table}[tb]
\centering
\begin{tabular}{|rcl|rcl|}\hline
 \(\scO_{GG}\) &\(=\)& \(g_s^2 \abs{H}^2G_{\mu \nu }^aG^{a,\mu \nu }\) & \(\scO_H\)   &\(=\)& \(\frac{1}{2}\big(\pd_{\mu} \abs{H}^2\big)^2\)\\
 \(\scO_{WW}\) &\(=\)& \(g^2  \abs{H}^2 W_{\mu \nu }^aW^{a,\mu \nu } \) &  \(\scO_T\)   &\(=\)& \(\frac{1}{2}\big( H^{\dag} \Dfbd H\big)^2\) \\
 \(\scO_{BB}\) &\(=\)& \(g'^2 \abs{H}^2 B_{\mu \nu }B^{\mu \nu }\) & \(\scO_R\)   &\(=\)& \(\abs{H}^2\abs{D_{\m}H}^2\) \\
 \(\scO_{WB}\) &\(=\)& \(2gg'H^\dag {\tau^a}H W_{\mu \nu }^a B^{\mu \nu }\) &  \(\scO_D\)   &\(=\)& \(\abs{D^2H}^2\) \\
 \(\scO_W\)   &\(=\)& \(ig\big(H^\dag \tau^a \Dfb H\big)D^\nu W_{\mu \nu }^a\) &  \(\scO_6\)   &\(=\)& \(\abs{H}^6\) \\
 \(\scO_B\)   &\(=\)& \(ig'Y_H\big(H^\dag \Dfb H\big)\pd^\nu B_{\mu \nu }\)  &  \(\scO_{2G}\) &\(=\)& \(-\frac{1}{2} \big(D^\mu G_{\mu \nu }^a\big)^2\) \\
 \(\scO_{3G}\) &\(=\)& \(\frac{1}{3!}g_sf^{abc}G_\rho ^{a\mu }G_\mu ^{b\nu }G_\nu ^{c\rho }\) &  \(\scO_{2W}\) &\(=\)& \(-\frac{1}{2} \big(D^\mu W_{\mu \nu }^a\big)^2\) \\
 \(\scO_{3W}\) &\(=\)& \(\frac{1}{3!}g \e^{abc}W_\rho ^{a\mu }W_\mu ^{b\nu }W_\nu ^{c\rho }\) & \(\scO_{2B}\) &\(=\)& \(-\frac{1}{2} \big(\pd^{\mu} B_{\mu \nu }\big)^2\) \\
  \hline
\end{tabular}
\caption{Dimension-six bosonic operators for our mapping analysis.} \label{tbl:operators2}
\vspace{-10pt}
\end{table}
\renewcommand\arraystretch{0}

There are four categories of precision observables on which present and near future precision programs will be able to reach a per mille level sensitivity: (1) Electroweak Precision Observables (EWPO), (2) Triple Gauge Couplings (TGC), (3) Higgs decay widths, and (4) Higgs production cross sections. In the mapping calculation, we can keep only up to linear order of Wilson coefficients and we only include tree-level diagrams of the Wilson coefficients. This is because the near future precision experiments will only be sensitive to one-loop physics, and we practically consider each power of $\frac{1}{\Lambda^2}c_i$ as one-loop size, since it is already known that the SM is a very good theoretical description and the deviations should be small. Although in some UV models Wilson coefficients can arise at tree-level, the corresponding $\frac{1}{\Lambda^2}$ must be small enough to be consistent with the current constraints. So considering $\frac{1}{\Lambda^2}c_i$ as one-loop size is practically appropriate.

Our convention when expanding the Higgs doublet around the EW breaking vacuum is to take \(H = \begin{pmatrix} 0 & v + h/\sqrt{2} \end{pmatrix}^T\) where \(v \approx 174\) GeV.

\subsection{Electroweak precision observables}

\renewcommand\arraystretch{2.0}
\begin{table}[tb]
\centering
\begin{tabular}{|rl|}
  \hline
  $S=$ & \hspace{-0.4cm} $ -\dfrac{4c_Z^{} s_Z^{}}{\alpha} \Pi'_{3B}(0)$ \\
  $X=$ & \hspace{-0.4cm} $ -\dfrac{1}{2} m_W^2 \Pi''_{3B}(0)$ \\
  \hline
  $T=$ & \hspace{-0.4cm} $ \dfrac{1}{\alpha}\dfrac{1}{m_W^2} \left[ \Pi_{WW}(0) - \Pi_{33}(0) \right]$ \\
  $U=$ & \hspace{-0.4cm} $ \dfrac{{4s_Z^2}}{\alpha }\left[ \Pi'_{WW}(0) - \Pi'_{33}(0) \right]$ \\
  $V=$ & \hspace{-0.4cm} $ \dfrac{1}{2} m_W^2 \left[ \Pi''_{WW}(0) - \Pi''_{33}(0) \right]$ \hspace{2mm} \\
  \hline
  \hspace{2mm} $W=$ & \hspace{-0.4cm} $ -\dfrac{1}{2} m_W^2 \Pi''_{33}(0)$ \\
  \hline
  $Y=$ & \hspace{-0.4cm} $ -\dfrac{1}{2} m_W^2 \Pi''_{BB}(0)$ \\
  \hline
\end{tabular}
\caption{Definitions of the EWPO parameters, where the single/double prime denotes the first/second derivative of the transverse vacuum polarization functions.} \label{tbl:EWPOdef}
\vspace{-5pt}
\end{table}
\renewcommand\arraystretch{0}

Electroweak precision observables represent the oblique corrections to the propagators of electroweak gauge bosons. Specifically, there are four transverse vacuum polarization functions: $\Pi_{WW}(p^2)$, $\Pi_{ZZ}(p^2)$, $\Pi_{\gamma\gamma}(p^2)$, and $\Pi_{\gamma Z}(p^2)$,\footnote{Throughout this paper, we use $\Pi(p^2)$ to denote the additional part of the transverse vacuum polarization function due to the Wilson coefficients. In a more precise notation, one should use $\Pi^\text{new}(p^2)$ as in \cite{Beringer:1900zz} or $\delta\Pi(p^2)$ as in~\cite{Maksymyk:1993zm,*Burgess:1993mg} for it, but we simply use $\Pi(p^2)$ to reduce the clutter. That said, our $\Pi(p^2)$ at leading order is linear in $c_i$.} each of which can be expanded in $p^2$
\begin{equation}
\Pi(p^2)=a_0+a_2 p^2+a_4 p^4+{\cal O}(p^6) \label{eqn:PiExpand}.
\end{equation}
Two out of these expansion coefficients are fixed to zero by the masslessness of the photon: $\Pi_{\gamma\gamma}(0)=\Pi_{\gamma Z}(0)=0$. Another three combinations are fixed (absorbed) by the definition of the three free parameters $g$, $g'$, and $v$ in electroweak theory. So up to $p^2$ order, there are three left-over parameters that can be used to test the predictions of the model. These are the Peskin-Takeuchi parameters $S$, $T$, and $U$~\cite{Peskin:1991sw,Beringer:1900zz}, which capture all possible non-decoupling electroweak oblique corrections. As higher energy scales were probed at LEP II, it was proposed to also include the coefficients of $p^4$ terms, which brings us four additional parameters $W, Y, X, V$~\cite{Maksymyk:1993zm,*Burgess:1993mg,Kundu:1996ah,Barbieri:2004qk}.


So in total, we have seven EWPO parameters in consideration, $S, T, U, W, Y, X, V$. In this paper, we take the definitions of them as listed in Table~\ref{tbl:EWPOdef},\footnote{Our definitions in Table~\ref{tbl:EWPOdef} agree with~\cite{Peskin:1991sw} and~\cite{Barbieri:2004qk}. Many other popular definitions are in common use as well ({\it e.g.} see~\cite{Maksymyk:1993zm,Beringer:1900zz,Baak:2012kk}). The main differences lie in the choice of using derivatives of $\Pi(p^2)$ evaluated at $p^2=0$, such as $\Pi'_{WW}(0)$, {\it etc.}, versus using some form of finite distance subtraction, such as $\frac{\Pi_{WW}(m_W^2)-\Pi_{WW}(0)}{m_W^2}$, {\it etc}. Up to $p^4$ order in $\Pi(p^2)$, this discrepancy would only cause a disagreement in the result of $U$. For example, the definition in~\cite{Beringer:1900zz} would result in nonzero $U$ parameter from the custodial preserving operator ${\cal O}_{2W}$: $U=\frac{s_Z^4}{\alpha}\frac{4m_Z^2}{\Lambda^2}c_{2W}^{} \ne 0$. In this paper, we stick to the definition in~\cite{Peskin:1991sw} to make $U$ a purely custodial violating parameter. Under our definition, $U=0$ at dim-6 level.} where for the purpose of conciseness, we use the alternative set $\left\{\Pi_{33}, \Pi_{BB}, \Pi_{3B}\right\}$ instead of $\left\{\Pi_{ZZ}, \Pi_{\gamma\gamma}, \Pi_{\gamma Z}\right\}$.\footnote{One may also be concerned that these definitions through the transverse polarization functions $\Pi(p^2)$ are not generically gauge invariant. In principle, these $\Pi(p^2)$ functions can be promoted to gauge invariant ones ${\overline\Pi}(p^2)$ by a ``pinch technique'' prescription. (For examples, see discussions in~\cite{Degrassi:1993kn,Chen:2013kfa,Binosi:2009qm}.)} And due to the relation $W^3=c_Z^{}Z+s_Z^{}A$ and $B=-s_Z^{}Z+c_Z^{}A$,\footnote{Throughout this paper, we adopt the notation $c_Z^{}\equiv \cos\theta_Z$ {\it etc.}, with $\theta_Z$ denoting the weak mixing angle. We do not use $\theta_W$ in order to avoid clash with the Wilson coefficient for the operator ${\cal O}_W$.} the two set are simply related by the transformations
\begin{eqnarray}
\Pi_{33} &=& c_Z^2\Pi_{ZZ} + s_Z^2\Pi_{\gamma\gamma} + 2c_Z^{} s_Z^{}\Pi_{\gamma Z} , \label{eqn:Pi33} \\
\Pi_{BB} &=& s_Z^2\Pi_{ZZ} + c_Z^2\Pi_{\gamma\gamma} - 2c_Z^{} s_Z^{}\Pi_{\gamma Z} , \label{eqn:PiBB} \\
\Pi_{3B} &=& -c_Z^{} s_Z^{}\Pi_{ZZ} + c_Z^{} s_Z^{}\Pi_{\gamma\gamma} + (c_Z^2-s_Z^2)\Pi_{\gamma Z} . \label{eqn:Pi3B}
\end{eqnarray}

Table~\ref{tbl:EWPOresults} summarizes the mapping results of the seven EWPO parameters, {\it i.e.} each of them as a linear function (to leading order) of the Wilson coefficients $c_i$. These results are straightforward to calculate. First, we calculate $\Pi_{WW}(p^2)$, $\Pi_{ZZ}(p^2)$, $\Pi_{\gamma\gamma}(p^2)$, and $\Pi_{\gamma Z}(p^2)$ in terms of $c_i$. This can be done by expanding out the dim-6 operators in Table~\ref{tbl:operators2}, identifying the relevant Lagrangian terms, and reading off the two-point Feynman rules. The details of these steps together with the results of $\Pi_{WW}(p^2)$, $\Pi_{ZZ}(p^2)$, $\Pi_{\gamma\gamma}(p^2)$, and $\Pi_{\gamma Z}(p^2)$ (Table~\ref{tbl:Pis}) are shown in Appendix~\ref{subsec:FeynmanRules}. Next, we compute the alternative combinations $\Pi_{WW}(p^2)-\Pi_{33}(p^2)$, $\Pi_{33}(p^2)$, $\Pi_{BB}(p^2)$, $\Pi_{3B}(p^2)$ using the transformation relations Eq.~\eqref{eqn:Pi33}-Eq.~\eqref{eqn:Pi3B}, the results of which are also summarized in Appendix~\ref{subsec:FeynmanRules} (Table~\ref{tbl:AlternativePis}). Finally, we combine Table~\ref{tbl:AlternativePis} with the definitions of EWPO parameters (Table~\ref{tbl:EWPOdef}) to obtain the results in Table~\ref{tbl:EWPOresults}.

\renewcommand\arraystretch{2.0}
\begin{table}[tb]
\centering
\begin{tabular}{|rl|rl|}
  \hline
  $S=$ & \hspace{-0.4cm} $\dfrac{c_Z^2s_Z^2}{\alpha}\dfrac{4m_Z^2}{\Lambda^2}(4c_{WB}^{}+c_W^{}+c_B^{})$ & $W=$ & \hspace{-0.4cm} $\dfrac{m_W^2}{\Lambda^2}c_{2W}^{}$ \\
  $T=$ & \hspace{-0.4cm} $\dfrac{1}{\alpha}\dfrac{2{v^2}}{\Lambda^2}c_T^{}$ & $Y=$ & \hspace{-0.4cm} $\dfrac{m_W^2}{\Lambda^2}c_{2B}^{}$ \\
  $U=$ & \hspace{-0.4cm} $0$ & $X=$ & \hspace{-0.4cm} $V=0$ \\
  \hline
\end{tabular}
\caption{EWPO parameters in terms of Wilson coefficients.}\label{tbl:EWPOresults}
\vspace{-5pt}
\end{table}
\renewcommand\arraystretch{0}

We would like to emphasize the importance of $W$ and $Y$ parameters. It should be clear from the definitions Table~\ref{tbl:EWPOdef} that the seven EWPO parameters fall into four different classes: $\{S, X\}$, $\{T, U, V\}$, $\{W\}$, and $\{Y\}$. Therefore $W$ and $Y$ out of the four $p^4$ order EWPO parameters supplement the classes formed by $S, T, U$ (see also the discussions in \cite{Barbieri:2004qk}). Our mapping results in Table~\ref{tbl:EWPOresults} also show that $W$ and $Y$ are practically more important compared to $X$ and $V$, for $W$ and $Y$ are nonzero while $X$ and $V$ vanish at dim-6 level.

\subsection{Triple gauge couplings}

The TGC parameters can be described by the a phenomenological Lagrangian \cite{Gaemers:1978hg,Hagiwara:1986vm,*Hagiwara:1992eh,Gounaris:1996rz}
\begin{eqnarray}
{{\cal L}_{{\rm{TGC}}}} &=& ig{c_Z^{}}{Z^\mu } \cdot g_1^Z(\hat W_{\mu \nu }^ - {W^{ + \nu }} - \hat W_{\mu \nu }^ + {W^{ - \nu }}) + igW_\mu ^ + W_\nu ^ - ({\kappa _Z} \cdot {c_Z^{}}{{\hat Z}^{\mu \nu }} + {\kappa _\gamma } \cdot {s_Z^{}}{{\hat A}^{\mu \nu }}) \nonumber \\
&& + \frac{{ig}}{{m_W^2}}\hat W_\mu ^{ - \rho }\hat W_{\rho \nu }^ + ({\lambda _Z} \cdot {c_Z^{}}{{\hat Z}^{\mu \nu }} + {\lambda _\gamma } \cdot {s_Z^{}}{{\hat A}^{\mu \nu }}) , \label{eqn:LagTGC}
\end{eqnarray}
where ${\hat V}_{\mu\nu} \equiv {\partial_\mu}{V_\nu}-{\partial_\nu}{V_\mu}$. Among the five parameters above, there are two relations due to an accidental custodial symmetry. We take $g_1^Z$, $\kappa_\gamma$, and $\lambda_\gamma$ as the three independent parameters. The other two can be expressed as \cite{Gounaris:1996rz}
\begin{eqnarray}
 {\kappa _Z} &=& g_1^Z - \frac{{s_Z^2}}{{c_Z^2}}({\kappa _\gamma } - 1) , \\
 {\lambda _Z} &=& {\lambda _\gamma } .
\end{eqnarray}
The SM values of TGC parameters are $g_{1,\rm{SM}}^Z=\kappa_{\gamma,\rm{SM}}=1, \lambda_{\gamma,\rm{SM}}=0$. Their deviations from SM are currently constrained at percent level \cite{Schael:2013ita}, and will be improved to $10^{-4}$ level at ILC500 (see the second reference in~\cite{Behnke:2013xla}). Their mapping results are summarized in Table~\ref{tbl:TGCresults}.\footnote{These results are also obtained in \cite{Elias-Miro:2013eta}.}

\renewcommand\arraystretch{2.0}
\begin{table}[tb]
\centering
\begin{tabular}{|rl|}
  \hline
  $\delta g_1^Z=$ & \hspace{-0.4cm} $-\dfrac{m_Z^2}{\Lambda^2}c_W^{}$ \\
  $\delta \kappa_\gamma=$ & \hspace{-0.4cm} $\dfrac{4m_W^2}{\Lambda^2}c_{WB}^{}$ \\
  $\lambda_\gamma=$ & \hspace{-0.4cm} $-\dfrac{m_W^2}{\Lambda^2}c_{3W}^{}$ \\
  \hline
\end{tabular}
\caption{TGC parameters in terms of Wilson coefficients.} \label{tbl:TGCresults}
\vspace{-5pt}
\end{table}
\renewcommand\arraystretch{0}

\subsection{Deviations in Higgs decay widths}

The dim-6 operators bring deviations in the Higgs decay widths from the Standard Model. In this paper, we study all the SM Higgs decay modes that near future linear colliders can have sub-percent sensitivity on, {\it i.e.} $\Gamma \in \left\{\Gamma_{h \to f\bar{f}}, \Gamma_{h \to gg}, \Gamma_{h \to \gamma\gamma}, \Gamma_{h \to \gamma Z} ,\Gamma_{h \to WW^*}, \Gamma_{h \to ZZ^*}\right\}$. Our analysis for the decay modes through off-shell vector gauge bosons $h \to WW^*$ and $h \to ZZ^*$ apply to all their fermionic modes, namely that $h \to WW^* \to W l \bar\nu / W d {\bar u}$ and $h \to ZZ^* \to Z f {\bar f}$.

For each decay width $\Gamma$ above, we define its deviation from the SM
\begin{equation}
\epsilon \equiv \frac{\Gamma}{\Gamma _\text{SM}} - 1. \label{eqn:epsdef}
\end{equation}
It turns out that at leading order (linear power) in $c_i$, this deviation is generically a sum of three parts, (1) the ``interference correction'' $\epsilon_I^{}$, (2) the ``residue correction'' $\epsilon_R^{}$, and (3) the ``parametric correction'' $\epsilon_P^{}$:
\begin{equation}
\epsilon = \epsilon_I^{} + \epsilon_R^{} + \epsilon_P^{} .
\end{equation}
In the following, we will first give a brief description of the meaning and the mapping results of each part, and then explain in detail how to derive these results.

\renewcommand\arraystretch{2.0}
\begin{table}[tb]
\centering
\begin{tabular}{|rl|}
  \hline
  $\epsilon_{hf\bar f,I}^{}=$ & \hspace{-4mm} $0$ \\
  $\epsilon_{hgg,I}^{}=$ & \hspace{-4mm} $\dfrac{(4\pi)^2} {{\mathop{\rm Re}\nolimits} A_{hgg}^\text{SM}} \dfrac{16v^2}{\Lambda^2}{c_{GG}^{}}$ \\
  $\epsilon_{h\gamma\gamma,I}^{}=$ & \hspace{-4mm} $\dfrac{(4\pi)^2}{{\mathop{\rm Re}\nolimits} A_{h\gamma\gamma}^\text{SM}} \dfrac{8v^2}{\Lambda^2}\left(c_{WW}^{}+ c_{BB}^{}-c_{WB}^{}\right)$ \\
  $\epsilon_{h\gamma Z}^{}=$ & \hspace{-4mm} $\dfrac{(4\pi)^2}{{\mathop{\rm Re}\nolimits} A_{h\gamma Z}^\text{SM}}\dfrac{4v^2}{\Lambda^2}\dfrac{1}{c_Z^{}}\Big[ 2\left(c_Z^2c_{WW}^{}-s_Z^2c_{BB}^{}\right)-\left(c_Z^2-s_Z^2\right)c_{WB}^{} \Big]$ \\
  $\epsilon_{hWW^*,I}^{}=$ & \hspace{-4mm} $\Big[ 2I_a(\beta_W)-I_b(\beta_W) \Big] \dfrac{m_W^2}{\Lambda^2} c_{2W}^{} - \Big[ 2I_b(\beta_W)-I_c(\beta_W) \Big] \dfrac{4m_W^2}{\Lambda^2}c_{WW}^{}$ \\
   & \hspace{-4mm} $-I_a(\beta_W) \dfrac{2m_W^2}{\Lambda^2} c_W^{} - I_b(\beta_W) \dfrac{v^2}{\Lambda^2} c_R^{} + \dfrac{2m_h^2}{\Lambda^2} c_D^{}$ \\
  $\epsilon_{hZZ^*,I}^{}=$ & \hspace{-4mm} $ +\Big[ 2I_a(\beta_Z)-I_b(\beta_Z) \Big] \dfrac{m_Z^2}{\Lambda^2}\left(c_Z^2 c_{2W}^{} + s_Z^2 c_{2B}^{}\right)$ \\
   & \hspace{-4mm} $-\Big[ 2I_b(\beta_Z)-I_c(\beta_Z) \Big] \dfrac{4m_Z^2}{\Lambda^2}\left(c_Z^4 c_{WW}^{} + s_Z^4 c_{BB}^{} + c_Z^2 s_Z^2 c_{WB}^{}\right)$ \\
   & \hspace{-4mm} $-I_a(\beta_Z)\dfrac{2m_Z^2}{\Lambda^2}\left(c_Z^2 c_W^{}+s_Z^2 c_B^{}\right) +I_b(\beta_Z) \dfrac{v^2}{\Lambda^2} \left(2c_T^{}-c_R^{}\right) + \dfrac{2m_h^2}{\Lambda^2} c_D^{}$ \\
   & \hspace{-4mm} $+\dfrac{e Q_f 2c_Z^2 s_Z^{}}{g(T_f^3-s_Z^2 Q_f)} \left\{\begin{array}{l}
\Big[ I_a(\beta_Z)-I_b(\beta_Z)-1 \Big] \dfrac{m_Z^2}{\Lambda^2}\left(c_{2W}^{} - c_{2B}^{} - c_W^{} + c_B^{}\right) \\
 + I_d(\beta_Z) \dfrac{m_Z^2}{\Lambda^2}\Big[2c_Z^2 c_{WW}^{} - 2s_Z^2 c_{BB}^{} - \left(c_Z^2-s_Z^2\right) c_{WB}^{}\Big]
\end{array} \right\} $ \\
  \hline
\end{tabular}
\caption{Interference corrections $\epsilon_I^{}$ to Higgs decay widths, with $\beta_W\equiv\frac{m_W}{m_h}$, $\beta_Z\equiv\frac{m_Z}{m_h}$, and the auxiliary integrals $I_a(\beta)$, $I_b(\beta)$, $I_c(\beta)$, $I_d(\beta)$ listed in Eq.~\eqref{eqn:Iadef}-\eqref{eqn:Iddef} of the appendix. The $A_{hgg}^\text{SM}$, $A_{h\gamma\gamma}^\text{SM}$, and $A_{h\gamma Z}^\text{SM}$ are the standard form factors, whose expressions are listed in Eq.~\eqref{eqn:Ahgg}-\eqref{eqn:AhAZ} of Appendix~\ref{subsec:GammaIdetails}.} \label{tbl:epsI_width}
\end{table}
\renewcommand\arraystretch{0}

\renewcommand\arraystretch{2.0}
\begin{table}[tb]
\centering
\begin{tabular}{|c|c|c|}
  \hline
   & $\epsilon_R^{}$ & $\epsilon_P^{}$ \\
  \hline
  $\Gamma_{hf\bar f}$ & $\Delta r_h^{}$ & $\Delta w_{y_f^2}$ \\
  $\Gamma_{hgg}$ & $0$ & $0$ \\
  $\Gamma_{h\gamma\gamma}$ & $0$ & $0$ \\
  $\Gamma_{h\gamma Z}$ & $0$ & $0$ \\
  $\Gamma_{hWW^*}$ & $\Delta r_h^{}+\Delta r_W^{}$ & $3\Delta w_{g^2}+\Delta w_{v^2}$ \\
  $\Gamma_{hZZ^*}$ & $\Delta r_h^{}+\Delta r_Z^{}$ & $3\Delta w_{g^2}+\Delta w_{v^2}+\left(3\dfrac{s_Z^2}{c_Z^2}-\dfrac{2s_Z^2 Q_f}{T_f^3-s_Z^2 Q_f}\right)\Delta w_{s_Z^2}$ \\
  \hline
\end{tabular}
\caption{Residue corrections $\epsilon_R^{}$ and parametric corrections $\epsilon_P^{}$ to Higgs decay widths. The explicit results in terms of the dim-6 Wilson coefficients of the residue modifications $\Delta r_h^{}, \Delta r_W^{}, \Delta r_Z^{}$ and parameter modifications $\Delta w_{g^2}, \Delta w_{v^2}, \Delta w_{s_Z^2}, \Delta w_{y_f^2}$ are listed, respectively, in Tables~\ref{tbl:Deltars} and~\ref{tbl:Deltaws} of Appendix~\ref{sec:app_mapping}.} \label{tbl:epsRP_width}
\end{table}
\renewcommand\arraystretch{0}

\subsubsection{Brief description of the results}\label{sec:map_brief_descrip}

\begin{itemize}
  \item ``Interference Correction'' $\epsilon_I^{}$

  $\epsilon_I^{}$ captures the effects of new, amputated Feynman diagrams $iM_\text{AD,new}(c_i)$ introduced by the dim-6 effective operators. This modifies the value of the total amputated diagram
  \begin{equation}
  iM_\text{AD} = iM_\text{AD,SM} + iM_\text{AD,new}(c_i) .
  \end{equation}
  Upon modulus square, the cross term, namely the interference between the new amplitude and the SM amplitude, gives the leading order contribution to the deviation:
  \begin{equation}
  \epsilon_I^{} = \frac{\int{d\Pi_f \overline{M_\text{AD,SM}^*M_\text{AD,new}(c_i)+c.c.}}}{\int{d\Pi_f \overline{\left| M_\text{AD,SM} \right|^2}}} , \label{eqn:epsIdef}
  \end{equation}
  where $\int d\Pi_f$ denotes the phase space integral, and the overscore denotes any step needed for getting the unpolarized result, namely a sum of final spins and/or an average over the initial spins, if any. The results of $\epsilon_I$ are summarized in Table~\ref{tbl:epsI_width}. Details of the calculation are relegated to an appendix. Specifically, in Appendix~\ref{subsec:FeynmanRules} we list out the new set of Feynman rules generated by the dim-6 operators; in Appendix~\ref{subsec:GammaIdetails} we list out all the relevant new amputated diagrams involved in each $\epsilon_I^{}$. Due to the phase space integral, there are some complicated auxiliary integrals involved in the results. The definitions and values of these auxiliary integrals are given in Eq.~\eqref{eqn:Iadef}-\eqref{eqn:Iddef}. The $A_{hgg}^\text{SM}$, $A_{h\gamma\gamma}^\text{SM}$, and $A_{h\gamma Z}^\text{SM}$ in Table~\ref{tbl:epsI_width} are the standard form factors, detailed expressions of which are shown in Eq.~\eqref{eqn:Ahgg}-\eqref{eqn:AhAZ} of the appendix.

  \item ``Residue Correction'' $\epsilon_R^{}$

 $\epsilon_R^{}$ captures the effects of residue modifications at the pole mass, \textit{i.e.} wavefunction corrections, by the dim-6 effective operators. We know from the LSZ reduction formula that the invariant amplitude $iM$ equals the value of amputated diagram $iM_\text{AD}$ multiplied by the square root of the mass pole residue $r_k$ of each external leg particle $k$
  \begin{equation}
  iM = \left( \prod\limits_{k \in \{ \text{external legs} \} } r_k^{1/2} \right) \cdot iM_\text{AD} .
  \end{equation}
Besides the corrections to $iM_\text{AD}$ discussed before, a mass pole residue modification $\Delta r_k$ of an external leg particle $k$ also feeds into the decay width deviation. Upon modulus square, this part of deviation is
  \begin{equation}
  \epsilon_R^{} = \sum\limits_{k \in \{ \text{external legs} \} } \Delta r_k^{}.
  \end{equation}
  The results of $\epsilon_R^{}$ for each decay width are summarized in the second column of Table~\ref{tbl:epsRP_width}. The values of the relevant residue modifications $\Delta r_k$ are listed in Appendix~\ref{subsec:RMdetails} (Table~\ref{tbl:Deltars}). Note that, unlike the interference correction $\epsilon_I^{}$, the residue correction $\epsilon_R^{}$ corresponds to a contribution with the size of $\Gamma_\text{SM} \times c_i$. But for $\Gamma_{hgg}$, $\Gamma_{h\gamma\gamma}$, and $\Gamma_{h\gamma Z}$, the SM value $\Gamma_\text{SM}$ is already of one-loop size. So $\epsilon_{hgg,R}^{}$, $\epsilon_{h\gamma\gamma,R}^{}$, and $\epsilon_{h\gamma Z,R}^{}$ should be one-loop size in Wilson coefficients, namely that $\frac{1}{16\pi^2}\times c_i$. Therefore, to our order of approximation, this size should be neglected for consistency, hence why \(\e_R = 0\) for $\Gamma_{hgg}$, $\Gamma_{h\gamma\gamma}$, and $\Gamma_{h\gamma Z}$ in Table~\ref{tbl:epsRP_width}.

  \item ``Parametric Correction'' $\epsilon_P^{}$

  $\epsilon_P^{}$ captures how the dim-6 effective operators modify the parameters of the SM Lagrangian. When computing the decay width $\Gamma$, one usually writes it in terms of a set of Lagrangian parameters $\{\rho\}$, which in our case are $\{\rho\}=\{g^2, v^2, s_Z^2, y_f^2\}$. So $\Gamma=\Gamma(\rho,c_i)$ is what one usually calculates. However, the deviation $\epsilon$ is supposed to be a physical observable that describes the change of the relation between $\Gamma$ and other physical observables $\{obs\}$, which in our case can be taken as $\{obs\}=\{\hat\alpha, \hat{G}_F, \hat{m}_Z^2, \hat{m}_f^2\}$. So one should eliminate $\{\rho\}$ in terms of $\{obs\}$. This elimination brings additional dependence on $\{c_i\}$, because the Wilson coefficients also modify the relation between $\{\rho\}$ and $\{obs\}$ through $\rho=\rho(obs,c_i)$. Therefore, to include the full dependence on $c_i$, one should write the decay width as
  \begin{equation}
  \Gamma=\Gamma\big(\rho(obs,c_i),c_i\big) .
  \end{equation}
  The $\epsilon_I^{}$ and $\epsilon_R^{}$ discussed previously only take into account of the explicit dependence on $c_i$, with $\{\rho\}$ held fixed. The implicit dependence on $c_i$ through modifying the Lagrangian parameter $\rho$ is what we call ``parametric correction'':
  \begin{equation}
  \epsilon_P^{} = \sum\limits_{\rho  \in \{ {g^2},{v^2},s_Z^2,y_f^2\} } {\frac{{\partial \ln \Gamma (\rho ,{c_i})}}{{\partial \ln \rho }}\Delta \ln \rho}=\sum\limits_{\rho  \in \{ {g^2},{v^2},s_Z^2,y_f^2\} } {\frac{{\partial \ln \Gamma (\rho ,{c_i})}}{{\partial \ln \rho }}\Delta {w_\rho }} ,
  \end{equation}
  where $\Delta w_\rho$ denotes the Lagrangian parameter modification
  \begin{equation}
  \Delta w_\rho = \Delta\ln\rho  = \frac{\Delta\rho}{\rho} .
  \end{equation}
  The parametric correction $\epsilon_P^{}$ in terms of $\Delta w_\rho$ are summarized in the third column of Table~\ref{tbl:epsRP_width}. And a detailed calculation of $\Delta w_\rho$ is in Appendix~\ref{subsec:LPMdetails}, with the results summarized in Table~\ref{tbl:Deltaws}. As with the residue correction case, $\epsilon_{hgg,P}^{}$, $\epsilon_{h\gamma\gamma,P}^{}$, and $\epsilon_{h\gamma Z,P}^{}$ are one-loop size in Wilson coefficients and hence neglected for consistency.
\end{itemize}

\subsubsection{Detailed derivation}

Clearly from Eq.~\eqref{eqn:LeffSM}, the SM EFT goes back to the SM when all $c_i=0$. Thus, up to linear power of $c_i$, the deviation defined in Eq.~\eqref{eqn:epsdef} is
\begin{equation}
\epsilon \equiv \frac{\Gamma }{{{\Gamma _{{\rm{SM}}}}}} - 1 = \frac{{\Gamma ({c_i})}}{{\Gamma ({c_i} = 0)}} - 1 = {\left. {\frac{{d\ln \Gamma }}{{d{c_i}}}} \right|_{{c_i} = 0}}{c_i} . \label{eqn:epsdefext}
\end{equation}
As explained before, this function $\Gamma(c_i)$ in Eq.~\eqref{eqn:epsdefext} should be understood as the dependence of $\Gamma$ on $\{c_i\}$ with the values of $\{obs\}$ held fixed. Practically, it is most convenient to first compute both $\Gamma$ and $\{obs\}$ in terms of the Lagrangian parameters $\{\rho\}$:
\begin{eqnarray}
\Gamma &=& \Gamma(\rho,c_i) , \\
obs &=& obs(\rho,c_i) ,
\end{eqnarray}
One can then plug the inverse of the second function $\rho=\rho(obs,c_i)$ into the first to get
\begin{equation}
\Gamma(c_i) = \Gamma \big(\rho(obs,c_i),c_i\big) .
\end{equation}
This makes it clear that in addition to the explicit dependence on $c_i$, $\Gamma$ also has an implicit dependence on $c_i$ through the Lagrangian parameters $\rho(obs,c_i)$:
\begin{equation}
\frac{{d\ln \Gamma }}{{d{c_i}}} = \frac{{\partial \ln \Gamma (\rho ,{c_i})}}{{\partial {c_i}}} + \sum\limits_\rho  {\frac{{\partial \ln \Gamma (\rho ,{c_i})}}{{\partial \ln \rho }}\frac{{\partial \ln \rho (obs,{c_i})}}{{\partial {c_i}}}} . \label{eqn:dependence}
\end{equation}
Putting it another way, the first term in the above shows the deviation when $\rho$ are fixed numbers. But $\rho$ are not fixed numbers. They are a set of Lagrangian parameters determined by a set of experimental measurements $obs$ through relations that get modified by $c_i$ as well. So the truly fixed numbers are the experimental inputs $obs$. By adding the second piece in Eq.~\eqref{eqn:dependence}, we get the full amount of deviation with $obs$ as fixed input numbers. By putting $obs$ in the place of $\ln\Gamma$, one can also explicitly check that Eq.~\eqref{eqn:dependence} keeps $obs$ fixed. Making use of the fact
\begin{equation}
{\frac{{\partial \ln \rho (obs,{c_i})}}{{\partial {c_i}}} = {{\left. {\frac{{d\ln \rho }}{{d{c_i}}}} \right|}_{obs = {\rm{const}}}} =  - \frac{{{{\left. {\frac{{\partial (obs)}}{{\partial {c_i}}}} \right|}_\rho }}}{{{{\left. {\frac{{\partial (obs)}}{{\partial \ln \rho }}} \right|}_{{c_i}}}}}} ,
\end{equation}
we clearly see that
\begin{equation}
{\frac{{d(obs)}}{{d{c_i}}} = \frac{{\partial (obs)}}{{\partial {c_i}}} + \sum\limits_\rho  {{{\left. {\frac{{\partial (obs)}}{{\partial \ln \rho }}} \right|}_{{c_i}}}\frac{{\partial \ln \rho (obs,{c_i})}}{{\partial {c_i}}}}  = 0} .
\end{equation}

Because of Eq.~\eqref{eqn:dependence}, the deviation Eq.~\eqref{eqn:epsdefext} is split into two parts
\begin{eqnarray}
\epsilon &=& {\left. {\frac{{\partial \ln \Gamma (\rho ,{c_i})}}{{\partial {c_i}}}} \right|_{{c_i} = 0}}{c_i} + \sum\limits_\rho  {\left[ {{{\left. {\frac{{\partial \ln \Gamma (\rho ,{c_i})}}{{\partial \ln \rho }}} \right|}_{{c_i} = 0}}\left( {{{\left. {\frac{{\partial \ln \rho (obs,{c_i})}}{{\partial {c_i}}}} \right|}_{{c_i} = 0}}{c_i}} \right)} \right]} \nonumber \\
&=& {\left. {\frac{{\partial \ln \Gamma (\rho ,{c_i})}}{{\partial {c_i}}}} \right|_{{c_i} = 0}}{c_i} + \epsilon_P^{} ,
\end{eqnarray}
where the implicit dependence part is defined as the parametric correction $\epsilon_P^{}$
\begin{equation}
\epsilon_P^{} \equiv \sum\limits_\rho {{\left. {\frac{{\partial \ln \Gamma (\rho ,{c_i})}}{{\partial \ln \rho }}} \right|}_{{c_i} = 0}}\Delta w_\rho ,
\end{equation}
with the parameter modifications $\Delta w_\rho$ defined as
\begin{equation}
{\Delta {w_\rho } \equiv {{\left. {\frac{{\partial \ln \rho (obs,{c_i})}}{{\partial {c_i}}}} \right|}_{{c_i} = 0}}{c_i} = \Delta \ln \rho  = \frac{{\Delta \rho }}{\rho }} . \label{eqn:Deltawdef}
\end{equation}

The explicit dependence part can be further split by noting that
\begin{eqnarray}
i{M_{{\rm{AD}}}} &=& i{M_{{\rm{AD,SM}}}} + i{M_{{\rm{AD,new}}}}({c_i}) , \\
iM &=& \left( {\prod\limits_{k \in \{ {\text{external legs\} }}} {r_k^{1/2}} } \right) \cdot i{M_{{\rm{AD}}}} , \\
\Gamma (\rho ,{c_i}) &=& \frac{1}{{2{m_h}}}\int {d{\Pi _f}\overline {{{\left| M \right|}^2}} }  = \frac{1}{{2{m_h}}}\left( {\prod\limits_{k \in \{ {\text{external legs\} }}} {{r_k^{}}} } \right) \cdot \int {d{\Pi _f}\overline {{{\left| {{M_{{\rm{AD}}}}} \right|}^2}} } .
\end{eqnarray}
Therefore we have
\begin{eqnarray}
{\left. {\frac{{\partial \ln \Gamma (\rho ,{c_i})}}{{\partial {c_i}}}} \right|_{{c_i} = 0}}{c_i} &=& {\left. {\frac{{\partial \ln \left[ {\int {d{\Pi _f}\overline {{{\left| {{M_{AD}}} \right|}^2}} } } \right]}}{{\partial {c_i}}}} \right|_{{c_i} = 0}}{c_i} + \sum\limits_{k \in \{ {\text{external legs\} }}} {{{\left. {\frac{{\partial \ln {r_k}}}{{\partial {c_i}}}} \right|}_{{c_i} = 0}}{c_i}} \nonumber \\
 &=& {\left. {\frac{{\Delta \left( {\int {d{\Pi _f}\overline {{{\left| {{M_{{\rm{AD}}}}} \right|}^2}} } } \right)}}{{\int {d{\Pi _f}\overline {{{\left| {{M_{{\rm{AD}}}}} \right|}^2}} } }}} \right|_{{c_i} = 0}} + \sum\limits_{k \in \{ {\text{external legs\} }}} {{{\left. {\frac{{\Delta {r_k}}}{{{r_k}}}} \right|}_{{c_i} = 0}}} \nonumber \\
 &=& \frac{{\int {d{\Pi _f}\overline {M_{{\rm{AD,SM}}}^*{M_{{\rm{AD,new}}}}({c_i}) + c.c.} } }}{{\int {d{\Pi _f}\overline {{{\left| {{M_{{\rm{AD,SM}}}}} \right|}^2}} } }} + \sum\limits_{k \in \{ {\text{external legs\} }}} {\Delta {r_k}} \nonumber \\
 &=& \epsilon_I^{} + \epsilon_R^{} ,
\end{eqnarray}
with $\epsilon_I^{}$ and $\epsilon_R^{}$ defined as
\begin{eqnarray}
{\epsilon_I^{}} &\equiv& \frac{{\int {d{\Pi _f}\overline {M_{{\rm{AD,SM}}}^*{M_{{\rm{AD,new}}}}({c_i}) + c.c.} } }}{{\int {d{\Pi _f}\overline {{{\left| {{M_{{\rm{AD,SM}}}}} \right|}^2}} } }} , \\
{\epsilon_R^{}} &\equiv& \sum\limits_{i \in \{ {\text{external legs\} }}} {\Delta {r_i}} .
\end{eqnarray}

So in summary, the total deviation in decay width has three parts $\epsilon = \epsilon_I^{}+\epsilon_R^{}+\epsilon_P^{}$, with
\begin{eqnarray}
{\epsilon_I^{}} &=& \frac{{\int {d{\Pi _f}\overline {M_{{\rm{AD,SM}}}^*{M_{{\rm{AD,new}}}}({c_i}) + c.c.} } }}{{\int {d{\Pi _f}\overline {{{\left| {{M_{{\rm{AD,SM}}}}} \right|}^2}} } }} , \\
{\epsilon_R^{}} &=& \sum\limits_{i \in \{ {\text{external legs\} }}} {\Delta {r_i}} , \\
{\epsilon_P^{}} &=& \sum\limits_{\rho  \in \{ {g^2},{v^2},s_Z^2,y_f^2\} } {{{\left. {\frac{{\partial \ln \Gamma (\rho ,{c_i})}}{{\partial \ln \rho }}} \right|}_{{c_i} = 0}}\Delta {w_\rho }} ,
\end{eqnarray}
where
\begin{equation}
\Delta w_\rho \equiv \left. \frac{\partial\ln\rho(obs,c_i)}{\partial c_i} \right|_{c_i=0} c_i = \Delta\ln\rho = \frac{\Delta\rho}{\rho} .
\end{equation}
For each decay width in consideration, we computed these three parts of deviation. The results are summarized in Table~\ref{tbl:epsI_width} and Table~\ref{tbl:epsRP_width}. It is worth noting that this splitting is a convenient intermediate treatment of the calculation, but each of $\epsilon_I^{}$, $\epsilon_R^{}$, $\epsilon_P^{}$ alone would not be physical, because it depends on the renormalization scheme as well as the choice of operator basis. It is the total sum of the three that reflects the physical deviation in the decay widths.

\subsection{Deviations in Higgs production cross sections}

\begin{figure}[t]
 \centering
 \includegraphics[height=7cm]{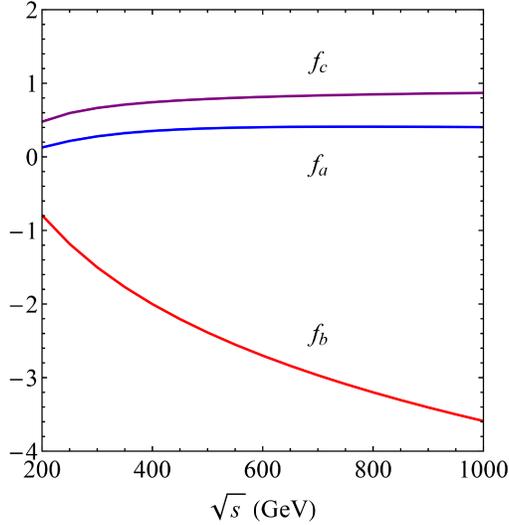}
 \caption{Numerical results of auxiliary functions $f_a(s)$, $f_b(s)$, and $f_c(s)$ in $\epsilon_{WWh,I}^{}(s)$. Mathematica code for these auxiliary functions can be found at~\href{http://hitoshi.berkeley.edu/HiggsEFT/auxiliary.html}{http://hitoshi.berkeley.edu/HiggsEFT/auxiliary.html}.} \label{fig:plotf}
\end{figure}

The dim-6 operators also induce deviations in the Higgs production cross sections. In this paper, we focus on the production modes $\sigma \in \big\{\sigma_{ggF}^{}, \sigma_{WWh}^{}, \sigma_{Wh}^{}, \sigma_{Zh}^{}\big\}$, which are the most important ones for both hadron colliders such as the LHC and possible future lepton colliders such as the ILC. As with the decay width case, we define the cross section deviation
\begin{equation}
\epsilon \equiv \frac{\sigma}{\sigma_\text{SM}} - 1 .
\end{equation}
Again, there are three types of corrections
\begin{equation}
\epsilon=\epsilon_I^{}+\epsilon_R^{}+\epsilon_P^{} .
\end{equation}
The mapping results are summarized in Table~\ref{tbl:epsI_sigma} and Table~\ref{tbl:epsRP_sigma}. Relevant new amputated Feynman diagrams for $\epsilon_I^{}$ are listed in Appendix~\ref{subsec:SigmaIdetails}. The calculation of the interference correction to $\sigma_{WWh}^{}$ turns out to be very involved. Its lengthy analytical expression $\epsilon_{WWh,I}^{}(s)$ does not help much, so we instead show its numerical results in Table~\ref{tbl:epsI_sigma}. The auxiliary functions $f_a(s)$, $f_b(s)$, $f_c(s)$ in $\epsilon_{WWh,I}^{}(s)$ are defined in Appendix~\ref{subsec:SigmaIdetails} (Eq.~\eqref{eqn:fadef}-\eqref{eqn:fcdef}), where more details of the phase space integral are also shown. The numerical values of $f_a(s)$, $f_b(s)$, $f_c(s)$ are plotted in Fig.~\ref{fig:plotf}. We also provide \href{http://hitoshi.berkeley.edu/HiggsEFT/auxiliary.html}{Mathematica code} so that one can make use of these auxiliary funcitons.\footnote{This code can be found at~\href{http://hitoshi.berkeley.edu/HiggsEFT/auxiliary.html}{http://hitoshi.berkeley.edu/HiggsEFT/auxiliary.html}}

\renewcommand\arraystretch{2.0}
\begin{table}[tb]
\centering
\begin{tabular}{|rl|}
  \hline
  $\epsilon_{ggF,I}^{}=$ & \hspace{-4mm} $\dfrac{(4\pi)^2}{{\mathop{\rm Re}\nolimits}(A_{hgg}^\text{SM})}\dfrac{16v^2}{\Lambda^2}c_{GG}^{}$ \\
  $\epsilon_{WWh,I}^{}(s)=$ & \hspace{-4mm} $\Big[-f_b(s)-f_c(s)\Big]\dfrac{2m_W^2}{\Lambda^2}c_{2W}^{} +\Big[-f_a(s)+2f_c(s)\Big] \dfrac{8m_W^2}{\Lambda^2}c_{WW}^{}$ \\
  & \hspace{-4mm} $+\Big[f_b(s)+2f_c(s)\Big] \dfrac{2m_W^2}{\Lambda^2}c_W^{} + f_c(s)\dfrac{2v^2}{\Lambda^2}c_R^{} + \dfrac{2m_h^2}{\Lambda^2}c_D^{}$ \\
  $\epsilon_{Wh,I}^{}=$ & \hspace{-4mm} $\dfrac{1}{{1 - \eta_W^2}}\left[ \begin{array}{l}
 - \dfrac{{2s}}{{{\Lambda ^2}}}{c_{2W}^{}} + {I_{VH}}({\eta_h^{}},{\eta_W^{}})\dfrac{{16m_W^2}}{{{\Lambda^2}}}{c_{WW}^{}}\\
 + (1 + 2\eta_W^2 - \eta_W^4)\dfrac{{2s}}{{{\Lambda^2}}}{c_W^{}} + (2 - \eta_W^2)\dfrac{{2{v^2}}}{{{\Lambda ^2}}}{c_R^{}}
\end{array} \right] + \dfrac{{2m_h^2}}{{{\Lambda^2}}}{c_D^{}}$ \\
  $\epsilon_{Zh,I}^{}=$ & \hspace{-4mm} $\dfrac{1}{{1 - \eta_Z^2}}\left[ \begin{array}{l}
 - \dfrac{{2s}}{{{\Lambda^2}}}\left(c_Z^2{c_{2W}^{}} + s_Z^2{c_{2B}^{}}\right)\\
 + {I_{VH}}({\eta_h^{}},{\eta_Z^{}})\dfrac{{16m_Z^2}}{{{\Lambda^2}}}\left(c_Z^4{c_{WW}^{}} + s_Z^4{c_{BB}^{}} + c_Z^2s_Z^2{c_{WB}^{}}\right)\\
 + \left(1 + 2\eta_Z^2 - \eta_Z^4\right)\dfrac{{2s}}{{{\Lambda^2}}}\left(c_Z^2{c_W^{}} + s_Z^2{c_B^{}}\right)\\
 + \left(2 - \eta_Z^2\right)\dfrac{{2{v^2}}}{{{\Lambda^2}}}\left( -2{c_T^{}} + {c_R^{}}\right)
\end{array} \right] + \dfrac{{2m_h^2}}{{{\Lambda^2}}}{c_D^{}}$ \\
   & \hspace{-4mm} $ + \dfrac{2e{Q_f}{c_Z^2}{s_Z^{}}}{{g(T_f^3 - s_Z^2{Q_f})}}\left\{ \begin{array}{l}
 - \dfrac{s}{{{\Lambda ^2}}}\left({c_{2W}^{}} - {c_{2B}^{}} - {c_W^{}} + {c_B^{}}\right)\\
 + {I_{VH}}({\eta_h^{}},{\eta_Z^{}})\dfrac{{4m_Z^2}}{\Lambda^2}\Big[2c_Z^2{c_{WW}^{}}-2s_Z^2{c_{BB}^{}}- \left(c_Z^2-s_Z^2\right){c_{WB}^{}}\Big]
\end{array} \right\}$ \\
  \hline
\end{tabular}
\caption{Interference corrections $\epsilon_I^{}$ to Higgs production cross sections, with $\eta_h^{} \equiv \frac{m_h}{\sqrt s}$, $\eta_Z^{} \equiv \frac{m_Z}{\sqrt s}$, and the auxiliary function defined as $I_{VH}(\eta_h^{}, \eta_V^{}) \equiv 1 + \frac{6(1-\eta_h^2+\eta_V^2)(1-\eta_V^2)}{(1 - \eta_h^2 + \eta_V^2)^2 + 8\eta_V^2}$. The numerical results of the auxiliary functions $f_a(s)$, $f_b(s)$, and $f_c(s)$ in $\epsilon_{WWh,I}^{}(s)$ are shown in Fig.~\ref{fig:plotf}.} \label{tbl:epsI_sigma}
\vspace{-5pt}
\end{table}
\renewcommand\arraystretch{1.0}

\renewcommand\arraystretch{2.0}
\begin{table}[tb]
\centering
\begin{tabular}{|c|c|c|}
  \hline
   & $\epsilon_R^{}$ & $\epsilon_P^{}$ \\
  \hline
  $\sigma_{ggF}^{}$ & $0$ & $0$ \\
  $\sigma_{WWh}^{}$ & $\Delta r_h^{}$ & $4\Delta w_{g^2}+\Delta w_{v^2}$ \\
  $\sigma_{Wh}^{}$ & $\Delta r_h^{}+\Delta r_W^{}$ & $3\Delta w_{g^2}+\Delta w_{v^2}$ \\
  $\sigma_{Zh}^{}$ & $\Delta r_h^{}+\Delta r_Z^{}$ & $3\Delta w_{g^2}+\Delta w_{v^2}+\left(3\dfrac{s_Z^2}{c_Z^2}-\dfrac{2s_Z^2 Q_f}{T_f^3 - s_Z^2 Q_f}\right)\Delta w_{s_Z^2}$ \\
  \hline
\end{tabular}
\caption{Residue corrections $\epsilon_R^{}$ and parametric corrections $\epsilon_P^{}$ to Higgs production cross sections. The results of residue modifications and parameter modifications are listed in Tables~\ref{tbl:Deltars} and~\ref{tbl:Deltaws} of Appendix~\ref{sec:app_mapping}.} \label{tbl:epsRP_sigma}
\vspace{-5pt}
\end{table}
\renewcommand\arraystretch{1.0}

\section{Summary of results} \label{sec:conclusions}


In the vein of studying how specific new models of physics affect precision observables, we have aimed in this work to provide tools to easily make use of the Standard Model effective field theory. As with any EFT, there is a practical three-step procedure that one makes use of: matching the UV theory onto the EFT at the scale where heavy states are integrated out, RG evolving the EFT down to the scale where measurements are made, and mapping the EFT onto observables at the measurement scale. While each of these steps is straightforward in the abstract, in practice they can be complicated for the SM EFT primarily due to the large number of higher dimension operators in the SM EFT. Here we provide a summary of some of the central results in this paper.

In section~\ref{sec:CDE} we developed the covariant derivative expansion, which allows one to compute the tree and one-loop effective action at the matching scale in a manifestly gauge covariant fashion. This calculation at tree-level is particularly obvious, and was explained in Sec.~\ref{sec:CDE_tree}. At one-loop, the effective action can be brought to the form (Eq.~\eqref{eqn:sum_CDE_lag})
\begin{equation*}
\D S_{\text{eff,1-loop}} = ic_s\Tr \log\big[-P^2 + m^2 +U(x)\big] =ic_s \int d^4x \, \frac{d^4q}{(2\pi)^4} \, \tr \log\left[-\left(q_{\m}+\tG_{\n\m}\frac{\pd}{\pd q_{\n}}\right)^2 +m^2 + \tU\right]
\end{equation*}
where \(\tG_{\n\m}\) and \(\tU\), Eq.~\eqref{eqn:sum_tilde_defs}, are expansions containing HDOs through commutators of the covariant derivative \(P_{\m}\) with itself and the low-energy (SM) fields in \(U(x)\), together with derivatives of the auxiliary momentum \(q_{\m}\). The general form of \(U(x)\) for scalars, fermions, and vector bosons is summarized in Sec.~\ref{sec:CDE_fermions}.

The above effective action is then evaluated in an inverse mass expansion, leading to universal formulas for the one-loop effective action. In the case that \(m^2\) commutes with \(U(x)\), we explicitly performed this covariant derivative expansion and the general results up through dimension-six operators is given in Eq.~\eqref{eqn:CDE_universal_lag}. With these results, in Sec.~\ref{sec:CDE_examples} we computed the Wilson coefficients of dimension-six operators for numerous physically interesting and non-trivial models of new physics. Besides the inherent physical interest of the UV models considered, these examples hopefully offer a pedagogical explanation of how the CDE can be used to easily obtain the effective action at the matching scale.

In section~\ref{sec:operatorbasis} we considered the step of RG running Wilson coefficients at the matching scale down to the observation scale. At leading order, this involves making use of the anomalous dimension matrix \(\g_{ij}\). In the past few years, there has been great progress on computing \(\g_{ij}\). Instead of examining the technical details of this calculation, we explored the questions of when are these results needed and how to make use of them. Due to the per-mille sensitivity of present and future precision measurements, as a general rule of thumb RG running needs to be considered only when Wilson coefficients are generated at tree-level. If one does need to make use of RG evolution, the most practical ingredient one needs to understand to make use of existing computations of \(\g_{ij}\) concerns RG closure and choice of an operator basis. We provided a brief explanation of the choice of operator sets as well as common operator bases in the literature and how one can go between these bases.

Finally, in section~\ref{sec:mapping} we studied how higher dimension operators impact precision observables. In particular, we computed the impact of all purely bosonic dimension-six operators (Table~\ref{tbl:operators2}) on electroweak precision observables, Higgs' decay widths, and Higgs production cross sections. This calculation was done to leading (linear) order in the Wilson coefficients. While various parts of these results have been computed in the literature previously, we believe our results offer the first complete and systematic results for the bosonic operators we considered.

The effect of the bosonic HDOs on the electroweak precision observables and triple gauge couplings can be found in Tables~\ref{tbl:EWPOresults} and~\ref{tbl:TGCresults}, respectively. For the Higgs decay widths and production cross sections, we considered the deviations that the HDOs lead to relative to the SM prediction, 
\begin{equation*}
\e = \frac{\G}{\G_{\text{SM}}} - 1 \text{ and } \e = \frac{\s}{\s_{\text{SM}}} - 1.
\end{equation*}
These deviations can be further refined into the impact of the HDOs in diagrammatic interference, residue (wavefunction) corrections, and changes to Lagrangian parameters (Sec~\ref{sec:map_brief_descrip}). In other words,
\begin{equation*}
\e = \e_I^{} + \e_R^{} + \e_P^{},
\end{equation*}
where \(\e_{I,R,P}^{}\) stand for interference, residue, and parametric corrections, respectively. The values of \(\e_{I,R,P}^{}\) in terms of the dimension-six Wilson coefficients can be found in Tables~\ref{tbl:epsI_width} and~\ref{tbl:epsRP_width} for Higgs decay widths and Tables~\ref{tbl:epsI_sigma} and~\ref{tbl:epsRP_sigma} for Higgs production cross sections.

Besides being the appropriate, model-independent framework to study precision observables, effective field theory provides great simplification to studying how specific new models of physics impact precision observables. We have outlined in detail the algorithmic procedure for doing this with the SM EFT. Given a UV model, one can easily match it onto the SM EFT using the covariant derivative expansion. One then decides if RG running down to the weak scale is of practical relevance; if it is, existing computations of the anomalous dimension matrix can be employed to do this step. At the weak scale, one then simply takes the Wilson coefficients of the bosonic operators and plugs them into Tables~\ref{tbl:EWPOresults}-\ref{tbl:epsRP_sigma} to study the deviations the UV model induces on electroweak and Higgs observables. We hope that the tools and results developed in this work not only highlight the utility of the SM EFT, but also demonstrate how one can use the SM EFT with relative ease.

\acknowledgments
\noindent We thank Sally Dawson and Matthew McCullough for useful discussions. BH is grateful to the Brantley-Tuttle fellowship for support while this work was completed. This work was supported by the U.S. DOE under Contract DE-AC03-76SF00098, by the NSF under grants PHY-1002399 and PHY-1316783. HM was also supported by the JSPS grant (C) 23540289, and by WPI, MEXT, Japan.

\newpage
\appendix

\section{Supplemental details for the CDE}

This appendix shows some details in using the CDE method. First, in appendix~\ref{sec:app_cde_ferm}, we present some details of the derivation of CDE for fermions and gauge bosons. Appendix~\ref{sec:app_cde_identity} then list out quite a bit useful identities that one frequently encounters while using CDE. Finally, appendix~\ref{sec:app_CDE_universal} shows intermediate steps in deriving the universal formula of the CDE.

\subsection{CDE for fermions and gauge bosons}\label{sec:app_cde_ferm}

\subsubsection*{Fermions}
We now consider the functional determinant for massive fermion fields and provide the formulas for the covariant derivative expansion for them. We work in the notation of Dirac fermions, denoting the gamma matrices by \(\g^{\m}\) and employing slashed notation, {\it e.g.} \(\slashed{D} = \g^{\m}D_{\m}\). This discussion is easily modified if one wants to consider Weyl fermions and use two-component notation.

Consider the Lagrangian containing the fermions to be
\begin{equation}
\scL[\ps,\ph]  = \bps\big(i\slashed{D} - m - M(x) \big) \ps,
\end{equation}
where \(m\) is the fermion mass and \(M(x)\) is in general dependent on the light fields \(\ph(x)\). Upon integrating over the Grassman valued fields in the path integral, the one-loop contribution to the effective action is given by
\begin{equation}
S_{\text{eff,1-loop}} \equiv \D\Seff = -i \Tr \log \big(\sP- m- M \big),
\end{equation}
where, as before, \(P_{\m} \equiv i D_{\m}\). Using \(\Tr \log AB = \Tr \log A + \Tr \log B\) and the fact that the trace is invariant under changing signs of gamma matrices we have
\begin{align}
\D\Seff &= -\frac{i}{2} \Big[ \Tr \log \big(-\sP -m - M \big) + \Tr \log \big(\sP - m - M \big)\Big] \nonumber \\
&=-\frac{i}{2} \Tr \log \Big( -\sP^{2} + m^2 + 2mM + M^2 + \sP M \Big). \label{eqn:ferm_sq}
\end{align}
where \(\sP M \equiv [\sP,M]\), as defined in Eq.~\eqref{eqn:notation_short}. With \(\g^{\m}\g^{\n} = (\{\g^{\m},\g^{\n}\} + [\g^{\m},\g^{\n}])/2 = g^{\m\n} - i\s^{\m\n}\),
\begin{equation}
\sP^2 = P^2 + \frac{i}{2} \s^{\m\n}[D_{\m},D_{\n}] = P^2 + \frac{i}{2} \s \cdot G',
\end{equation}
where \(G_{\m\n}' \equiv [D_{\m},D_{\n}]\), as defined in Eq.~\eqref{eqn:notation_short}.

We thus see that the trace for fermions,
\begin{equation}
\Tr \log \Big( -P^2 + m^2  -\frac{i}{2}\s^{\m\n}G'_{\m\n} + 2mM + M^2 + \sP M \Big),
\label{eqn:ferm_trace}
\end{equation}
is of the form \(\Tr \log (-P^2 + m^2 + U)\). Therefore, all the steps in evaluating the trace and shifting by the covariant derivative using \(e^{\pm P\cdot \pd/\pd q}\) are the same as previously considered and we can immediately write down the answer from Eq.~\eqref{eqn:sum_CDE_lag}. Defining
\begin{equation}
U_{\text{ferm}} \equiv -\frac{i}{2}\s^{\m\n}G'_{\m\n} + 2mM + M^2 + \sP M,
\label{eqn:U_ferm}
\end{equation}
the one-loop effective Lagrangian for fermions is then given by
\begin{equation}
\D \scL_{\text{eff,ferm}} = -\frac{i}{2} \int dq \, \tr \, \log\Big[ -\Big(q_{\m} + \tG_{\n\m}\pd_{\n}\Big)^2 + \tU_{\text{ferm}} \Big] ,
\label{eqn:Leff_ferm}
\end{equation}
where \(\tG\) and \(\tU_{\text{ferm}}\) are defined as in Eq.~\eqref{eqn:sum_tilde_short} with \(U \to U_{\text{ferm}}\).

We note that the result originally obtained in~\cite{Gaillard:1985} contains an error (see Eq.~(4.21) therein compared to our result Eq.~\eqref{eqn:Leff_ferm}). This mistake originates from an error in Eq.~(4.17) of~\cite{Gaillard:1985} where a term proportional to $[{\tilde G}_{\mu\nu}\pd_\nu, {\tilde G}_{\rho\sigma}\pd_\sigma] \ne 0$ was missing.

\subsubsection*{Massless gauge bosons}
Here we consider the one-loop contribution to the 1PI effective action from massless gauge fields. The spirit here is slightly different from our previous discussions involving massive scalars and fermions; we are not integrating the gauge bosons out of the theory but instead are evaluating the 1PI effective action. Nevertheless, the manipulations are exactly the same since the one-loop contribution to the 1PI effective action is still a functional trace of the form \(\Tr \log(D^2 + U)\).

In evaluating the 1PI effective action, we split the gauge boson into a background piece plus fluctuations around this background, \(A_{\m} = A_{B,\m} + Q_{\m}\), and perform the path integral over the fluctuations \(Q_{\m}\) while holding the background \(A_{B,\m}\) fixed. In order to do the path integral, one must gauge fix the \(Q_{\m}\) fields. At first glance, one might think that gauge fixing destroys the possibility of keeping gauge invariance manifest while evaluating the one-loop effective action. However, this turns out not to be the case. It is well known that there is a convenient gauge fixing condition that leaves the gauge symmetry of the background \(A_{B,\m}\) field manifest, {\it i.e.} it only gauge fixes \(Q_{\m}\) and not \(A_{B,\m}\). This technique is known as the background field method (for example, see~\cite{Abbott:1981} and references therein).\footnote{All techniques of evaluating effective actions are, by the definition of holding fields fixed while doing a path integral, background field methods. Nevertheless, the term ``background field method'' is usually taken to refer to employing this special gauge fixing condition while evaluating the 1PI effective action.} Because the gauge symmetry of the background \(A_{B,\m}\) field is not fixed, we will still be able to employ the techniques of the covariant derivative expansion, allowing a manifestly gauge invariant computation of the one-loop effective action.

The issues around gauge symmetry are actually quite distinct for the background field method versus the CDE. However, because similar words are used in both discussions, it is worth clarifying what aspects of gauge symmetry are handled in each case. The background field method makes it manifestly clear that the effective action of \(A_{B,\m}\) possesses a gauge symmetry by only gauge fixing the fluctuating field \(Q_{\m}\). This is an all orders statement. However, when evaluating the effective action order-by-order, one still works with the non-covariant quantities \(A_{B,\m}\), \(Q_{\m}\), and  \(\pd/\pd x^{\m}\) at intermediate steps.\footnote{To one-loop order, one only deals with \(A_{B,\m}\) and \(\pd_{\m}\).} The covariant derivative expansion, on the other hand, is a technique for evaluating the {\it one-loop} effective action that keeps gauge invariance manifest at all stages of the computation by working with gauge covariant quantities such as \(D_{\m}\). To understand this point more explicitly, one can compare the method of the CDE presented in this paper and in~\cite{Cheyette:1987} with the evaluation of the functional determinant using the component fields as presented in detail in Peskin and Schroeder~\cite{Peskin:1995}.

Now onto the calculation, we take pure \(SU(N)\) gauge theory,
\begin{equation}
\scL[A_{\m}] = - \frac{1}{2Ng^2} \tr \, F_{\m\n}^2 = - \frac{1}{4g^2} \big(F_{\m\n}^a\big)^2,
\end{equation}
where \(F_{\m\n} = F_{\m\n}^at^a\) and we take the \(t^a\) in the adjoint representation, \(\tr\, t^a t^b = N \d^{ab}, \ (t^b)_{ac} = if^{abc}\).
We denote the covariant derivative as \(\mathcal{D}_{\m} = \pd_{\m} - i A_{\m}\) with the field strength defined as usual, \(F_{\m\n} = i[\mathcal{D}_{\m},\mathcal{D}_{\n}]\). Note that we have normalized the gauge field such that the coupling constant does not appear in the covariant derivative.

Let \(\G[A_{B}]\) be the 1PI effective action. To find \(\G[A_B]\), we split the gauge field into a background piece and a fluctuating piece, \(A_{\m} = A_{B,\m} + Q_{\m}\), and integrate out the \(Q_{\m}\) fields.\footnote{To keep our discussion short, we are being slightly loose here. In particular, a source term \(J\) for the fluctuating fields needs to be introduced. After integrating out the fluctuating field, we obtain an effective action which is a functional of \(J\) and the background fields, \(W[J,A_B]\). The 1PI effective action, \(\G[A_B]\), is obtained by a Legendre transform of \(W\). For more details see, for example,~\cite{Abbott:1981}.} The one-loop contribution to \(\G\) comes from the quadratic terms in \(Q_{\m}\). We have
\begin{subequations}
\begin{align}
\mathcal{D}_{\m} &= \pd_{\m} - i (A_{B,\m} + Q_{\m}) \equiv D_{\m} - i Q_{\m} , \\
F_{\m\n} &= i[D_{\m},D_{\n}] + D_{\m}Q_{\n} - D_{\n}Q_{\m} - i[Q_{\m},Q_{\n}] \equiv G_{\m\n} + Q_{\m\n}  - i[Q_{\m},Q_{\n}] , \\
\scL &= -\frac{1}{2Ng^2} \Tr \big(G_{\m\n} + Q_{\m\n} - i[Q_{\m},Q_{\n}] \big)^2 . \label{eqn:YM_BFM}
\end{align}
\end{subequations}
Note that \(D_{\m} = \pd_{\m} - i A_{B,\m}\) and \(G_{\m\n} = i [D_{\m},D_{\n}]\) are the covariant derivative and field strength of the background field alone.

In order to get sensible results out of the path integral, we need to gauge fix. As in the background field method, we employ a gauge fixing condition which is covariant with respect to the background field \(A_{B,\m}\). Namely, the gauge-fixing condition \(G^a\) is taken to be \(G^a = D^{\m}Q_{\m}^a\). The resultant gauge-fixed Lagrangian---including ghosts to implement the Fadeev-Popov determinant---is, {\it e.g.}~\cite{Abbott:1981,Peskin:1995},
\begin{equation}
\scL_{g.f.} + \scL_{gh} = -\frac{1}{2g^2\x}\big(D^{\m}Q^a_{\m}\big)^2 +D^{\m}\xoverline{c}^a\big(D_{\m}c^a + f^{abc}Q_{\m}^bc^c\big),
\label{eqn:gauge_fix_L}
\end{equation}
where \(\x\) is the gauge-fixing parameter. The utility of this gauge fixing condition is that the fluctuating \(Q_{\m}^a\) is gauge fixed while the Lagrangian~\eqref{eqn:YM_BFM} together with \(\scL_{g.f.} + \scL_{gh}\) possesses a manifest gauge symmetry with gauge field \(A_{B,\m}\) that is not gauge fixed. Thus we can perform the path integral over \(Q_{\m}^a\) while leaving the gauge invariance of the effective action of \(A_{B,\m}\) manifest. Under a background gauge symmetry transformation, \(A_{B,\m}\) transforms as a gauge field, \(A_{B,\m} \to V(A_{B,\m} + i \pd_{\m})V^{\dag}\) while \(Q_{\m}\) (and the ghosts \(c\) and \(\xoverline{c}\)) transforms simply as a field in the adjoint representation, \(Q_{\m} \to V Q_{\m} V^{\dag}\).  Procedurally, when performing the path integral over \(Q\) and \(c\), one can simply think about these fields as regular scalar and fermion\footnote{Of course ghosts aren't fermions; they are anti-commuting scalars. We are speaking very loosely and by fermion we are referring to their anti-commuting properties.} fields in the adjoint of some gauge symmetry and with interactions dictated by the Lagrangians in~\eqref{eqn:YM_BFM} and~\eqref{eqn:gauge_fix_L}. 

The quadratic piece of the combined Yang-Mills, gauge-fixing, and ghost Lagrangian is
\begin{equation}
\scL = - \frac{1}{2g^2} Q_{\m}^a \Big[ - g^{\m\n} (D^2)^{ac} - \frac{1-\x}{\x} (D^{\m}D^{\n})^{ac} - 2 f^{abc}G^{b\, \m\n}\Big]Q_{\n}^c + \xoverline{c}^a\big[- (D^2)^{ac}\big]c^c.
\end{equation}
We will work in Feynman gauge with \(\x = 1\) so that we can drop the \(D^{\m}D^{\n}\) term. Note that everything inside the square brackets in the above is in the adjoint representation (recall, \(f^{abc} = -i (t^b)_{ac}\)). Using the generator for Lorentz transformations on four-vectors, \((\mathcal{J}_{\r\s})^{\m\n} = i (\d^{\m}_{\r}\d^{\n}_{\s} - \d^{\n}_{\r}\d^{\m}_{\s})\), we can write
\begin{equation*}
G^{\m\n} = -\frac{i}{2} \big(G^{\r\s}\mathcal{J}_{\r\s}\big)^{\m\n}.
\end{equation*}
The quadratic piece of the Lagrangian is then given by
\begin{equation}
\scL = - \frac{1}{2g^2} Q_{\m}^a \big[ -  D^2 \mathbf{1}_4  + G\cdot \mathcal{J} \big]^{\m,ac}_{\n}Q^{\n,c} + \xoverline{c}^a\big[- D^2\big]^{ac}c^c,
\label{eqn:massless_gauge_quadratic}
\end{equation}
where \(\mathbf{1}_4\) is the \(4\times 4\) identity matrix for the Lorentz indices, {\it i.e.} \((\mathbf{1}_4)^{\m}_{\n} = \d^{\m}_{\n}\). Performing the path-integral over the gauge and ghost fields we obtain
\begin{equation}
\G_{\text{1-loop}}[A_B] = \frac{i}{2} \Tr \log \big( D^2 \mathbf{1}_4 - G\cdot \mathcal{J}\big)- i \Tr \log\big( D^2 \big),
\label{eqn:Seff_massless_gauge}
\end{equation}
where the factor of \(1/2\) in the first term is because the \(Q_{\m}^a\) are real bosons, while the factor of \(-1\) in the second term is because the \(c^a\) are anti-commuting. Note that the functional traces makes totally transparent the role of the ghosts. The trace of the gauge boson term containing \(D^2\) picks up a factor of 4 from the trace over Lorentz indices, one for each \(Q_{\m}\) \(\m = 0,1,2,3\). Of course, the gauge boson only has two physical degrees of freedom; we see explicitly above that the ghost piece cancels the contribution of two of the degrees of freedom.

Each of the traces in the above are of the form \(\Tr( -P^2 + U)\), and thus we can immediately apply the transformations leading to the covariant derivative expansion. Switching to our notation \(G'_{\m\n} = [D_{\m},D_{\n}] = -i G_{\m\n}\) and defining
\begin{equation}
U_{\text{gauge}} = -i \mathcal{J}^{\m\n}G'_{\m\n},
\label{eqn:U_gauge}
\end{equation}
we have
\begin{align}
\G_{\text{1-loop}}[A_B] = \frac{i}{2} &\int dx \, dq \, \tr \, \log \Big[ -\Big(q_{\m} + \tG_{\n\m}\pd_{\n}\Big)^2 + \tU_{\text{gauge}} \Big] \nonumber \\
&- i \int dx \, dq \, \tr\, \log \Big[ -\Big(q_{\m} + \tG_{\n\m}\pd_{\n}\Big)^2\Big] ,
\label{eqn:Gamma_gauge}
\end{align}
where \(\tG\) and \(\tU_{\text{gauge}}\) are defined as in Eq.~\eqref{eqn:sum_tilde_short} with \(U \to U_{\text{ferm}}\). The first term in the above is from the fluctuating gauge fields, while the second is from the ghosts. Note also that the trace ``tr'' in the first term includes over the Lorentz indices, just as the trace for fermions in Eq.~\eqref{eqn:Leff_ferm} is over the Lorentz (spinor) indices. In fact, it should be clear that \(U_{\text{gauge}}\) is very similar to the first term in \(U_{\text{ferm}}\) (Eq.~\eqref{eqn:U_ferm}): \(U_{\text{ferm}} \supset -i (\s^{\m\n}/2) G'_{\m\n}\) where \(\s^{\m\n}/2\) is the generator for Lorentz transformations on spinors.

Note that the effective action~\eqref{eqn:Gamma_gauge} contains infrared divergences from the massless gauge and ghost fields that we integrated out. These divergences can be regulated by adding a mass term for \(Q_{\m}^a\) and \(c^a\) because these mass terms respect the gauge invariance of the background field \(A_{B,\m}\).\footnote{As stated previously, procedurally one can just think of \(Q_{\m}\) and \(c\) as scalars and fermions transforming in the adjoint of some gauge symmetry whose gauge field is \(A_{B,\m}\). Just as scalars and fermions can have mass terms without disturbing gauge-invariance, \(Q_{\m}\) and \(c\) can have mass terms without disturbing the background gauge-invariance.}

\subsubsection*{Massive gauge bosons}
With our understanding of the story for massless gauge bosons, it turns out to be simple to obtain the result for massive gauge bosons. We consider massive vector bosons \(Q_{\m}\) transforming under an unbroken, low-energy gauge group. As is well known, beyond tree-level perturbation theory, the Nambu-Goldstone bosons (NGBs) \(\chi^i\) ``eaten'' by the massive vector boson must be included, {\it i.e.} we cannot work in unitary gauge. By working in a generalized \(R_{\x}\) gauge, we will be able to maintain manifest covariance of the low-energy gauge group. As we will see, mathematically, the results are essentially the same as the the massless case in Eqs.~\eqref{eqn:massless_gauge_quadratic} and~\eqref{eqn:Seff_massless_gauge}, modified by the presence of mass terms for the \(Q_{\m}\) and ghosts as well as an additional term for the NGBs.

First, as we mentioned in the main text, the gauge-kinetic piece of the Lagrangian up to quadratic term in $Q_\mu^i$ is
\begin{equation}
{\cal L}_\text{g.k.} \supset \frac{1}{2} Q_\mu^i \left( D^2 g^{\mu\nu} - D^\nu D^\mu + [D^\mu,D^\nu] \right)^{ij} Q_\nu^j , \label{eqn:LgkAppA}
\end{equation}
where $D_\mu$ denotes the covariant derivative that contains only the unbroken gauge fields. {\it A priori}, one may think that the coefficient of the magnetic dipole term, \(Q_\mu^i [D^\mu,D^\nu]^{ij} Q_\nu^j\), could be a free parameter. However, tree-level unitarity forces it be universally unity in the above equation, regardless of the details of symmetry breaking~\cite{Weinberg:1970,Ferrara:1992yc}. In appendix~\ref{sec:app_gkpiece}, we provide a new, algebraic derivation of this universality and also explain it via the physical argument of tree-level unitarity.

Second, because we are integrating out the heavy gauge bosons $Q_\mu^i$ perturbatively, we need to fix the part of gauge transformation corresponding to $Q_\mu^i$. But we would also like to preserve the unbroken gauge symmetry. To achieve this, we can adopt a generalized $R_\xi$ gauge fixing term as following
\begin{equation}
{{\cal L}_\text{g.f.}} =  - \frac{1}{{2\xi }}{\left( {\xi {m_Q}{\chi ^i} + {D^\mu }Q_\mu ^i} \right)^2} , \label{eqn:LgfAppA}
\end{equation}
where $\partial^\mu Q_\mu^i$ from the usual $R_\xi$ gauge fixing is promoted to $D^\mu Q_\mu^i$ to preserve the unbroken gauge symmetry.

Now combining Eq.~\eqref{eqn:LgkAppA} and~\eqref{eqn:LgfAppA} with the appropriate ghost term
\begin{equation}
{{\cal L}_\text{ghost}} = {{\bar c}^i}{\left( { - {D^2} - \xi m_Q^2} \right)^{ij}}{c^j} ,
\end{equation}
the mass term of $Q_\mu^i$ due to the symmetry breaking,
\begin{equation}
{\cal L}_\text{mass} \supset \frac{1}{2}{\left( {{D_\mu }{\chi ^i} - {m_Q}Q_\mu ^i} \right)^2} ,
\end{equation}
and a generic interaction term quadratic in $Q_\mu^i$,
\begin{equation}
{\cal L}_\text{I} = \frac{1}{2} Q_\mu^i \left(M^{\mu\nu}\right)^{ij} Q_\nu^j,
\end{equation}
we find the full Lagrangian up to quadratic power in $Q_\mu^i$ to be
\begin{eqnarray}
\Delta {\cal L} &=& \frac{1}{2}Q_\mu^i \left(D^2 g^{\mu\nu} - D^\nu D^\mu + m_Q^2 g^{\mu\nu} + [D^\mu, D^\nu] + \frac{1}{\xi} D^\mu D^\nu + M^{\mu\nu} \right)^{ij} Q_\nu^j \nonumber \\
 && + \frac{1}{2}{\chi ^i}{\left( { - {D^2} - \xi m_Q^2} \right)^{ij}}{\chi ^j} + {{\bar c}^i}{\left( { - {D^2} - \xi m_Q^2} \right)^{ij}}{c^j}.
\end{eqnarray}
Taking Feynman gauge $\xi=1$, we get
\begin{eqnarray}
\Delta {\cal L} &=& \frac{1}{2}Q_\mu ^i{\left( {D^2}{g^{\mu \nu }} + m_Q^2{g^{\mu \nu }} + 2[{D^\mu },{D^\nu }] + M^{\mu\nu} \right)^{ij}}Q_\nu ^j \nonumber \\
 && + \frac{1}{2}{\chi ^i}{\left( { - {D^2} - m_Q^2} \right)^{ij}}{\chi ^j} + {{\bar c}^i}{\left( { - {D^2} - m_Q^2} \right)^{ij}}{c^j}.
\end{eqnarray}
This is what we presented in the main text, Eq.~\eqref{eqn:Sgauge}.

\subsection{Useful identities}\label{sec:app_cde_identity}

\noindent \textbf{Expansion of $\tG_{\nu\mu}$}

\begin{eqnarray}
\tG_{\n\m} &=& \sum_{n=0}^{\infty} \frac{(n+1)}{(n+2)!}\big(P_{\a_1}\dots P_{\a_n}G'_{\n\m}\big) \pd^n_{\a_1\a_2 \dots \a_{n}} \nonumber \\
 &=& \frac{1}{2}G'_{\n\m} + \frac{1}{3}(P_{\a}G'_{\n\m})\pd_{\a} + \frac{1}{8}(P_{\a_1}P_{\a_2},G_{\n\m})\pd^2_{\a_1\a_2} + \dots \hspace{2mm} .
\end{eqnarray}

\noindent \textbf{Commutators/anti-commutators}\footnote{Note that we are not distinguishing upper and lower indices, so in the following, $\delta_{\mu\nu}$ here should be understood as $g_{\mu\nu}$.}

\begin{eqnarray}
\{q_{\m},\pd_{\a}\} &=& 2 q_{\m}\pd_{\a} + \d_{\m\a}  , \\
\{q_{\m},\pd^2_{\a_1\a_2}\} &=&  2 q_{\m}\pd^2_{\a_1\a_2} + \d_{\m\a_1}\pd_{\a_2} + \d_{\m\a_2}\pd_{\a_1} , \\
\{q_{\m},\pd^3_{\a_1\a_2\a_3}\} &=&  2 q_{\m}\pd^3_{\a_1\a_2\a_3} + \d_{\m\a_1}\pd^2_{\a_2\a_3} + \d_{\m\a_2}\pd^2_{\a_1\a_3} + \d_{\m\a_3}\pd^2_{\a_1\a_2} , \\
\{q_{\m},\pd^n_{\a_1\dots\a_n}\} &=& 2q_{\m}\pd^n_{\a_1\dots\a_n} + \sum_{i=1}^n\d_{\m\a_i}\prod_{j\ne i} \pd_{\a_j} .
\end{eqnarray}
And hence we have
\begin{eqnarray}
\{q_{\m},\tG_{\n\m}\} &=& G'_{\n\m}q_{\m} + \frac{1}{3}(P_{\a}G'_{\n\m})\big( 2 q_{\m}\pd_{\a} + \d_{\m\a} \big) \nonumber \\
 && +\frac{1}{8}(P_{\a_1}P_{\a_2}G'_{\n\m}) \big(2 q_{\m}\pd^2_{\a_1\a_2} + \d_{\m\a_1}\pd_{\a_2} + \d_{\m\a_2}\pd_{\a_1} \big)+\dots \hspace{2mm} .
\end{eqnarray}

\noindent \textbf{Derivatives and integrals}

\renewcommand\arraystretch{2.0}
\begin{eqnarray}
\pd_{\a_1}\D &=& (-1)\cdot 2\cdot q_{\a_1}\D^2 , \\
\pd^2_{\a_1\a_2}\D &=& (-1)\cdot 2\cdot \d_{\a_1\a_2}\D^2 + (-1)^2\cdot2!\cdot2^2\cdot q_{\a_1}q_{\a_2}\D^3 , \\
\pd^3_{\a_1\a_2\a_3}\D &=& (-1)^2\cdot2!\cdot 2^2\big(\underbrace{\d_{\a_1\a_2}q_{\a_3} + \text{ perm}}_{3\text{ terms}}\big) \D^3 + (-1)^3 \cdot 3!\cdot 2^3\cdot q_{\a_1}q_{\a_2}q_{\a_3} \D^4 , \\
\pd^4_{\a_1\a_2\a_3\a_4}\D &=& (-1)^2 \cdot 2! \cdot 2^2 \big(\underbrace{\d_{\a_1\a_2}\d_{\a_3\a_4} + \text{ perm}}_{3\text{ terms}}\big) \D^3 \nonumber \\
&& + (-1)^3 \cdot 3!\cdot 2^3 \big( \underbrace{\d_{\a_1\a_2}q_{\a_3}q_{\a_4} + \text{ perm}}_{6\text{ terms}}\big) \D^4 \nonumber \\
&& + (-1)^4 \cdot 4! \cdot 2^4 \cdot q_{\a_1}q_{\a_2}q_{\a_3}q_{\a_4} \D^5 .
\end{eqnarray}
\renewcommand\arraystretch{1.0}

These derivatives, which are part of the integrand, take simplified forms under \(q\)-integration:

\renewcommand\arraystretch{2.0}
\begin{eqnarray}
\pd^2_{\a_1\a_2}\D &\to& 2\d_{\a_1\a_2}(-\D^2 + q^2\D^3) , \nonumber \\
q_{\a_4}\pd^3_{\a_1\a_2\a_3}\D &\to& 2\big(\d_{\a_1\a_2}\d_{\a_3\a_4} + \d_{\a_1\a_3}\d_{\a_2\a_4} + \d_{\a_1\a_4}\d_{\a_2\a_3}\big)\big[q^2\D^3 - (q^2)^2\D^4\big] , \nonumber \\
\pd^4_{\a_1\a_2\a_3\a_4}\D &\to& 4\big(\d_{\a_1\a_2}\d_{\a_3\a_4} + \d_{\a_1\a_3}\d_{\a_2\a_4} + \d_{\a_1\a_4}\d_{\a_2\a_3}\big)\big[2\D^3 - 6q^2\D^4+ 4(q^2)^2\D^5\big] , \nonumber \\
q_{\a_5}q_{\a_6}\pd^4_{\a_1\a_2\a_3\a_4}\D &\to& 2\d_{\a_5\a_6} \big(\d_{\a_1\a_2}\d_{\a_3\a_4} + \d_{\a_1\a_3}\d_{\a_2\a_4} + \d_{\a_1\a_4}\d_{\a_2\a_3}\big)\big[q^2\D^3 - (q^2)^2\D^4\big] \nonumber \\
&& + 2\big(\underbrace{\d_{\a_1\a_2}\d_{\a_3\a_4}\d_{\a_5\a_5} + \text{ perm}}_{15\text{ terms}}\big)\big[-(q^2)^2\D^4 + (q^2)^3\D^5\big] . \nonumber
\end{eqnarray}
\renewcommand\arraystretch{1.0}

The following are useful integrals. They are in Minkowski space, and the powers of the free propagator---\(n\) in \(\D^n\)---is assumed large enough to make the integral converge:
\begin{align*}
I_0^{(n)} &\equiv \int \frac{d^4q}{(2\pi)^4} \D^n = i\frac{(-1)^n}{(4\pi)^2} \frac{1}{(n-1)(n-2)} \frac{1}{(m^2)^{n-2}} , \\
I_2^{(n)} &\equiv \int \frac{d^4q}{(2\pi)^4} q^2\D^n = -i\frac{(-1)^n}{(4\pi)^2} \frac{2}{(n-1)(n-2)(n-3)} \frac{1}{(m^2)^{n-3}} , \\
I_4^{(n)} &\equiv \int \frac{d^4q}{(2\pi)^4} (q^2)^2\D^n = i\frac{(-1)^n}{(4\pi)^2} \frac{6}{(n-1)(n-2)(n-3)(n-4)} \frac{1}{(m^2)^{n-4}} , \\
I_6^{(n)} &\equiv \int \frac{d^4q}{(2\pi)^4} (q^2)^3\D^n = -i\frac{(-1)^n}{(4\pi)^2} \frac{24}{(n-1)(n-2)(n-3)(n-4)(n-5)} \frac{1}{(m^2)^{n-3}} .
\end{align*}

\noindent \textbf{Operator identities and trace computations}

Let us state some basics of covariant derivative calculus. Most of these are obvious, but we list them here because we make use of them over and over in calculations.
\begin{itemize}
\item The covariant derivative acting on a matrix is given by the commutator, \(D_{\m}A = [D_{\m},A]\).
\item The basic rules of calculus are the same. In particular, the chain rule holds: \(D (A B) = (DA)B + A(DB)\). This implies integration by parts holds, \(\int dx \tr\big[A(DB)\big] = \int dx \tr\big[-(DA)B\big]\).
\item The covariant derivative acting on a gauge invariant quantity is just the partial derivatve, \(D_{\m} \abs{H}^2 = \pd_{\m}\abs{H}^2\).
\end{itemize}

\begin{equation}
2 \abs{H^{\dag} D_{\m} H}^2 = \frac{1}{2} \big(\pd_{\m} \abs{H}^2\big)^2 - \frac{1}{2} \big( H^{\dag} \Dfb H \big)^2  \ \Leftrightarrow \ 2 \scO_{HD} = \scO_H - \scO_T .
\label{eqn:op_identity}
\end{equation}

A term that often shows up in calculations is
\begin{equation}
\big(H^{\dag}D_{\m} H\big)^2 +  \big((D_{\m}H)^{\dag}H\big)^2 = \scO_T + \scO_H .
\label{eqn:cdeA_opsq}
\end{equation}

\begin{equation}
\Tr\big[D_{\m}(HH^{\dag})\big]^2 = \big(H^{\dag}D H\big)^2 + \big((DH)^{\dag}H\big)^2 + 2 \abs{H}^2\abs{DH}^2 = \scO_T + \scO_H + 2\scO_r .
\end{equation}

\subsection{Evaluating terms in the CDE: results for the $\mathcal{I}_n$}\label{sec:app_CDE_universal}

\begin{equation*}
\frac{1}{A^{-1}(1-AB)} = \sum_{n=0}^{\infty}(AB)^nA ,
\end{equation*}

\begin{equation*}
\scI_n \equiv \int dq \, dm^2\, \tr \Big[ \D\Big( -\{q\tG\} -\tG^2 + \tU \Big)\Big]^n\D ,
\end{equation*}

\begin{equation*}
\D \scL_{\scI_n} = -i c_s \scI_n .
\end{equation*}
Breaking \(\scI_n\) into easier to work with pieces, we define integrals involving only \(\tG\) as \(J_n\) and integrals involving only \(\tU\) as \(K_n\),
\begin{align*}
J_n &\equiv \int dq \, dm^2\, \tr \Big[ \D\Big( -\{q\tG\} -\tG^2 \Big)\Big]^n\D , \\
K_n &\equiv \int dq \, dm^2\, \tr \Big[ \D\tU\Big]^n\D .
\end{align*}
We define \(L_n\) for integrals involving mixed \(\tG\) and \(\tU\) terms as \(L_n\); for example, \(L_2\) is given by
\begin{equation*}
L_2 \equiv  \int dq \, dm^2 \, \tr \Big[ -\D\Big(\{q\tG\} + \tG^2 \Big)\D \tU \D - \D \tU \D\Big( \{q\tG\} +\tG^2 \Big)\D  \Big].
\end{equation*}

\begin{equation*}
\D {\cal L}_{J_1 + J_2} = -\frac{1}{(4\pi)^2} \bigg[\frac{1}{6}\Big(\log \frac{m^2}{\m^2} - 1\Big) \cdot \Big(\frac{1}{2}\tr G'_{\m\n}G'_{\m\n}\Big) + \frac{1}{m^2} \cdot \frac{1}{60} \cdot \tr \big(P_{\m}G_{\m\n}'\big)^2 + \frac{1}{m^2} \cdot \frac{1}{90} \cdot \tr \big(G'_{\m\n}G'_{\n\s}G'_{\s\m}\big) \bigg] .
\end{equation*}

\begin{align*}
\D {\cal L}_{K_1} &= \frac{1}{(4\pi)^2}m^2\Big[-\log \frac{m^2}{\m^2}+1\Big] \cdot \tr U , \\
\D {\cal L}_{K_2} &= \frac{1}{(4\pi)^2} \bigg[-\frac{1}{2} \log \frac{m^2}{\m^2} \cdot \tr U^2 -\frac{1}{m^2} \cdot \frac{1}{12} \cdot \tr \Big( [P_{\m},U]^2 \Big) + \frac{1}{m^4}\cdot\frac{1}{120}\cdot \tr \Big( \big[P_{\m}[P_{\m},U]\big]\big[P_{\n}[P_{\n},U]\big] \Big)\bigg] , \\
\D {\cal L}_{K_3} &= \frac{1}{(4\pi)^2} \bigg[- \frac{1}{m^2} \cdot \frac{1}{6} \cdot \tr \big(U^3\big) +  \frac{1}{m^4} \cdot \frac{1}{12}\cdot \tr \Big( U [P_{\m},U][P_{\m},U] \Big) \bigg] , \\
\D {\cal L}_{K_4} &= \frac{1}{(4\pi)^2}\cdot \left[ \frac{1}{m^4}\cdot \frac{1}{24}\cdot \tr\big(U^4\big) - \frac{1}{m^6}\cdot \frac{1}{20}\cdot \tr\big( U^2 [P_{\m},U][P_{\m},U] \big) - \frac{1}{m^6}\cdot \frac{1}{30}\cdot \tr\big( U [P_{\m},U]U[P_{\m},U] \big) \right] , \\
\D {\cal L}_{K_5} &= -\frac{1}{(4\pi)^2}\cdot \frac{1}{m^6}\cdot \frac{1}{60}\cdot \tr\big(U^5\big) , \\
\D {\cal L}_{K_6} &= \frac{1}{(4\pi)^2}\cdot \frac{1}{m^8}\cdot \frac{1}{120} \cdot \tr\big(U^6\big) .
\end{align*}

\begin{align*}
\D {\cal L}_{L_2} &= - \frac{1}{(4\pi)^2} \cdot \frac{1}{m^2} \cdot \frac{1}{12} \cdot \tr \Big( U G'_{\m\n} G'_{\m\n} \Big) , \\
\D {\cal L}_{L_3} &= \frac{1}{(4\pi)^2}\cdot \frac{1}{m^4} \cdot \bigg[ \frac{1}{24} \cdot \Big( U^2 G'_{\m\n}G'_{\m\n} \Big) - \frac{1}{120} \cdot \tr \Big( \big[[P_{\m},U],[P_{\n},U]\big] G'_{\m\n} \Big) - \frac{1}{120}\cdot \tr \Big( \big[U[U,G'_{\m\n}]\big] G'_{\m\n} \Big) \bigg] .
\end{align*}

\newpage
\section{Universality of Magnetic Dipole Term}\label{sec:app_gkpiece}

Assuming that there is a weakly coupled renormalizable UV model, \footnote{In general, this need not be the case. For example, the \(Q_{\m}\) could be composite particles in the low-energy effective description of some strongly interacting theory. Another example is when additional massive vector bosons are needed to UV complete the theory. For example, an effective theory with a massive vector transforming as a doublet under a \(SU(2)\) gauge symmetry is non-renormalizable---a valid UV completion could be an \(SU(3)\) gauge symmetry broken to \(SU(2)\), but this requires an additional doublet and singlet vector.}${}^,$\footnote{As in all the other cases considered in this work, although never explicitly stated, we are also assuming the fields we integrate out are weakly coupled amongst themselves and the low-energy fields, so that it makes sense to integrate them out.}${}^,$\footnote{\(G\) itself may be contained in some larger group \(\mathcal{G}\) which also contains exact and approximate global symmetries and the same mechanism responsible for breaking \(G \to H\) may also break some of these global symmetries. These generalities do not affect our results below, which concern the transformation of \(Q_{\m}\) and its associated NGBs under \(H\). We therefore stick to our simplified picture for clarity.} we consider a general picture that the full gauge symmetry group $G$ of the UV model is spontaneously broken into a subgroup $H$. A set of gauge bosons $Q_\mu^i$ have ``eaten'' the Nambu-Goldstone bosons $\chi^i$ and obtained mass $m_Q$. For this setup, it turns out that $Q_\mu^i$ form a certain representation of the unbroken gauge group $H$, and under this representation, the general form of the gauge-kinetic piece of the Lagrangian up to quadratic term in $Q_\mu^i$ is given by Eq.~\eqref{eqn:Lgk}, which we reproduce here for convenience
\begin{equation}
{\cal L}_\text{g.k.} \supset \frac{1}{2} Q_\mu^i \left( D^2 g^{\mu\nu} - D^\nu D^\mu + [D^\mu,D^\nu] \right)^{ij} Q_\nu^j , \label{eqn:LgaugeAppB}
\end{equation}
with $D_\mu$ denoting the covariant derivative that contains only the massless gauge bosons. One remarkable feature of this general gauge-kinetic term is that the coefficient of the ``magnetic dipole term'' $\frac{1}{2} Q_\mu^i \left\{\left[D^\mu, D^\nu\right] \right\}^{ij} Q_\nu^j$ is universal, namely that its coefficient is fixed to $1$ relative to the ``curl'' terms $\frac{1}{2} Q_\mu^i \left\{ D^2 g^{\mu\nu} - D^\nu D^\mu\right\}^{ij} Q_\nu^j$, regardless of the details of the symmetry breaking. We use the word ``curl'' since the term comes from the quadratic piece in \((D_{\m}Q_{\n} - D_{\n}Q_{\m})^2\).

The universal coefficient of the magnetic dipole term is known to be a consequence of tree-level unitarity~\cite{Weinberg:1970,Ferrara:1992yc}. In this appendix, we present an additional, new way of proving Eq.~\eqref{eqn:LgaugeAppB} that is completely algebraic. We note that these algebraic methods developed may be useful for other purposes since they allow a very compact way of writing the gauge kinetic terms for multiple gauge groups with different coupling constants, see Eq.~\eqref{eqn:Lgaugeinner}. We also give the physical argument based on tree-level unitarity for the validity of Eq.~\eqref{eqn:LgaugeAppB}, similar to~\cite{Weinberg:1970,Ferrara:1992yc}.

\subsection{Algebraic Proof}

Let us first give an algebraic derivation of Eq.~\eqref{eqn:LgaugeAppB}, which we believe is new. Let $G$ have a general structure of product group
\begin{equation}
G = G_1 \times G_2 \times \cdots \times G_n . \label{eqn:Gdecompose}
\end{equation}
Let $T^A$ be the set of generators of $G$, with $A = 1,2, \ldots ,\dim (G)$. Due to Eq.~\eqref{eqn:Gdecompose}, the set of generators $T^A$ are composed by a number of subsets
\begin{equation}
\Big\{T^A\Big\}=\Big\{T_1^{A_1}\Big\} \cup \Big\{T_2^{A_2}\Big\} \cup \cdots \cup \Big\{T_n^{A_n}\Big\} , \label{eqn:TAdecompose}
\end{equation}
with ${{A_i} = 1,2, \ldots ,\dim ({G_i})}$. Let $f_G^{ABC}$ denote the structure constant of $G$ :
\begin{equation}
\left[ {{T^A},{T^B}} \right] = if_G^{ABC}{T^C} . \label{eqn:Galgebra}
\end{equation}
Obviously $f_G^{ABC}=0$ if any two indices belong to different subsets in Eq.~\eqref{eqn:TAdecompose}.

The full covariant derivative ${\bar D}$ of the UV model and its commutator is
\begin{eqnarray}
{{\bar D}_\mu } &=& {\partial _\mu } - i{g^A}G_\mu^A{T^A} , \label{eqn:fullD} \\
\left[ {{{\bar D}_\mu },{{\bar D}_\nu }} \right] &=&  - i{g^A}G_{\mu \nu }^A{T^A} ,
\end{eqnarray}
where $G_\mu^A$ denote the gauge fields, $G_{\mu\nu}^A$ the field strengths, and $g^A$ the gauge couplings that could be arbitrarily different for $T^A$ of different subsets in Eq.~\eqref{eqn:TAdecompose}. Here we emphasize that the above expression of the full covariant derivative holds for any representation of $G$.

Because we have put the arbitrary gauge couplings into the covariant derivative, the gauge boson kinetic term of the UV Lagrangian is simply
\begin{equation}
{\cal L}_\text{g.k.} =  - \frac{1}{4}{\left( {G_{\mu \nu }^{{A_1}}} \right)^2} - \frac{1}{4}{\left( {G_{\mu \nu }^{{A_2}}} \right)^2} -  \cdots  - \frac{1}{4}{\left( {G_{\mu \nu }^{{A_n}}} \right)^2} . \label{eqn:LgaugeUV}
\end{equation}
In order to write this kinetic term in terms of the full covariant derivative ${\bar D}_\mu$, let us define an inner product in the generator space $\{T^A\}$:
\begin{equation}
\left\langle {{T^A},{T^B}} \right\rangle  \equiv \frac{1}{{2{{({g^A})}^2}}}{\delta ^{AB}} , \label{eqn:innerproduct}
\end{equation}
which just looks like a scaled version of trace. However, we emphasize that, although it should be quite clear from definition, this inner product is essentially very different from the trace. The inner product can only be taken over two vectors in the generator space, while a trace action can be taken over arbitrary powers of generators. Nevertheless, the inner product defined in Eq.~\eqref{eqn:innerproduct} has many similar properties as the trace action. For example, if one of the two vectors is given in a form of a commutator of two other generators, a cyclic permutation is allowed
\begin{eqnarray}
\left\langle {{T^A},\left[ {{T^B},{T^C}} \right]} \right\rangle  &=& \left\langle {{T^A},if_G^{BCD}{T^D}} \right\rangle  = if_G^{BCD}\frac{1}{{2{{({g^A})}^2}}}{\delta ^{AD}} \nonumber \\
 &=& if_G^{ABC}\frac{1}{{2{{({g^A})}^2}}} = if_G^{ABC}\frac{1}{{2{{({g^C})}^2}}} \nonumber \\
 &=& if_G^{CAB}\frac{1}{{2{{({g^C})}^2}}} = \left\langle {{T^C},\left[ {{T^A},{T^B}} \right]} \right\rangle . \label{eqn:cyclic}
\end{eqnarray}
Note that the second line above is true because for the case $g^A \ne g^C$, $f^{ABC}=0$. As we shall see shortly, this cyclic permutation property will play a very important role in our derivation. With the inner product defined in Eq.~\eqref{eqn:innerproduct}, the gauge boson kinetic term Eq.~\eqref{eqn:LgaugeUV} can be very conveniently written as
\begin{equation}
{\cal L}_\text{g.k.} = \frac{1}{2}\left\langle {\left[ {{{\bar D}_\mu },{{\bar D}_\nu }} \right],\left[ {{{\bar D}^\mu },{{\bar D}^\nu }} \right]} \right\rangle . \label{eqn:Lgaugeinner}
\end{equation}

Now let us consider the subgroup $H$ of $G$. Let $t^a$ be the generators of $H$, which span a subspace of the full group generator space, and have closed algebra
\begin{equation}
\left[ {{t^a},{t^b}} \right] = if_H^{abc}{t^c} , \label{eqn:Halgebra}
\end{equation}
with $f_H^{abc}$ denotes the structure constant of $H$, and $a=1,2,...,\dim(H)$. Once the full group $G$ is spontaneously broken into $H$, it is obviously convenient to divide the full generator space into the unbroken generators $t^a$ and the broken generators $X^i$, $i=1,2,...,\dim(G)-\dim(H)$, with the corresponding massless gauge fields $A_\mu^a$ and massive gauge bosons $Q_\mu^i$
\begin{equation}
\begin{array}{*{20}{c}}
{\left( {{t^A}} \right) = \left( {\begin{array}{*{20}{c}}
{{g_H^a}{t^a}}\\
{{X^i}}
\end{array}} \right)}&,\hspace{1cm}&{\left( {W_\mu ^A} \right) = \left( {\begin{array}{*{20}{c}}
{A_\mu ^a}\\
{Q_\mu ^i}
\end{array}} \right)}
\end{array} . \label{eqn:decomposition}
\end{equation}
In the above, we write $t^A$ instead of $T^A$, and $W_\mu^A$ instead of $G_\mu^A$, because $t^a$ is generically a linear combination of $T^A$, and there is a linear transformation between $t^A$ and $T^A$, as well as between $W_\mu^A$ and $G_\mu^A$ in accordance. This linear transformation is typically chosen to be orthogonal between gauge field~\footnote{Other linear transformations will lead to equivalent theories upon field redefinition.}, in order to preserve the universal coefficients structure in Eq.~\eqref{eqn:LgaugeUV}. Then we have
\begin{equation}
W_\mu ^A = {O^{AB}}G_\mu ^B \hspace{3mm} , \hspace{3mm} \text{with } {O^T}O = 1 . \label{eqn:rotation}
\end{equation}
The full covariant derivative Eq.~\eqref{eqn:fullD} can be rewritten as
\begin{eqnarray}
{{\bar D}_\mu } &=& {\partial _\mu } - iW_\mu ^A{t^A} = {\partial _\mu } - ig_H^aA_\mu ^a{t^a} - iQ_\mu ^i{X^i} = {D_\mu } - iQ_\mu ^i{X^i} , \\
{t^A} &=& {O^{AB}}{g^B}{T^B} , \label{eqn:tTrelation}
\end{eqnarray}
where the second line serves as the definition of $t^A$ in terms of $T^A$. Note that a factor $g_H^a$ is needed in Eq.~\eqref{eqn:decomposition} to make Eqs.~\eqref{eqn:Galgebra},~\eqref{eqn:Halgebra} and~\eqref{eqn:tTrelation} consistent. This is how one determines the gauge coupling constant $g_H^a$ of the unbroken gauge group. We have also used $D_\mu$ to denote the covariant derivative that contains only the massless gauge bosons $A_\mu^a$. The above definition of $t^A$ preserves the orthogonality of them under the inner product defined in Eq.~\eqref{eqn:innerproduct}
\begin{equation}
\left\langle {{t^A},{t^B}} \right\rangle  = \left\langle {{O^{AC}}{g^C}{T^C},{O^{BD}}{g^D}{T^D}} \right\rangle  = \frac{1}{{2{{({g^C})}^2}}}{O^{AC}}{O^{BD}}{g^C}{g^D}{\delta ^{CD}} = \frac{1}{2}{\delta ^{AB}} ,
\end{equation}
which specifically means that
\begin{equation}
{\left\langle {{t^a},{t^b}} \right\rangle  = \frac{1}{2\left(g_H^a\right)^2}{\delta ^{ab}}}, \hspace{0.5cm} {\left\langle {{X^i},{X^j}} \right\rangle  = \frac{1}{2}{\delta ^{ij}}}, \hspace{0.5cm} {\left\langle {{t^a},{X^i}} \right\rangle  = 0} .
\end{equation}

Let us first prove that $Q_\mu^i$ defined through Eq.~\eqref{eqn:decomposition} and Eq.~\eqref{eqn:rotation} form a representation under the unbroken gauge group $H$. This is essentially to prove that the commutator between $t^a$ and $X^i$ is only a linear combination of $X^i$
\begin{equation}
\left[ {{t^a},{X^i}} \right] =  - {(t_Q^a)^{ij}}{X^j} , \label{eqn:tXcommutator}
\end{equation}
with a certain set of matrices $\left(t_Q^a\right)^{ij}$ that also need to be antisymmetric between $i,j$. Both points can be easily proven by making use of our inner product defined in Eq.~\eqref{eqn:innerproduct} and its cyclic permutation property Eq.~\eqref{eqn:cyclic}. Eq.~\eqref{eqn:tXcommutator} is obvious from
\begin{equation}
\left\langle {{t^b},\left[ {{t^a},{X^i}} \right]} \right\rangle = \left\langle {{X^i},\left[ {{t^b},{t^a}} \right]} \right\rangle  = 0 ,
\end{equation}
and the antisymmetry is clear from
\begin{equation}
{\left( {t_Q^a} \right)^{ij}} =  - 2\left\langle {{X^j},\left[ {{t^a},{X^i}} \right]} \right\rangle  =  - 2\left\langle {{t^a},\left[ {{X^i},{X^j}} \right]} \right\rangle .
\end{equation}
Once Eq.~\eqref{eqn:tXcommutator} is proven, it follows that
\begin{equation}
\left[ {{t^a},Q_\mu ^i{X^i}} \right] =  - Q_\mu ^i{\left( {t_Q^a} \right)^{ij}}{X^j} = {\left( {t_Q^a} \right)^{ij}}Q_\mu ^j{X^i} ,
\end{equation}
where we see that $t_Q^a$ serves as the generator matrix or ``charge'' of $Q_\mu^i$. And therefore
\begin{eqnarray}
\left[ {{D_\mu },Q_\nu ^i{X^i}} \right] &=& ({\partial _\mu }Q_\nu ^i){X^i} - ig_H^aA_\mu ^a\left[ {{t^a},Q_\nu ^i{X^i}} \right] = ({\partial _\mu }Q_\nu ^i){X^i} - ig_H^aA_\mu ^a{\left( {t_Q^a} \right)^{ij}}Q_\mu ^j{X^i} \nonumber \\
 &=& \left[ {\left( {{\partial _\mu }Q_\nu ^i} \right) - ig_H^aA_\mu ^a{{\left( {t_Q^a} \right)}^{ij}}Q_\mu ^j} \right]{X^i} = \left( {{D_\mu }Q_\nu ^i} \right){X^i} . \label{eqn:Qcharge}
\end{eqnarray}

With all the above preparations, we are eventually ready to decompose the full gauge boson kinetic term in Eq.~\eqref{eqn:LgaugeUV}. First, the commutator of the full covariant derivative is
\begin{eqnarray}
\left[ {{{\bar D}_\mu },{{\bar D}_\nu }} \right] &=& \left[ {{D_\mu } - iQ_\mu ^i{X^i},{D_\nu } - iQ_\nu ^j{X^j}} \right] \nonumber \\
 &=& \left[ {{D_\mu },{D_\nu }} \right] - i\left\{ {\left[ {{{\bar D}_\mu },Q_\nu ^i{X^i}} \right] - \left[ {{{\bar D}_\nu },Q_\mu ^i{X^i}} \right]} \right\} - \left[ {Q_\mu ^i{X^i},Q_\nu ^j{X^j}} \right] \nonumber \\
 &=& \left[ {{D_\mu },{D_\nu }} \right] - i\left[ {({D_\mu }Q_\nu ^i) - ({D_\nu }Q_\mu ^i)} \right]{X^i} - \left[ {Q_\mu ^i{X^i},Q_\nu ^j{X^j}} \right] .
\end{eqnarray}
Keeping only terms relevant and up to quadratic power for $Q_\mu^i$, it follows from Eq.~\eqref{eqn:Lgaugeinner} that
\begin{eqnarray}
{\cal L}_\text{g.k.} &=& \frac{1}{2}\left\langle {\left[ {\bar D}_\mu,{\bar D}_\nu \right],\left[{\bar D}^\mu,{\bar D}^\nu \right]} \right\rangle \nonumber \\
 &\supset& - \frac{1}{4}{\left[ {({D_\mu }Q_\nu ^i) - ({D_\nu }Q_\mu ^i)} \right]^2} - \left\langle {\left[ {{D^\mu },{D^\nu }} \right],\left[ {Q_\mu ^i{X^i},Q_\nu ^j{X^j}} \right]} \right\rangle \nonumber \\
 &=& \frac{1}{2}Q_\mu ^i{\left( {{D^2}{g^{\mu \nu }} - {D^\nu }{D^\mu }} \right)^{ij}}Q_\nu ^j - \left\langle {Q_\mu ^i{X^i},\left[ {Q_\nu ^j{X^j},\left[ {{D^\mu },{D^\nu }} \right]} \right]} \right\rangle \nonumber \\
 &=& \frac{1}{2}Q_\mu ^i{\left( {{D^2}{g^{\mu \nu }} - {D^\nu }{D^\mu }} \right)^{ij}}Q_\nu ^j + \left\langle {Q_\mu ^i{X^i},\left[ {{D^\mu },\left[ {{D^\nu },Q_\nu ^j{X^j}} \right]} \right]} \right\rangle  - \left\langle {Q_\mu ^i{X^i},\left[ {{D^\nu },\left[ {{D^\mu },Q_\nu ^j{X^j}} \right]} \right]} \right\rangle \nonumber \\
 &=& \frac{1}{2}Q_\mu ^i{\left( {{D^2}{g^{\mu \nu }} - {D^\nu }{D^\mu }} \right)^{ij}}Q_\nu ^j + Q_\mu ^i\left\langle {{X^i},\left( {{D^\mu }{D^\nu }Q_\nu ^j} \right){X^j}} \right\rangle  - Q_\mu ^i\left\langle {{X^i},\left( {{D^\nu }{D^\mu }Q_\nu ^j} \right){X^j}} \right\rangle \nonumber \\
 &=& \frac{1}{2}Q_\mu ^i{\left( {{D^2}{g^{\mu \nu }} - {D^\nu }{D^\mu }} \right)^{ij}}Q_\nu ^j + \frac{1}{2}Q_\mu ^i{\left( {{D^\mu }{D^\nu } - {D^\nu }{D^\mu }} \right)^{ij}}Q_\nu ^j \nonumber \\
 &=& \frac{1}{2}Q_\mu ^i{\left\{ {{D^2}{g^{\mu \nu }} - {D^\nu }{D^\mu } + \left[ {{D^\mu },{D^\nu }} \right]} \right\}^{ij}}Q_\nu ^j ,
\end{eqnarray}
where from the second line to the third line, we have used the cyclic permutation property of the inner product, and the fourth line follows from the third line due to Jacobi identity. This finishes our algebraic derivation of Eq.~\eqref{eqn:LgaugeAppB}.

We would like to stress that in spite of the allowance of arbitrary gauge couplings for each simple group $G_i$, the end gauge-interaction piece of the Lagrangian of the heavy vector boson $Q_\mu^i$ has the above universal form, especially that the coefficient of the magnetic dipole term $\frac{1}{2} Q_\mu^i\left[D^\mu, D^\nu\right]^{ij} Q_\nu^j$ is fixed at to unity relative to the curl terms $\frac{1}{2} Q_\mu^i \left\{ D^2 g^{\mu\nu} - D^\nu D^\mu\right\}^{ij} Q_\nu^j$.

\subsection{Physical Proof}

Now let us give a physical argument to explain this universality, which is from the tree-level unitarity. This argument is known~\cite{Weinberg:1970,Ferrara:1992yc}, but we provide it here for completeness. Let us consider one component of the massless background gauge boson and call it a ``photon'' $A_\mu$ with its coupling constant $e$ and generator $Q$. It is helpful to use a complex linear combination of generators $X^i$ to form $X^\alpha$ and $X^{\alpha\dagger}$ that are ``eigenstates'' of the generator $Q$, $[Q, X^\alpha] = q^\alpha X^\alpha$ and $[Q, X^{\alpha\dagger}] = -q^\alpha X^{\alpha\dagger}$. We also define $Q_\mu^\alpha$ and $Q_\mu^{\alpha\dagger}$ to keep $Q_\mu^i X^i=Q_\mu^\alpha X^\alpha+Q_\mu^{\alpha\dagger} X^{\alpha\dagger}$. Note that $Q_\mu^i$ are real, but $Q_\mu^\alpha$ are complex fields. The normalization of $Q_\mu^\alpha$ is chosen such that $\frac{1}{2} Q_\mu^i Q^{\mu i} = Q_\mu^{\alpha\dagger} Q^{\mu\alpha}$. It should be clear that in this part of the appendix where we discuss integrating out a heavy gauge boson, indices $\alpha,\beta$ are used to denote the complex generators $X^\alpha$, $X^{\alpha\dagger}$, and their accordingly defined complex gauge fields $Q^\alpha$, $Q^{\alpha\dagger}$. Lorentz indices are denoted by $\mu$, $\nu$, $\rho$, {\it etc}.

First, one can check that the ``curl'' terms in Eq.~\eqref{eqn:LgaugeAppB} written in terms of $Q_\mu^i$ gives the correct kinetic term for $Q_\mu^\alpha$ coupled to photon according to its charge $q^\alpha$, because from Eq.~\eqref{eqn:Qcharge} we have
\begin{eqnarray}
\left( {{D_\mu }Q_\nu ^i} \right){X^i} &=& \left[ {{D_\mu },Q_\nu ^i{X^i}} \right] = \left[ {{D_\mu },Q_\nu ^\alpha{X^\alpha} + Q_\nu ^{\alpha\dag }{X^{\alpha\dag }}} \right] \nonumber \\
 &\supset& \left( {{\partial _\mu }Q_\nu ^\alpha} \right){X^\alpha} + \left( {{\partial _\mu }Q_\nu ^{\alpha\dag }} \right){X^{\alpha\dag }} - ie{A_\mu }\left[ {Q,Q_\nu ^\alpha{X^\alpha} + Q_\nu ^{\alpha\dag }{X^{\alpha\dag }}} \right] \nonumber \\
 &=& \left( {{\partial _\mu }Q_\nu ^\alpha} \right){X^\alpha} + \left( {{\partial _\mu }Q_\nu ^{\alpha\dag }} \right){X^{\alpha\dag }} - ie{q^\alpha}{A_\mu }\left( {Q_\nu ^\alpha{X^\alpha} - Q_\nu ^{\alpha\dag }{X^{\alpha\dag }}} \right) \nonumber \\
 &=& \left( {{\partial _\mu }Q_\nu ^\alpha - ie{q^\alpha}{A_\mu }Q_\nu ^\alpha} \right){X^\alpha} + \left( {{\partial _\mu }Q_\nu ^{\alpha\dag } + ie{q^\alpha}{A_\mu }Q_\nu ^{\alpha\dag }} \right){X^{\alpha\dag }} ,
\end{eqnarray}
and the ``curl'' form derives from the original Yang-Mills Lagrangian in the UV theory
\begin{equation}
{{\cal L}_{{\text{Yang - Mills}}}} \supset  - \frac{1}{4}{\left( {{D_\mu }Q_\nu ^i - {D_\nu }Q_\mu ^i} \right)^2} = \frac{1}{2}Q_\mu ^i{\left\{ {{D^2}{g^{\mu \nu }} - {D^\nu }{D^\mu }} \right\}^{ij}}Q_\nu ^j .
\end{equation}

What is the least obvious is the universal coefficient for the ``magnetic dipole term''
\begin{equation}
\frac{1}{2} Q_\mu^i\left[D^\mu, D^\nu\right]^{ij} Q_\nu^j = -\frac{1}{2\mu(R)} \text{tr} \left( [Q_\mu, Q_\nu] [D^\mu, D^\nu] \right) ,
\end{equation}
where $Q_\mu \equiv Q_\mu^i X^i = Q_\mu^\alpha X^\alpha + Q_\mu^{\alpha\dagger}X^{\alpha\dagger}$, and $\text{tr}(X^i X^j)=\mu(R)\delta^{ij}$. This term is gauge invariant under the unbroken gauge symmetry and one may wonder whether the coefficient can be arbitrary and model dependent. Focusing on the ``photon'' coupling piece, this term contains
\begin{eqnarray}
 - \frac{1}{{2\mu (R)}}{\rm{tr}}\left( {\left[ {{Q_\mu },{Q_\nu }} \right]\left[ {{D^\mu },{D^\nu }} \right]} \right) &\supset& \frac{{ie{{\hat A}^{\mu \nu }}}}{{2\mu (R)}}{\rm{tr}}\left( {\left[ {{Q_\mu },{Q_\nu }} \right]Q} \right) \nonumber \\
 &=& \frac{{ie{{\hat A}^{\mu \nu }}}}{{2\mu (R)}}{\rm{tr}}\left( {\left[ {Q,{Q_\mu }} \right]{Q_\nu }} \right) \nonumber \\
 &=& \frac{{ie{{\hat A}^{\mu \nu }}}}{{2\mu (R)}}{\rm{tr}}\left\{ {\left( {{q^\alpha}Q_\mu ^\alpha{X^\alpha} - {q^\alpha}Q_\mu ^{\alpha\dag }{X^{\alpha\dag }}} \right)\left( {Q_\nu ^\beta{X^\beta} + Q_\nu ^{\beta\dag }{X^{\beta\dag }}} \right)} \right\} \nonumber \\
 &=& \frac{{ - ie{{\hat A}^{\mu \nu }}}}{2}{q^\alpha}\left( {Q_\mu ^{\alpha\dag }Q_\nu ^\alpha - Q_\nu ^{\alpha\dag }Q_\mu ^\alpha} \right) \nonumber \\
 &=&  - ie{{\hat A}^{\mu \nu }}{q^\alpha}Q_\mu ^{\alpha\dag }Q_\nu ^\alpha , \label{eqn:DipoleTGC}
\end{eqnarray}
where ${\hat A}_{\mu\nu}\equiv \partial_\mu A_\nu-\partial_\nu A_\mu$, and we have used the property $\text{tr}\left(X^\alpha X^{\beta\dagger}\right)=\mu(R)\delta^{\alpha\beta}$, $\text{tr}\left(X^\alpha X^\beta\right)= \text{tr}\left(X^{\alpha\dagger}X^{\beta\dagger}\right)=0$. So it is clear that the coefficient of this ``magnetic dipole term'' is exactly the ``triple gauge coupling'' between the heavy gauge boson $Q_\mu^\alpha$ and the massless gauge bosons $A_\mu$. One can make it more transparent by taking the SM analog of Eq.~\eqref{eqn:DipoleTGC}. In the case of SM electroweak symmetry breaking, one recognizes $q^\alpha=-1$, $Q_\nu^\alpha=W_\nu^-$, and $Q_\mu^{\alpha\dagger}=W_\mu^+$, then Eq.~\eqref{eqn:DipoleTGC} is nothing but the $\kappa_\gamma$ term in Eq.~\eqref{eqn:LagTGC}. It is well known that the amplitude for $\gamma\gamma \rightarrow W^+ W^-$ would grow as $E_W^2$ in the Standard Model if the magnetic dipole moment $\kappa_\gamma \neq 1$.  The quadratic part of the Lagrangian ({\it i.e.} Eq.~\eqref{eqn:LgaugeAppB}) is sufficient to determine the tree-level amplitude, and the diagrams are exactly the same as those in the Standard Model.  Unless $\kappa_\gamma = 1$, it violates perturbative unitarity at high energies.  Because the amplitude does not involve the Higgs or other heavy vector bosons, the amplitude is exactly the same as that in the UV theory, which is unitary.  Therefore, the perturbative unitarity for this amplitude needs to be satisfied with the quadratic Lagrangian, which requires the dipole moment to have this value.

\newpage

\section{Supplemental details for mapping the $c_i$ to physical observables} \label{sec:app_mapping}

This appendix shows the calculational details of the mapping step described in Section~\ref{sec:mapping}. We first list out in Appendix~\ref{subsec:FeynmanRules} all the relevant two-point and three-point Feynman rules from the set of dimension-six operators in Table~\ref{tbl:operators2}. Transverse vacuum polarization functions, that can be readily read off from the two-point Feynman rules, are also tabulated. Then in Appendix~\ref{subsec:GammaIdetails} and~\ref{subsec:SigmaIdetails} we present details in calculating the ``interference correction'' $\epsilon_I^{}$ for Higgs decay widths and Higgs production cross sections, respectively. We list out relevant Feynman diagrams, definitions of auxiliary functions, and conventional form factors. Finally, in Appendix~\ref{subsec:RMdetails} and~\ref{subsec:LPMdetails} we show our calculation steps of the residue modifications and the parameter modifications, which are related to the ``residue correction'' $\epsilon_R^{}$ and the ``parametric correction'' $\epsilon_P^{}$, respectively.

\subsection{Additional Feynman rules from dim-6 effective operators} \label{subsec:FeynmanRules}

\subsubsection{Feynman rules for vacuum polarization functions}

\renewcommand\arraystretch{2.0}
\begin{table}[tb]
\centering
\begin{tabular}{|rl|}
  \hline
  $\Pi_{WW}(p^2) =$ & \hspace{-0.4cm} $p^4\left(-\dfrac{1}{\Lambda^2}c_{2W}^{}\right) +p^2\dfrac{2m_W^2}{\Lambda^2}(4c_{WW}^{}+c_W^{}) +m_W^2\dfrac{v^2}{\Lambda^2}c_R^{}$ \\
  $\Pi_{ZZ}(p^2) =$ & \hspace{-0.4cm} $p^4\left[-\dfrac{1}{\Lambda^2}\left(c_Z^2 c_{2W}^{}+s_Z^2 c_{2B}^{}\right)\right] +p^2\dfrac{2m_Z^2}{\Lambda^2}\Big[ 4\left(c_Z^4 c_{WW}^{}+s_Z^4 c_{BB}^{}+c_Z^2 s_Z^2 c_{WB}^{}\right)$ \\
   & \hspace{-0.4cm} $ +\left(c_Z^2 c_W^{}+s_Z^2 c_B^{}\right) \Big] + m_Z^2\dfrac{v^2}{\Lambda^2}\left(-2c_T^{}+c_R^{}\right)$ \\
  $\Pi_{\gamma\gamma}(p^2) =$ & \hspace{-0.4cm} $p^4\left[-\dfrac{1}{\Lambda^2}\left(s_Z^2 c_{2W}^{}+c_Z^2 c_{2B}^{}\right)\right] +p^2\dfrac{8m_Z^2}{\Lambda^2}c_Z^2 s_Z^2(c_{WW}^{}+c_{BB}^{}-c_{WB}^{})$ \\
  $\Pi_{\gamma Z}(p^2) =$ & \hspace{-0.4cm} $p^4\Big[-\dfrac{1}{\Lambda^2}c_Z^{}s_Z^{}(c_{2W}^{}-c_{2B}^{})\Big] +p^2\dfrac{m_Z^2}{\Lambda^2}c_Z^{} s_Z^{} \Big[8\left(c_Z^2 c_{WW}^{}-s_Z^2 c_{BB}^{}\right)$ \\
   & \hspace{-0.4cm} $-4\left(c_Z^2-s_Z^2\right) c_{WB}^{}+(c_W^{}-c_B^{})\Big]$ \\
  $\Sigma(p^2) =$ & \hspace{-0.4cm} $p^4\left(-\dfrac{1}{\Lambda^2}c_D^{}\right) +p^2\left[-\dfrac{v^2}{\Lambda^2}(2c_H^{}+c_R^{})\right]$ \\
  \hline
\end{tabular}
\caption{Transverse Vacuum polarization functions in terms of Wilson coefficients.} \label{tbl:Pis}
\vspace{-5pt}
\end{table}
\renewcommand\arraystretch{1.0}

\renewcommand\arraystretch{2.0}
\begin{table}[tb]
\centering
\begin{tabular}{|rl|}
  \hline
  $\Pi_{WW}(p^2)-\Pi_{33}(p^2) =$ & \hspace{-0.4cm} $m_W^2\dfrac{2v^2}{\Lambda^2}c_T^{}$ \\
  $\Pi_{33}(p^2) =$ & \hspace{-0.4cm} $p^4\left(-\dfrac{1}{\Lambda^2}c_{2W}^{}\right) +p^2\dfrac{2m_W^2}{\Lambda^2}(4c_{WW}^{}+c_W^{}) +m_W^2\dfrac{v^2}{\Lambda^2}(-2c_T^{}+c_R^{})$ \\
  $\Pi_{BB}(p^2) =$ & \hspace{-0.4cm} $p^4\left(-\dfrac{1}{\Lambda^2}c_{2B}^{}\right) +p^2\dfrac{2m_Z^2s_Z^2}{\Lambda^2}(4c_{BB}^{}+c_B^{}) +m_Z^2s_Z^2\dfrac{v^2}{\Lambda^2}(-2c_T^{}+c_R^{})$ \\
  $\Pi_{3B}(p^2) =$ & \hspace{-0.4cm} $p^2\left(-\dfrac{m_Z^2}{\Lambda^2}c_Z^{} s_Z^{}\right)(4c_{WB}^{}+c_W^{}+c_B^{}) + m_Z^2\dfrac{v^2}{\Lambda^2}c_Z^{} s_Z^{} (2c_T^{}-c_R^{})$ \\
  \hline
\end{tabular}
\caption{Alternative set of transverse vacuum polarization functions that are used in our definitions of EWPO parameters Table~\ref{tbl:EWPOdef}.} \label{tbl:AlternativePis}
\vspace{-5pt}
\end{table}
\renewcommand\arraystretch{1.0}

\begin{figure}[t]
\centering
 \subfigure[]{
 \centering
\begin{fmffile}{VV}
\begin{fmfgraph*}(25,25)
\fmfleft{i}
\fmfright{o}
\fmflabel{$\mu$}{i}
\fmflabel{$\nu$}{o}
\fmf{photon}{i,v,o}
\fmfv{decor.shape=circle,decor.filled=full,decor.size=4thick,l=$i\Pi_{VV}^{\mu\nu}(p^2)$,l.a=-90,l.d=5mm}{v}
\end{fmfgraph*}
\end{fmffile}
 }\hspace{2cm}
 \subfigure[]{
 \centering
\begin{fmffile}{hh}
\begin{fmfgraph*}(25,25)
\fmfleft{i}
\fmfright{o}
\fmflabel{$h$}{i}
\fmflabel{$h$}{o}
\fmf{dashes}{i,v,o}
\fmfv{decor.shape=circle,decor.filled=full,decor.size=4thick,l=$-i\Sigma(p^2)$,l.a=-90,l.d=5mm}{v}
\end{fmfgraph*}
\end{fmffile}
 }
 \caption{Feynman rules for vacuum polarization functions.} \label{fig:TwoPointFR}
\end{figure}
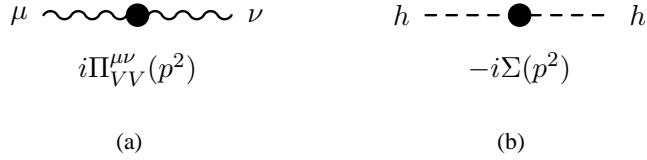

Throughout the calculations in the paper, the relevant vacuum polarization functions are those of the vector bosons $i\Pi_{VV}^{\mu\nu}(p^2) \in \left\{i\Pi_{WW}^{\mu\nu}(p^2), i\Pi_{ZZ}^{\mu\nu}(p^2), i\Pi_{\gamma\gamma}^{\mu\nu}(p^2), i\Pi_{\gamma Z}^{\mu\nu}(p^2)\right\}$ and that of the Higgs boson $-i\Sigma(p^2)$. It is straightforward to expand out the dim-6 effective operators listed in Table~\ref{tbl:operators2}, identify the relevant Lagrangian pieces, and obtain the Feynman rules. The relevant Lagrangian pieces are shown in Eq.~\eqref{eqn:LWW}-Eq.~\eqref{eqn:Lhh}. The resulting Feynman rules of the vacuum polarization functions are drawn in Fig.~\ref{fig:TwoPointFR}, with the detailed values listed in Eq.~\eqref{eqn:RuleWW}-Eq.~\eqref{eqn:Rulehh}. In the diagrams, we use a big solid dot to denote the interactions due to the dim-6 effective operators ({\it i.e.} due to Wilson coefficients $c_i$), while a simple direct connecting would represent the SM interaction.

For vector bosons, one can easily identify the transverse part of the vacuum polarization functions $\Pi_{VV}(p^2)$ from
\begin{equation}
i\Pi_{VV}^{\mu\nu}(p^2) = i\left(g^{\mu\nu}-\frac{p^\mu p^\nu}{p^2}\right)\Pi_{VV}(p^2) + \left( i\frac{p^\mu p^\nu}{p^2} \text{ term} \right) .
\end{equation}
These transverse vacuum polarization functions $\left\{\Pi_{WW}(p^2), \Pi_{ZZ}(p^2), \Pi_{\gamma\gamma}(p^2), \Pi_{\gamma Z}(p^2)\right\}$ together with $-i\Sigma(p^2)$ are summarized in Table~\ref{tbl:Pis}. In some occasions, such as defining the EWPO parameters, it is more concise to use the alternative set $\left\{\Pi_{33}, \Pi_{BB}, \Pi_{3B}\right\}$ instead of $\left\{\Pi_{ZZ}, \Pi_{\gamma\gamma}, \Pi_{\gamma Z}\right\}$. Due to the relation $W^3=c_Z^{}Z+s_Z^{}A$ and $B=-s_Z^{}Z+c_Z^{}A$, there is a simple transformation between these two sets
\begin{eqnarray}
\Pi_{33} &=& c_Z^2\Pi_{ZZ} + s_Z^2\Pi_{\gamma\gamma} + 2c_Z^{} s_Z^{}\Pi_{\gamma Z} , \\
\Pi_{BB} &=& s_Z^2\Pi_{ZZ} + c_Z^2\Pi_{\gamma\gamma} - 2c_Z^{} s_Z^{}\Pi_{\gamma Z} , \\
\Pi_{3B} &=&  -c_Z^{} s_Z^{}\Pi_{ZZ} + c_Z^{} s_Z^{}\Pi_{\gamma\gamma} + \big(c_Z^2-s_Z^2\big)\Pi_{\gamma Z} ,
\end{eqnarray}
where we have adopted the notation $c_Z^{} \equiv \cos\theta_Z$ {\it etc.}, with $\theta_Z$ denoting the weak mixing angle. This alternative set of vector boson transverse vacuum polarization functions are summarized in Table~\ref{tbl:AlternativePis}.

\begin{align}
{\cal L}_{WW} &= W_\mu^ + \left(\partial^4 g^{\mu\nu} -\partial^2 \partial^\mu \partial^\nu \right) W_\nu^- \cdot \left(-\frac{1}{\Lambda^2}c_{2W}^{} \right) \nonumber \\
 & \qquad+ W_\mu^+ \left( -\partial^2 g^{\mu\nu} +\partial^\mu \partial^\nu \right) W_\nu^- \cdot \frac{2m_W^2}{\Lambda^2}(4c_{WW}^{} +c_W^{}) \nonumber \\
 & \qquad+ m_W^2 W_\mu^+ W^{-\mu} \cdot \frac{v^2}{\Lambda^2} c_R^{} -W^{-\mu}\left(\partial_\mu \partial_\nu\right) W^{+\nu} \cdot \frac{m_W^2}{\Lambda^2}c_D^{} , \label{eqn:LWW} \\
{{\cal L}_{ZZ}} &= \frac{1}{2}{Z_\mu }\left( {{\partial ^4}{g^{\mu \nu }} - {\partial ^2}{\partial ^\mu }{\partial ^\nu }} \right){Z_\nu } \cdot \left[ { - \frac{1}{{{\Lambda ^2}}}\left( {c_Z^2{c_{2W}^{}} + s_Z^2{c_{2B}^{}}} \right)} \right] \nonumber \\
 &\qquad +\frac{1}{2}{Z_\mu }\left( { - {\partial ^2}{g^{\mu \nu }} + {\partial ^\mu }{\partial ^\nu }} \right){Z_\nu } \cdot \frac{{2m_Z^2}}{{{\Lambda ^2}}}\left[ \begin{array}{l}
4\left( {c_Z^4{c_{WW}^{}} + s_Z^4{c_{BB}^{}} + c_Z^2s_Z^2{c_{WB}^{}}} \right)\\
 + \left( {c_Z^2{c_W^{}} + s_Z^2 c_B^{}} \right)
\end{array} \right] \nonumber \\
 &\qquad +\frac{1}{2}m_Z^2{Z_\mu }{Z^\mu } \cdot \frac{{{v^2}}}{{{\Lambda ^2}}}\left( { - 2{c_T^{}} + {c_R^{}}} \right) , \label{eqn:LZZ}  \\
{{\cal L}_{\gamma \gamma }} &= \frac{1}{2}{A_\mu }\left( {{\partial ^4}{g^{\mu \nu }} - {\partial ^2}{\partial ^\mu }{\partial ^\nu }} \right){A_\nu } \cdot \left[ { - \frac{1}{{{\Lambda ^2}}}\left( {s_Z^2{c_{2W}^{}} + c_Z^2{c_{2B}^{}}} \right)} \right] \nonumber \\
 &\qquad +\frac{1}{2}{A_\mu }\left( { - {\partial ^2}{g^{\mu \nu }} + {\partial ^\mu }{\partial ^\nu }} \right){A_\nu } \cdot \frac{{8m_Z^2}}{{{\Lambda ^2}}}c_Z^2s_Z^2\left( {{c_{WW}^{}} + {c_{BB}^{}} - {c_{WB}^{}}} \right) , \label{eqn:LAA} \\
{{\cal L}_{\gamma Z}} &= {A_\mu }\left( {{\partial ^4}{g^{\mu \nu }} - {\partial ^2}{\partial ^\mu }{\partial ^\nu }} \right){Z_\nu } \cdot \left[ { - \frac{1}{{{\Lambda ^2}}}{c_Z^{}}{s_Z^{}}\left( {{c_{2W}^{}} - {c_{2B}^{}}} \right)} \right] \nonumber \\
 &\qquad +{A_\mu }\left( { - {\partial ^2}{g^{\mu \nu }} + {\partial ^\mu }{\partial ^\nu }} \right){Z_\nu } \cdot \frac{{m_Z^2}}{{{\Lambda ^2}}}{c_Z^{}}{s_Z^{}}\left[ \begin{array}{l}
8\left( {c_Z^2{c_{WW}^{}} - s_Z^2{c_{BB}^{}}} \right)\\
 - 4\left( {c_Z^2 - s_Z^2} \right){c_{WB}^{}} + \left( {{c_W^{}} - {c_B^{}}} \right)
\end{array} \right] , \label{eqn:LAZ} \\
{{\cal L}_{hh}} &= \frac{1}{2}h\left( {{\partial ^4}} \right)h \cdot \frac{1}{{{\Lambda ^2}}}{c_D^{}} + \frac{1}{2}h\left( { - {\partial ^2}} \right)h \cdot \frac{{{v^2}}}{{{\Lambda ^2}}}\left( {2{c_H^{}} + {c_R^{}}} \right) . \label{eqn:Lhh}
\end{align}

\begin{align}
i\Pi _{WW}^{\mu \nu }({p^2}) &= i\left( {{g^{\mu \nu }} - \frac{{{p^\mu }{p^\nu }}}{{{p^2}}}} \right) \cdot \bigg[ {p^4}\left( { - \frac{1}{{{\Lambda ^2}}}{c_{2W}^{}}} \right) + {p^2}\frac{{2m_W^2}}{{{\Lambda ^2}}}\left( {4{c_{WW}^{}} + {c_W^{}}} \right) \nonumber \\
&\qquad + m_W^2\frac{{{v^2}}}{{{\Lambda ^2}}}{c_R^{}} \bigg] + i\frac{{{p^\mu }{p^\nu }}}{{{p^2}}} \cdot \left( {{p^2}\frac{{m_W^2}}{{{\Lambda ^2}}}{c_D^{}} + m_W^2\frac{{{v^2}}}{{{\Lambda ^2}}}{c_R^{}}} \right) , \label{eqn:RuleWW} \\
i\Pi _{ZZ}^{\mu \nu }({p^2}) &= i\left( {{g^{\mu \nu }} - \frac{{{p^\mu }{p^\nu }}}{{{p^2}}}} \right) \cdot \Bigg\{
{p^4}\left[ { - \frac{1}{{{\Lambda ^2}}}\left( {c_Z^2{c_{2W}^{}} + {c_{2B}^{}}s_Z^2} \right)} \right] \nonumber \\
 &\qquad +{p^2}\frac{{2m_Z^2}}{{{\Lambda ^2}}}\left[ 4\left( {c_Z^4{c_{WW}^{}} + s_Z^4{c_{BB}^{}} + c_Z^2s_Z^2{c_{WB}^{}}} \right) + \left( {c_Z^2{c_W^{}} + s_Z^2{c_B^{}}} \right) \right] \nonumber \\
 &\qquad + m_Z^2\frac{{{v^2}}}{{{\Lambda ^2}}}\left( { - 2{c_T^{}} + {c_R^{}}} \right) \Bigg\} + i\frac{{{p^\mu }{p^\nu }}}{{{p^2}}} \cdot m_Z^2\frac{{{v^2}}}{{{\Lambda ^2}}}\left( { - 2{c_T^{}} + {c_R^{}}} \right) , \label{eqn:RuleZZ} \\
i\Pi _{\gamma \gamma }^{\mu \nu }({p^2}) &= i\left( {{g^{\mu \nu }} - \frac{{{p^\mu }{p^\nu }}}{{{p^2}}}} \right) \cdot \Bigg\{ {p^4}\left[ { - \frac{1}{{{\Lambda ^2}}}\left( {s_Z^2{c_{2W}^{}} + c_Z^2{c_{2B}^{}}} \right)} \right] \nonumber \\
 &\qquad + {p^2}\frac{{8m_Z^2}}{{{\Lambda ^2}}}c_Z^2s_Z^2\left( {{c_{WW}^{}} + {c_{BB}^{}} - {c_{WB}^{}}} \right) \Bigg\} , \label{eqn:RuleAA} \displaybreak \\
i\Pi _{\gamma Z}^{\mu \nu }({p^2}) &= i\left( {{g^{\mu \nu }} - \frac{{{p^\mu }{p^\nu }}}{{{p^2}}}} \right) \cdot \Bigg\{
{p^4}\left[ { - \frac{1}{{{\Lambda ^2}}}{c_Z^{}}{s_Z^{}}\left( {{c_{2W}^{}} - {c_{2B}^{}}} \right)} \right] \nonumber \\
 &\qquad +{p^2}\frac{{m_Z^2}}{{{\Lambda ^2}}}{c_Z^{}}{s_Z^{}}\left[ 8\left( {c_Z^2{c_{WW}^{}} - s_Z^2{c_{BB}^{}}} \right) -4\left( {c_Z^2 - s_Z^2} \right){c_{WB}^{}} + {c_W^{}} - {c_B^{}} \right] \Bigg\} , \hspace{0.4cm} \label{eqn:RuleAZ} \\
 - i\Sigma ({p^2}) &= i{p^4}\frac{1}{{{\Lambda ^2}}}{c_D^{}} + i{p^2}\frac{{{v^2}}}{{{\Lambda^2}}}\left( {2{c_H^{}} + {c_R^{}}} \right) . \label{eqn:Rulehh}
\end{align}

\subsubsection{Feynman rules for three-point vertices}

\begin{figure}[t]
\centering
 \subfigure[]{
 \centering\label{subfig:RulehWW}
\begin{fmffile}{hWW}
\begin{fmfgraph*}(25,35)
\fmfleft{i}
\fmfright{d6,o2,d5,d4,d3,d2,o1,d1}
\fmflabel{$h$}{i}
\fmflabel{$W^+,\mu$}{o1}
\fmflabel{$W^-,\nu$}{o2}
\fmfv{decor.shape=circle,decor.filled=full,decor.size=4thick,l=$iM_{hWW}^{\mu\nu}(p_1,,p_2)$,l.a=0,l.d=5mm}{v}
\fmf{dashes}{i,v}
\fmf{photon,label=$p_1$,label.side=left}{v,o1}
\fmf{photon,label=$p_2$,label.side=right}{v,o2}
\end{fmfgraph*}
\end{fmffile}
 }\hspace{3cm}
 \subfigure[]{
 \centering\label{subfig:RulehZZ}
\begin{fmffile}{hZZ}
\begin{fmfgraph*}(25,35)
\fmfleft{i}
\fmfright{d6,o2,d5,d4,d3,d2,o1,d1}
\fmflabel{$h$}{i}
\fmflabel{$Z,\mu$}{o1}
\fmflabel{$Z,\nu$}{o2}
\fmfv{decor.shape=circle,decor.filled=full,decor.size=4thick,l=$iM_{hZZ}^{\mu\nu}(p_1,,p_2)$,l.a=0,l.d=5mm}{v}
\fmf{dashes}{i,v}
\fmf{photon,label=$p_1$,label.side=left}{v,o1}
\fmf{photon,label=$p_2$,label.side=right}{v,o2}
\end{fmfgraph*}
\end{fmffile}
 }\hspace{3cm}
 \subfigure[]{
 \centering\label{subfig:RulehAZ}
\begin{fmffile}{hAZ}
\begin{fmfgraph*}(25,35)
\fmfleft{i}
\fmfright{d6,o2,d5,d4,d3,d2,o1,d1}
\fmflabel{$h$}{i}
\fmflabel{$\gamma,\mu$}{o1}
\fmflabel{$Z,\nu$}{o2}
\fmfv{decor.shape=circle,decor.filled=full,decor.size=4thick,l=$iM_{h\gamma Z}^{\mu\nu}(p_1,,p_2)$,l.a=0,l.d=5mm}{v}
\fmf{dashes}{i,v}
\fmf{photon,label=$p_1$,label.side=left}{v,o1}
\fmf{photon,label=$p_2$,label.side=right}{v,o2}
\end{fmfgraph*}
\end{fmffile}
 }

\vspace{0.5cm}
 \subfigure[]{
 \centering\label{subfig:RulehAA}
\begin{fmffile}{hAA}
\begin{fmfgraph*}(25,35)
\fmfleft{i}
\fmfright{d6,o2,d5,d4,d3,d2,o1,d1}
\fmflabel{$h$}{i}
\fmflabel{$\gamma,\mu$}{o1}
\fmflabel{$\gamma,\nu$}{o2}
\fmfv{decor.shape=circle,decor.filled=full,decor.size=4thick,l=$iM_{h\gamma\gamma}^{\mu\nu}(p_1,,p_2)$,l.a=0,l.d=5mm}{v}
\fmf{dashes}{i,v}
\fmf{photon,label=$p_1$,label.side=left}{v,o1}
\fmf{photon,label=$p_2$,label.side=right}{v,o2}
\end{fmfgraph*}
\end{fmffile}
 }\hspace{3cm}
 \subfigure[]{
 \centering\label{subfig:Rulehgg}
\begin{fmffile}{hgg}
\begin{fmfgraph*}(25,35)
\fmfleft{i}
\fmfright{d6,o2,d5,d4,d3,d2,o1,d1}
\fmflabel{$h$}{i}
\fmflabel{$g,\mu$}{o1}
\fmflabel{$g,\nu$}{o2}
\fmfv{decor.shape=circle,decor.filled=full,decor.size=4thick,l=$iM_{hgg}^{\mu\nu}(p_1,,p_2)$,l.a=0,l.d=5mm}{v}
\fmf{dashes}{i,v}
\fmf{gluon,label=$p_1$,label.side=left}{v,o1}
\fmf{gluon,label=$p_2$,label.side=right}{v,o2}
\end{fmfgraph*}
\end{fmffile}
 }
 \caption{Feynman rules for three-point vertices.} \label{fig:ThreePointFR}
\end{figure}
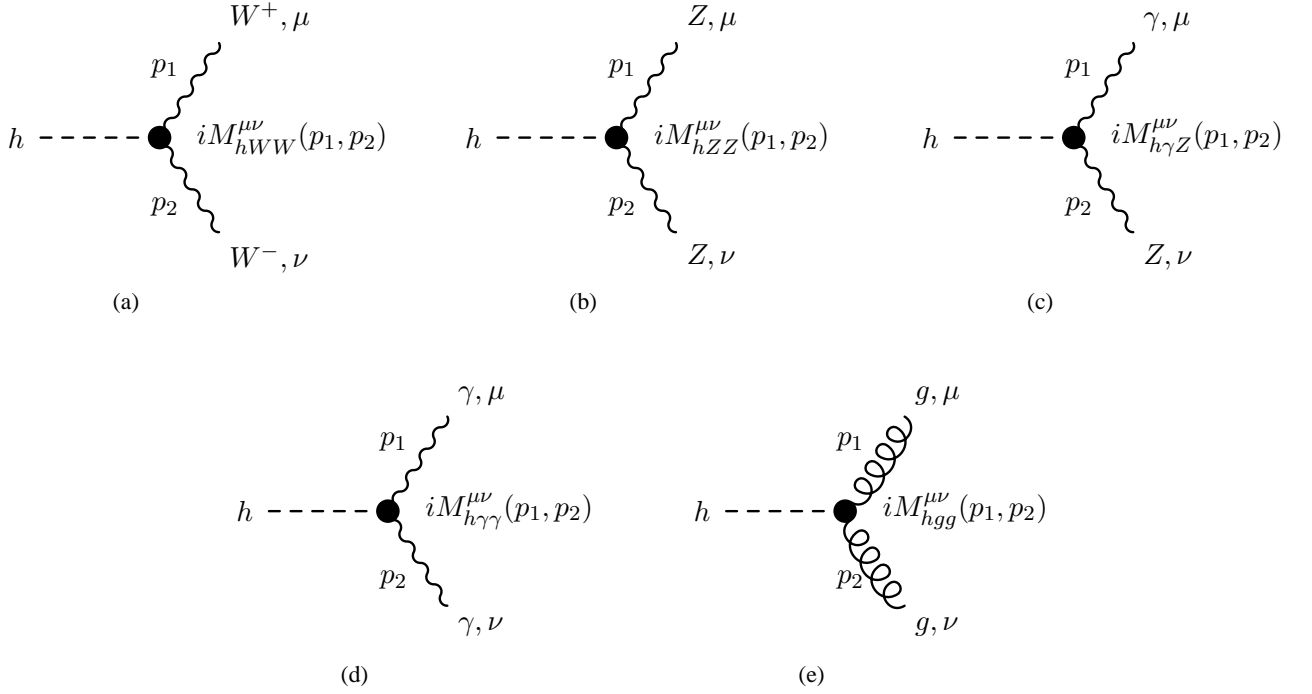

In this paper, the relevant three-point vertices are $hWW$, $hZZ$, $h\gamma Z$, $h\gamma\gamma$, and $hgg$ vertices. As with the vacuum polarization functions case, we expand out the dim-6 effective operators in Table~\ref{tbl:operators2} and identify the relevant Lagrangian pieces (Eq.~\eqref{eqn:LhWW}-Eq.~\eqref{eqn:Lhgg}). These Lagrangian pieces generate the Feynman rules shown in Fig.~\ref{fig:ThreePointFR}, with detailed values listed in Eq.~\eqref{eqn:RulehWW}-Eq.~\eqref{eqn:Rulehgg}.

\renewcommand\arraystretch{1.2}
\begin{eqnarray}
{{\cal L}_{hWW}} &=& \frac{{\sqrt 2 m_W^2}}{v}\Bigg\{ \frac{1}{2}h\hat W_{\mu \nu }^ + {{\hat W}^{ - \mu \nu }} \cdot \frac{1}{{{\Lambda ^2}}}8{c_{WW}} + h\left[ \begin{array}{l}
W_\mu ^ + \left( { - {\partial ^2}{g^{\mu \nu }} + {\partial ^\mu }{\partial ^\nu }} \right)W_\nu ^ - \\
 + W_\mu ^ - \left( { - {\partial ^2}{g^{\mu \nu }} + {\partial ^\mu }{\partial ^\nu }} \right)W_\nu ^ +
\end{array} \right] \cdot \frac{1}{{{\Lambda ^2}}}{c_W} \nonumber \\
 && + \left[ \begin{array}{l}
 - \left( {{\partial ^2}h} \right)W_\mu ^ - {W^{ + \mu }} - h\left( {{\partial _\mu }{W^{ - \mu }}} \right)\left( {{\partial _\nu }{W^{ + \nu }}} \right)\\
 - h{W^{ + \mu }}\left( {{\partial _\mu }{\partial _\nu }} \right){W^{ - \nu }} - h{W^{ - \mu }}\left( {{\partial _\mu }{\partial _\nu }} \right){W^{ + \nu }}
\end{array} \right] \cdot \frac{1}{{{\Lambda ^2}}}{c_D} + hW_\mu ^ + {W^{ - \mu }} \cdot \frac{{2{v^2}}}{{{\Lambda ^2}}}{c_R} \Bigg\} , \hspace{1cm} \label{eqn:LhWW} \\
{{\cal L}_{hZZ}} &=& \frac{{\sqrt 2 m_Z^2}}{v}\Bigg\{
\frac{1}{4}h{{\hat Z}_{\mu \nu }}{{\hat Z}^{\mu \nu }} \cdot \frac{1}{{{\Lambda ^2}}}8\left( {c_Z^4{c_{WW}} + s_Z^4{c_{BB}} + c_Z^2s_Z^2{c_{WB}}} \right) \nonumber \\
 && + \frac{1}{2}h{Z_\mu }\left( { - {\partial ^2}{g^{\mu \nu }} + {\partial ^\mu }{\partial ^\nu }} \right){Z_\nu } \cdot \frac{1}{{{\Lambda ^2}}}2\left( {c_Z^2{c_W} + s_Z^2{c_B}} \right) \nonumber \\
 && + \left[ { - \frac{1}{2}\left( {{\partial ^2}h} \right)\left( {{Z_\mu }{Z^\mu }} \right) - \frac{1}{2}h\left( {{\partial _\mu }{Z^\mu }} \right)\left( {{\partial _\nu }{Z^\nu }} \right) - h{Z^\mu }\left( {{\partial _\mu }{\partial _\nu }} \right){Z^\nu }} \right] \cdot \frac{1}{{{\Lambda ^2}}}{c_D} \nonumber \\
 && + \frac{1}{2}h{Z_\mu }{Z^\mu } \cdot \frac{{2{v^2}}}{{{\Lambda ^2}}}\left( { - 2{c_T} + {c_R}} \right) \Bigg\} , \label{eqn:LhZZ} \\
{{\cal L}_{h\gamma Z}} &=& \frac{{\sqrt 2 m_Z^2}}{v}\Bigg\{ \frac{1}{2}h{{\hat Z}_{\mu \nu }}{A^{\mu \nu }} \cdot \frac{1}{{{\Lambda ^2}}}4{c_Z^{}}{s_Z^{}}\Big[ {2\left( {c_Z^2{c_{WW}^{}} - s_Z^2{c_{BB}^{}}} \right) - \left( {c_Z^2 - s_Z^2} \right){c_{WB}^{}}} \Big] \nonumber \\
 && + h{Z_\mu }\left( { - {\partial ^2}{g^{\mu \nu }} + {\partial ^\mu }{\partial ^\nu }} \right){A_\nu } \cdot \frac{1}{{{\Lambda ^2}}}{c_Z^{}}{s_Z^{}}({c_W^{}} - {c_B^{}}) \Bigg\} , \label{eqn:LhAZ} \\
{{\cal L}_{h\gamma \gamma }} &=& \frac{{\sqrt 2 m_Z^2}}{v}\frac{1}{4}h{A_{\mu \nu }}{A^{\mu \nu }} \cdot \frac{1}{{{\Lambda ^2}}}8c_Z^2s_Z^2\left( {{c_{WW}^{}} + {c_{BB}^{}} - {c_{WB}^{}}} \right) , \label{eqn:LhAA} \\
{{\cal L}_{hgg}} &=& \frac{{\sqrt 2 g_s^2{v^2}}}{{2v}}\frac{1}{4}hG_{\mu \nu }^a{G^{a,\mu \nu }} \cdot \frac{1}{{{\Lambda ^2}}}8{c_{GG}^{}} . \label{eqn:Lhgg}
\end{eqnarray}
\renewcommand\arraystretch{1.0}

\begin{eqnarray}
iM_{hWW}^{\mu\nu}(p_1, p_2) &=& i\frac{\sqrt 2 m_W^2}{v}\Bigg\{-\left(p_1 p_2 g^{\mu\nu}-p_1^\nu p_2^\mu\right)\frac{1}{\Lambda^2}8c_{WW}^{} +\Big[ \left(p_1^2 g^{\mu\nu}-p_1^\mu p_1^\nu \right) + \left(p_2^2 g^{\mu\nu} -p_2^\mu p_2^\nu \right) \Big]\frac{1}{\Lambda^2} c_W^{} \nonumber \\
 && +\Big[(p_1+p_2)^2 g^{\mu\nu}+p_1^\mu p_2^\nu+p_1^\mu p_1^\nu+p_2^\mu p_2^\nu \Big]\frac{1}{\Lambda^2}c_D^{} +g^{\mu\nu} \frac{2v^2}{\Lambda^2} c_R^{} \Bigg\} , \label{eqn:RulehWW} \\
iM_{hZZ}^{\mu\nu}(p_1, p_2) &=& i\frac{\sqrt{2} m_Z^2}{v} \Bigg\{ -\big(p_1 p_2 g^{\mu\nu} - p_1^\nu p_2^\mu \big)\frac{1}{\Lambda^2} 8\big(c_Z^4 c_{WW}^{} + s_Z^4 c_{BB}^{} + c_Z^2 s_Z^2 c_{WB}^{} \big) \nonumber \\
 && +\Big[\big(p_1^2 g^{\mu\nu} - p_1^\mu p_1^\nu \big) + \big(p_2^2 g^{\mu\nu}- p_2^\mu p_2^\nu \big) \Big]\frac{1}{\Lambda^2} \big(c_Z^2 c_W^{} + s_Z^2 c_B^{}\big) \nonumber \\
 && +\Big[ (p_1+p_2)^2 g^{\mu\nu} + p_1^\mu p_2^\nu + p_1^\mu p_1^\nu + p_2^\mu p_2^\nu \Big] \frac{1}{\Lambda^2} c_D^{} + g^{\mu\nu} \frac{2v^2}{\Lambda^2} (-2c_T^{}+c_R^{}) \Bigg\} , \label{eqn:RulehZZ} \\
iM_{h\gamma Z}^{\mu\nu}(p_1, p_2) &=& i\frac{\sqrt{2} m_Z^2}{v} \Bigg\{ -\big(p_1 p_2 g^{\mu\nu} - p_1^\nu p_2^\mu\big) \frac{4c_Z^{}s_Z^{}}{\Lambda^2}\Big[2\big(c_Z^2 c_{WW}^{} - s_Z^2 c_{BB}^{}\big) - \big(c_Z^2-s_Z^2\big) c_{WB}^{} \Big] \nonumber \\
 && +\big(p_1^2 g^{\mu\nu} - p_1^\mu p_1^\nu \big)\frac{1}{\Lambda^2} c_Z^{}s_Z^{}(c_W^{}-c_B^{}) \Bigg\} , \label{eqn:RulehAZ} \\
iM_{h\gamma\gamma}^{\mu\nu}(p_1, p_2) &=& -i\frac{\sqrt{2} m_Z^2}{v} \big(p_1 p_2 g^{\mu\nu} -p_1^\nu p_2^\mu \big) \frac{1}{\Lambda^2} 8c_Z^2 s_Z^2(c_{WW}^{} + c_{BB}^{} - c_{WB}^{}) , \label{eqn:RulehAA} \\
iM_{hgg}^{\mu\nu}(p_1, p_2) &=& -i\frac{\sqrt{2} g_s^2 v^2}{2v}\big(p_1 p_2 g^{\mu\nu} - p_1^\nu p_2^\mu \big)\frac{1}{\Lambda^2}8c_{GG}^{} . \label{eqn:Rulehgg}
\end{eqnarray}

\subsection{Details on interference corrections to the Higgs decay widths} \label{subsec:GammaIdetails}

\begin{figure}[t]
\centering
 \subfigure[]{
 \centering
\begin{fmffile}{GammahWW1}
\begin{fmfgraph*}(40,30)
\fmfleft{i}
\fmfright{o4,o3,o1}
\fmfv{decor.shape=circle,decor.filled=full,decor.size=4thick}{v1}
\fmfv{l=$h$,l.a=180,l.d=3mm}{i}
\fmfv{l=$W^+$,l.a=0,l.d=3mm}{o1}
\fmfv{l=$l / d$,l.a=0,l.d=3mm}{o3}
\fmfv{l=$\bar\nu / {\bar u}$,l.a=0,l.d=3mm}{o4}
\fmf{dashes,tension=1.2}{i,v1}
\fmf{photon,tension=0.35}{v1,o1}
\fmf{photon,label=$W^-$,label.side=right}{v1,v2}
\fmf{fermion}{o4,v2,o3}
\end{fmfgraph*}
\end{fmffile}
 }\hspace{2cm}
 \subfigure[]{
 \centering
\begin{fmffile}{GammahWW2}
\begin{fmfgraph*}(45,35)
\fmfleft{i}
\fmfright{o4,o3,o1}
\fmfv{decor.shape=circle,decor.filled=full,decor.size=4thick}{v3}
\fmfv{l=$h$,l.a=180,l.d=3mm}{i}
\fmfv{l=$W^+$,l.a=0,l.d=3mm}{o1}
\fmfv{l=$l / d$,l.a=0,l.d=3mm}{o3}
\fmfv{l=$\bar\nu / {\bar u}$,l.a=0,l.d=3mm}{o4}
\fmf{dashes,tension=0.8}{i,v1}
\fmf{photon,tension=0.2}{v1,o1}
\fmf{photon,label=$W^-$,label.side=right}{v1,v3}
\fmf{photon,label=$W^-$,label.side=right}{v3,v2}
\fmf{fermion}{o4,v2,o3}
\end{fmfgraph*}
\end{fmffile}
 }
 \caption{New amputated Feynman diagrams for $\Gamma_{hWW^*}$.} \label{fig:epsIhWW}
\end{figure}
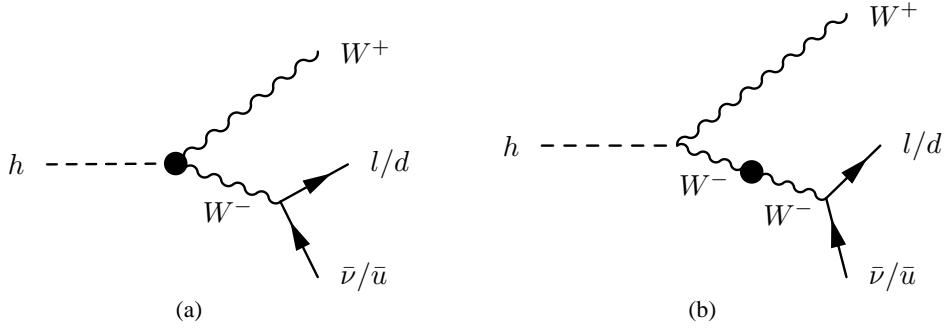

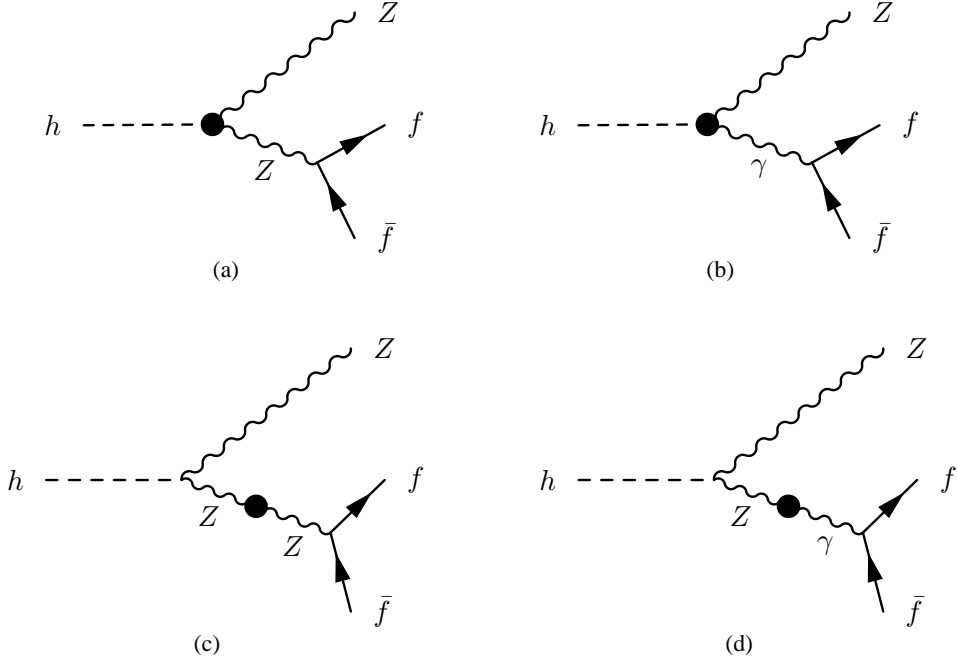
\begin{figure}[t]
\centering
 \subfigure[]{
 \centering
\begin{fmffile}{GammahZZ1}
\begin{fmfgraph*}(40,30)
\fmfleft{i}
\fmfright{o4,o3,o1}
\fmfv{decor.shape=circle,decor.filled=full,decor.size=4thick}{v1}
\fmfv{l=$h$,l.a=180,l.d=3mm}{i}
\fmfv{l=$Z$,l.a=0,l.d=3mm}{o1}
\fmfv{l=$f$,l.a=0,l.d=3mm}{o3}
\fmfv{l=${\bar f}$,l.a=0,l.d=3mm}{o4}
\fmf{dashes,tension=1.2}{i,v1}
\fmf{photon,tension=0.35}{v1,o1}
\fmf{photon,label=$Z$,label.side=right}{v1,v2}
\fmf{fermion}{o4,v2,o3}
\end{fmfgraph*}
\end{fmffile}
 }\hspace{2cm}
 \subfigure[]{
 \centering
\begin{fmffile}{GammahZZ2}
\begin{fmfgraph*}(40,30)
\fmfleft{i}
\fmfright{o4,o3,o1}
\fmfv{decor.shape=circle,decor.filled=full,decor.size=4thick}{v1}
\fmfv{l=$h$,l.a=180,l.d=3mm}{i}
\fmfv{l=$Z$,l.a=0,l.d=3mm}{o1}
\fmfv{l=$f$,l.a=0,l.d=3mm}{o3}
\fmfv{l=${\bar f}$,l.a=0,l.d=3mm}{o4}
\fmf{dashes,tension=1.2}{i,v1}
\fmf{photon,tension=0.35}{v1,o1}
\fmf{photon,label=$\gamma$,label.side=right}{v1,v2}
\fmf{fermion}{o4,v2,o3}
\end{fmfgraph*}
\end{fmffile}
 }

\vspace{0.5cm}
 \subfigure[]{
 \centering
\begin{fmffile}{GammahZZ3}
\begin{fmfgraph*}(45,35)
\fmfleft{i}
\fmfright{o4,o3,o1}
\fmfv{decor.shape=circle,decor.filled=full,decor.size=4thick}{v3}
\fmfv{l=$h$,l.a=180,l.d=3mm}{i}
\fmfv{l=$Z$,l.a=0,l.d=3mm}{o1}
\fmfv{l=$f$,l.a=0,l.d=3mm}{o3}
\fmfv{l=${\bar f}$,l.a=0,l.d=3mm}{o4}
\fmf{dashes,tension=0.8}{i,v1}
\fmf{photon,tension=0.2}{v1,o1}
\fmf{photon,label=$Z$,label.side=right}{v1,v3}
\fmf{photon,label=$Z$,label.side=right}{v3,v2}
\fmf{fermion}{o4,v2,o3}
\end{fmfgraph*}
\end{fmffile}
 }\hspace{2cm}
 \subfigure[]{
 \centering
\begin{fmffile}{GammahZZ4}
\begin{fmfgraph*}(45,35)
\fmfleft{i}
\fmfright{o4,o3,o1}
\fmfv{decor.shape=circle,decor.filled=full,decor.size=4thick}{v3}
\fmfv{l=$h$,l.a=180,l.d=3mm}{i}
\fmfv{l=$Z$,l.a=0,l.d=3mm}{o1}
\fmfv{l=$f$,l.a=0,l.d=3mm}{o3}
\fmfv{l=${\bar f}$,l.a=0,l.d=3mm}{o4}
\fmf{dashes,tension=0.8}{i,v1}
\fmf{photon,tension=0.2}{v1,o1}
\fmf{photon,label=$Z$,label.side=right}{v1,v3}
\fmf{photon,label=$\gamma$,label.side=right}{v3,v2}
\fmf{fermion}{o4,v2,o3}
\end{fmfgraph*}
\end{fmffile}
 }
 \caption{New amputated Feynman diagrams for $\Gamma_{hZZ^*}$.} \label{fig:epsIhZZ}
\end{figure}

There is no new amputated diagrams for $h \to f{\bar f}$ decay modes up to leading order (linear power and tree level) in Wilson coefficients, because we are considering only the bosonic dim-6 effective operators (Table~\ref{tbl:operators2}). The $h \to gg$, $h \to \gamma\gamma$, and $h \to \gamma Z$ decay widths are already at one-loop order in the SM, so the only new amputated diagram up to leading order in Wilson coefficients is given by the new three-point vertices $iM_{hgg}^{\mu\nu}(p_1,p_2)$, $iM_{h\gamma\gamma}^{\mu\nu}(p_1,p_2)$, and $iM_{h\gamma Z}^{\mu\nu}(p_1,p_2)$ (Fig.~\ref{subfig:RulehAA}, Fig.~\ref{subfig:Rulehgg}, and Fig.~\ref{subfig:RulehAZ}) multiplied by appropriate polarization vectors
\begin{eqnarray}
iM_{hgg,\text{ AD,new}} &=& iM_{hgg}^{\mu\nu}(p_1,p_2) \epsilon_\mu^*(p_1) \epsilon_\nu^*(p_2) , \\
iM_{h\gamma\gamma,\text{ AD,new}} &=& iM_{h\gamma\gamma}^{\mu\nu}(p_1,p_2) \epsilon_\mu^*(p_1) \epsilon_\nu^*(p_2) , \\
iM_{h\gamma Z,\text{ AD,new}} &=& iM_{h\gamma Z}^{\mu\nu}(p_1,p_2) \epsilon_\mu^*(p_1) \epsilon_\nu^*(p_2) .
\end{eqnarray}
The $h \to WW^*$ and $h \to ZZ^*$ modes are a little more complicated, because they are at tree level in the SM. It turns out that there are two new amputated diagrams for $h \to WW^*$ mode as shown in Fig.~\ref{fig:epsIhWW}, and four new amputated diagrams for $h \to ZZ^*$ mode as shown in Fig.~\ref{fig:epsIhZZ}.

It is straightforward to evaluate these relevant new diagrams using the new Feynman rules listed in Section~\ref{subsec:FeynmanRules} (together with the SM Feynman rules). One can then compute the interference correction $\epsilon_I^{}$ for each decay mode from its definition (Eq.~\eqref{eqn:epsIdef}). The three-body phase space integrals are analytically manageable, albeit a little bit tedious. We summarize the final results of $\epsilon_I^{}$ in Table~\ref{tbl:epsI_width}, where the auxiliary integrals $I_a(\beta)$, $I_b(\beta)$, $I_c(\beta)$, and $I_d(\beta)$ are defined as

\renewcommand\arraystretch{1.2}
\begin{eqnarray}
{I_\text{SM}}(\beta ) &\equiv& \frac{1}{{8{\beta ^2}}}\left[ {{I_2}(\beta ) + 2(1 - 6{\beta ^2}){I_1}(\beta ) + (1 - 4{\beta ^2} + 12{\beta ^4}){I_0}(\beta )} \right] , \label{eqn:ISMdef} \\
{I_a}(\beta ) &\equiv& \frac{1}{{8{\beta ^4}{I_\text{SM}}(\beta )}}\left[ \begin{array}{l}
{I_3}(\beta ) + (1 - 16{\beta ^2}){I_2}(\beta ) + (1 - 12{\beta ^2} + 62{\beta ^4}){I_1}(\beta )\\
 - 4({\beta ^2} - 5{\beta ^4} + 18{\beta ^6}){I_0}(\beta ) + 2({\beta ^4} - 4{\beta ^6} + 12{\beta ^8}){I_{ - 1}}(\beta )
\end{array} \right] , \label{eqn:Iadef} \\
{I_b}(\beta ) &\equiv& \frac{1}{{4{\beta ^2}{I_\text{SM}}(\beta )}}\left[ \begin{array}{l}
 - 2{I_2}(\beta ) - (4 - 25{\beta ^2}){I_1}(\beta ) - 2(1 - 5{\beta ^2} + 18{\beta ^4}){I_0}(\beta )\\
 + {\beta ^2}(1 - 4{\beta ^2} + 12{\beta ^4}){I_{ - 1}}(\beta )
\end{array} \right] , \label{eqn:Ibdef} \\
{I_c}(\beta ) &\equiv& \frac{{5{I_2}(\beta ) + 2(2 - 3{\beta ^2}){I_1}(\beta ) - (1 + 2{\beta ^2}){I_0}(\beta )}}{{2{\beta ^2}{I_\text{SM}}(\beta )}} , \label{eqn:Icdef} \\
{I_d}(\beta ) &\equiv& \frac{{7{I_2}(\beta ) + 8(1 - 3{\beta ^2}){I_1}(\beta ) + (1 - 4{\beta ^2} + 12{\beta ^4}){I_0}(\beta )}}{{2{\beta ^2}{I_\text{SM}}(\beta )}} , \label{eqn:Iddef}
\end{eqnarray}
\renewcommand\arraystretch{1.0}
where another set of auxiliary integrals $I_0(\beta)$, $I_1(\beta)$, $I_2(\beta)$, $I_3(\beta)$, $I_{-1}(\beta)$ are defined as follows, with $\beta \in (\frac{1}{2},1)$
\begin{eqnarray}
{I_0}(\beta ) &\equiv& \int_{2\beta  - 1}^{{\beta ^2}} {\frac{{dy\sqrt {{{(y + 1)}^2} - 4{\beta ^2}} }}{{{y^2}}}}  = 1 - \frac{1}{{{\beta ^2}}} - \ln \beta  + \frac{{\frac{\pi }{2} - \arcsin \frac{{3{\beta ^2} - 1}}{{2{\beta ^3}}}}}{{\sqrt {4{\beta ^2} - 1} }} , \nonumber \\
{I_1}(\beta ) &\equiv& \int_{2\beta  - 1}^{{\beta ^2}} {\frac{{dy\sqrt {{{(y + 1)}^2} - 4{\beta ^2}} }}{{{y^2}}}y}  = 1 - {\beta ^2} - \ln \beta  - \frac{{\frac{\pi }{2} - \arcsin \frac{{3{\beta ^2} - 1}}{{2{\beta ^3}}}}}{{\sqrt {4{\beta ^2} - 1} }}(4{\beta ^2} - 1) , \nonumber \\
{I_2}(\beta ) &\equiv& \int_{2\beta  - 1}^{{\beta ^2}} {\frac{{dy\sqrt {{{(y + 1)}^2} - 4{\beta ^2}} }}{{{y^2}}}{y^2}}  = \frac{1}{2}(1 - {\beta ^4}) + 2{\beta ^2}\ln \beta , \nonumber \\
{I_3}(\beta ) &\equiv& \int_{2\beta  - 1}^{{\beta ^2}} {\frac{{dy\sqrt {{{(y + 1)}^2} - 4{\beta ^2}} }}{{{y^2}}}({y^3} + {y^2})}  = \frac{1}{3}{(1 - {\beta ^2})^3} , \nonumber \\
{I_{ - 1}}(\beta ) &\equiv& \int_{2\beta  - 1}^{{\beta ^2}} {\frac{{dy\sqrt {{{(y + 1)}^2} - 4{\beta ^2}} }}{{{y^3}}}}  = \frac{{2{\beta ^2}\Big(\frac{\pi }{2} - \arcsin \frac{{3{\beta ^2} - 1}}{{2{\beta ^3}}}\Big)}}{{{{(4{\beta ^2} - 1)}^{\frac{3}{2}}}}} - \frac{{(1 - {\beta ^2})(3{\beta ^2} - 1)}}{{2{\beta ^4}(4{\beta ^2} - 1)}} . \nonumber
\end{eqnarray}
The $A_{hgg}^\text{SM}$, $A_{h\gamma\gamma}^\text{SM}$, and $A_{h\gamma Z}^\text{SM}$ in Table~\ref{tbl:epsI_width} are the standard form factors
\begin{eqnarray}
A_{hgg}^\text{SM} &=& \sum\limits_Q {{A_{1/2}}({\tau _Q})} , \label{eqn:Ahgg} \\
A_{h\gamma\gamma}^\text{SM} &=& {A_1}({\tau _W}) + \sum\limits_f {{N_C}Q_f^2{A_{1/2}}({\tau _f})} , \label{eqn:AhAA} \\
A_{h\gamma Z}^\text{SM} &=& {A_1}({\tau _W},{\lambda _W}) + \sum\limits_f {{N_C}\frac{{2{Q_f}}}{{{c_Z}}}\big(T_f^3 - 2s_Z^2{Q_f}\big){A_{1/2}}({\tau _f},{\lambda _f})} , \label{eqn:AhAZ}
\end{eqnarray}
with $\tau_i\equiv\frac{4m_i^2}{m_h^2}$, $\lambda_i\equiv\frac{4m_i^2}{m_Z^2}$, and $A_{1/2}(\tau)$, $A_1(\tau)$, $A_{1/2}(\tau,\lambda)$, $A_1(\tau,\lambda)$ being the conventional form factors (for example see~\cite{Djouadi:2005gi})
\begin{eqnarray}
{A_{1/2}}\left(\tau \right) &=& 2{\tau ^{ - 2}}\Big[ {\tau  + \left(\tau  - 1\right)f\left(\tau \right)} \Big] , \\
{A_1}\left(\tau \right) &=&  - {\tau ^{ - 2}}\Big[ {2{\tau ^2} + 3\tau  + 3\left(2\tau  - 1\right)f\left(\tau \right)} \Big] , \\
{A_{1/2}}\left(\tau ,\lambda \right) &=& {B_1}\left(\tau ,\lambda \right) - {B_2}\left(\tau ,\lambda \right) , \\
{A_1}\left(\tau ,\lambda \right) &=& {c_Z}\Bigg\{ {4\left(3 - \frac{{s_Z^2}}{{c_Z^2}}\right){B_2}\left(\tau ,\lambda \right) + \left[ {\left(1 + \frac{2}{\tau }\right)\frac{{s_Z^2}}{{c_Z^2}} - \left(5 + \frac{2}{\tau }\right)} \right]{B_1}\left(\tau ,\lambda \right)} \Bigg\} ,
\end{eqnarray}
with
\begin{eqnarray}
B_1(\tau, \lambda) &\equiv& \frac{\tau\lambda}{2(\tau-\lambda)} + \frac{\tau^2 \lambda^2}{2(\tau-\lambda)^2} \left[ f\left(\frac{1}{\tau}\right) - f\left(\frac{1}{\lambda}\right) \right] + \frac{\tau^2 \lambda}{(\tau-\lambda)^2} \left[ g\left(\frac{1}{\tau}\right) - g\left(\frac{1}{\lambda}\right) \right] , \nonumber \\
B_2(\tau, \lambda) &\equiv& -\frac{\tau\lambda}{2(\tau-\lambda)} \left[ f\left(\frac{1}{\tau}\right) - f\left(\frac{1}{\lambda}\right) \right] , \nonumber
\end{eqnarray}
and
\begin{eqnarray}
f(\tau ) &=& \left\{ \begin{array}{ll}
{\arcsin ^2}\sqrt \tau & \hspace{1cm} \tau \le 1 \\
 - \dfrac{1}{4}{\left[ {\log \dfrac{{1 + \sqrt {1 - {\tau ^{ - 1}}} }}{{1 - \sqrt {1 - {\tau ^{ - 1}}} }} - i\pi } \right]^2} & \hspace{1cm} \tau > 1
\end{array} \right. , \\
g(\tau ) &=& \left\{ \begin{array}{ll}
\sqrt {{\tau ^{ - 1}} - 1} \arcsin \sqrt \tau & \hspace{1cm} \tau \le 1 \\
\dfrac{{\sqrt {1 - {\tau ^{ - 1}}} }}{2}\left[ {\log \dfrac{{1 + \sqrt {1 - {\tau ^{ - 1}}} }}{{1 - \sqrt {1 - {\tau ^{ - 1}}} }} - i\pi } \right] & \hspace{1cm} \tau > 1
\end{array} \right. .
\end{eqnarray}

\subsection{Details on interference corrections to Higgs production cross section} \label{subsec:SigmaIdetails}

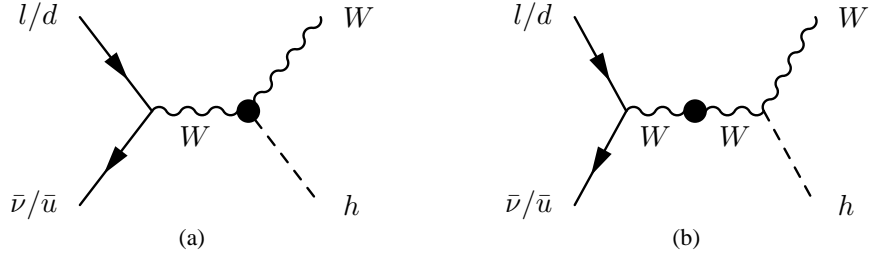
\begin{figure}[t]
\centering
 \subfigure[]{
 \centering
\begin{fmffile}{sigmaWh1}
\begin{fmfgraph*}(40,25)
\fmfleft{i2,i1}
\fmfright{o2,o1}
\fmfv{decor.shape=circle,decor.filled=full,decor.size=4thick}{v2}
\fmfv{l=$l / d$,l.a=180,l.d=3mm}{i1}
\fmfv{l=${\bar\nu / {\bar u}}$,l.a=180,l.d=3mm}{i2}
\fmfv{l=$W$,l.a=0,l.d=3mm}{o1}
\fmfv{l=$h$,l.a=0,l.d=3mm}{o2}
\fmf{fermion}{i1,v1,i2}
\fmf{photon,tension=1.5,label=$W$,label.side=right}{v1,v2}
\fmf{photon}{v2,o1}
\fmf{dashes}{v2,o2}
\end{fmfgraph*}
\end{fmffile}
 }\hspace{2cm}
 \subfigure[]{
 \centering
\begin{fmffile}{sigmaWh2}
\begin{fmfgraph*}(40,25)
\fmfleft{i2,i1}
\fmfright{o2,o1}
\fmfv{decor.shape=circle,decor.filled=full,decor.size=4thick}{v3}
\fmfv{l=$l / d$,l.a=180,l.d=3mm}{i1}
\fmfv{l=${\bar\nu / {\bar u}}$,l.a=180,l.d=3mm}{i2}
\fmfv{l=$W$,l.a=0,l.d=3mm}{o1}
\fmfv{l=$h$,l.a=0,l.d=3mm}{o2}
\fmf{fermion}{i1,v1,i2}
\fmf{photon,tension=1.5,label=$W$,label.side=right}{v1,v3}
\fmf{photon,tension=1.5,label=$W$,label.side=right}{v3,v2}
\fmf{photon}{v2,o1}
\fmf{dashes}{v2,o2}
\end{fmfgraph*}
\end{fmffile}
 }
 \caption{New amputated Feynman diagrams for $\sigma_{Wh}$.} \label{fig:epsIWh}
\end{figure}

The $ggF$ Higgs production mode is just the time reversal of the $h \to gg$ decay. Again as it is already at one-loop order in the SM, the only new amputated diagram up to leading order in Wilson coefficients is given by the new three-point vertex $iM_{hgg}^{\mu\nu}(p_1,p_2)$ (Fig.~\ref{subfig:Rulehgg}) multiplied by the polarization vectors
\begin{equation}
iM_{ggF,\text{ AD,new}} = iM_{hgg}^{\mu\nu}(p_1,p_2) \epsilon_\mu(p_1) \epsilon_\nu(p_2) .
\end{equation}
Obviously, the interference correction to $ggF$ production cross section is the same as that to $h \to gg$ decay width
\begin{equation}
\epsilon_{ggF,I}^{} = \epsilon_{hgg,I}^{} = \dfrac{(4\pi)^2}{{\mathop{\rm Re}\nolimits}(A_{hgg}^\text{SM})}\dfrac{16v^2}{\Lambda^2}c_{GG}^{} .
\end{equation}
The vector boson fusion production mode $\sigma_{WWh}^{}$ has three new amputated diagrams as shown in Fig.~\ref{fig:epsIWWh} (in which one of the fermion lines can be inverted to take account of production mode in lepton colliders such as the ILC). For the vector boson associate production modes, there are two new diagrams for $\sigma_{Wh}^{}$ (Fig.~\ref{fig:epsIWh}) and four for $\sigma_{Zh}^{}$ (Fig.~\ref{fig:epsIZh}).

Again from the definition (Eq.~\eqref{eqn:epsIdef}), we compute the interference correction $\epsilon_I^{}$ for each Higgs production mode. The final results are summarized in Table~\ref{tbl:epsI_sigma}. For $\sigma_{Wh}^{}$ and $\sigma_{Zh}^{}$, the final states phase space integral is only two-body and quite simple. On the other hand, $\sigma_{WWh}^{}$ requires to integrate over a three-body phase space, which turns out to be quite involved. The analytical result $\epsilon_{WWh,I}^{}(s)$ is several pages long and hence would not be that useful. Instead, we provide numerical results of it in Table~\ref{tbl:epsI_sigma}, where three auxiliary functions $f_a(s)$, $f_b(s)$, and $f_c(s)$ are defined. We provide the numerical results of these auxiliary functions (Fig.\ref{fig:plotf}) as well as mathematica code of their calculations.

To show the definition of $f_a(s)$, $f_b(s)$, and $f_c(s)$, we need to describe the three-body phase space integral of $\sigma_{WWh}^{}$. We take the center of mass frame of the colliding fermions and setup the spherical coordinates with the positive $z$-axis being the direction of ${\mathord{\buildrel{\lower3pt\hbox{$\scriptscriptstyle\rightharpoonup$}}\over p}}_a$. Then the various momenta labeled in Fig.~\ref{fig:epsIWWh} can be expressed as
\begin{eqnarray}
{p_a} &=& \frac{{\sqrt s }}{2}(1,0,0,1) , \\
{p_b} &=& \frac{{\sqrt s }}{2}(1,0,0, - 1) , \\
{p_3} &=& \frac{{\sqrt s }}{2}{x_3}(1,{s_3},0,{c_3}) , \\
{p_4} &=& \frac{{\sqrt s }}{2}{x_4}(1,{s_4}\cos \phi ,{s_4}\sin \phi ,{c_4}) .
\end{eqnarray}
where we have defined $x_3\equiv\frac{2E_3}{\sqrt{s}}$, $x_4\equiv\frac{2E_4}{\sqrt{s}}$, and adopted the notation $c_3\equiv\cos\theta_3$ {\it etc}. Due to the axial symmetry around the $z$-axis, we have also taken the parametrization $\phi_3=0$ and $\phi_4=\phi$ without loss of generality. For further convenience, let us also define $\eta_h\equiv\frac{m_h}{\sqrt{s}}$, $\eta_W\equiv\frac{m_W}{\sqrt{s}}$, and $\alpha_\phi\equiv\frac{1}{2}(1-c_3c_4-s_3s_4\cos\phi)$. The three-body phase space has nine variables to integrate over. But the axial symmetry and the $\delta$-function of 4-momentum make five of them trivial, leaving us with four nontrivial ones, which we choose to be $x_3$, $c_3$, $c_4$, and $\phi$. Sometimes, we will still use the quantity $x_4$ to make the expression short, but it has been fixed by the energy $\delta$-function and should be understood as a function of the other four
\begin{equation}
x_4(x_3,c_3,c_4,\phi) = \frac{{1 - \eta _h^2 - {x_3}}}{{1 - {\alpha _\phi }{x_3}}} .
\end{equation}
Now the phase space integral can be written as
\begin{eqnarray}
\frac{1}{{2s}}\int {d{\Pi _3}(1,3,4)} &=& \frac{1}{{2s}}\int {\frac{{{d^3}{{\mathord{\buildrel{\lower3pt\hbox{$\scriptscriptstyle\rightharpoonup$}}
\over p} }_3}}}{{{{(2\pi )}^3}}}\frac{1}{{2{E_3}}}\frac{{{d^3}{{\mathord{\buildrel{\lower3pt\hbox{$\scriptscriptstyle\rightharpoonup$}}
\over p} }_4}}}{{{{(2\pi )}^3}}}\frac{1}{{2{E_4}}}\frac{{{d^3}{{\mathord{\buildrel{\lower3pt\hbox{$\scriptscriptstyle\rightharpoonup$}}
\over p} }_1}}}{{{{(2\pi )}^3}}}\frac{1}{{2{E_1}}}{{(2\pi )}^4}{\delta ^4}({p_1} + {p_3} + {p_4} - p)} \nonumber \\
 &=& \frac{1}{{2048{\pi ^4}}}\int_0^{1 - \eta _h^2} {d{x_3}} \int_{ - 1}^1 {d{c_3}d{c_4}\int_0^{2\pi } {d\phi \frac{{(1 - \eta _h^2 - {x_3}){x_3}}}{{{{(1 - {\alpha _\phi }{x_3})}^2}}}} } .
\end{eqnarray}
The modulus square of the SM invariant amplitude is
\begin{eqnarray}
\overline {{{\left| {{M_{WWh,SM}}} \right|}^2}} &=& {\left( {\frac{g}{{\sqrt 2 }}} \right)^4}\frac{{2m_W^4}}{{{v^2}}}\frac{{{g^{\mu \nu }}{g^{\alpha \beta }}\frac{1}{4}\tr\big({{\slashed p}_a}{\gamma _\alpha }{{\slashed p}_3}{\gamma _\mu }{P_L}\big)\tr\big({{\slashed p}_b}{\gamma _\nu }{{\slashed p}_4}{\gamma _\beta }{P_L}\big)}}{{{{(k_1^2 - m_W^2)}^2}{{(k_2^2 - m_W^2)}^2}}} \nonumber \\
 &=& \frac{{m_W^4}}{{{v^6}}}2\eta _W^4\frac{{4{x_3}{x_4}(1 + {c_3})(1 - {c_4})}}{{{{\left[ {{x_3}(1 - {c_3}) + 2\eta _W^2} \right]}^2}{{\left[ {{x_4}(1 + {c_4}) + 2\eta _W^2} \right]}^2}}} .
\end{eqnarray}
Now we are about ready to show the definition of $f_a(s)$, $f_b(s)$, and $f_c(s)$. Let us introduce an ``average'' definition of $A$ as
\begin{equation}
\left\langle A \right\rangle  \equiv \frac{{\frac{1}{{2s}}\int {d{\Pi _3}(1,3,4)\overline {{{\left| {{M_{WWh,SM}}} \right|}^2}} A} }}{{\frac{1}{{2s}}\int {d{\Pi _3}(1,3,4)\overline {{{\left| {{M_{WWh,SM}}} \right|}^2}} } }} .
\end{equation}
Then $f_a(s)$, $f_b(s)$, and $f_c(s)$ are defined as
\begin{eqnarray}
{f_a}(s) &\equiv& \left\langle {\frac{{\big({k_1}{k_2}{g^{\mu \nu }} - k_1^\nu k_2^\mu \big){g^{\alpha \beta }}\frac{1}{4}\tr\big({{\slashed p}_a}{\gamma _\alpha }{{\slashed p}_3}{\gamma _\mu }{P_L}\big)\tr\big({{\slashed p}_b}{\gamma _\nu }{{\slashed p}_4}{\gamma _\beta }{P_L}\big) + c.c.}}{{m_W^2{g^{\mu \nu }}{g^{\alpha \beta }}\frac{1}{4}\tr\big({{\slashed p}_a}{\gamma _\alpha }{{\slashed p}_3}{\gamma _\mu }{P_L}\big)\tr\big({{\slashed p}_b}{\gamma _\nu }{{\slashed p}_4}{\gamma _\beta }{P_L}\big)}}} \right\rangle \nonumber \\
 &=& \left\langle { - \frac{1}{{2\eta _W^2}}\Big(\frac{{{x_4}}}{{1 + {c_3}}} + \frac{{{x_3}}}{{1 - {c_4}}}\Big){s_3}{s_4}\cos \phi } \right\rangle , \label{eqn:fadef} \\
{f_b}(s) &\equiv& \left\langle {\frac{{k_1^2 + k_2^2}}{{m_W^2}}} \right\rangle  = \left\langle { - \frac{1}{{2\eta _W^2}}\Big[ {{x_3}(1 - {c_3}) + {x_4}(1 + {c_4})} \Big]} \right\rangle , \label{eqn:fbdef} \\
{f_c}(s) &\equiv& \left\langle {\frac{{k_1^2}}{{k_1^2 - m_W^2}} + \frac{{k_2^2}}{{k_2^2 - m_W^2}}} \right\rangle  = \left\langle {\frac{{{x_3}(1 - {c_3})}}{{{x_3}(1 - {c_3}) + 2\eta _W^2}} + \frac{{{x_4}(1 + {c_4})}}{{{x_4}(1 + {c_4}) + 2\eta _W^2}}} \right\rangle , \label{eqn:fcdef}
\end{eqnarray}
where various momenta are as labeled in Fig.~\ref{fig:epsIWWh}, and $P_L=\frac{1-\gamma^5}{2}$, with the $\gamma$ matrices defined as usual.

\begin{figure}[t]
\centering
 \subfigure[]{
 \centering
\begin{fmffile}{sigmaZh1}
\begin{fmfgraph*}(40,25)
\fmfleft{i2,i1}
\fmfright{o2,o1}
\fmfv{decor.shape=circle,decor.filled=full,decor.size=4thick}{v2}
\fmfv{l=$f$,l.a=180,l.d=3mm}{i1}
\fmfv{l=${\bar f}$,l.a=180,l.d=3mm}{i2}
\fmfv{l=$Z$,l.a=0,l.d=3mm}{o1}
\fmfv{l=$h$,l.a=0,l.d=3mm}{o2}
\fmf{fermion}{i1,v1,i2}
\fmf{photon,tension=1.5,label=$Z$,label.side=right}{v1,v2}
\fmf{photon}{v2,o1}
\fmf{dashes}{v2,o2}
\end{fmfgraph*}
\end{fmffile}
 }\hspace{2cm}
 \subfigure[]{
 \centering
\begin{fmffile}{sigmaZh2}
\begin{fmfgraph*}(40,25)
\fmfleft{i2,i1}
\fmfright{o2,o1}
\fmfv{decor.shape=circle,decor.filled=full,decor.size=4thick}{v3}
\fmfv{l=$f$,l.a=180,l.d=3mm}{i1}
\fmfv{l=${\bar f}$,l.a=180,l.d=3mm}{i2}
\fmfv{l=$Z$,l.a=0,l.d=3mm}{o1}
\fmfv{l=$h$,l.a=0,l.d=3mm}{o2}
\fmf{fermion}{i1,v1,i2}
\fmf{photon,tension=1.5,label=$Z$,label.side=right}{v1,v3}
\fmf{photon,tension=1.5,label=$Z$,label.side=right}{v3,v2}
\fmf{photon}{v2,o1}
\fmf{dashes}{v2,o2}
\end{fmfgraph*}
\end{fmffile}
 }

\vspace{0.5cm}
 \subfigure[]{
 \centering
\begin{fmffile}{sigmaZh3}
\begin{fmfgraph*}(40,25)
\fmfleft{i2,i1}
\fmfright{o2,o1}
\fmfv{decor.shape=circle,decor.filled=full,decor.size=4thick}{v2}
\fmfv{l=$f$,l.a=180,l.d=3mm}{i1}
\fmfv{l=${\bar f}$,l.a=180,l.d=3mm}{i2}
\fmfv{l=$Z$,l.a=0,l.d=3mm}{o1}
\fmfv{l=$h$,l.a=0,l.d=3mm}{o2}
\fmf{fermion}{i1,v1,i2}
\fmf{photon,tension=1.5,label=$\gamma$,label.side=right}{v1,v2}
\fmf{photon}{v2,o1}
\fmf{dashes}{v2,o2}
\end{fmfgraph*}
\end{fmffile}
 }\hspace{2cm}
 \subfigure[]{
 \centering
\begin{fmffile}{sigmaZh4}
\begin{fmfgraph*}(40,25)
\fmfleft{i2,i1}
\fmfright{o2,o1}
\fmfv{decor.shape=circle,decor.filled=full,decor.size=4thick}{v3}
\fmfv{l=$f$,l.a=180,l.d=3mm}{i1}
\fmfv{l=${\bar f}$,l.a=180,l.d=3mm}{i2}
\fmfv{l=$Z$,l.a=0,l.d=3mm}{o1}
\fmfv{l=$h$,l.a=0,l.d=3mm}{o2}
\fmf{fermion}{i1,v1,i2}
\fmf{photon,tension=1.5,label=$\gamma$,label.side=right}{v1,v3}
\fmf{photon,tension=1.5,label=$Z$,label.side=right}{v3,v2}
\fmf{photon}{v2,o1}
\fmf{dashes}{v2,o2}
\end{fmfgraph*}
\end{fmffile}
 }
 \caption{New amputated Feynman diagrams for $\sigma_{Zh}$.} \label{fig:epsIZh}
\end{figure}
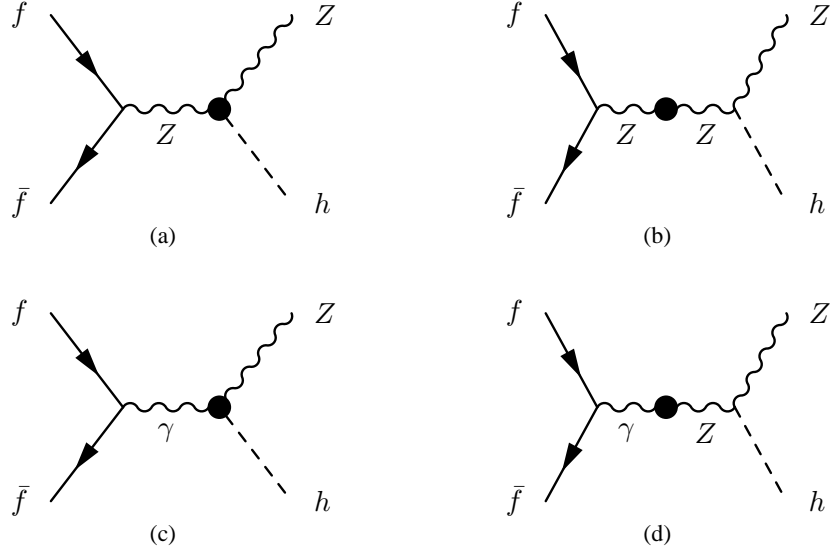

\begin{figure}[t]
\centering
 \subfigure[]{
 \centering
\begin{fmffile}{sigmaWWh1}
\begin{fmfgraph*}(32,28)
\fmfleft{i2,i1}
\fmfright{o2,o,o1}
\fmfv{decor.shape=circle,decor.filled=full,decor.size=4thick}{v}
\fmfv{l=$l / d$,l.a=180,l.d=3mm}{i1}
\fmfv{l=$\nu / u$,l.a=180,l.d=3mm}{i2}
\fmfv{l=$\nu / u$,l.a=0,l.d=3mm}{o1}
\fmfv{l=$l / d$,l.a=0,l.d=3mm}{o2}
\fmfv{l=$h$,l.a=0,l.d=3mm}{o}
\fmf{fermion,tension=1.5}{i1,v1}
\fmf{fermion}{v1,o1}
\fmf{fermion,tension=1.5}{i2,v2}
\fmf{fermion}{v2,o2}
\fmf{photon,label=$W$,label.side=left}{v1,v}
\fmf{photon,label=$W$,label.side=right}{v2,v}
\fmf{dashes}{v,o}
\end{fmfgraph*}
\end{fmffile}
 }\hspace{2cm}
 \subfigure[]{
 \centering
\begin{fmffile}{sigmaWWh2}
\begin{fmfgraph*}(36,32)
\fmfleft{i2,i1}
\fmfright{o2,o,o1}
\fmfv{decor.shape=circle,decor.filled=full,decor.size=4thick}{v3}
\fmfv{l=$l / d$,l.a=180,l.d=3mm}{i1}
\fmfv{l=$\nu / u$,l.a=180,l.d=3mm}{i2}
\fmfv{l=$\nu / u$,l.a=0,l.d=3mm}{o1}
\fmfv{l=$l / d$,l.a=0,l.d=3mm}{o2}
\fmfv{l=$h$,l.a=0,l.d=3mm}{o}
\fmf{fermion,tension=1.5}{i1,v1}
\fmf{fermion}{v1,o1}
\fmf{fermion,tension=1.5}{i2,v2}
\fmf{fermion}{v2,o2}
\fmf{photon,label=$W$,label.side=right}{v1,v3}
\fmf{photon,label=$W$,label.side=right}{v3,v}
\fmf{photon,tension=0.5,label=$W$,label.side=right}{v2,v}
\fmf{dashes}{v,o}
\end{fmfgraph*}
\end{fmffile}
 }\hspace{2cm}
 \subfigure[]{
 \centering
\begin{fmffile}{sigmaWWh3}
\begin{fmfgraph*}(36,32)
\fmfleft{i2,i1}
\fmfright{o2,o,o1}
\fmfv{decor.shape=circle,decor.filled=full,decor.size=4thick}{v3}
\fmfv{l=$l / d$,l.a=180,l.d=3mm}{i1}
\fmfv{l=$\nu / u$,l.a=180,l.d=3mm}{i2}
\fmfv{l=$\nu / u$,l.a=0,l.d=3mm}{o1}
\fmfv{l=$l / d$,l.a=0,l.d=3mm}{o2}
\fmfv{l=$h$,l.a=0,l.d=3mm}{o}
\fmf{fermion,tension=1.5}{i1,v1}
\fmf{fermion}{v1,o1}
\fmf{fermion,tension=1.5}{i2,v2}
\fmf{fermion}{v2,o2}
\fmf{photon,tension=0.5,label=$W$,label.side=left}{v1,v}
\fmf{photon,label=$W$,label.side=left}{v2,v3}
\fmf{photon,label=$W$,label.side=left}{v3,v}
\fmf{dashes}{v,o}
\end{fmfgraph*}
\end{fmffile}
 }
 \caption{New amputated Feynman diagrams for $\sigma_{WWh}$.} \label{fig:epsIWWh}
\end{figure}

\subsection{Calculation of residue modifications} \label{subsec:RMdetails}

The mass pole residue modification $\Delta r_k^{}$ of each external leg $k$ can be computed using the corresponding vacuum polarization function. In our paper, the relevant mass pole residue modifications are
\begin{eqnarray}
\Delta r_h^{} &=& {\left. {\frac{{d\Sigma ({p^2})}}{{d{p^2}}}} \right|_{{p^2} = m_h^2}} , \\
\Delta r_W^{} &=& {\left. {\frac{{d{\Pi _{WW}}({p^2})}}{{d{p^2}}}} \right|_{{p^2} = m_W^2}} , \\
\Delta r_Z^{} &=& {\left. {\frac{{d{\Pi _{ZZ}}({p^2})}}{{d{p^2}}}} \right|_{{p^2} = m_Z^2}} ,
\end{eqnarray}
where $-i\Sigma(p^2)$ denotes the vacuum polarization function of the physical Higgs field $h$. With all the vacuum polarization functions listed in Table~\ref{tbl:Pis}, it is straightforward to calculate $\Delta r$. The results are summarized in Table~\ref{tbl:Deltars}.

\renewcommand\arraystretch{2.0}
\begin{table}[tb]
\centering
\begin{tabular}{|rl|}
  \hline
  $\Delta r_h^{}=$ & \hspace{-4mm} $-\dfrac{v^2}{\Lambda^2}\left(2c_H^{} + c_R^{}\right) - \dfrac{2m_h^2}{\Lambda^2}c_D^{}$ \\
  $\Delta r_Z^{}=$ & \hspace{-4mm} $\dfrac{2m_Z^2}{\Lambda^2}\Big[ - c_Z^2 c_{2W}^{} - s_Z^2 c_{2B}^{} + 4\left(c_Z^4 c_{WW}^{} + s_Z^4 c_{BB}^{} + c_Z^2 s_Z^2 c_{WB}^{}\right) + c_Z^2 c_W^{} + s_Z^2 c_B^{} \Big]$ \\
  $\Delta r_W^{}=$ & \hspace{-4mm} $\dfrac{2m_W^2}{\Lambda^2}\left(-c_{2W}^{} + 4c_{WW}^{} + c_W^{}\right)$ \\
  \hline
\end{tabular}
\caption{Residue modifications $\Delta r$ in terms of Wilson coefficients.} \label{tbl:Deltars}
\vspace{-10pt}
\end{table}
\renewcommand\arraystretch{1.0}

\subsection{Calculation of Lagrangian parameter modifications} \label{subsec:LPMdetails}

The set of Lagrangian parameters relevant for us are $\{\rho\}=\{g^2,v^2,s_Z^2,y_f^2\}$. We would like to compute them in terms of the physical observables and the Wilson coefficients $\rho=\rho(obs,c_i)$, where the set of observables relevant to us can be taken as $\{obs\}=\{\hat\alpha,\hat{G}_F,\hat{m}_Z^2,\hat{m}_f^2\}$. We put a hat on the quantities to denote that it is a physical observable measured from the experiments. On the other hand, for notation convenience, we also define the following auxiliary Lagrangian parameters that are related to the basic ones $\{\rho\}=\{g^2,v^2,s_Z^2,y_f^2\}$:
\begin{eqnarray}
m_W^2 &\equiv& \frac{1}{2} g^2 v^2 , \\
m_Z^2 &\equiv& \frac{1}{2} g^2 v^2\frac{1}{1-s_Z^2} .
\end{eqnarray}
These auxiliary Lagrangian parameters are not hatted.

As explained in Section~\ref{sec:mapping}, in order to obtain $\rho=\rho(obs,c_i)$, we first need to compute the function $obs=obs(\rho,c_i)$, which up to linear order in $c_i$ are
\begin{eqnarray}
\hat \alpha &=& \frac{{{g^2}s_Z^2}}{{4\pi }}{\left. {\frac{{{p^2}}}{{{p^2} - {\Pi _{\gamma \gamma }}({p^2})}}} \right|_{{p^2} \to 0}} = \frac{{{g^2}s_Z^2}}{{4\pi }}\left[ {1 + {{\Pi '}_{\gamma \gamma }}(0)} \right] , \\
{{\hat G}_F} &=& \frac{{\sqrt 2 {g^2}}}{8}{\left. {\frac{{ - 1}}{{{p^2} - m_W^2 - {\Pi _{WW}}({p^2})}}} \right|_{{p^2} = 0}} = \frac{1}{{2\sqrt 2 {v^2}}}\left[ {1 - \frac{1}{{m_W^2}}{\Pi _{WW}}(0)} \right] , \\
\hat m_Z^2 &=& m_Z^2 + {\Pi _{ZZ}}(m_Z^2) = \frac{1}{2}{g^2}{v^2}\frac{1}{{1 - s_Z^2}}\left[ {1 + \frac{1}{{m_Z^2}}{\Pi _{ZZ}}(m_Z^2)} \right] , \\
\hat m_f^2 &=& y_f^2{v^2} .
\end{eqnarray}
Note that the vacuum polarization functions are linear in $c_i$ and hence only kept up to first order. Next we need to take the inverse of these to get the function $\rho=\rho(obs,c_i)$. Again, because the vacuum polarization functions are already linear in $c_i$, one can neglect the modification of the Lagrangian parameters multiplying them when taking the inverse at the leading order. This gives
\begin{eqnarray}
{g^2}s_Z^2 &=& 4\pi \hat \alpha \left[ {1 - {{\Pi '}_{\gamma \gamma }}(0)} \right] , \\
{v^2} &=& \frac{1}{{2\sqrt 2 {{\hat G}_F}}}\left[ {1 - \frac{1}{{m_W^2}}{\Pi _{WW}}(0)} \right] , \\
s_Z^2(1 - s_Z^2) &=& \frac{{\pi \hat \alpha }}{{\sqrt 2 {{\hat G}_F}\hat m_Z^2}}\left[ {1 - {{\Pi '}_{\gamma \gamma }}(0)} \right]\left[ {1 - \frac{1}{{m_W^2}}{\Pi _{WW}}(0)} \right]\left[ {1 + \frac{1}{{m_Z^2}}{\Pi _{ZZ}}(m_Z^2)} \right] , \\
y_f^2 &=& 2\sqrt 2 {{\hat G}_F}\hat m_f^2\left[ {1 + \frac{1}{{m_W^2}}{\Pi _{WW}}(0)} \right] .
\end{eqnarray}
Then taking $\log$ and derivative on both sides, we obtain
\begin{eqnarray}
\Delta {w_{{g^2}}} + \Delta {w_{s_Z^2}} &=&  - {{\Pi '}_{\gamma \gamma }}(0) , \\
\Delta {w_{{v^2}}} &=& - \frac{1}{{m_W^2}}{\Pi _{WW}}(0) , \\
\frac{c_Z^2-s_Z^2}{c_Z^2}\Delta {w_{s_Z^2}} &=& - {{\Pi '}_{\gamma \gamma }}(0) - \frac{1}{{m_W^2}}{\Pi _{WW}}(0) + \frac{1}{{m_Z^2}}{\Pi _{ZZ}}(m_Z^2) , \\
\Delta {w_{y_f^2}} &=& \frac{1}{{m_W^2}}{\Pi _{WW}}(0) ,
\end{eqnarray}
which leads us to the final results
\begin{eqnarray}
\Delta {w_{{g^2}}} &=&  - {{\Pi '}_{\gamma \gamma }}(0) - \frac{{c_Z^2}}{c_Z^2-s_Z^2}\left[ { - {{\Pi '}_{\gamma \gamma }}(0) - \frac{1}{{m_W^2}}{\Pi _{WW}}(0) + \frac{1}{{m_Z^2}}{\Pi _{ZZ}}(m_Z^2)} \right] , \\
\Delta {w_{{v^2}}} &=&  - \frac{1}{{m_W^2}}{\Pi _{WW}}(0) , \\
\Delta {w_{s_Z^2}} &=& \frac{{c_Z^2}}{c_Z^2-s_Z^2}\left[ { - {{\Pi '}_{\gamma \gamma }}(0) - \frac{1}{{m_W^2}}{\Pi _{WW}}(0) + \frac{1}{{m_Z^2}}{\Pi _{ZZ}}(m_Z^2)} \right] , \\
\Delta {w_{y_f^2}} &=& \frac{1}{{m_W^2}}{\Pi _{WW}}(0) .
\end{eqnarray}
Plugging in the vacuum polarization functions listed in Table~\ref{tbl:Pis}, one can get the Lagrangian parameter modifications $\Delta w_\rho$ summarized in Table~\ref{tbl:Deltaws}.

\renewcommand\arraystretch{2.0}
\begin{table}[tb]
\centering
\begin{tabular}{|rl|}
  \hline
  $\Delta w_{g^2} =$ & \hspace{-0.4cm} $\dfrac{m_W^2}{\Lambda^2}\dfrac{1}{c_Z^2-s_Z^2}\bigg\{ \big(c_Z^2 c_{2W}^{}+s_Z^2 c_{2B}^{}\big) - 8\Big[ \big(c_Z^2-s_Z^2\big) c_{WW}^{} + s_Z^2 c_{WB}^{}\Big]$ \\
  & \hspace{-0.4cm} $-2\big(c_Z^2 c_W^{} + s_Z^2 c_B^{}\big) \bigg\} +\dfrac{c_Z^2}{c_Z^2-s_Z^2}\dfrac{2v^2}{\Lambda^2}c_T^{}$ \\
  $\Delta w_{v^2} =$ & \hspace{-0.4cm} $-\dfrac{v^2}{\Lambda^2}c_R^{}$ \\
  $\Delta w_{s_Z^2} =$ & \hspace{-0.4cm} $\dfrac{m_W^2}{\Lambda^2}\dfrac{1}{c_Z^2-s_Z^2}\bigg\{ -\big(c_Z^2 c_{2W}^{}+s_Z^2 c_{2B}^{}\big) + 8\Big[ \big(c_Z^2-s_Z^2\big)\big(c_Z^2 c_{WW}^{}-s_Z^2 c_{BB}^{}\big) +2c_Z^2s_Z^2 c_{WB}^{}\Big]$ \\
  & \hspace{-0.4cm} $+2\big(c_Z^2 c_W^{} + s_Z^2 c_B^{}\big)\bigg\} -\dfrac{c_Z^2}{c_Z^2-s_Z^2}\dfrac{2v^2}{\Lambda^2}c_T^{}$ \\
  $\Delta w_{y_f^2} =$ & \hspace{-0.4cm} $\dfrac{v^2}{\Lambda^2}c_R^{}$ \\
  \hline
\end{tabular}
\caption{Parameter modifications $\Delta w_\rho$ in terms of Wilson coefficients.} \label{tbl:Deltaws}
\vspace{-10pt}
\end{table}
\renewcommand\arraystretch{1.0}










\bibliography{HiggsPrecision}
\bibliographystyle{kickass_mciteplus}

\end{document}